\documentclass[twocolumn,showkeys,showpacs,preprintnumbers,prd,superscriptaddress,nofootinbib]{revtex4-1}
\usepackage{graphicx,epsf,bm,amsmath,amsfonts,amssymb,epstopdf,natbib,hyperref,color,verbatim,multirow,bm}
\hypersetup{colorlinks=true,urlcolor=blue,citecolor=blue,linkcolor=blue,menucolor=blue,anchorcolor=blue,filecolor=blue}

\date{\today}
            
\begin{document}

\title{Inflationary interpretation of the stochastic gravitational wave background signal detected by pulsar timing array experiments}

\author{Sunny Vagnozzi}
\email{sunny.vagnozzi@unitn.it}
\affiliation{Department of Physics, University of Trento, Via Sommarive 14, 38123 Povo (TN), Italy}
\affiliation{Trento Institute for Fundamental Physics and Applications (TIFPA)-INFN, Via Sommarive 14, 38123 Povo (TN), Italy}

\begin{abstract}
\noindent Various pulsar timing array (PTA) experiments (NANOGrav, EPTA, PPTA, CPTA, including data from InPTA) very recently reported evidence for excess red common-spectrum signals in their latest datasets, with inter-pulsar correlations following the Hellings-Downs pattern, pointing to a stochastic gravitational wave background (SGWB) origin. Focusing for concreteness on the NANOGrav signal (given that all signals are in good agreement between each other), I inspect whether it supports an inflationary SGWB explanation, finding that such an interpretation calls for an extremely blue tensor spectrum, with spectral index $n_T \simeq 1.8 \pm 0.3$, while Big Bang Nucleosynthesis limits require a very low reheating scale, $T_{\rm rh} \lesssim 10\,{\rm GeV}$. While not impossible, an inflationary origin for the PTA signal is barely tenable: within well-motivated inflationary models it is hard to achieve such a blue tilt, whereas models who do tend to predict sizeable non-Gaussianities, excluded by observations. Intriguingly, ekpyrotic models naturally predict a SGWB with spectral index $n_T=2$, although with an amplitude too suppressed to be able to explain the signal detected by PTA experiments. Finally, I provide explicit expressions for a bivariate Gaussian approximation to the joint posterior distribution for the intrinsic-noise amplitude and spectral index of the NANOGrav signal, which can facilitate extending similar analyses to different theoretical signals.
\end{abstract}

\maketitle

\section{Introduction}
\label{sec:introduction}

The existence of a stochastic gravitational wave background (SGWB) covering a wide range of frequencies is a robust prediction of many motivated physical scenarios~\cite{Caprini:2018mtu,Renzini:2022alw}. Possible sources include but are not limited to cosmological relics associated to phase transitions~\cite{Siemens:2006yp,Caprini:2010xv,Ramberg:2019dgi,Caprini:2019egz,Ellis:2020awk}, and astrophysical processes such as merging supermassive black hole binaries (SMBHBs)~\cite{Rajagopal:1994zj,Jaffe:2002rt,Wyithe:2002ep,Sesana:2004sp,Burke-Spolaor:2018bvk}. In particular, the incoherent superposition of the gravitational radiation produced during the slow adiabatic inspiral phase by all SMBHBs adds up to a broadband SGWB signal peaking in the ${\cal O}({\rm nHz})$ frequency range~\cite{Burke-Spolaor:2018bvk}. A well-known detection channel for ${\rm nHz}$ GWs is via pulsar timing arrays (PTAs), exploiting the fact that millisecond pulsars behave as extremely stable clocks~\cite{Sazhin:1978ghw,Detweiler:1979wn}. PTAs search for spatially correlated fluctuations in the pulse arrival-time measurements of pulsars in a widely distributed array, induced by passing GWs perturbing the space-time metric along the line of sight to each pulsar~\cite{Hobbs:2017oam}.

In 2020, the North American Nanohertz Observatory for Gravitational Waves (NANOGrav) PTA collaboration reported strong evidence for a red-stochastic common-spectrum process in the timing residuals of the 47 pulsars in their 12.5-year dataset~\cite{NANOGrav:2020bcs}. This signal was later confirmed by the Parkes PTA (PPTA)~\cite{Goncharov:2021oub} and European PTA (EPTA)~\cite{Chen:2021rqp} collaborations, as well as the combined International PTA (IPTA)~\cite{Antoniadis:2022pcn}. While an excess residual power with consistent amplitude and spectral shape across all pulsars is the first expected SGWB sign in a PTA~\cite{NANOGrav:2020spf,Romano:2020sxq}, a GW origin for such signal can only be attributed in the presence of phase-coherent inter-pulsar correlations following the Hellings-Downs (HD) pattern~\cite{Hellings:1983fr}, which would exclude more mundane explanations such as intrinsic pulsar processes~\cite{Goncharov:2022ktc,Zic:2022sxd} or common systematic noise~\cite{Tiburzi:2015kqa}.

Very recently, in June 2023, various PTA experiments, including NANOGrav, EPTA, PPTA, and the Chinese Pulsar Timing Array (CPTA), in the case of EPTA including also data from the Indian PTA (InPTA), reported on the analyses of their latest datasets, which all confirm the presence of excess red common-spectrum signals, with strain amplitude of order ${\cal O}(10^{-15})$ at the reference frequency $f=1\,{\rm yr}^{-1}$~\cite{NANOGrav:2023gor,Antoniadis:2023ott,Reardon:2023gzh,Xu:2023wog,NANOGrav:2023hde,NANOGrav:2023ctt,NANOGrav:2023hvm,NANOGrav:2023hfp,NANOGrav:2023ghw,NANOGrav:2023pdq,NANOGrav:2023icp,Antoniadis:2023lym,Antoniadis:2023puu,Antoniadis:2023aac,Antoniadis:2023xlr,Smarra:2023ljf,Reardon:2023zen,Zic:2023gta}. Importantly, all analyses report evidence (with varying strength) for HD correlations, which point to a genuine GW origin for the signals, in turn making these the first convincing detections of a SGWB signal in the ${\rm nHz}$ range. For instance, focusing on the NANOGrav 15-year results which for concreteness I will consider in the remainder of this work, the inferred amplitude of the excess red common-spectrum signal is $A \sim 6.3 \times 10^{-15}$ at the reference frequency $f=1\,{\rm yr}^{-1}$, whereas a model including a HD-correlated power-law SGWB was found to be preferred over a spatially uncorrelated common-spectrum power-law SGWB with Bayes factor of up to $1000$.

An important cosmological SGWB source are GWs produced during inflation, a theorized stage of quasi-de Sitter expansion in the very early Universe, introduced to solve the flatness, horizon, and entropy problems, alongside the apparent lack of topological defects: during inflation, tiny (quantum) tensor and scalar fluctuations to the metric are stretched outside the causal horizon, eventually re-entering much later~\cite{Kazanas:1980tx,Starobinsky:1980te,Sato:1981ds,Guth:1980zm,Mukhanov:1981xt,Linde:1981mu,Albrecht:1982wi}. Despite a few potential foundational problems pointed out in recent years (see e.g.\ Refs.~\cite{Ijjas:2013vea,Ijjas:2014nta,Obied:2018sgi,Agrawal:2018own,Achucarro:2018vey,Garg:2018reu,Kehagias:2018uem,Kinney:2018nny,Ooguri:2018wrx,Palti:2019pca,Bedroya:2019tba,Geng:2019phi,Trivedi:2020wxf,Anchordoqui:2021eox,Gashti:2022hey}), the inflationary paradigm remains in very good health and is consistent with a large number of precision cosmological observations (see e.g.\ Refs.~\cite{Martin:2013tda,Benetti:2013cja,Martin:2013nzq,Creminelli:2014oaa,Dai:2014jja,Rinaldi:2014gua,Rinaldi:2014gha,Myrzakulov:2015fra,Rinaldi:2015yoa,Escudero:2015wba,Benetti:2016tvm,Benetti:2016ycg,Benetti:2016jhf,Guo:2017qjt,Campista:2017ovq,Ni:2017jxw,SantosdaCosta:2017ctv,Park:2017xbl,Guo:2018uic,DiValentino:2018wum,Chowdhury:2019otk,Benetti:2019kgw,Haro:2019peq,Guo:2019dui,Li:2019ipk,Aich:2019obd,Braglia:2020fms,Cicoli:2020bao,Keeley:2020rmo,SantosdaCosta:2020dyl,Rodrigues:2020fle,Vagnozzi:2020rcz,Neves:2020anh,Vagnozzi:2020dfn,Ye:2021nej,Dhawan:2021mel,Stein:2021uge,Forconi:2021que,dosSantos:2021vis,Cabass:2022wjy,Ye:2022afu,Antony:2022ert,Cabass:2022ymb,Ye:2022efx,Ghoshal:2022qxk,Gangopadhyay:2022vgh,Montefalcone:2022owy,Stein:2022cpk,Cabass:2022oap,Montefalcone:2022jfw}), and further tests thereof are among the key science drivers of a number of upcoming cosmological surveys~\cite{CMB-S4:2016ple,SimonsObservatory:2018koc,SimonsObservatory:2019qwx}. Nevertheless, the ``smoking gun'' detection of the inflationary SGWB remains to be achieved: while this signal has typically been sought at very low frequencies ($f \lesssim 10^{-15}\,{\rm Hz}$)~\cite{Kamionkowski:2015yta}, based on expectations within the simplest inflationary models, such a search needs not be limited to these low frequencies, as models beyond the minimal ones can naturally predict rich features at higher frequencies, including potentially the ${\cal O}({\rm nHz})$ range. In the earlier work of Ref.~\cite{Vagnozzi:2020gtf} I examined whether the NANOGrav 12.5-year signal could be due to an inflationary SGWB, finding the answer to be potentially positive. In this work I revisit this question in light of the signals very recently detected by PTA experiments, prompted by their convincing GW origin. For concreteness and simplicity, I will focus on the NANOGrav 15-year signal, given that the signals observed in all four PTA experiments are mutually consistent, and that NANOGrav achieved the most precise determination of the signal's spectral index. Overall, I find that an inflationary interpretation of these signals is now significantly less tenable, when compared to the previous explanation of the NANOGrav 12.5-year signal.

The rest of this paper is organized as follows. In Sec.~\ref{sec:inflation} I briefly review the physics of inflationary GWs. In Sec.~\ref{sec:methods} I discuss my analysis methods, which are applied to obtain the results discussed in Sec.~\ref{sec:results}, before drawing finishing remarks in Sec.~\ref{sec:conclusions}.

\section{Inflationary gravitational waves and pulsar timing arrays}
\label{sec:inflation}

Here I very briefly review the inflationary GW spectrum, focusing on scales relevant for PTA experiments. My discussion will closely follow Refs.~\cite{Zhao:2013bba,Liu:2015psa,Vagnozzi:2020gtf}, modulo a few minor numerical updates, and I encourage the reader to refer to these works for further details. I work in synchronous gauge where, denoting by $a$ and $\eta$ respectively the scale factor and conformal time, and by $h_{ij}$ a transverse, traceless (i.e.\ $\partial^ih_{ij}=g^{ij}h_{ij}=0$) symmetric $3 \times 3$ matrix describing GWs, the line element of the perturbed FLRW metric reads:
\begin{eqnarray}
ds^2 = a^2(\eta) \left [ d\eta^2 - (\delta_{ij}+h_{ij})dx^idx^j \right ]\,.
\label{eq:flrw}
\end{eqnarray}
Moving to Fourier space (where $k$ denotes the mode wavenumber) and assuming isotropy, the evolution of the GW field $h_k$ (for both the $+$ and $\times$ polarizations) is described by the following equation:
\begin{eqnarray}
h_k^{''}+2{\cal H}h_k^{'}+k^2h_k=0\,.
\label{eq:hdotdot}
\end{eqnarray}
where $^{'}$ denotes a derivative with respect to conformal time, and ${\cal H} \equiv a'/a$ denotes the conformal Hubble rate. The evolution of a GW field given by $h_k(\eta_i)$ at an initial conformal time $\eta_i$, and characterized by its primordial (tensor) spectrum ${\cal P}^{\rm prim}_T(k) = 2k^3 \vert h_k(\eta_i) \vert^2/\pi^2$, can be obtained by determining the GW transfer function ${\cal T}(\eta\,,k) = h_k(\eta)/h_k(\eta_i)$, with $h_k(\eta)$ evaluated at conformals time $\eta \gg \eta_i$ and obtained by solving Eq.~(\ref{eq:hdotdot}). The relevant quantity for GW direct detection experiments (such as PTAs) is the spectral energy density parameter $\Omega_{\rm gw}(k)$, i.e.\ the logarithmic derivative with respect to wavenumber of the present GW energy density $\rho_{\rm gw}$, divided by the critical density $\rho_c$:
\begin{eqnarray}
\Omega_{\rm gw}(k) \equiv \frac{1}{\rho_c}\frac{d\rho_{\rm gw}}{d\ln k}=\frac{k^2}{12H_0^2}{\cal T}^2(\eta_0\,,k){\cal P}^{\rm prim}_T(k)\,,
\label{eq:omegagw}
\end{eqnarray}
with $\eta_0$ and $H_0$ denoting the present conformal time and the Hubble constant respectively (see e.g.\ Refs.~\cite{Kuroyanagi:2008ye,Kuroyanagi:2010mm,Kuroyanagi:2011fy}).~\footnote{For reviews deriving in detail all the above equations, I refer the reader to Refs.~\cite{Giovannini:2019oii,Odintsov:2022cbm}.} In order to compare the expected theoretical signal to what is observed in PTA experiments, it is more convenient to work with frequencies $f$ rather than wavenumbers $k$, and the two are related by:
\begin{eqnarray}
f=\frac{ck}{2\pi} \implies f \simeq 1.54 \times 10^{-15} \left ( \frac{k}{{\rm Mpc}^{-1}} \right ) \,{\rm Hz}\,,
\label{eq:k}
\end{eqnarray}
where $c$ denotes the speed of light, and I have made the commonly used units of measure explicit.

In order to connect all the above quantities to inflation, one needs to specify the primordial tensor power spectrum ${\cal P}^{\rm prim}_T(k)$. Although this is clearly an approximation (to be discussed more in detail later), it is customary to approximate the primordial tensor spectrum as being described by a pure power-law:
\begin{eqnarray}
{\cal P}^{\rm prim}_T(k) = rA_s \left ( \frac{k}{k_{\star}} \right )^{n_T}\,,
\label{eq:p}
\end{eqnarray}
where $r$ denotes the tensor-to-scalar ratio, $A_s$ is the amplitude of primordial scalar perturbations at the Cosmic Microwave Background (CMB) pivot scale $k_{\star}$ (here I assume the \textit{Planck} 2018 pivot scale $k_{\star}=0.05\,{\rm Mpc}^{-1}$), and $n_T$ is the tensor spectral index. Note that a scale-invariant primordial tensor spectrum is described by $n_T=0$, whereas a blue [red] spectrum is described by $n_T>0$ [$n_T<0$]. Within single-field slow-roll inflationary models, the so-called ``consistency relation'' $n_T=-r/8$ is satisfied~\cite{Copeland:1993ie}: given that $r>0$, within these models one expects the tensor spectrum to be slightly red-tilted. Note, in addition, that the tensor spectral index is related to the equation of state of the dominant component during the inflationary stage, $w$, via the following relation (see e.g.\ Ref.~\cite{Zhao:2013bba}):
\begin{eqnarray}
n_T \approx \frac{4}{1+3w}+2\,.
\label{eq:ntw}
\end{eqnarray}
From the above we note that a pure de Sitter expansion stage (which clearly cannot be a completely realistic case given that inflation has to end at some point), i.e.\ one for which $w=-1$, leads to a scale-invariant tensor spectrum where $n_T=0$. Within the simplest single-field slow-roll inflationary models based on canonical scalar fields, the effective equation of state is $w>-1$, leading to $n_T<0$, and therefore a slight red tilt in the tensor power spectrum. On the other hand, within ``phantom'' inflationary models where $w<-1$, one expects $n_T>0$ and thereby a blue-tilted spectrum. When focusing on the range $w<-1/3$, necessary for the Universe to undergo an accelerated expansion and thereby for successful inflation to occur, it is worth noting that $n_T$ in Eq.~(\ref{eq:ntw}) is bounded from above by $n_T<2$.

If one assumes that inflation is followed by instantaneous reheating and the standard radiation, matter, and dark energy eras, it is well known that the transfer function evaluated today, ${\cal T}(\eta_0\,,k)$, admits relatively simple analytical approximations~\cite{Turner:1993vb,Chongchitnan:2006pe,Watanabe:2006qe,Zhao:2006mm,Giovannini:2009kg} (see also Refs.~\cite{Zhao:2013bba,Kite:2021yoe}). For our purposes, we are interested in $\Omega_{\rm gw}(f)$ at PTA scales $f \sim {\cal O}(10^{-9})\,{\rm Hz}$, whose corresponding wavenumbers $k \sim {\cal O}(10^6)\,{\rm Mpc}^{-1}$ are much larger than the wavenumber associated to a mode crossing the horizon at matter-radiation equality, $k_{\rm eq} \simeq 0.073\Omega_mh^2\,{\rm Mpc}^{-1} \sim {\cal O}(10^{-2})\,{\rm Mpc}^{-1}$, with $\Omega_m$ and $h$ the matter density parameter and reduced Hubble constant respectively. In other words, modes observed on PTA scales crossed the horizon deep in the radiation era, well before equality. In the $k \gg k_{\rm eq}$ regime, the GW spectral energy density [Eq.~(\ref{eq:omegagw})] reduces to~\cite{Zhao:2013bba}:
\begin{eqnarray}
\Omega_{\rm gw}(f) \approx \frac{15}{16}\frac{\Omega_m^2rA_s}{H_0^2\eta_0^4k_{\rm eq}^2}\left ( \frac{f}{f_{\star}} \right )^{n_T}\,,
\label{eq:omega1}
\end{eqnarray}
where $f_{\star} \simeq 7.7 \times 10^{-17}\,{\rm Hz}$ is the frequency corresponding to the pivot wavenumber $k_{\star}=0.05\,{\rm Mpc}^{-1}$, and $\eta_0$ is the present-day conformal time.

On the other hand, the GW spectral energy density associated to a PTA signal is conventionally expressed in the following way:
\begin{eqnarray}
\Omega^{\rm PTA}_{\rm gw}(f) = \frac{2\pi^2}{3H_0^2}f^2h_c^2(f)\,,
\label{eq:omegagwhc}
\end{eqnarray}
with $h_f(c)$ the power spectrum of the GW strain, which at the reference frequency of $f_{\rm yr}=1\,{\rm yr}^{-1} \approx 3.17 \times 10^{-8}\,{\rm Hz}$ is usually approximated as a power law with amplitude $A$ and spectral index $\alpha$:
\begin{eqnarray}
h_c(f) = A \left ( \frac{f}{f_{\rm yr}} \right ) ^{\alpha}\,,
\label{eq:hc}
\end{eqnarray}
where the spectral index $\alpha$ is further related to the spectral index $\gamma$ for the pulsar timing-residual cross-power spectral density $S(f) \propto f^{-\gamma}$ by:
\begin{eqnarray}
\alpha = \frac{3-\gamma}{2}\,.
\label{eq:alphagamma}
\end{eqnarray}
The expectation for the SGWB signal from merging SMBHBs is $\gamma \simeq 13/3$~\cite{Phinney:2001di}, and therefore $\alpha \simeq -2/3$. Combining Eqs.~(\ref{eq:omegagwhc}--\ref{eq:alphagamma}), one finds that the GW spectrum associated to PTA signals can be expressed as:
\begin{eqnarray}
\Omega^{\rm PTA}_{\rm gw}(f) = A^2\frac{2\pi^2}{3H_0^2}\frac{f^{5-\gamma}}{{\rm yr}^{\gamma-3}}\,.
\label{eq:omegagwhcpta}
\end{eqnarray}
Note that in comparison to earlier works on the same subject, in order to simplify the notation, I have dropped the subscript $_{\rm CP}$ (standing for ``common-process'') from the amplitude $A$ and spectral index $\gamma$. The results of PTA GW searches are therefore characterized by the inferred values of the intrinsic-noise amplitude $A$ and spectral index $\gamma$, and in fact these results are often reported in the form of joint $A$-$\gamma$ posteriors, or alternatively as the posterior of $A$ at a fiducial value of $\gamma$ (typically the merging SMBHBs value $\gamma=13/3$).

To connect the results of PTA GW searches with inflationary parameters, one can equate $\Omega^{\rm PTA}_{\rm gw}(f)$ given in Eq.~(\ref{eq:omegagwhcpta}) with $\Omega_{\rm gw}(f)$ given in Eq.~(\ref{eq:omega1}), the latter being an excellent approximation within the regime $f \gg f_{\rm eq}$ we are interested in, with $f_{\rm eq} \simeq 1.54 \times 10^{-17}\,{\rm Hz}$ the frequency associated to a mode crossing the horizon at matter-radiation equality. Doing so one finds that the tensor spectral index is related to $\gamma$ via:
\begin{eqnarray}
n_T = 5-\gamma\,.
\label{eq:ntgamma}
\end{eqnarray}
Further equating Eq.~(\ref{eq:omega1}) and Eq.~(\ref{eq:omegagwhcpta}) while imposing Eq.~(\ref{eq:ntgamma}) gives a relation between the amplitude of the PTA signal $A$ appearing in Eq.~(\ref{eq:omegagwhcpta}) and the relevant cosmological parameters (including the inflationary ones). Doing so, I find (see also Ref.~\cite{Zhao:2013bba}):
\begin{eqnarray}
A = \sqrt{\frac{45\Omega_m^2A_s}{32\pi^2(\eta_0k_{\rm eq})^2}}c\frac{{\rm yr}}{\eta_0} \left ( \frac{{\rm yr}^{-1}}{f_{\star}} \right ) ^{\frac{n_T}{2}}\sqrt{r}\,.
\label{eq:rnta}
\end{eqnarray}
which carries a dependence on $n_T$ reflecting the large ``lever arm'' between the CMB pivot frequency at which $A_s$ is constrained, and the frequency of the PTA signal [$1\,{\rm yr}^{-1}$]. Inserting into Eq.~(\ref{eq:rnta}) the best-fit values for the cosmological parameters as per the \textit{Planck} 2018 results~\cite{Planck:2018vyg}, i.e.\ $\Omega_m=0.315$, $h=0.6736$ (required to compute $\eta_0 \simeq 1.38 \times 10^4\,{\rm Mpc}$), $A_s=2.1 \times 10^{-9}$ at $k_{\star}=0.05\,{\rm Mpc}^{-1}$, and $k_{\rm eq} \simeq 0.01\,{\rm Mpc}^{-1}$, I find:
\begin{eqnarray}
A \approx 0.875\sqrt{r} \times 10^{4.3n_T-18}\,,
\label{eq:amplitude}
\end{eqnarray}
from which one clearly sees that for allowed values of $r \lesssim 0.06$, unless the spectrum is blue ($n_T>0$), the amplitude of the inflationary SGWB on PTA scales will be $A \lesssim {\cal O}(10^{-20})$, and hence beyond any hope of detection.~\footnote{Note that Eq.~(\ref{eq:amplitude}) differs slightly from the analogous Eq.~(16) in Ref.~\cite{Zhao:2013bba}, where a factor of $5$ instead of $4.3$ appears in the exponent. The results are completely consistent, and this difference in exponent is entirely due to the different choice of pivot scale, which was $k_{\star}=0.002\,{\rm Mpc}^{-1}$ in Ref.~\cite{Zhao:2013bba}. This corresponds to $f_{\star} \simeq 3.08 \times 10^{-18}\,{\rm Hz}$, leading to the factor of $({\rm yr}^{-1}/f_{\star})$ in Eq.~(\ref{eq:rnta}) being fortuitously close to $10^{10}$, and thereby $({\rm yr}^{-1}/f_{\star})^{n_T/2} \approx 10^{5n_T}$. With the current choice of pivot scale instead, $f_{\star} \approx 7.7 \times 10^{-17}\,{\rm Hz}$ and therefore $({\rm yr}^{-1}/f_{\star}) \approx 4.1 \times 10^8 \approx 10^{8.6}$, explaining the factor of $(10^{8.6})^{n_T/2}=10^{4.3n_T}$ in Eq.~(\ref{eq:amplitude}).} From Eq.~(\ref{eq:amplitude}) one sees that the amplitude of the inflationary SGWB on PTA scales is controlled by the cosmological parameters $r$, $n_T$, $\Omega_m$, $A_s$, and $H_0$ (or equivalently, thinking in terms of the 6 $\Lambda$CDM parameters, the acoustic scale $\theta_s$), whereas $k_{\rm eq}$ and $\eta_0$ are known once $\Omega_m$ and $H_0$ are specified. However it is also easy to check that, given the current level of extremely low relative uncertainty in the determination of $\Omega_m$, $A_s$, and $H_0$,~\footnote{Modulo uncertainty floors in $H_0$ due to the Hubble tension~\cite{DiValentino:2021izs}, as well as the model-dependence of inferred constraints on $\Omega_m$ and $A_s$, the size of both of which however remain subdominant with respect to the effects of $r$ and $n_T$.} the value of $A$ is virtually controlled only by $r$ and $n_T$, where essentially all of the theoretical uncertainty is condensed (particularly for what concerns $r$, given the lack of detection of inflationary B-modes, and therefore our complete ignorance regarding even just the order of magnitude of the inflationary SGWB signal).

Just as any other massless degrees of freedom, GWs contribute to the radiation energy density of the early Universe, with potentially important effects at various epochs, including Big Bang Nucleosynthesis (BBN) and the time of recombination when the CMB formed. The SGWB contribution to the radiation energy budget can be characterized by $\Delta N_{\rm eff}$, i.e.\ its contribution to the effective number of relativistic species $N_{\rm eff}$.~\footnote{The standard value of $N_{\rm eff}$ has been subject to several increasingly precise revisions over the past decades (see e.g.\ Refs.~\cite{Mangano:2001iu,Mangano:2005cc,Bennett:2019ewm,Akita:2020szl,Froustey:2020mcq,Bennett:2020zkv,Cielo:2023bqp}), but lies safely between $3.043$ and $3.047$.} This contribution is given by (see e.g.\ Refs.~\cite{Allen:1997ad,Smith:2006nka,Boyle:2007zx,Kuroyanagi:2014nba,Vagnozzi:2022qmc}):~\footnote{See also Ref.~\cite{Giare:2022wxq} for a recent work examining caveats related to this calculation.}
\begin{eqnarray}
\Delta N_{\rm eff} \approx 1.8 \times 10^5\int^{f_{\max}}_{f_{\min}}df\,\frac{\Omega_{\rm gw}(f)h^2}{f^2}\,,
\label{eq:neff}
\end{eqnarray}
where the values of $f_{\min}$ and $f_{\max}$ depend respectively on the epoch of interest, and the maximum temperature reached in the hot Big Bang era. In the following, I will be interested in the epoch of BBN, during which light elements were synthesized. This sets $f_{\min} \simeq 10^{-10}\,{\rm Hz}$, corresponding to the frequency of a mode crossing the horizon during the time of BBN, when the temperature of the radiation bath was $T \sim {\cal O}({\rm MeV})$. The reason is that only sub-horizon modes contribute to the radiation energy at any given time, as it is these modes that oscillate and propagate as massless modes, thereby contributing to the local energy density. As a consequence, frequencies associated to modes which have yet to cross the horizon at the epoch of interest should be discarded, which in turn explains the appearance of the lower cutoff in the integral in Eq.~(\ref{eq:neff}). On the other hand, assuming instantaneous reheating and a standard thermal history following inflation, $f_{\rm max}$ is directly related to $T_{\rm rh}$, the temperature at which the Universe reheats following inflation. For instance, GUT-scale reheating corresponds to $f_{\max} \simeq 10^8\,{\rm Hz}$, with lower values of $T_{\rm rh}$ corresponding to lower values of $f_{\max}$. In order not to spoil successful BBN predictions, reheating should occur at temperatures $T_{\rm rh} \gtrsim 5\,{\rm MeV}$~\cite{deSalas:2015glj}. Finally, $\Delta N_{\rm eff}$ is constrained by BBN and CMB probes, where an indicative $2\sigma$ upper limit is $\Delta N_{\rm eff} \lesssim 0.4$~\cite{Aver:2015iza,Cooke:2017cwo,Planck:2018vyg,Vagnozzi:2019ezj,Hsyu:2020uqb,ACT:2020gnv,Mossa:2020gjc}.~\footnote{Although BBN and CMB probes constrain $\Delta N_{\rm eff}$ to a similar level of precision, formally speaking in this work I will only be considering BBN constraints. The reason, as discussed in footnote~6 of Ref.~\cite{Kuroyanagi:2020sfw}, is that in most CMB analyses the extra radiation components are implemented as a neutrino-like (free-streaming) fluid (see e.g.\ Refs.~\cite{Vagnozzi:2017ovm,Vagnozzi:2018jhn,RoyChoudhury:2019hls,Giare:2020vzo,Gariazzo:2022ahe}). However, perturbations in the radiation originating from GWs are expected to behave quite differently, and hence it is unclear to what extent the CMB constraints on $N_{\rm eff}$ can be directly applied to GWs. To be as conservative as possible, I will thus only consider BBN constraints, therefore setting $f_{\rm min}=10^{-10}\,{\rm Hz}$ in Eq.~(\ref{eq:neff}) -- nevertheless, since a posteriori I will be considering a (very) blue spectrum, the integral in Eq.~(\ref{eq:neff}) ends up being for all intents and purposes insensitive to the choice of $f_{\min}$, whereas it depends strongly on the chosen value of $f_{\max}$.}

\section{Data and methods}
\label{sec:methods}

In June 2023 the NANOGrav, EPTA, PPTA, and CPTA PTA experiments reported evidence for excess red common-spectrum signals in their latest datasets (NANOGrav 15-year, EPTA DR2 complemented with InPTA DR1, PPTA DR3, and CPTA DR1), with inter-pulsar correlations following the HD pattern~\cite{NANOGrav:2023gor,Antoniadis:2023ott,Reardon:2023gzh,Xu:2023wog,NANOGrav:2023hde,NANOGrav:2023ctt,NANOGrav:2023hvm,NANOGrav:2023hfp,NANOGrav:2023ghw,NANOGrav:2023pdq,NANOGrav:2023icp,Antoniadis:2023lym,Antoniadis:2023puu,Antoniadis:2023aac,Antoniadis:2023xlr,Smarra:2023ljf,Reardon:2023zen,Zic:2023gta}. Considering that all four signals are consistent with each other, and the fact that NANOGrav achieved the most precise determination of the spectral index (which not all experiments managed to infer), for concreteness and simplicity I will focus on the NANOGrav 15-year signal in the remainder of this work, but I expect my main conclusions to broadly carry over to the results of the other three PTA experiments as well.

The NANOGrav 15-year dataset includes timing observations for 68 pulsars, collected between July 2004 and August 2020 by the Arecibo Observatory, the Green Bank Telescope, and the Very Large Array~\cite{NANOGrav:2020gpb,NANOGrav:2020qll,NANOGrav:2023gor}. Compared to the previous 12.5-year dataset, the current dataset contains 21 additional pulsars, whereas 2.9 more years of timing data are available for the 47 pulsars present previously. The consensus NANOGrav analysis of their 15-year dataset focused only on narrowband time-of-arrival data for the 67 pulsars with a timing baseline of at least 3 years~\cite{NANOGrav:2023gor}. The analysis makes use of the TT(BIPM2019) timescale and JPL DE440 ephemeris~\cite{Park:2021ghw}, the latter leading to significant improvements in the determination of the Jovian orbit compared to the previously used ephemeris. For further details on the analysis assumptions adopted by the NANOGrav collaboration, in particular for what concerns the underlying model fit to the time-of-arrival data, I invite the reader to refer to Sec.~2 of Ref.~\cite{NANOGrav:2023gor}.

Analyzing their 15-year dataset, the NANOGrav collaboration reported very strong evidence for a time-correlated stochastic signal with common amplitude and spectrum across all pulsars~\cite{NANOGrav:2023gor}. The analysis also provides positive Bayesian evidence that the common-spectrum signal includes HD inter-pulsar correlations: the latter case is preferred over a model with common-spectrum spatially-uncorrelated red noise (CURN), with Bayes factors as large as 1000. A model with common-spectrum spatially-uncorrelated red noise is itself strongly preferred over a model with intrinsic red noise only, confirming the earlier 12.5-year dataset findings. The HD versus CURN Bayes factors can be converted into detection statistics, with the corresponding $p$-values being of order $10^{-3}$~\cite{NANOGrav:2023gor}. Overall, together with the latest EPTA (+InPTA), PPTA, and CPTA datasets, the NANOGrav 15-year dataset therefore provides the first compelling evidence for the presence of HD correlations in pulsar timing residuals, convincingly pointing to a SGWB origin.

Fitting a model describing an isotropic SGWB characterized by HD correlations, the NANOGrav collaboration inferred the values of the intrinsic-noise amplitude $A$ and spectral index $\gamma$ appearing in Eq.~(\ref{eq:omegagwhcpta}) to be $A=6.4_{-2.7}^{+4.2} \times 10^{-15}$ and $\gamma=3.2 \pm 0.6$ respectively (both reported at $2\sigma$), at a reference frequency of $f_{\rm yr}=1\,{\rm yr}^{-1}$~\cite{NANOGrav:2023gor}. The joint $\log_{10}A$-$\gamma$ posterior is reported in the lower left panel of Fig.~1 of Ref.~\cite{NANOGrav:2023gor} (blue contours, reference frequency $1\,{\rm yr}^{-1}$). It is worth noting that, compared to the 12.5-year signal, the value of $A$ is a factor of $\approx 5$ higher, whereas $\gamma$ is significantly lower (as discussed in more detail in the next paragraph), corresponding to a bluer spectrum given the relation between $n_T$ and $\gamma$ in Eq.~(\ref{eq:ntgamma}). Therefore, one can already expect quantitative differences in the inflationary interpretation compared to my earlier results of Ref.~\cite{Vagnozzi:2020gtf}, going in the direction of a bluer spectrum due to both the smaller value of $\gamma$ and the larger value of $A$. This expectation will in fact be confirmed in Sec.~\ref{sec:results}.

As noted in Ref.~\cite{NANOGrav:2023gor}, the inferred value of $\gamma$ is low relative to (and in moderate tension with) the value $\gamma=13/3$ expected from merging SMBHBs, which lies at the upper boundary of the $3\sigma$ region. Interestingly a similar ``anomaly'', albeit reversed in direction, was observed in the 12.5-year dataset, where $\gamma \sim 5.52$ was inferred~\cite{NANOGrav:2020bcs}. In that case, the value $\gamma=13/3$ lay just outside the lower boundary of the $1\sigma$ region. However, astrophysical effects can alter the value of $\gamma$ arising from merging SMBHBs, and the estimation of $\gamma$ itself is sensitive to details in the modeling of intrinsic red noise and of interstellar medium timing delays in a few pulsars. The NANOGrav collaboration performed several robustness and validation tests, and found the inferred values of $A$ and $\gamma$ to be quite stable against changes in the underlying analysis assumptions: I shall therefore take the consensus results reported in the lower left panel of Fig.~1 of Ref.~\cite{NANOGrav:2023gor} as the baseline signal for which the viability of an inflationary interpretation will be examined.

\begin{figure}[!ht]
\includegraphics[width=1.0\linewidth]{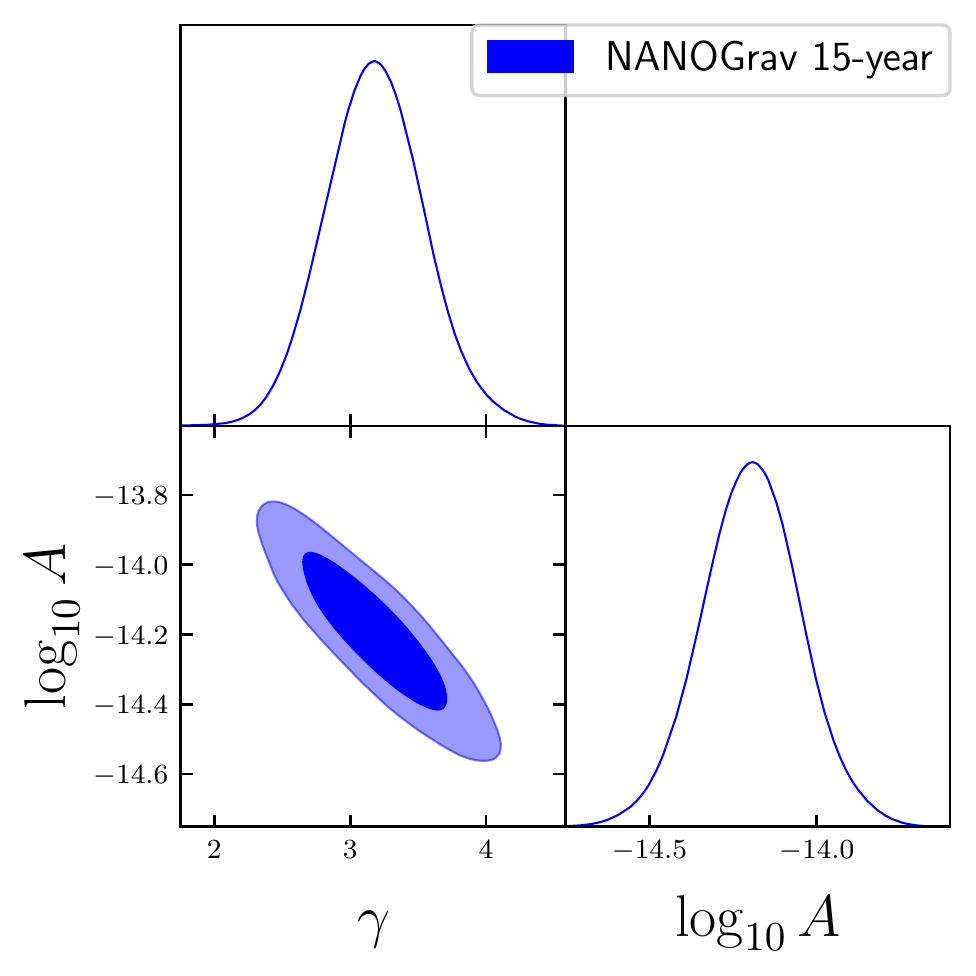}
\caption{Triangular plot showing the 2D joint and 1D marginalized posterior probability distributions for the logarithm of the intrinsic-noise SGWB amplitude $\log_{10}A$ and spectral index $\gamma$. These are in excellent agreement with the posteriors reported in the lower left panel of Fig.~1 of Ref.~\cite{NANOGrav:2023gor} (blue contours, reference frequency $1\,{\rm yr}^{-1}$), confirming the goodness of the approximation for the NANOGrav posterior discussed in Sec.~\ref{sec:methods}.}
\label{fig:nanograv_gammalogA}
\end{figure}

In my earlier work of Ref.~\cite{Vagnozzi:2020gtf} I translated the inferred NANOGrav 12.5-year constraints on $A$ and $\gamma$ into constraints on $r$ and $n_T$ through a simple grid scan. Although this simple method was more than sufficiently accurate to identify the benchmark region of inflationary parameter space required to explain the signal, here I improve the robustness and portability of the analysis method by performing a Bayesian analysis, through a Markov Chain Monte Carlo (MCMC) scan of the parameter space.~\footnote{A posteriori, I have cross-checked that repeating the analysis of Ref.~\cite{Vagnozzi:2020gtf} through an MCMC rather than a grid scan recovers essentially the same results.} Specifically, my goal is to derive constraints on the cosmological parameters, with focus on the inflationary parameters $r$ and $n_T$, treating the $A$-$\gamma$ posterior distribution inferred by NANOGrav as a likelihood.~\footnote{See Ref.~\cite{Visinelli:2018utg} where a similar simplified approach was adopted and justified by means of Bayes' theorem, and Appendix~A of Ref.~\cite{Giare:2023kiv} which confirms the goodness of the Gaussian approximation I have adopted.} To this end, I seek a closed-form approximation for the NANOGrav $A$-$\gamma$ joint posterior distribution. I find that the latter is well approximated by a bivariate Gaussian in $\log_{10}A$ and $\gamma$, with mean vector $\boldsymbol{\mu}_{15}$ and covariance matrix $\boldsymbol{\Sigma}_{15}$ given by:
\begin{eqnarray}
\boldsymbol{\mu}_{15} & \approx & (-14.20\,,3.20)\,,\nonumber \\
\boldsymbol{\Sigma}_{15} & \approx & \begin{pmatrix}
0.127 & -0.045 \\
-0.045 & 0.021
\end{pmatrix}
\,.
\label{eq:musigma}
\end{eqnarray}
To test the goodness of this approximation, I run a test MCMC on the 2 parameters $\log_{10}A$ and $\gamma$, using the bivariate Gaussian described above as likelihood, and assuming flat wide priors on both parameters. As the priors are flat and no other likelihoods are used, the recovered joint $\log_{10}A$-$\gamma$ posterior directly traces the input bivariate Gaussian distribution, and by extension the NANOGrav 15-year joint $\log_{10}A$-$\gamma$ posterior for which I am seeking a closed-form approximation. The resulting joint $\log_{10}A$-$\gamma$ posterior, alongside the 1D marginalized distributions in $\log_{10}A$ and $\gamma$, are shown in Fig.~\ref{fig:nanograv_gammalogA}. Comparing this to the lower left panel of Fig.~1 of Ref.~\cite{NANOGrav:2023gor} (blue contours, reference frequency $1\,{\rm yr}^{-1}$), I find excellent agreement between the two, confirming the goodness of my approximation, especially for the purpose of pinpointing the (a posteriori rather implausible) benchmark region of inflationary parameter space required to explain the NANOGrav signal.

Having made these considerations, I therefore approximate the NANOGrav log-likelihood as:
\begin{eqnarray}
\ln{\cal L}(\boldsymbol{\theta}) = -\frac{(\boldsymbol{x}(\boldsymbol{\theta})-\boldsymbol{\mu}_{15})^T\boldsymbol{\Sigma}_{15}^{-1}(\boldsymbol{x}(\boldsymbol{\theta})-\boldsymbol{\mu}_{15})}{2}\,,
\label{eq:loglikelihoodnanograv}
\end{eqnarray}
where $\boldsymbol{x}(\boldsymbol{\theta})$ is the vector of derived parameters $\boldsymbol{x}(\boldsymbol{\theta}) \equiv \{\log_{10}A(\boldsymbol{\theta})\,,\gamma(\boldsymbol{\theta})\}$, where $\boldsymbol{\theta}$ indicates the vector of cosmological parameters, $^T$ denotes the transpose operation, and $\boldsymbol{\mu}_{15}$ and $\boldsymbol{\Sigma}_{15}$ are respectively the vector of mean values and covariance matrix for the derived parameters $\log_{10}A$ and $\gamma$ and the covariance matrix discussed earlier and given in Eq.~(\ref{eq:musigma}). In principle $\boldsymbol{\theta}$ should include all the relevant cosmological parameters governing the amplitude and shape of the inflationary SGWB: $r$, $n_T$, $\Omega_m$, $A_s$, and $H_0$ (or equivalently $\theta_s$). However, as argued earlier in Sec.~\ref{sec:inflation}, in practice only $r$ and $n_T$ have an important effect on the inflationary SGWB signal, whereas the effect of the other parameters is completely subdominant given the level of precision to which they have been determined by current cosmological observations, in particular the \textit{Planck} satellite. Thus, in what follows I assume $\boldsymbol{\theta} \equiv \{r,n_T\}$, i.e.\ that the independent cosmological parameters are $r$ and $n_T$. I connect $r$ and $n_T$ to the derived parameters $\log_{10} A$ and $\gamma$ entering the NANOGrav likelihood of Eq.~(\ref{eq:loglikelihoodnanograv}) through Eqs.~(\ref{eq:ntgamma},\ref{eq:amplitude}), fixing all other parameters to their best-fit values as per the \textit{Planck} 2018 results~\cite{Planck:2018vyg} quoted in Sec.~\ref{sec:inflation}.

Given the fact that the order of magnitude of the tensor-to-scalar ratio $r$ is presently unknown, from a statistical point of view it proves more convenient to work with $\log_{10}r$ rather than $r$ itself. To sample the joint $\log_{10}r$-$n_T$ posterior distribution in light of the NANOGrav signal, whose information is included by means of the log-likelihood of Eq.~(\ref{eq:loglikelihoodnanograv}), I generate MCMC chains through the cosmological sampler \texttt{MontePython}~\cite{Audren:2012wb,Brinckmann:2018cvx}, which I have configured to act as a generic sampler. I impose wide flat priors on both $\log_{10}r$ and $n_T$, and have verified a posteriori that the prior ranges do not cut the posteriors of the two parameters where these are significantly non-zero. In addition, I impose a hard prior on $\log_{10}r$ such that $r<0.036$, to reflect the current $2\sigma$ upper limit on $r$ (at the same pivot scale at which $A_s$ has been reported) obtained by a joint analysis of \textit{Planck}, WMAP, BICEP2, \textit{Keck Array}, and BICEP3 data~\cite{BICEP:2021xfz} (I have check a posteriori that the choice of including this limit has very little effect on my results). I assess the convergence of the generated MCMC chains by using the Gelman-Rubin parameter $R-1$~\cite{Gelman:1992zz}, and require $R-1<10^{-3}$ in order for the MCMC chains to be considered converged.

\section{Results and discussions}
\label{sec:results}

\begin{figure}[!ht]
\includegraphics[width=1.0\linewidth]{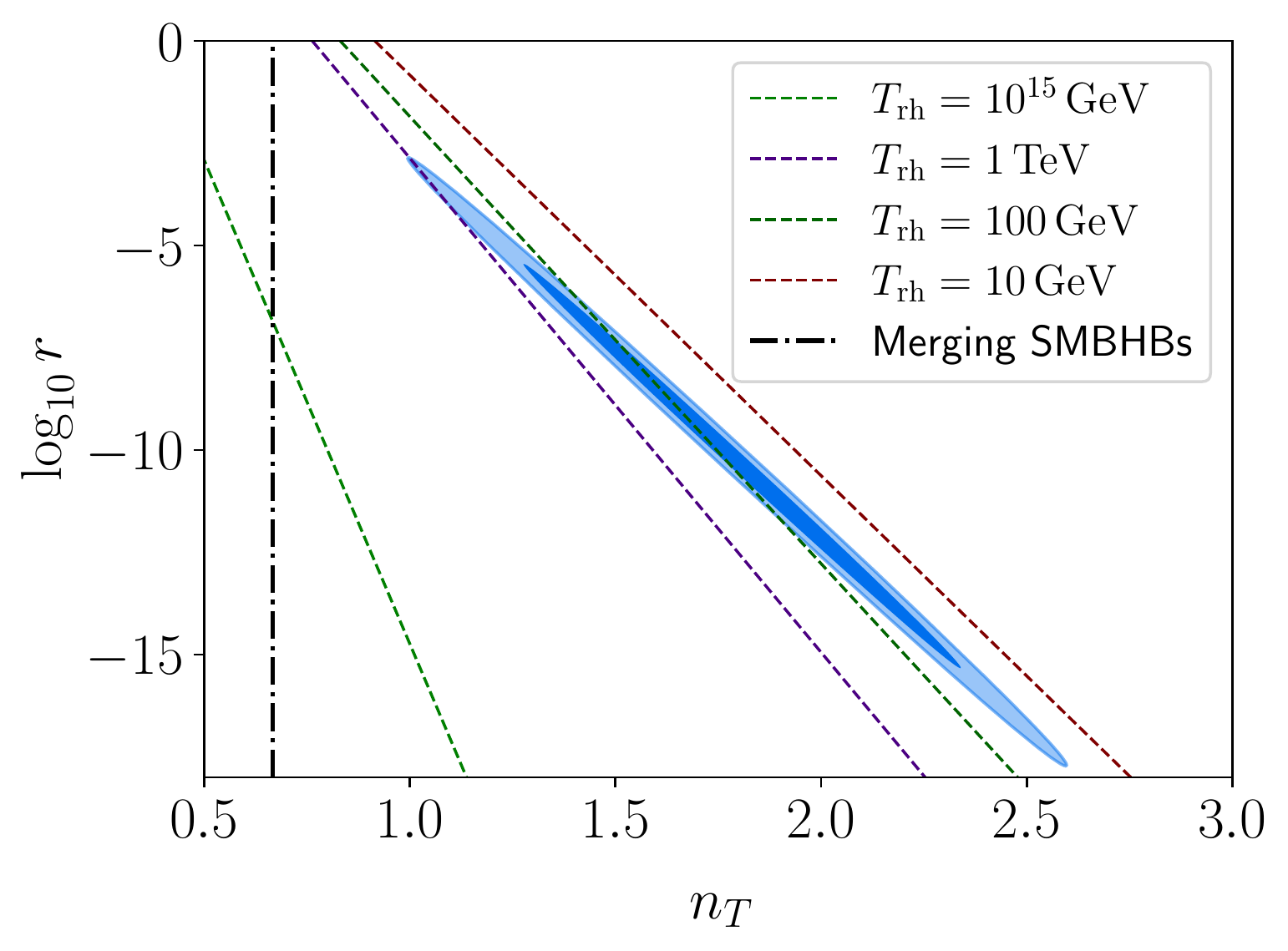}
\caption{2D joint posterior probability distribution for the tensor spectral index $n_T$ and the logarithm of the tensor-to-scalar ratio $\log_{10}r$. The dashed curves correspond to isocontours of constant $\Delta N_{\rm eff}=0.4$ for different values of the reheating temperature $T_{\rm rh}$ (see the color coding), so the regions to the right of the curves are excluded for the specific value of $T_{\rm rh}$, given the indicative BBN/CMB $2\sigma$ upper limit $\Delta N_{\rm eff} \lesssim 0.4$. The black dot-dashed vertical line at $n_T=2/3$ corresponds to $\gamma=13/3$, the value of the intrinsic-noise SGWB spectral index expected for the SGWB arising from a population of inspiraling supermassive BH binaries.}
\label{fig:nanograv_ntlogr_15yr_2D_reheating}
\end{figure}

I perform an MCMC run as described above to identify which region of inflationary parameter space can, at face value, explain the NANOGrav 15-year signal.~\footnote{Besides the inflationary interpretation I put forward in Ref.~\cite{Vagnozzi:2020gtf}, the NANOGrav 12.5-year signal (and the related PPTA, EPTA, and IPTA signals) stimulated a lot of work devoted to possible theoretical interpretations: for an inevitably limited selection of examples, see e.g.\ Refs.~\cite{Ellis:2020ena,Blasi:2020mfx,Vaskonen:2020lbd,DeLuca:2020agl,Buchmuller:2020lbh,Nakai:2020oit,Addazi:2020zcj,Kohri:2020qqd,Ratzinger:2020koh,Samanta:2020cdk,Bian:2020urb,Namba:2020kij,Neronov:2020qrl,Li:2020cjj,Sugiyama:2020roc,Liu:2020mru,Paul:2020wbz,Zhou:2020kkf,Domenech:2020ers,Bhattacharya:2020lhc,Abe:2020sqb,Kitajima:2020rpm,Middleton:2020asl,Inomata:2020xad,Tahara:2020fmn,Chigusa:2020rks,Pandey:2020gjy,Bigazzi:2020avc,Ramberg:2020oct,Cai:2020qpu,Barman:2020jrf,Chiang:2020aui,Atal:2020yic,Datta:2020bht,Chen:2021wdo,Gao:2021dfi,Li:2021qer,Gorghetto:2021fsn,Kawasaki:2021ycf,Blanco-Pillado:2021ygr,Sharma:2021rot,Brandenburg:2021tmp,Hindmarsh:2021mnl,Lazarides:2021uxv,Zhou:2021cfu,Sakharov:2021dim,NANOGrav:2021flc,Moore:2021ibq,Borah:2021ocu,Yi:2021lxc,Wu:2021zta,Haque:2021dha,Liu:2021svg,Lewicki:2021xku,Chang:2021afa,Buchmuller:2021mbb,Masoud:2021prr,Casey-Clyde:2021xro,Li:2021htg,Cai:2021wzd,Spanos:2021hpk,Khodadi:2021ees,Wu:2021kmd,Izquierdo-Villalba:2021prf,Chen:2021ncc,Borah:2021ftr,NANOGrav:2021ini,Gao:2021lno,Lin:2021vwc,Benetti:2021uea,Dunsky:2021tih,Zhang:2021rqs,Cai:2022nqv,RoperPol:2022iel,Wang:2022wwj,Ashoorioon:2022raz,Ahmed:2022rwy,Afzal:2022vjx,Wang:2022rjz,Cheong:2022gfc,Bernardo:2022vlj,Ghoshal:2022ruy,Datta:2022tab,ElBourakadi:2022anr,Blasi:2022ayo,Borah:2022iym,Saad:2022mzu,Bian:2022qbh,Cai:2022lec,NANOGrav:2023bts,Berbig:2023yyy,Wu:2023pbt,Dandoy:2023jot,Bernardo:2023mxc,An:2023idh,Ferrer:2023uwz,Ghoshal:2023sfa,Ferrante:2023bgz,Bringmann:2023opz,Madge:2023cak,Baldes:2023fsp}.} The resulting joint $\log_{10}r$-$n_T$ posterior is shown in Fig.~\ref{fig:nanograv_ntlogr_15yr_2D_reheating}. Ignoring for the moment the dashed lines associated to different reheating temperatures, I find $n_T=1.81 \pm 0.34$, completely in line with expectations given the inferred values of $\gamma$. The vertical dot-dashed line at $n_T=2/3$ indicates the expected value from merging SMBHBs corresponding to $\gamma=13/3$: not surprisingly, it lies out of the $2\sigma$ region. In addition, I find a very strong negative degeneracy between $\log_{10}r$ and $n_T$, with a correlation coefficient $\sim -0.90$, also not unexpected given that a lower value of $r$ can be compensated with a bluer spectrum, and viceversa. From Fig.~\ref{fig:nanograv_ntlogr_15yr_2D_reheating} one also sees that the NANOGrav signal actually points towards rather low values of $r$, as a direct consequence of the very high value of $n_T$: larger values of $r$ would actually lead to the resulting SGWB exceeding the amplitude of the signal detected by NANOGrav. Within this scenario, the resulting inflationary SGWB would be barely detectable by next-generation CMB experiments~\cite{CMB-S4:2016ple,SimonsObservatory:2018koc,SimonsObservatory:2019qwx}. As additionally anticipated earlier, the value of the tensor spectral index $n_T$ required to explain the NANOGrav 15-year signal is larger than that required to explain the 12.5-year signal~\cite{Vagnozzi:2020gtf}, as a direct consequence of both the higher value of $A$, as well as the significantly lower value of $\gamma$.

The key point to take away from these results is that explaining the NANOGrav 15-year signal requires a very blue spectrum, which very strongly violates the consistency relation $n_T=-r/8$~\cite{Copeland:1993ie}. Obviously, this cannot be realized within the simplest single-field slow-roll inflationary models. Nevertheless, a number of inflationary (and non-inflationary) models leading to a blue spectrum (or more generally significant deviations from the simplest expectations) have been discussed in recent years: with no claims as to completeness, these models include inflationary models based on phantom fields~\cite{Piao:2004tq,Liu:2010dh,Liu:2012iba,Dinda:2014zta,Richarte:2016qqm}, modifications to gravity~\cite{Kobayashi:2010cm,Myrzakulov:2015qaa,Fujita:2018ehq,Odintsov:2019evb,Nojiri:2019riz,Nojiri:2019fft,Odintsov:2020nwm,Nojiri:2020pqr,Odintsov:2020vjb,Kawai:2021edk,Oikonomou:2021kql,Odintsov:2021kup,Odintsov:2021urx,Odintsov:2022sdk,Oikonomou:2022xoq,Odintsov:2022hxu,Oikonomou:2022pdf,Oikonomou:2022yle,Oikonomou:2022ijs,Oikonomou:2022irx} or non-commutative space-times~\cite{Calcagni:2004as,Calcagni:2013lya}, breaking of spatial or temporal diffeomorphism invariance~\cite{Endlich:2012pz,Cannone:2014uqa,Graef:2015ova,Ricciardone:2016lym,Graef:2017cfy}, violations of the null energy condition~\cite{Baldi:2005gk}, couplings to axionic, gauge, and spin-2 fields~\cite{Maleknejad:2011sq,Adshead:2012kp,Maleknejad:2016qjz,Dimastrogiovanni:2016fuu,Adshead:2016omu,Obata:2016oym,Nojiri:2019nar,Iacconi:2020yxn,Oikonomou:2023bah} and particle production~\cite{Cook:2011hg,Pajer:2013fsa,Mukohyama:2014gba}, initial states other than the Bunch-Davies one~\cite{Ashoorioon:2014nta}, elastic media~\cite{Gruzinov:2004ty}, higher order effective gravitational action corrections~\cite{Giare:2020plo}, second-order effects~\cite{Biagetti:2013kwa} (including those associated to the formation of primordial BHs), sound speed resonance effects~\cite{Cai:2016ldn,Cai:2020ovp} and finite-time singularities~\cite{Kleidis:2016vmd}, alternatives to inflation such as string gas cosmology~\cite{Brandenberger:1988aj,Brandenberger:2006vv,Brandenberger:2006xi,Stewart:2007fu,Brandenberger:2014faa}, ekpyrotic~\cite{Khoury:2001wf,Hipolito-Ricaldi:2016kqq} and bouncing~\cite{Brandenberger:2009jq} scenarios, and many others (see Ref.~\cite{Wang:2014kqa} for a more detailed overview of these scenarios). It is worth noting that, in principle, the inferred value of $n_T$ is not inconsistent with CMB observations alone, as these are unable to place strong constraints on $n_T$ in the absence of a detection of non-zero $r$, or without including additional data (e.g.\ interferometer-scale limits on the the SGWB amplitude).~\footnote{See e.g.\ Refs.~\cite{Meerburg:2015zua,Cabass:2015jwe,Wang:2016tbj,Graef:2018fzu,Giare:2019snj} for examples of works examining constraints on $n_T$ from CMB data, which in the absence of a detection of primordial B-modes are extremely weak due to the degeneracy between $r$ and $n_T$. An alternative approach explored by Ref.~\cite{Galloni:2022mok} is to re-parameterize the tensor power spectrum using two different tensor-to-scalar ratios defined at two pivot scales: this ``two-scales approach'' was shown to improve the constraints on $n_T$, although when using CMB data alone these constraints remain comparatively weak. See however Ref.~\cite{Giare:2020vss} for a different work which argued that constraints on $n_T$ from CMB data might actually be stronger than previously thought.}

More importantly, such a blue spectrum has a very strong impact on $N_{\rm eff}$, and hence on BBN. In Fig.~\ref{fig:nanograv_ntlogr_15yr_2D_reheating}, the dashed curves correspond to isocontours of $\Delta N_{\rm eff}=0.4$ for different values of the reheating temperature $T_{\rm rh}$ (see color coding) and therefore of the frequency cutoff $f_{\max}$ appearing in Eq.~(\ref{eq:neff}). The regions to the right of the curves are excluded by the requirement $\Delta N_{\rm eff} \lesssim 0.4$ from BBN and CMB considerations (indicative and rather generous $2\sigma$ upper limit, with the caveats for the CMB discussed in footnote~6). From Fig.~\ref{fig:nanograv_ntlogr_15yr_2D_reheating} it is clear that the inflationary interpretation of the NANOGrav signal is in strong tension with limits on $N_{\rm eff}$ unless the reheating temperature is as low as $T_{\rm rh} \lesssim 10\,{\rm GeV}$. Unsurprisingly, this upper limit on the reheating temperature is even stronger than the limit I obtained previously from the 12.5-year dataset, $T_{\rm rh} \lesssim 100\,{\rm GeV}-1\,{\rm TeV}$, as a direct result of the bluer spectrum. While low-scale reheating models have been studied for quite a while~\cite{Kawasaki:2000en,Giudice:2000ex,Hannestad:2004px}, they are generally less easier to construct, and require e.g.\ an unusually long period of slow-roll or the existence of long-lived massive particles. It is worth noting that with such a low reheating scale, cosmic neutrinos may not have had time to thermalize completely by the time they decouple -- as a consequence, the current relic neutrino energy and number densities may be lower compared to standard expectations~\cite{deSalas:2015glj,Gerbino:2016sgw}.~\footnote{Given  the recent results from the Atacama Cosmology Telescope (ACT) which have inferred a value of $N_{\rm eff}$ lower than expected~\cite{ACT:2020gnv}, one may in principle be tempted to speculate as to whether an underlying low-reheating scenario may explain these results. See, however, Refs.~\cite{Handley:2020hdp,Giare:2022rvg,Zhai:2023yny,DiValentino:2023fei,Giare:2023wzl} for discussions on the tension between ACT and \textit{Planck} for what concerns the determination of parameters such as $N_{\rm eff}$, $\omega_b$, and $n_s$.}

An important caveat to all these results is my choice of treating ${\cal P}_T(k)$ as a pure power-law from CMB frequencies down to PTA frequencies. While this approximation is widespread (just by way of recent well-known examples, see e.g.\ Refs.~\cite{Meerburg:2015zua,Planck:2018jri,Campeti:2020xwn}), it might not be entirely justified when considering probes leading to a wide lever arm in frequency space~\cite{Giare:2020vhn,Kinney:2021nje}. For instance, there are about $9$ decades in frequency between CMB and PTA scales, and an additional $10$ decades in frequency between PTA and current interferometer scales. To address this point, what would ultimately be required is to estimate the theoretical uncertainties in the predictions for inflationary parameters in a way as potential-agnostic as possible (for instance through Monte Carlo inflation flow potential reconstruction, see e.g.\ Refs.~\cite{Kinney:2005in,Caligiuri:2014ola}). Nevertheless, for what concerns the present work, treating the primordial tensor spectrum as being a pure power-law between CMB and PTA scales might actually not be too bad an approximation: in this respect, see for example Fig.~5 in Ref.~\cite{Kinney:2021nje}, where PTA scales correspond to $k \sim {\cal O}(10^6)\,{\rm Mpc}^{-1}$, where the relative deviation from the pure power-law extrapolation for an example inflationary potential is actually rather small. Nevertheless, extending this treatment down to interferometer scales and beyond is clearly not reasonable (as is clear from the same Figure). Overall this question, while important, goes well beyond the scope of my work, and I therefore defer it to future studies.

A somewhat observationally more relevant point related to the above is that the blue spectrum described earlier, if na\"{i}vely extrapolated down to interferometer scales (which, \textit{repetita juvant}, should not be done in any case!), would inevitably violate upper limits on the SGWB amplitude on interferometer scales, where Advanced LIGO/Advanced Virgo set the $2\sigma$ upper limit $\Omega_{\rm gw} \lesssim 6.6 \times 10^{-9}$ at $f \sim 25\,{\rm Hz}$~\cite{KAGRA:2021kbb}. An obvious and well-motivated way to address this problem is to envisage the existence of a break in the tensor power spectrum, which in the simplest case would therefore be described by a broken power-law. Such an approach was discussed in detail in Ref.~\cite{Benetti:2021uea}, to which I refer the reader for further details. Here, I simply note that a wide variety of well-motivated scenarios naturally give rise to a break in the tensor power spectrum: such scenarios include but are not limited to a non-instantaneous reheating process (possibly with a low reheating scale), an early period of non-standard expansion (e.g.\ an early matter domination phase, or a kination phase), late-time entropy injection, models where the inflaton is coupled to gauge fields, and alternatives to inflation. Whichever mechanism one chooses to invoke, the break needs to take place at frequencies $f>f_{\rm yr}$, and the post-break spectrum needs to be sufficiently red in order for the SGWB amplitude to be low enough on interferometer scales (see Ref.~\cite{Benetti:2021uea} for further details on this phenomenological approach).

\begin{figure*}[!ht]
\centering
\includegraphics[width=0.8\linewidth]{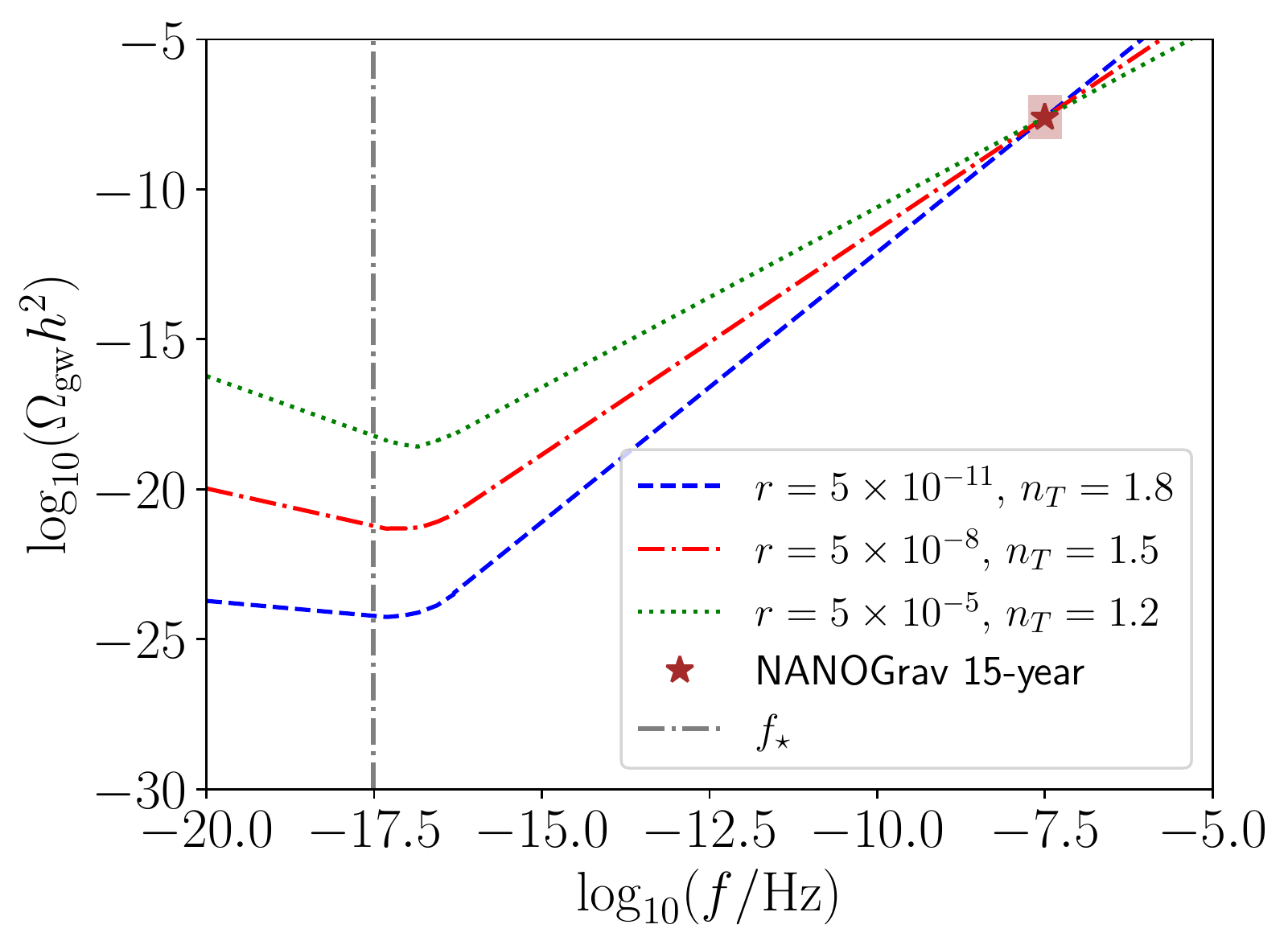}
\caption{Benchmark examples (consistent with data indicatively at the $2\sigma$ level) of SGWB spectra required to explain the NANOGrav signal while remaining consistent with upper limits on the tensor-to-scalar ratio on CMB scales. The three examples are characterized by different values of the tensor-to-scalar ratio $r$ and tensor spectral index $n_T$, namely $r=5 \times 10^{-11}$ and $n_T=1.8$ (blue dashed curve), $r=5 \times 10^{-8}$ and $n_T=1.5$ (red dot-dashed curve), $r=5 \times 10^{-5}$ and $n_T=1.2$ (green dotted curve). The NANOGrav signal is denoted by a brown star, with the shaded brown region representative of observational uncertainties. The vertical grey dot-dashed line indicates the pivot frequency $f_{\star} \approx 7.7 \times 10^{-17}\,{\rm Hz}$, which corresponds to the pivot wavenumber $k_{\star}=0.05\,{\rm Mpc}^{-1}$ at which $r$ is constrained by CMB observations.}
\label{fig:nanograv_spectrum_benchmark}
\end{figure*}

The required reheating scale required for an inflationary interpretation of the NANOGrav 15-year signal to be viable at face value is extremely low, below the electroweak scale. While this is in principle allowed by cosmological probes, which only require $T_{\rm rh} \gtrsim 5\,{\rm MeV}$~\cite{deSalas:2015glj}, realizing it in practice from the particle point of view is more challenging (for instance, it is unclear whether electroweak symmetry breaking can proceed successfully as usual, although on the other hand such a scenario may facilitate the generation of a baryon asymmetry). Finally, it is worth noting that the reheating scale can be lowered even in the absence of new physics, through effects such as ``Higgs blocking''~\cite{Freese:2017ace,Litsa:2020rsm,Litsa:2020mvj}.

The biggest challenge for an inflationary interpretation of the NANOGrav 15-year results, however, is the extremely blue spectrum required to fit the signal, with $n_T$ close to $2$. As one sees from Eq.~(\ref{eq:ntw}), even within effective phantom inflationary scenarios, the value $n_T=2$ can only be reached asymptotically if one requires $w<-1/3$ (it can be reached for positive values of $w$, but these are not of interest to an inflationary discussion). Nevertheless, even just getting close to $n_T=2$ within phantom inflationary models would require considering very deep phantom models: needless to say, such models are theoretically problematic and hard to achieve in a controlled setting. Just by way of example, from Eq.~(\ref{eq:ntw}) one sees that obtaining $n_T=1$ requires $w=-5/3$, whereas obtaining $n_T=1.5$ requires $w=-7/3$ -- both correspond to values deep in the phantom regime, and therefore theoretically very problematic. Motivated by the 2014 BICEP-2 signal, Ref.~\cite{Wang:2014kqa} discussed a number of inflationary scenarios leading to blue tensor spectra, including the ones discussed earlier in this Section: for most of these models, it was found that reaching values of the tensor spectral index $n_T \sim 2$ is extremely hard if not impossible altogether. In a more generic setting, this is also further supported by the findings of Ref.~\cite{Giare:2022wxq}, where based on an extension of the flow approach it was shown that most of the mechanisms proposed to obtain a blue spectrum (especially those which can be cast in terms of additional effective field theory operators) actually only support $n_T>0$ for few e-folds, which is quite different from what is required to explain the NANOGrav signal. This further supports the untenability of an inflationary interpretation of the NANOGrav signal.

An explicit phantom inflationary model is just one way of violating the null energy condition (NEC), which is a well-motivated possibility to obtain a blue spectrum. Besides explicit phantom inflationary models, the NEC can be violated within models based on modified gravity, and an interesting model in this sense is the G-inflation model~\cite{Kobayashi:2010cm}, based on a Galileon field (see also Refs.~\cite{Deffayet:2009mn,He:2016uiy}). This allows for NEC violation at the cosmological background level without introducing a ghost in the scalar sector. However, in its simplest incarnation this model is problematic because \textit{a)} the scalar and tensor spectra tilt in the same direction, so a very blue tensor spectrum would require a very blue scalar spectrum, strongly excluded by observations, and \textit{b)} the bluer the tensor spectrum, the larger the level of equilateral non-Gaussianity predicted (due to non-linear self-interactions of the Galileon field), which in the case of $n_T \sim 2$ are safely ruled out. Another possibility involves perturbations generated from a non-Bunch-Davies vacuum, although again achieving a large tensor spectral index comes at the price of introducing a significant level of (folded) non-Gaussianity both in the scalar and tensor sectors, with the former excluded by observations~\cite{Wang:2014kqa}. Remaining within the inflationary realm, and in the context of the non-standard scenarios discussed earlier in this Section, one finds that either $n_T$ is positive but too small (certainly not $n_T \sim 2$ as required by the NANOGrav signal), or achieving large $n_T$ comes at the price of a significant level of non-Gaussianities, already excluded by observations.~\footnote{One possible interesting exception is the non-local extension of Starobinsky inflation, itself motivated by attempts to construct a 4-dimensional theory of quantum gravity, and whose phenomenology was recently studied in Refs.~\cite{Koshelev:2016xqb,Koshelev:2017tvv,Koshelev:2020foq,Koshelev:2020xby,Koshelev:2022olc,Koshelev:2022bvg,Koshelev:2023elc}.} All these considerations bring me to the general conclusion that an inflationary interpretation of the NANOGrav signal, and by extension of the signal recently observed by EPTA+InPTA, PPTA, and CPTA (given the broad agreement between all four signals) is hardly tenable.

Before closing, I briefly discuss predictions for $n_T$ within alternatives to inflation. It is worth noting that one of the first predictions for a blue tensor spectrum came from string gas cosmology~\cite{Brandenberger:1988aj,Brandenberger:2006vv}, wherein the universe emerged from a string Hagedorn phase at a nearly constant temperature, until the decay of the string winding modes allowed for the expansion of the Universe. Within string gas cosmology, one finds~\cite{Brandenberger:2014faa,Wang:2014kqa}:
\begin{eqnarray}
n_T \approx 1-n_s\,,
\label{eq:stringgas}
\end{eqnarray}
which is not suppressed by the slow-roll parameters, and therefore clearly has a hard time explaining the ${\cal O}(1)$ value of $n_T$ required by the NANOGrav signal, given that constraints on the scalar spectral index from \textit{Planck} imply $\vert n_s-1 \vert \sim {\cal O}(10^{-2})$.~\footnote{A related scenario was studied in Ref.~\cite{Biswas:2014kva}, where the possibility of producing primordial fluctuations from statistical thermal fluctuations within string gas cosmology was explored. In particular, blue GWs are produced during a stringy thermal contracting phase at temperatures higher than the Hagedorn temperature. However, this scenario predicts $n_T \sim 3$, which is at odds with the NANOGrav signal.}

A more interesting possibility is the so-called ``old ekpyrotic'' scenario~\cite{Khoury:2001wf} (see also Refs.~\cite{Khoury:2001zk,Buchbinder:2007ad,Lehners:2008vx,Oikonomou:2014yua,Odintsov:2015uca}), wherein the Universe starts in a cold, very slowly contracting state, after which the collision of a brane in the bulk space with a bounding orbifold plane starts the hot Big Bang expansion era. The ekpyrotic model predicts a quasi-scale-invariant spectrum of scalar perturbations and a very blue tensor spectrum with $n_T=2$~\cite{Boyle:2003km}. However, on cosmological scales the amplitude of tensor modes is highly suppressed~\cite{Khoury:2001wf,Boyle:2003km}, so that despite the large blue tilt, the resulting SGWB remains well below the detection threshold even on PTA scales~\cite{Boyle:2003km,Calcagni:2020tvw}. While the ``new ekpyrotic'' scenario~\cite{Brandenberger:2020tcr,Brandenberger:2020eyf,Brandenberger:2020wha} avoids these problems, it also predicts a relation between $n_T$ and $n_s$ of the same form as Eq.~(\ref{eq:stringgas}), and therefore a tensor spectrum which does not carry a strong enough blue tilt to explain the signal observed by PTA experiments.

\section{Conclusions}
\label{sec:conclusions}

A number of pulsar timing array (PTA) experiments (NANOGrav, EPTA+InPTA, PPTA, and CPTA) have recently confirmed the detection of a common-spectrum low-frequency stochastic signal across the pulsars in their latest datasets, with the additional compelling evidence of Hellings-Downs inter-pulsar correlations, achieving the first convincing detection of a stochastic gravitational wave background (SGWB) in the ${\rm nHz}$ frequency range~\cite{NANOGrav:2023gor,Antoniadis:2023ott,Reardon:2023gzh,Xu:2023wog,NANOGrav:2023hde,NANOGrav:2023ctt,NANOGrav:2023hvm,NANOGrav:2023hfp,NANOGrav:2023ghw,NANOGrav:2023pdq,NANOGrav:2023icp,Antoniadis:2023lym,Antoniadis:2023puu,Antoniadis:2023aac,Antoniadis:2023xlr,Smarra:2023ljf,Reardon:2023zen,Zic:2023gta}. This has officially broadened the horizons of GW astronomy, and the quest is now open for determining the origin of these signals. An unresolved background of merging supermassive black hole binaries is a natural candidate explanation, whereas another exciting possibility entails a cosmological origin. While other types of observations (for instance the detection of anisotropies~\cite{Bartolo:2019oiq,Bartolo:2019zvb,LISACosmologyWorkingGroup:2022kbp,Galloni:2022rgg,Schulze:2023ich}) are ultimately required to discern between these two possibilities, it is interesting to examine whether cosmological explanations are consistent with the currently reported characteristics of the signal. In this work, focusing for concreteness and simplicity on the NANOGrav 15-year signal (given that the signals observed in all four PTA experiments are mutually consistent), I have tested the viability of an inflationary SGWB interpretation, given the uttermost theoretical importance of an inflationary era as a source of cosmological GWs.

The main outcomes of my analysis are summarized in Fig.~\ref{fig:nanograv_ntlogr_15yr_2D_reheating} and Fig.~\ref{fig:nanograv_spectrum_benchmark}. Compared to the earlier NANOGrav 12.5-year results, for which an inflationary interpretation was in principle not unreasonable~\cite{Vagnozzi:2020gtf}, an inflationary origin for the NANOGrav 15-year signal appears much less tenable. Firstly, the spectrum required to explain the signal is extremely blue ($n_T \sim 2$), to an extent which, while theoretically not impossible to obtain, presents huge challenges within inflationary settings. As discussed in Sec.~\ref{sec:results}, obtaining $n_T \sim 2$ within inflationary models is extremely challenging (and at the very least requires moving far from the single-field slow-roll paradigm), and the least unlikely realizations all tend to predict sizeable non-Gaussianities, which have not been detected. Alternatives to inflation can also provide a source for such a blue-tilted SGWB spectrum, with an ekpyrotic phase being a particularly intriguing possibility in this sense, given the fact that such scenarios naturally predict $n_T \sim 2$, although in this case the expected amplitude of the signal is far too small. Returning to the inflationary realm, another difficulty concerns the fact that the required reheating scale has to be extremely low, below the electroweak scale, posing additional difficulties on the model-building side.~\footnote{Even if an inflationary model which overcomes such difficulties can be constructed, the blue spectrum would anyhow require a break at frequencies above PTA scales, in order not to violate upper limits on the SGWB amplitude on interferometer scales (see e.g.\ the model-agnostic approach in Ref.~\cite{Benetti:2021uea}).} Finally, I have provided explicit expressions for a bivariate Gaussian approximation to the joint posterior distribution for the intrinsic-noise amplitude and spectral index inferred from the NANOGrav signal. This can facilitate extending similar simplified likelihood analyses to different underlying theoretical signals for which a prediction for the amplitude and spectral index can easily be obtained.

Overall, it is safe to say that an inflationary interpretation of the signals reported in June 2023 by various PTA experiments appears exceptionally unlikely. Other cosmological sources, such as GWs generated by a network of cosmic strings or associated to the formation of primordial black holes, may provide a better explanation for the signal, and I expect that such possibilities will be examined in significant detail elsewhere in the literature.~\footnote{Besides the NANOGrav~\cite{NANOGrav:2023hvm} and EPTA~\cite{Antoniadis:2023xlr} official publications on searches for new physics, see e.g.\ Refs.~\cite{Liu:2023hte,King:2023cgv,Zu:2023olm,Han:2023olf,Lambiase:2023pxd,Ellis:2023dgf,Guo:2023hyp,Megias:2023kiy,Fujikura:2023lkn,Yang:2023aak,Li:2023yaj,Deng:2023btv,Franciolini:2023wjm,Shen:2023pan,Kitajima:2023cek,Ellis:2023tsl,Franciolini:2023pbf,Wang:2023len,Ghoshal:2023fhh,Bai:2023cqj,Addazi:2023jvg,Athron:2023mer,Oikonomou:2023qfz,Kitajima:2023vre,Huang:2023chx,Eichhorn:2023gat,Buchmueller:2023nll,Lazarides:2023ksx,Broadhurst:2023tus,Cai:2023dls,Yang:2023qlf,Blasi:2023sej,Inomata:2023zup,Depta:2023qst,Gouttenoire:2023ftk,Borah:2023sbc,Wang:2023ost,Murai:2023gkv,Datta:2023vbs,Barman:2023fad,Bi:2023tib,Lu:2023mcz,Xiao:2023dbb,Li:2023bxy,Zhang:2023lzt,Anchordoqui:2023tln,Liu:2023ymk,Konoplya:2023fmh,Chowdhury:2023opo,Niu:2023bsr}, for examples of works connecting the PTA signals to a variety of fundamental physics models.} Looking beyond, it will be interesting to see whether the characteristics of the signal will be confirmed by future PTA observations, e.g.\ the forthcoming IPTA Data Release 3, which will include a total of 80 pulsars observed for up to 24 years, with a considerably improved sensitivity to spatial correlations and spectral resolutions compared to single-PTA datasets such as the ones considered here. These datasets will help clarify the origin of the SGWB, settling the question of whether the latter is astrophysical or cosmological in origin. Regardless of the outcome, it is beyond doubt that PTA experiments are opening a new, exciting window onto GWs produced in the early Universe (including potentially during inflation), at energy scales and physical regimes which are completely inaccessible down on Earth: the era of nHz GW astronomy is upon us.

\section*{Note added}
\noindent I have independently started this work in preparation for the much anticipated public announcement of NANOGrav in coordination with other PTA experiment on June 29, 2023. Inevitably, by then NANOGrav~\cite{NANOGrav:2023hvm} and EPTA~\cite{Antoniadis:2023xlr} had already explored an inflationary interpretation of their results. However, they considered slightly different parametrizations and ingredients (e.g.\ non-instantaneous reheating), and conversely my work discusses in more detail the difficulties associated to an inflationary origin of the signals, which are only alluded to by NANOGrav. Our results are in good overall agreement, as we all agree on the need for a very blue spectrum and a very low reheating temperature. This provides a valuable and independent cross-check, whereas our discussions of the results nicely complement each other.

\begin{acknowledgments}
\noindent I thank William Giar\`{e} and Will Kinney for many useful discussions around the subject of this work. This publication is based upon work from the COST Action CA21136 ``Addressing observational tensions in cosmology with systematics and fundamental physics (CosmoVerse), supported by COST (European Cooperation in Science and Technology).
\end{acknowledgments}

\bibliography{PTA2023}

\begin{thebibliography}{435}%
\makeatletter
\providecommand \@ifxundefined [1]{%
 \@ifx{#1\undefined}
}%
\providecommand \@ifnum [1]{%
 \ifnum #1\expandafter \@firstoftwo
 \else \expandafter \@secondoftwo
 \fi
}%
\providecommand \@ifx [1]{%
 \ifx #1\expandafter \@firstoftwo
 \else \expandafter \@secondoftwo
 \fi
}%
\providecommand \natexlab [1]{#1}%
\providecommand \enquote  [1]{``#1''}%
\providecommand \bibnamefont  [1]{#1}%
\providecommand \bibfnamefont [1]{#1}%
\providecommand \citenamefont [1]{#1}%
\providecommand \href@noop [0]{\@secondoftwo}%
\providecommand \href [0]{\begingroup \@sanitize@url \@href}%
\providecommand \@href[1]{\@@startlink{#1}\@@href}%
\providecommand \@@href[1]{\endgroup#1\@@endlink}%
\providecommand \@sanitize@url [0]{\catcode `\\12\catcode `\$12\catcode
  `\&12\catcode `\#12\catcode `\^12\catcode `\_12\catcode `\%12\relax}%
\providecommand \@@startlink[1]{}%
\providecommand \@@endlink[0]{}%
\providecommand \url  [0]{\begingroup\@sanitize@url \@url }%
\providecommand \@url [1]{\endgroup\@href {#1}{\urlprefix }}%
\providecommand \urlprefix  [0]{URL }%
\providecommand \Eprint [0]{\href }%
\providecommand \doibase [0]{http://dx.doi.org/}%
\providecommand \selectlanguage [0]{\@gobble}%
\providecommand \bibinfo  [0]{\@secondoftwo}%
\providecommand \bibfield  [0]{\@secondoftwo}%
\providecommand \translation [1]{[#1]}%
\providecommand \BibitemOpen [0]{}%
\providecommand \bibitemStop [0]{}%
\providecommand \bibitemNoStop [0]{.\EOS\space}%
\providecommand \EOS [0]{\spacefactor3000\relax}%
\providecommand \BibitemShut  [1]{\csname bibitem#1\endcsname}%
\let\auto@bib@innerbib\@empty
\bibitem [{\citenamefont {Caprini}\ and\ \citenamefont
  {Figueroa}(2018)}]{Caprini:2018mtu}%
  \BibitemOpen
  \bibfield  {author} {\bibinfo {author} {\bibfnamefont {C.}~\bibnamefont
  {Caprini}}\ and\ \bibinfo {author} {\bibfnamefont {D.~G.}\ \bibnamefont
  {Figueroa}},\ }\href {\doibase 10.1088/1361-6382/aac608} {\bibfield
  {journal} {\bibinfo  {journal} {Class. Quant. Grav.}\ }\textbf {\bibinfo
  {volume} {35}},\ \bibinfo {pages} {163001} (\bibinfo {year} {2018})},\
  \Eprint {http://arxiv.org/abs/1801.04268} {arXiv:1801.04268 [astro-ph.CO]}
  \BibitemShut {NoStop}%
\bibitem [{\citenamefont {Renzini}\ \emph {et~al.}(2022)\citenamefont
  {Renzini}, \citenamefont {Goncharov}, \citenamefont {Jenkins},\ and\
  \citenamefont {Meyers}}]{Renzini:2022alw}%
  \BibitemOpen
  \bibfield  {author} {\bibinfo {author} {\bibfnamefont {A.~I.}\ \bibnamefont
  {Renzini}}, \bibinfo {author} {\bibfnamefont {B.}~\bibnamefont {Goncharov}},
  \bibinfo {author} {\bibfnamefont {A.~C.}\ \bibnamefont {Jenkins}}, \ and\
  \bibinfo {author} {\bibfnamefont {P.~M.}\ \bibnamefont {Meyers}},\ }\href
  {\doibase 10.3390/galaxies10010034} {\bibfield  {journal} {\bibinfo
  {journal} {Galaxies}\ }\textbf {\bibinfo {volume} {10}},\ \bibinfo {pages}
  {34} (\bibinfo {year} {2022})},\ \Eprint {http://arxiv.org/abs/2202.00178}
  {arXiv:2202.00178 [gr-qc]} \BibitemShut {NoStop}%
\bibitem [{\citenamefont {Siemens}\ \emph {et~al.}(2007)\citenamefont
  {Siemens}, \citenamefont {Mandic},\ and\ \citenamefont
  {Creighton}}]{Siemens:2006yp}%
  \BibitemOpen
  \bibfield  {author} {\bibinfo {author} {\bibfnamefont {X.}~\bibnamefont
  {Siemens}}, \bibinfo {author} {\bibfnamefont {V.}~\bibnamefont {Mandic}}, \
  and\ \bibinfo {author} {\bibfnamefont {J.}~\bibnamefont {Creighton}},\ }\href
  {\doibase 10.1103/PhysRevLett.98.111101} {\bibfield  {journal} {\bibinfo
  {journal} {Phys. Rev. Lett.}\ }\textbf {\bibinfo {volume} {98}},\ \bibinfo
  {pages} {111101} (\bibinfo {year} {2007})},\ \Eprint
  {http://arxiv.org/abs/astro-ph/0610920} {arXiv:astro-ph/0610920} \BibitemShut
  {NoStop}%
\bibitem [{\citenamefont {Caprini}\ \emph {et~al.}(2010)\citenamefont
  {Caprini}, \citenamefont {Durrer},\ and\ \citenamefont
  {Siemens}}]{Caprini:2010xv}%
  \BibitemOpen
  \bibfield  {author} {\bibinfo {author} {\bibfnamefont {C.}~\bibnamefont
  {Caprini}}, \bibinfo {author} {\bibfnamefont {R.}~\bibnamefont {Durrer}}, \
  and\ \bibinfo {author} {\bibfnamefont {X.}~\bibnamefont {Siemens}},\ }\href
  {\doibase 10.1103/PhysRevD.82.063511} {\bibfield  {journal} {\bibinfo
  {journal} {Phys. Rev. D}\ }\textbf {\bibinfo {volume} {82}},\ \bibinfo
  {pages} {063511} (\bibinfo {year} {2010})},\ \Eprint
  {http://arxiv.org/abs/1007.1218} {arXiv:1007.1218 [astro-ph.CO]} \BibitemShut
  {NoStop}%
\bibitem [{\citenamefont {Ramberg}\ and\ \citenamefont
  {Visinelli}(2019)}]{Ramberg:2019dgi}%
  \BibitemOpen
  \bibfield  {author} {\bibinfo {author} {\bibfnamefont {N.}~\bibnamefont
  {Ramberg}}\ and\ \bibinfo {author} {\bibfnamefont {L.}~\bibnamefont
  {Visinelli}},\ }\href {\doibase 10.1103/PhysRevD.99.123513} {\bibfield
  {journal} {\bibinfo  {journal} {Phys. Rev. D}\ }\textbf {\bibinfo {volume}
  {99}},\ \bibinfo {pages} {123513} (\bibinfo {year} {2019})},\ \Eprint
  {http://arxiv.org/abs/1904.05707} {arXiv:1904.05707 [astro-ph.CO]}
  \BibitemShut {NoStop}%
\bibitem [{\citenamefont {Caprini}\ \emph {et~al.}(2020)\citenamefont {Caprini}
  \emph {et~al.}}]{Caprini:2019egz}%
  \BibitemOpen
  \bibfield  {author} {\bibinfo {author} {\bibfnamefont {C.}~\bibnamefont
  {Caprini}} \emph {et~al.},\ }\href {\doibase 10.1088/1475-7516/2020/03/024}
  {\bibfield  {journal} {\bibinfo  {journal} {JCAP}\ }\textbf {\bibinfo
  {volume} {03}},\ \bibinfo {pages} {024} (\bibinfo {year} {2020})},\ \Eprint
  {http://arxiv.org/abs/1910.13125} {arXiv:1910.13125 [astro-ph.CO]}
  \BibitemShut {NoStop}%
\bibitem [{\citenamefont {Ellis}\ \emph {et~al.}(2020)\citenamefont {Ellis},
  \citenamefont {Lewicki},\ and\ \citenamefont {No}}]{Ellis:2020awk}%
  \BibitemOpen
  \bibfield  {author} {\bibinfo {author} {\bibfnamefont {J.}~\bibnamefont
  {Ellis}}, \bibinfo {author} {\bibfnamefont {M.}~\bibnamefont {Lewicki}}, \
  and\ \bibinfo {author} {\bibfnamefont {J.~M.}\ \bibnamefont {No}},\ }\href
  {\doibase 10.1088/1475-7516/2020/07/050} {\bibfield  {journal} {\bibinfo
  {journal} {JCAP}\ }\textbf {\bibinfo {volume} {07}},\ \bibinfo {pages} {050}
  (\bibinfo {year} {2020})},\ \Eprint {http://arxiv.org/abs/2003.07360}
  {arXiv:2003.07360 [hep-ph]} \BibitemShut {NoStop}%
\bibitem [{\citenamefont {Rajagopal}\ and\ \citenamefont
  {Romani}(1995)}]{Rajagopal:1994zj}%
  \BibitemOpen
  \bibfield  {author} {\bibinfo {author} {\bibfnamefont {M.}~\bibnamefont
  {Rajagopal}}\ and\ \bibinfo {author} {\bibfnamefont {R.~W.}\ \bibnamefont
  {Romani}},\ }\href {\doibase 10.1086/175813} {\bibfield  {journal} {\bibinfo
  {journal} {Astrophys. J.}\ }\textbf {\bibinfo {volume} {446}},\ \bibinfo
  {pages} {543} (\bibinfo {year} {1995})},\ \Eprint
  {http://arxiv.org/abs/astro-ph/9412038} {arXiv:astro-ph/9412038} \BibitemShut
  {NoStop}%
\bibitem [{\citenamefont {Jaffe}\ and\ \citenamefont
  {Backer}(2003)}]{Jaffe:2002rt}%
  \BibitemOpen
  \bibfield  {author} {\bibinfo {author} {\bibfnamefont {A.~H.}\ \bibnamefont
  {Jaffe}}\ and\ \bibinfo {author} {\bibfnamefont {D.~C.}\ \bibnamefont
  {Backer}},\ }\href {\doibase 10.1086/345443} {\bibfield  {journal} {\bibinfo
  {journal} {Astrophys. J.}\ }\textbf {\bibinfo {volume} {583}},\ \bibinfo
  {pages} {616} (\bibinfo {year} {2003})},\ \Eprint
  {http://arxiv.org/abs/astro-ph/0210148} {arXiv:astro-ph/0210148} \BibitemShut
  {NoStop}%
\bibitem [{\citenamefont {Wyithe}\ and\ \citenamefont
  {Loeb}(2003)}]{Wyithe:2002ep}%
  \BibitemOpen
  \bibfield  {author} {\bibinfo {author} {\bibfnamefont {J.~S.~B.}\
  \bibnamefont {Wyithe}}\ and\ \bibinfo {author} {\bibfnamefont
  {A.}~\bibnamefont {Loeb}},\ }\href {\doibase 10.1086/375187} {\bibfield
  {journal} {\bibinfo  {journal} {Astrophys. J.}\ }\textbf {\bibinfo {volume}
  {590}},\ \bibinfo {pages} {691} (\bibinfo {year} {2003})},\ \Eprint
  {http://arxiv.org/abs/astro-ph/0211556} {arXiv:astro-ph/0211556} \BibitemShut
  {NoStop}%
\bibitem [{\citenamefont {Sesana}\ \emph {et~al.}(2004)\citenamefont {Sesana},
  \citenamefont {Haardt}, \citenamefont {Madau},\ and\ \citenamefont
  {Volonteri}}]{Sesana:2004sp}%
  \BibitemOpen
  \bibfield  {author} {\bibinfo {author} {\bibfnamefont {A.}~\bibnamefont
  {Sesana}}, \bibinfo {author} {\bibfnamefont {F.}~\bibnamefont {Haardt}},
  \bibinfo {author} {\bibfnamefont {P.}~\bibnamefont {Madau}}, \ and\ \bibinfo
  {author} {\bibfnamefont {M.}~\bibnamefont {Volonteri}},\ }\href {\doibase
  10.1086/422185} {\bibfield  {journal} {\bibinfo  {journal} {Astrophys. J.}\
  }\textbf {\bibinfo {volume} {611}},\ \bibinfo {pages} {623} (\bibinfo {year}
  {2004})},\ \Eprint {http://arxiv.org/abs/astro-ph/0401543}
  {arXiv:astro-ph/0401543} \BibitemShut {NoStop}%
\bibitem [{\citenamefont {Burke-Spolaor}\ \emph {et~al.}(2019)\citenamefont
  {Burke-Spolaor} \emph {et~al.}}]{Burke-Spolaor:2018bvk}%
  \BibitemOpen
  \bibfield  {author} {\bibinfo {author} {\bibfnamefont {S.}~\bibnamefont
  {Burke-Spolaor}} \emph {et~al.},\ }\href {\doibase 10.1007/s00159-019-0115-7}
  {\bibfield  {journal} {\bibinfo  {journal} {Astron. Astrophys. Rev.}\
  }\textbf {\bibinfo {volume} {27}},\ \bibinfo {pages} {5} (\bibinfo {year}
  {2019})},\ \Eprint {http://arxiv.org/abs/1811.08826} {arXiv:1811.08826
  [astro-ph.HE]} \BibitemShut {NoStop}%
\bibitem [{\citenamefont {{Sazhin}}(1978)}]{Sazhin:1978ghw}%
  \BibitemOpen
  \bibfield  {author} {\bibinfo {author} {\bibfnamefont {M.~V.}\ \bibnamefont
  {{Sazhin}}},\ }\href@noop {} {\bibfield  {journal} {\bibinfo  {journal}
  {Soviet Ast.}\ }\textbf {\bibinfo {volume} {22}},\ \bibinfo {pages} {36}
  (\bibinfo {year} {1978})}\BibitemShut {NoStop}%
\bibitem [{\citenamefont {Detweiler}(1979)}]{Detweiler:1979wn}%
  \BibitemOpen
  \bibfield  {author} {\bibinfo {author} {\bibfnamefont {S.~L.}\ \bibnamefont
  {Detweiler}},\ }\href {\doibase 10.1086/157593} {\bibfield  {journal}
  {\bibinfo  {journal} {Astrophys. J.}\ }\textbf {\bibinfo {volume} {234}},\
  \bibinfo {pages} {1100} (\bibinfo {year} {1979})}\BibitemShut {NoStop}%
\bibitem [{\citenamefont {Hobbs}\ and\ \citenamefont
  {Dai}(2017)}]{Hobbs:2017oam}%
  \BibitemOpen
  \bibfield  {author} {\bibinfo {author} {\bibfnamefont {G.}~\bibnamefont
  {Hobbs}}\ and\ \bibinfo {author} {\bibfnamefont {S.}~\bibnamefont {Dai}},\
  }\href {\doibase 10.1093/nsr/nwx126} {\bibfield  {journal} {\bibinfo
  {journal} {Natl. Sci. Rev.}\ }\textbf {\bibinfo {volume} {4}},\ \bibinfo
  {pages} {707} (\bibinfo {year} {2017})},\ \Eprint
  {http://arxiv.org/abs/1707.01615} {arXiv:1707.01615 [astro-ph.IM]}
  \BibitemShut {NoStop}%
\bibitem [{\citenamefont {Arzoumanian}\ \emph {et~al.}(2020)\citenamefont
  {Arzoumanian} \emph {et~al.}}]{NANOGrav:2020bcs}%
  \BibitemOpen
  \bibfield  {author} {\bibinfo {author} {\bibfnamefont {Z.}~\bibnamefont
  {Arzoumanian}} \emph {et~al.} (\bibinfo {collaboration} {NANOGrav}),\ }\href
  {\doibase 10.3847/2041-8213/abd401} {\bibfield  {journal} {\bibinfo
  {journal} {Astrophys. J. Lett.}\ }\textbf {\bibinfo {volume} {905}},\
  \bibinfo {pages} {L34} (\bibinfo {year} {2020})},\ \Eprint
  {http://arxiv.org/abs/2009.04496} {arXiv:2009.04496 [astro-ph.HE]}
  \BibitemShut {NoStop}%
\bibitem [{\citenamefont {Goncharov}\ \emph {et~al.}(2021)\citenamefont
  {Goncharov} \emph {et~al.}}]{Goncharov:2021oub}%
  \BibitemOpen
  \bibfield  {author} {\bibinfo {author} {\bibfnamefont {B.}~\bibnamefont
  {Goncharov}} \emph {et~al.},\ }\href {\doibase 10.3847/2041-8213/ac17f4}
  {\bibfield  {journal} {\bibinfo  {journal} {Astrophys. J. Lett.}\ }\textbf
  {\bibinfo {volume} {917}},\ \bibinfo {pages} {L19} (\bibinfo {year}
  {2021})},\ \Eprint {http://arxiv.org/abs/2107.12112} {arXiv:2107.12112
  [astro-ph.HE]} \BibitemShut {NoStop}%
\bibitem [{\citenamefont {Chen}\ \emph
  {et~al.}(2021{\natexlab{a}})\citenamefont {Chen} \emph
  {et~al.}}]{Chen:2021rqp}%
  \BibitemOpen
  \bibfield  {author} {\bibinfo {author} {\bibfnamefont {S.}~\bibnamefont
  {Chen}} \emph {et~al.},\ }\href {\doibase 10.1093/mnras/stab2833} {\bibfield
  {journal} {\bibinfo  {journal} {Mon. Not. Roy. Astron. Soc.}\ }\textbf
  {\bibinfo {volume} {508}},\ \bibinfo {pages} {4970} (\bibinfo {year}
  {2021}{\natexlab{a}})},\ \Eprint {http://arxiv.org/abs/2110.13184}
  {arXiv:2110.13184 [astro-ph.HE]} \BibitemShut {NoStop}%
\bibitem [{\citenamefont {Antoniadis}\ \emph {et~al.}(2022)\citenamefont
  {Antoniadis} \emph {et~al.}}]{Antoniadis:2022pcn}%
  \BibitemOpen
  \bibfield  {author} {\bibinfo {author} {\bibfnamefont {J.}~\bibnamefont
  {Antoniadis}} \emph {et~al.},\ }\href {\doibase 10.1093/mnras/stab3418}
  {\bibfield  {journal} {\bibinfo  {journal} {Mon. Not. Roy. Astron. Soc.}\
  }\textbf {\bibinfo {volume} {510}},\ \bibinfo {pages} {4873} (\bibinfo {year}
  {2022})},\ \Eprint {http://arxiv.org/abs/2201.03980} {arXiv:2201.03980
  [astro-ph.HE]} \BibitemShut {NoStop}%
\bibitem [{\citenamefont {Pol}\ \emph {et~al.}(2021)\citenamefont {Pol} \emph
  {et~al.}}]{NANOGrav:2020spf}%
  \BibitemOpen
  \bibfield  {author} {\bibinfo {author} {\bibfnamefont {N.~S.}\ \bibnamefont
  {Pol}} \emph {et~al.} (\bibinfo {collaboration} {NANOGrav}),\ }\href
  {\doibase 10.3847/2041-8213/abf2c9} {\bibfield  {journal} {\bibinfo
  {journal} {Astrophys. J. Lett.}\ }\textbf {\bibinfo {volume} {911}},\
  \bibinfo {pages} {L34} (\bibinfo {year} {2021})},\ \Eprint
  {http://arxiv.org/abs/2010.11950} {arXiv:2010.11950 [astro-ph.HE]}
  \BibitemShut {NoStop}%
\bibitem [{\citenamefont {Romano}\ \emph {et~al.}(2021)\citenamefont {Romano},
  \citenamefont {Hazboun}, \citenamefont {Siemens},\ and\ \citenamefont
  {Archibald}}]{Romano:2020sxq}%
  \BibitemOpen
  \bibfield  {author} {\bibinfo {author} {\bibfnamefont {J.~D.}\ \bibnamefont
  {Romano}}, \bibinfo {author} {\bibfnamefont {J.~S.}\ \bibnamefont {Hazboun}},
  \bibinfo {author} {\bibfnamefont {X.}~\bibnamefont {Siemens}}, \ and\
  \bibinfo {author} {\bibfnamefont {A.~M.}\ \bibnamefont {Archibald}},\ }\href
  {\doibase 10.1103/PhysRevD.103.063027} {\bibfield  {journal} {\bibinfo
  {journal} {Phys. Rev. D}\ }\textbf {\bibinfo {volume} {103}},\ \bibinfo
  {pages} {063027} (\bibinfo {year} {2021})},\ \Eprint
  {http://arxiv.org/abs/2012.03804} {arXiv:2012.03804 [gr-qc]} \BibitemShut
  {NoStop}%
\bibitem [{\citenamefont {Hellings}\ and\ \citenamefont
  {Downs}(1983)}]{Hellings:1983fr}%
  \BibitemOpen
  \bibfield  {author} {\bibinfo {author} {\bibfnamefont {R.~w.}\ \bibnamefont
  {Hellings}}\ and\ \bibinfo {author} {\bibfnamefont {G.~s.}\ \bibnamefont
  {Downs}},\ }\href {\doibase 10.1086/183954} {\bibfield  {journal} {\bibinfo
  {journal} {Astrophys. J. Lett.}\ }\textbf {\bibinfo {volume} {265}},\
  \bibinfo {pages} {L39} (\bibinfo {year} {1983})}\BibitemShut {NoStop}%
\bibitem [{\citenamefont {Goncharov}\ \emph {et~al.}(2022)\citenamefont
  {Goncharov} \emph {et~al.}}]{Goncharov:2022ktc}%
  \BibitemOpen
  \bibfield  {author} {\bibinfo {author} {\bibfnamefont {B.}~\bibnamefont
  {Goncharov}} \emph {et~al.},\ }\href {\doibase 10.3847/2041-8213/ac76bb}
  {\bibfield  {journal} {\bibinfo  {journal} {Astrophys. J.}\ }\textbf
  {\bibinfo {volume} {932}},\ \bibinfo {pages} {L22} (\bibinfo {year}
  {2022})},\ \Eprint {http://arxiv.org/abs/2206.03766} {arXiv:2206.03766
  [gr-qc]} \BibitemShut {NoStop}%
\bibitem [{\citenamefont {Zic}\ \emph {et~al.}(2022)\citenamefont {Zic} \emph
  {et~al.}}]{Zic:2022sxd}%
  \BibitemOpen
  \bibfield  {author} {\bibinfo {author} {\bibfnamefont {A.}~\bibnamefont
  {Zic}} \emph {et~al.},\ }\href {\doibase 10.1093/mnras/stac2100} {\bibfield
  {journal} {\bibinfo  {journal} {Mon. Not. Roy. Astron. Soc.}\ }\textbf
  {\bibinfo {volume} {516}},\ \bibinfo {pages} {410} (\bibinfo {year}
  {2022})},\ \Eprint {http://arxiv.org/abs/2207.12237} {arXiv:2207.12237
  [astro-ph.HE]} \BibitemShut {NoStop}%
\bibitem [{\citenamefont {Tiburzi}\ \emph {et~al.}(2016)\citenamefont
  {Tiburzi}, \citenamefont {Hobbs}, \citenamefont {Kerr}, \citenamefont
  {Coles}, \citenamefont {Dai}, \citenamefont {Manchester}, \citenamefont
  {Possenti}, \citenamefont {Shannon},\ and\ \citenamefont
  {You}}]{Tiburzi:2015kqa}%
  \BibitemOpen
  \bibfield  {author} {\bibinfo {author} {\bibfnamefont {C.}~\bibnamefont
  {Tiburzi}}, \bibinfo {author} {\bibfnamefont {G.}~\bibnamefont {Hobbs}},
  \bibinfo {author} {\bibfnamefont {M.}~\bibnamefont {Kerr}}, \bibinfo {author}
  {\bibfnamefont {W.}~\bibnamefont {Coles}}, \bibinfo {author} {\bibfnamefont
  {S.}~\bibnamefont {Dai}}, \bibinfo {author} {\bibfnamefont {R.}~\bibnamefont
  {Manchester}}, \bibinfo {author} {\bibfnamefont {A.}~\bibnamefont
  {Possenti}}, \bibinfo {author} {\bibfnamefont {R.}~\bibnamefont {Shannon}}, \
  and\ \bibinfo {author} {\bibfnamefont {X.}~\bibnamefont {You}},\ }\href
  {\doibase 10.1093/mnras/stv2143} {\bibfield  {journal} {\bibinfo  {journal}
  {Mon. Not. Roy. Astron. Soc.}\ }\textbf {\bibinfo {volume} {455}},\ \bibinfo
  {pages} {4339} (\bibinfo {year} {2016})},\ \Eprint
  {http://arxiv.org/abs/1510.02363} {arXiv:1510.02363 [astro-ph.IM]}
  \BibitemShut {NoStop}%
\bibitem [{\citenamefont {Agazie}\ \emph
  {et~al.}(2023{\natexlab{a}})\citenamefont {Agazie} \emph
  {et~al.}}]{NANOGrav:2023gor}%
  \BibitemOpen
  \bibfield  {author} {\bibinfo {author} {\bibfnamefont {G.}~\bibnamefont
  {Agazie}} \emph {et~al.} (\bibinfo {collaboration} {NANOGrav}),\ }\href
  {\doibase 10.3847/2041-8213/acdac6} {\bibfield  {journal} {\bibinfo
  {journal} {Astrophys. J. Lett.}\ }\textbf {\bibinfo {volume} {951}},\
  \bibinfo {pages} {L8} (\bibinfo {year} {2023}{\natexlab{a}})},\ \Eprint
  {http://arxiv.org/abs/2306.16213} {arXiv:2306.16213 [astro-ph.HE]}
  \BibitemShut {NoStop}%
\bibitem [{\citenamefont {Antoniadis}\ \emph
  {et~al.}(2023{\natexlab{a}})\citenamefont {Antoniadis} \emph
  {et~al.}}]{Antoniadis:2023ott}%
  \BibitemOpen
  \bibfield  {author} {\bibinfo {author} {\bibfnamefont {J.}~\bibnamefont
  {Antoniadis}} \emph {et~al.},\ }\href@noop {} {\  (\bibinfo {year}
  {2023}{\natexlab{a}})},\ \Eprint {http://arxiv.org/abs/2306.16214}
  {arXiv:2306.16214 [astro-ph.HE]} \BibitemShut {NoStop}%
\bibitem [{\citenamefont {Reardon}\ \emph
  {et~al.}(2023{\natexlab{a}})\citenamefont {Reardon} \emph
  {et~al.}}]{Reardon:2023gzh}%
  \BibitemOpen
  \bibfield  {author} {\bibinfo {author} {\bibfnamefont {D.~J.}\ \bibnamefont
  {Reardon}} \emph {et~al.},\ }\href {\doibase 10.3847/2041-8213/acdd02}
  {\bibfield  {journal} {\bibinfo  {journal} {Astrophys. J. Lett.}\ }\textbf
  {\bibinfo {volume} {951}},\ \bibinfo {pages} {L6} (\bibinfo {year}
  {2023}{\natexlab{a}})},\ \Eprint {http://arxiv.org/abs/2306.16215}
  {arXiv:2306.16215 [astro-ph.HE]} \BibitemShut {NoStop}%
\bibitem [{\citenamefont {Xu}\ \emph {et~al.}(2023)\citenamefont {Xu} \emph
  {et~al.}}]{Xu:2023wog}%
  \BibitemOpen
  \bibfield  {author} {\bibinfo {author} {\bibfnamefont {H.}~\bibnamefont {Xu}}
  \emph {et~al.},\ }\href {\doibase 10.1088/1674-4527/acdfa5} {\bibfield
  {journal} {\bibinfo  {journal} {Res. Astron. Astrophys.}\ }\textbf {\bibinfo
  {volume} {23}},\ \bibinfo {pages} {075026} (\bibinfo {year} {2023})},\
  \Eprint {http://arxiv.org/abs/2306.16216} {arXiv:2306.16216 [astro-ph.HE]}
  \BibitemShut {NoStop}%
\bibitem [{\citenamefont {Agazie}\ \emph
  {et~al.}(2023{\natexlab{b}})\citenamefont {Agazie} \emph
  {et~al.}}]{NANOGrav:2023hde}%
  \BibitemOpen
  \bibfield  {author} {\bibinfo {author} {\bibfnamefont {G.}~\bibnamefont
  {Agazie}} \emph {et~al.} (\bibinfo {collaboration} {NANOGrav}),\ }\href
  {\doibase 10.3847/2041-8213/acda9a} {\bibfield  {journal} {\bibinfo
  {journal} {Astrophys. J. Lett.}\ }\textbf {\bibinfo {volume} {951}},\
  \bibinfo {pages} {L9} (\bibinfo {year} {2023}{\natexlab{b}})},\ \Eprint
  {http://arxiv.org/abs/2306.16217} {arXiv:2306.16217 [astro-ph.HE]}
  \BibitemShut {NoStop}%
\bibitem [{\citenamefont {Agazie}\ \emph
  {et~al.}(2023{\natexlab{c}})\citenamefont {Agazie} \emph
  {et~al.}}]{NANOGrav:2023ctt}%
  \BibitemOpen
  \bibfield  {author} {\bibinfo {author} {\bibfnamefont {G.}~\bibnamefont
  {Agazie}} \emph {et~al.} (\bibinfo {collaboration} {NANOGrav}),\ }\href
  {\doibase 10.3847/2041-8213/acda88} {\bibfield  {journal} {\bibinfo
  {journal} {Astrophys. J. Lett.}\ }\textbf {\bibinfo {volume} {951}},\
  \bibinfo {pages} {L10} (\bibinfo {year} {2023}{\natexlab{c}})},\ \Eprint
  {http://arxiv.org/abs/2306.16218} {arXiv:2306.16218 [astro-ph.HE]}
  \BibitemShut {NoStop}%
\bibitem [{\citenamefont {Afzal}\ \emph {et~al.}(2023)\citenamefont {Afzal}
  \emph {et~al.}}]{NANOGrav:2023hvm}%
  \BibitemOpen
  \bibfield  {author} {\bibinfo {author} {\bibfnamefont {A.}~\bibnamefont
  {Afzal}} \emph {et~al.} (\bibinfo {collaboration} {NANOGrav}),\ }\href
  {\doibase 10.3847/2041-8213/acdc91} {\bibfield  {journal} {\bibinfo
  {journal} {Astrophys. J. Lett.}\ }\textbf {\bibinfo {volume} {951}},\
  \bibinfo {pages} {L11} (\bibinfo {year} {2023})},\ \Eprint
  {http://arxiv.org/abs/2306.16219} {arXiv:2306.16219 [astro-ph.HE]}
  \BibitemShut {NoStop}%
\bibitem [{\citenamefont {Agazie}\ \emph
  {et~al.}(2023{\natexlab{d}})\citenamefont {Agazie} \emph
  {et~al.}}]{NANOGrav:2023hfp}%
  \BibitemOpen
  \bibfield  {author} {\bibinfo {author} {\bibfnamefont {G.}~\bibnamefont
  {Agazie}} \emph {et~al.} (\bibinfo {collaboration} {NANOGrav}),\ }\href@noop
  {} {\  (\bibinfo {year} {2023}{\natexlab{d}})},\ \Eprint
  {http://arxiv.org/abs/2306.16220} {arXiv:2306.16220 [astro-ph.HE]}
  \BibitemShut {NoStop}%
\bibitem [{\citenamefont {Agazie}\ \emph
  {et~al.}(2023{\natexlab{e}})\citenamefont {Agazie} \emph
  {et~al.}}]{NANOGrav:2023ghw}%
  \BibitemOpen
  \bibfield  {author} {\bibinfo {author} {\bibfnamefont {G.}~\bibnamefont
  {Agazie}} \emph {et~al.} (\bibinfo {collaboration} {NANOGrav}),\ }\href@noop
  {} {\  (\bibinfo {year} {2023}{\natexlab{e}})},\ \Eprint
  {http://arxiv.org/abs/2306.16221} {arXiv:2306.16221 [astro-ph.HE]}
  \BibitemShut {NoStop}%
\bibitem [{\citenamefont {Agazie}\ \emph
  {et~al.}(2023{\natexlab{f}})\citenamefont {Agazie} \emph
  {et~al.}}]{NANOGrav:2023pdq}%
  \BibitemOpen
  \bibfield  {author} {\bibinfo {author} {\bibfnamefont {G.}~\bibnamefont
  {Agazie}} \emph {et~al.} (\bibinfo {collaboration} {NANOGrav}),\ }\href@noop
  {} {\  (\bibinfo {year} {2023}{\natexlab{f}})},\ \Eprint
  {http://arxiv.org/abs/2306.16222} {arXiv:2306.16222 [astro-ph.HE]}
  \BibitemShut {NoStop}%
\bibitem [{\citenamefont {Johnson}\ \emph {et~al.}(2023)\citenamefont {Johnson}
  \emph {et~al.}}]{NANOGrav:2023icp}%
  \BibitemOpen
  \bibfield  {author} {\bibinfo {author} {\bibfnamefont {A.~D.}\ \bibnamefont
  {Johnson}} \emph {et~al.} (\bibinfo {collaboration} {NANOGrav}),\ }\href@noop
  {} {\  (\bibinfo {year} {2023})},\ \Eprint {http://arxiv.org/abs/2306.16223}
  {arXiv:2306.16223 [astro-ph.HE]} \BibitemShut {NoStop}%
\bibitem [{\citenamefont {Antoniadis}\ \emph
  {et~al.}(2023{\natexlab{b}})\citenamefont {Antoniadis} \emph
  {et~al.}}]{Antoniadis:2023lym}%
  \BibitemOpen
  \bibfield  {author} {\bibinfo {author} {\bibfnamefont {J.}~\bibnamefont
  {Antoniadis}} \emph {et~al.},\ }\href {\doibase 10.1051/0004-6361/202346841}
  {\  (\bibinfo {year} {2023}{\natexlab{b}}),\ 10.1051/0004-6361/202346841},\
  \Eprint {http://arxiv.org/abs/2306.16224} {arXiv:2306.16224 [astro-ph.HE]}
  \BibitemShut {NoStop}%
\bibitem [{\citenamefont {Antoniadis}\ \emph
  {et~al.}(2023{\natexlab{c}})\citenamefont {Antoniadis} \emph
  {et~al.}}]{Antoniadis:2023puu}%
  \BibitemOpen
  \bibfield  {author} {\bibinfo {author} {\bibfnamefont {J.}~\bibnamefont
  {Antoniadis}} \emph {et~al.},\ }\href@noop {} {\  (\bibinfo {year}
  {2023}{\natexlab{c}})},\ \Eprint {http://arxiv.org/abs/2306.16225}
  {arXiv:2306.16225 [astro-ph.HE]} \BibitemShut {NoStop}%
\bibitem [{\citenamefont {Antoniadis}\ \emph
  {et~al.}(2023{\natexlab{d}})\citenamefont {Antoniadis} \emph
  {et~al.}}]{Antoniadis:2023aac}%
  \BibitemOpen
  \bibfield  {author} {\bibinfo {author} {\bibfnamefont {J.}~\bibnamefont
  {Antoniadis}} \emph {et~al.},\ }\href@noop {} {\  (\bibinfo {year}
  {2023}{\natexlab{d}})},\ \Eprint {http://arxiv.org/abs/2306.16226}
  {arXiv:2306.16226 [astro-ph.HE]} \BibitemShut {NoStop}%
\bibitem [{\citenamefont {Antoniadis}\ \emph
  {et~al.}(2023{\natexlab{e}})\citenamefont {Antoniadis} \emph
  {et~al.}}]{Antoniadis:2023xlr}%
  \BibitemOpen
  \bibfield  {author} {\bibinfo {author} {\bibfnamefont {J.}~\bibnamefont
  {Antoniadis}} \emph {et~al.},\ }\href@noop {} {\  (\bibinfo {year}
  {2023}{\natexlab{e}})},\ \Eprint {http://arxiv.org/abs/2306.16227}
  {arXiv:2306.16227 [astro-ph.CO]} \BibitemShut {NoStop}%
\bibitem [{\citenamefont {Smarra}\ \emph {et~al.}(2023)\citenamefont {Smarra}
  \emph {et~al.}}]{Smarra:2023ljf}%
  \BibitemOpen
  \bibfield  {author} {\bibinfo {author} {\bibfnamefont {C.}~\bibnamefont
  {Smarra}} \emph {et~al.},\ }\href@noop {} {\  (\bibinfo {year} {2023})},\
  \Eprint {http://arxiv.org/abs/2306.16228} {arXiv:2306.16228 [astro-ph.HE]}
  \BibitemShut {NoStop}%
\bibitem [{\citenamefont {Reardon}\ \emph
  {et~al.}(2023{\natexlab{b}})\citenamefont {Reardon} \emph
  {et~al.}}]{Reardon:2023zen}%
  \BibitemOpen
  \bibfield  {author} {\bibinfo {author} {\bibfnamefont {D.~J.}\ \bibnamefont
  {Reardon}} \emph {et~al.},\ }\href {\doibase 10.3847/2041-8213/acdd03}
  {\bibfield  {journal} {\bibinfo  {journal} {Astrophys. J. Lett.}\ }\textbf
  {\bibinfo {volume} {951}},\ \bibinfo {pages} {L7} (\bibinfo {year}
  {2023}{\natexlab{b}})},\ \Eprint {http://arxiv.org/abs/2306.16229}
  {arXiv:2306.16229 [astro-ph.HE]} \BibitemShut {NoStop}%
\bibitem [{\citenamefont {Zic}\ \emph {et~al.}(2023)\citenamefont {Zic} \emph
  {et~al.}}]{Zic:2023gta}%
  \BibitemOpen
  \bibfield  {author} {\bibinfo {author} {\bibfnamefont {A.}~\bibnamefont
  {Zic}} \emph {et~al.},\ }\href@noop {} {\  (\bibinfo {year} {2023})},\
  \Eprint {http://arxiv.org/abs/2306.16230} {arXiv:2306.16230 [astro-ph.HE]}
  \BibitemShut {NoStop}%
\bibitem [{\citenamefont {Kazanas}(1980)}]{Kazanas:1980tx}%
  \BibitemOpen
  \bibfield  {author} {\bibinfo {author} {\bibfnamefont {D.}~\bibnamefont
  {Kazanas}},\ }\href {\doibase 10.1086/183361} {\bibfield  {journal} {\bibinfo
   {journal} {Astrophys. J. Lett.}\ }\textbf {\bibinfo {volume} {241}},\
  \bibinfo {pages} {L59} (\bibinfo {year} {1980})}\BibitemShut {NoStop}%
\bibitem [{\citenamefont {Starobinsky}(1980)}]{Starobinsky:1980te}%
  \BibitemOpen
  \bibfield  {author} {\bibinfo {author} {\bibfnamefont {A.~A.}\ \bibnamefont
  {Starobinsky}},\ }\href {\doibase 10.1016/0370-2693(80)90670-X} {\bibfield
  {journal} {\bibinfo  {journal} {Phys. Lett. B}\ }\textbf {\bibinfo {volume}
  {91}},\ \bibinfo {pages} {99} (\bibinfo {year} {1980})}\BibitemShut {NoStop}%
\bibitem [{\citenamefont {Sato}(1981)}]{Sato:1981ds}%
  \BibitemOpen
  \bibfield  {author} {\bibinfo {author} {\bibfnamefont {K.}~\bibnamefont
  {Sato}},\ }\href {\doibase 10.1016/0370-2693(81)90805-4} {\bibfield
  {journal} {\bibinfo  {journal} {Phys. Lett. B}\ }\textbf {\bibinfo {volume}
  {99}},\ \bibinfo {pages} {66} (\bibinfo {year} {1981})}\BibitemShut {NoStop}%
\bibitem [{\citenamefont {Guth}(1981)}]{Guth:1980zm}%
  \BibitemOpen
  \bibfield  {author} {\bibinfo {author} {\bibfnamefont {A.~H.}\ \bibnamefont
  {Guth}},\ }\href {\doibase 10.1103/PhysRevD.23.347} {\bibfield  {journal}
  {\bibinfo  {journal} {Phys. Rev. D}\ }\textbf {\bibinfo {volume} {23}},\
  \bibinfo {pages} {347} (\bibinfo {year} {1981})}\BibitemShut {NoStop}%
\bibitem [{\citenamefont {Mukhanov}\ and\ \citenamefont
  {Chibisov}(1981)}]{Mukhanov:1981xt}%
  \BibitemOpen
  \bibfield  {author} {\bibinfo {author} {\bibfnamefont {V.~F.}\ \bibnamefont
  {Mukhanov}}\ and\ \bibinfo {author} {\bibfnamefont {G.~V.}\ \bibnamefont
  {Chibisov}},\ }\href@noop {} {\bibfield  {journal} {\bibinfo  {journal} {JETP
  Lett.}\ }\textbf {\bibinfo {volume} {33}},\ \bibinfo {pages} {532} (\bibinfo
  {year} {1981})}\BibitemShut {NoStop}%
\bibitem [{\citenamefont {Linde}(1982)}]{Linde:1981mu}%
  \BibitemOpen
  \bibfield  {author} {\bibinfo {author} {\bibfnamefont {A.~D.}\ \bibnamefont
  {Linde}},\ }\href {\doibase 10.1016/0370-2693(82)91219-9} {\bibfield
  {journal} {\bibinfo  {journal} {Phys. Lett. B}\ }\textbf {\bibinfo {volume}
  {108}},\ \bibinfo {pages} {389} (\bibinfo {year} {1982})}\BibitemShut
  {NoStop}%
\bibitem [{\citenamefont {Albrecht}\ and\ \citenamefont
  {Steinhardt}(1982)}]{Albrecht:1982wi}%
  \BibitemOpen
  \bibfield  {author} {\bibinfo {author} {\bibfnamefont {A.}~\bibnamefont
  {Albrecht}}\ and\ \bibinfo {author} {\bibfnamefont {P.~J.}\ \bibnamefont
  {Steinhardt}},\ }\href {\doibase 10.1103/PhysRevLett.48.1220} {\bibfield
  {journal} {\bibinfo  {journal} {Phys. Rev. Lett.}\ }\textbf {\bibinfo
  {volume} {48}},\ \bibinfo {pages} {1220} (\bibinfo {year}
  {1982})}\BibitemShut {NoStop}%
\bibitem [{\citenamefont {Ijjas}\ \emph {et~al.}(2013)\citenamefont {Ijjas},
  \citenamefont {Steinhardt},\ and\ \citenamefont {Loeb}}]{Ijjas:2013vea}%
  \BibitemOpen
  \bibfield  {author} {\bibinfo {author} {\bibfnamefont {A.}~\bibnamefont
  {Ijjas}}, \bibinfo {author} {\bibfnamefont {P.~J.}\ \bibnamefont
  {Steinhardt}}, \ and\ \bibinfo {author} {\bibfnamefont {A.}~\bibnamefont
  {Loeb}},\ }\href {\doibase 10.1016/j.physletb.2013.05.023} {\bibfield
  {journal} {\bibinfo  {journal} {Phys. Lett. B}\ }\textbf {\bibinfo {volume}
  {723}},\ \bibinfo {pages} {261} (\bibinfo {year} {2013})},\ \Eprint
  {http://arxiv.org/abs/1304.2785} {arXiv:1304.2785 [astro-ph.CO]} \BibitemShut
  {NoStop}%
\bibitem [{\citenamefont {Ijjas}\ \emph {et~al.}(2014)\citenamefont {Ijjas},
  \citenamefont {Steinhardt},\ and\ \citenamefont {Loeb}}]{Ijjas:2014nta}%
  \BibitemOpen
  \bibfield  {author} {\bibinfo {author} {\bibfnamefont {A.}~\bibnamefont
  {Ijjas}}, \bibinfo {author} {\bibfnamefont {P.~J.}\ \bibnamefont
  {Steinhardt}}, \ and\ \bibinfo {author} {\bibfnamefont {A.}~\bibnamefont
  {Loeb}},\ }\href {\doibase 10.1016/j.physletb.2014.07.012} {\bibfield
  {journal} {\bibinfo  {journal} {Phys. Lett. B}\ }\textbf {\bibinfo {volume}
  {736}},\ \bibinfo {pages} {142} (\bibinfo {year} {2014})},\ \Eprint
  {http://arxiv.org/abs/1402.6980} {arXiv:1402.6980 [astro-ph.CO]} \BibitemShut
  {NoStop}%
\bibitem [{\citenamefont {Obied}\ \emph {et~al.}(2018)\citenamefont {Obied},
  \citenamefont {Ooguri}, \citenamefont {Spodyneiko},\ and\ \citenamefont
  {Vafa}}]{Obied:2018sgi}%
  \BibitemOpen
  \bibfield  {author} {\bibinfo {author} {\bibfnamefont {G.}~\bibnamefont
  {Obied}}, \bibinfo {author} {\bibfnamefont {H.}~\bibnamefont {Ooguri}},
  \bibinfo {author} {\bibfnamefont {L.}~\bibnamefont {Spodyneiko}}, \ and\
  \bibinfo {author} {\bibfnamefont {C.}~\bibnamefont {Vafa}},\ }\href@noop {}
  {\  (\bibinfo {year} {2018})},\ \Eprint {http://arxiv.org/abs/1806.08362}
  {arXiv:1806.08362 [hep-th]} \BibitemShut {NoStop}%
\bibitem [{\citenamefont {Agrawal}\ \emph {et~al.}(2018)\citenamefont
  {Agrawal}, \citenamefont {Obied}, \citenamefont {Steinhardt},\ and\
  \citenamefont {Vafa}}]{Agrawal:2018own}%
  \BibitemOpen
  \bibfield  {author} {\bibinfo {author} {\bibfnamefont {P.}~\bibnamefont
  {Agrawal}}, \bibinfo {author} {\bibfnamefont {G.}~\bibnamefont {Obied}},
  \bibinfo {author} {\bibfnamefont {P.~J.}\ \bibnamefont {Steinhardt}}, \ and\
  \bibinfo {author} {\bibfnamefont {C.}~\bibnamefont {Vafa}},\ }\href {\doibase
  10.1016/j.physletb.2018.07.040} {\bibfield  {journal} {\bibinfo  {journal}
  {Phys. Lett. B}\ }\textbf {\bibinfo {volume} {784}},\ \bibinfo {pages} {271}
  (\bibinfo {year} {2018})},\ \Eprint {http://arxiv.org/abs/1806.09718}
  {arXiv:1806.09718 [hep-th]} \BibitemShut {NoStop}%
\bibitem [{\citenamefont {Ach\'ucarro}\ and\ \citenamefont
  {Palma}(2019)}]{Achucarro:2018vey}%
  \BibitemOpen
  \bibfield  {author} {\bibinfo {author} {\bibfnamefont {A.}~\bibnamefont
  {Ach\'ucarro}}\ and\ \bibinfo {author} {\bibfnamefont {G.~A.}\ \bibnamefont
  {Palma}},\ }\href {\doibase 10.1088/1475-7516/2019/02/041} {\bibfield
  {journal} {\bibinfo  {journal} {JCAP}\ }\textbf {\bibinfo {volume} {02}},\
  \bibinfo {pages} {041} (\bibinfo {year} {2019})},\ \Eprint
  {http://arxiv.org/abs/1807.04390} {arXiv:1807.04390 [hep-th]} \BibitemShut
  {NoStop}%
\bibitem [{\citenamefont {Garg}\ and\ \citenamefont
  {Krishnan}(2019)}]{Garg:2018reu}%
  \BibitemOpen
  \bibfield  {author} {\bibinfo {author} {\bibfnamefont {S.~K.}\ \bibnamefont
  {Garg}}\ and\ \bibinfo {author} {\bibfnamefont {C.}~\bibnamefont
  {Krishnan}},\ }\href {\doibase 10.1007/JHEP11(2019)075} {\bibfield  {journal}
  {\bibinfo  {journal} {JHEP}\ }\textbf {\bibinfo {volume} {11}},\ \bibinfo
  {pages} {075} (\bibinfo {year} {2019})},\ \Eprint
  {http://arxiv.org/abs/1807.05193} {arXiv:1807.05193 [hep-th]} \BibitemShut
  {NoStop}%
\bibitem [{\citenamefont {Kehagias}\ and\ \citenamefont
  {Riotto}(2018)}]{Kehagias:2018uem}%
  \BibitemOpen
  \bibfield  {author} {\bibinfo {author} {\bibfnamefont {A.}~\bibnamefont
  {Kehagias}}\ and\ \bibinfo {author} {\bibfnamefont {A.}~\bibnamefont
  {Riotto}},\ }\href {\doibase 10.1002/prop.201800052} {\bibfield  {journal}
  {\bibinfo  {journal} {Fortsch. Phys.}\ }\textbf {\bibinfo {volume} {66}},\
  \bibinfo {pages} {1800052} (\bibinfo {year} {2018})},\ \Eprint
  {http://arxiv.org/abs/1807.05445} {arXiv:1807.05445 [hep-th]} \BibitemShut
  {NoStop}%
\bibitem [{\citenamefont {Kinney}\ \emph {et~al.}(2019)\citenamefont {Kinney},
  \citenamefont {Vagnozzi},\ and\ \citenamefont {Visinelli}}]{Kinney:2018nny}%
  \BibitemOpen
  \bibfield  {author} {\bibinfo {author} {\bibfnamefont {W.~H.}\ \bibnamefont
  {Kinney}}, \bibinfo {author} {\bibfnamefont {S.}~\bibnamefont {Vagnozzi}}, \
  and\ \bibinfo {author} {\bibfnamefont {L.}~\bibnamefont {Visinelli}},\ }\href
  {\doibase 10.1088/1361-6382/ab1d87} {\bibfield  {journal} {\bibinfo
  {journal} {Class. Quant. Grav.}\ }\textbf {\bibinfo {volume} {36}},\ \bibinfo
  {pages} {117001} (\bibinfo {year} {2019})},\ \Eprint
  {http://arxiv.org/abs/1808.06424} {arXiv:1808.06424 [astro-ph.CO]}
  \BibitemShut {NoStop}%
\bibitem [{\citenamefont {Ooguri}\ \emph {et~al.}(2019)\citenamefont {Ooguri},
  \citenamefont {Palti}, \citenamefont {Shiu},\ and\ \citenamefont
  {Vafa}}]{Ooguri:2018wrx}%
  \BibitemOpen
  \bibfield  {author} {\bibinfo {author} {\bibfnamefont {H.}~\bibnamefont
  {Ooguri}}, \bibinfo {author} {\bibfnamefont {E.}~\bibnamefont {Palti}},
  \bibinfo {author} {\bibfnamefont {G.}~\bibnamefont {Shiu}}, \ and\ \bibinfo
  {author} {\bibfnamefont {C.}~\bibnamefont {Vafa}},\ }\href {\doibase
  10.1016/j.physletb.2018.11.018} {\bibfield  {journal} {\bibinfo  {journal}
  {Phys. Lett. B}\ }\textbf {\bibinfo {volume} {788}},\ \bibinfo {pages} {180}
  (\bibinfo {year} {2019})},\ \Eprint {http://arxiv.org/abs/1810.05506}
  {arXiv:1810.05506 [hep-th]} \BibitemShut {NoStop}%
\bibitem [{\citenamefont {Palti}(2019)}]{Palti:2019pca}%
  \BibitemOpen
  \bibfield  {author} {\bibinfo {author} {\bibfnamefont {E.}~\bibnamefont
  {Palti}},\ }\href {\doibase 10.1002/prop.201900037} {\bibfield  {journal}
  {\bibinfo  {journal} {Fortsch. Phys.}\ }\textbf {\bibinfo {volume} {67}},\
  \bibinfo {pages} {1900037} (\bibinfo {year} {2019})},\ \Eprint
  {http://arxiv.org/abs/1903.06239} {arXiv:1903.06239 [hep-th]} \BibitemShut
  {NoStop}%
\bibitem [{\citenamefont {Bedroya}\ \emph {et~al.}(2020)\citenamefont
  {Bedroya}, \citenamefont {Brandenberger}, \citenamefont {Loverde},\ and\
  \citenamefont {Vafa}}]{Bedroya:2019tba}%
  \BibitemOpen
  \bibfield  {author} {\bibinfo {author} {\bibfnamefont {A.}~\bibnamefont
  {Bedroya}}, \bibinfo {author} {\bibfnamefont {R.}~\bibnamefont
  {Brandenberger}}, \bibinfo {author} {\bibfnamefont {M.}~\bibnamefont
  {Loverde}}, \ and\ \bibinfo {author} {\bibfnamefont {C.}~\bibnamefont
  {Vafa}},\ }\href {\doibase 10.1103/PhysRevD.101.103502} {\bibfield  {journal}
  {\bibinfo  {journal} {Phys. Rev. D}\ }\textbf {\bibinfo {volume} {101}},\
  \bibinfo {pages} {103502} (\bibinfo {year} {2020})},\ \Eprint
  {http://arxiv.org/abs/1909.11106} {arXiv:1909.11106 [hep-th]} \BibitemShut
  {NoStop}%
\bibitem [{\citenamefont {Geng}(2020)}]{Geng:2019phi}%
  \BibitemOpen
  \bibfield  {author} {\bibinfo {author} {\bibfnamefont {H.}~\bibnamefont
  {Geng}},\ }\href {\doibase 10.1016/j.physletb.2020.135430} {\bibfield
  {journal} {\bibinfo  {journal} {Phys. Lett. B}\ }\textbf {\bibinfo {volume}
  {805}},\ \bibinfo {pages} {135430} (\bibinfo {year} {2020})},\ \Eprint
  {http://arxiv.org/abs/1910.14047} {arXiv:1910.14047 [hep-th]} \BibitemShut
  {NoStop}%
\bibitem [{\citenamefont {Trivedi}(2023)}]{Trivedi:2020wxf}%
  \BibitemOpen
  \bibfield  {author} {\bibinfo {author} {\bibfnamefont {O.}~\bibnamefont
  {Trivedi}},\ }\href {\doibase 10.1142/S0218271822501309} {\bibfield
  {journal} {\bibinfo  {journal} {Int. J. Mod. Phys. D}\ }\textbf {\bibinfo
  {volume} {32}},\ \bibinfo {pages} {2250130} (\bibinfo {year} {2023})},\
  \Eprint {http://arxiv.org/abs/2008.05474} {arXiv:2008.05474 [hep-th]}
  \BibitemShut {NoStop}%
\bibitem [{\citenamefont {Anchordoqui}\ \emph {et~al.}(2021)\citenamefont
  {Anchordoqui}, \citenamefont {Antoniadis}, \citenamefont {L\"ust},\ and\
  \citenamefont {Soriano}}]{Anchordoqui:2021eox}%
  \BibitemOpen
  \bibfield  {author} {\bibinfo {author} {\bibfnamefont {L.~A.}\ \bibnamefont
  {Anchordoqui}}, \bibinfo {author} {\bibfnamefont {I.}~\bibnamefont
  {Antoniadis}}, \bibinfo {author} {\bibfnamefont {D.}~\bibnamefont {L\"ust}},
  \ and\ \bibinfo {author} {\bibfnamefont {J.~F.}\ \bibnamefont {Soriano}},\
  }\href {\doibase 10.1103/PhysRevD.103.123537} {\bibfield  {journal} {\bibinfo
   {journal} {Phys. Rev. D}\ }\textbf {\bibinfo {volume} {103}},\ \bibinfo
  {pages} {123537} (\bibinfo {year} {2021})},\ \Eprint
  {http://arxiv.org/abs/2103.07982} {arXiv:2103.07982 [hep-th]} \BibitemShut
  {NoStop}%
\bibitem [{\citenamefont {Gashti}\ \emph {et~al.}(2022)\citenamefont {Gashti},
  \citenamefont {Sadeghi},\ and\ \citenamefont {Pourhassan}}]{Gashti:2022hey}%
  \BibitemOpen
  \bibfield  {author} {\bibinfo {author} {\bibfnamefont {S.~N.}\ \bibnamefont
  {Gashti}}, \bibinfo {author} {\bibfnamefont {J.}~\bibnamefont {Sadeghi}}, \
  and\ \bibinfo {author} {\bibfnamefont {B.}~\bibnamefont {Pourhassan}},\
  }\href {\doibase 10.1016/j.astropartphys.2022.102703} {\bibfield  {journal}
  {\bibinfo  {journal} {Astropart. Phys.}\ }\textbf {\bibinfo {volume} {139}},\
  \bibinfo {pages} {102703} (\bibinfo {year} {2022})},\ \Eprint
  {http://arxiv.org/abs/2202.06381} {arXiv:2202.06381 [astro-ph.CO]}
  \BibitemShut {NoStop}%
\bibitem [{\citenamefont {Martin}\ \emph
  {et~al.}(2014{\natexlab{a}})\citenamefont {Martin}, \citenamefont
  {Ringeval},\ and\ \citenamefont {Vennin}}]{Martin:2013tda}%
  \BibitemOpen
  \bibfield  {author} {\bibinfo {author} {\bibfnamefont {J.}~\bibnamefont
  {Martin}}, \bibinfo {author} {\bibfnamefont {C.}~\bibnamefont {Ringeval}}, \
  and\ \bibinfo {author} {\bibfnamefont {V.}~\bibnamefont {Vennin}},\ }\href
  {\doibase 10.1016/j.dark.2014.01.003} {\bibfield  {journal} {\bibinfo
  {journal} {Phys. Dark Univ.}\ }\textbf {\bibinfo {volume} {5-6}},\ \bibinfo
  {pages} {75} (\bibinfo {year} {2014}{\natexlab{a}})},\ \Eprint
  {http://arxiv.org/abs/1303.3787} {arXiv:1303.3787 [astro-ph.CO]} \BibitemShut
  {NoStop}%
\bibitem [{\citenamefont {Benetti}(2013)}]{Benetti:2013cja}%
  \BibitemOpen
  \bibfield  {author} {\bibinfo {author} {\bibfnamefont {M.}~\bibnamefont
  {Benetti}},\ }\href {\doibase 10.1103/PhysRevD.88.087302} {\bibfield
  {journal} {\bibinfo  {journal} {Phys. Rev. D}\ }\textbf {\bibinfo {volume}
  {88}},\ \bibinfo {pages} {087302} (\bibinfo {year} {2013})},\ \Eprint
  {http://arxiv.org/abs/1308.6406} {arXiv:1308.6406 [astro-ph.CO]} \BibitemShut
  {NoStop}%
\bibitem [{\citenamefont {Martin}\ \emph
  {et~al.}(2014{\natexlab{b}})\citenamefont {Martin}, \citenamefont {Ringeval},
  \citenamefont {Trotta},\ and\ \citenamefont {Vennin}}]{Martin:2013nzq}%
  \BibitemOpen
  \bibfield  {author} {\bibinfo {author} {\bibfnamefont {J.}~\bibnamefont
  {Martin}}, \bibinfo {author} {\bibfnamefont {C.}~\bibnamefont {Ringeval}},
  \bibinfo {author} {\bibfnamefont {R.}~\bibnamefont {Trotta}}, \ and\ \bibinfo
  {author} {\bibfnamefont {V.}~\bibnamefont {Vennin}},\ }\href {\doibase
  10.1088/1475-7516/2014/03/039} {\bibfield  {journal} {\bibinfo  {journal}
  {JCAP}\ }\textbf {\bibinfo {volume} {03}},\ \bibinfo {pages} {039} (\bibinfo
  {year} {2014}{\natexlab{b}})},\ \Eprint {http://arxiv.org/abs/1312.3529}
  {arXiv:1312.3529 [astro-ph.CO]} \BibitemShut {NoStop}%
\bibitem [{\citenamefont {Creminelli}\ \emph {et~al.}(2014)\citenamefont
  {Creminelli}, \citenamefont {L\'opez~Nacir}, \citenamefont {Simonovi\'c},
  \citenamefont {Trevisan},\ and\ \citenamefont
  {Zaldarriaga}}]{Creminelli:2014oaa}%
  \BibitemOpen
  \bibfield  {author} {\bibinfo {author} {\bibfnamefont {P.}~\bibnamefont
  {Creminelli}}, \bibinfo {author} {\bibfnamefont {D.}~\bibnamefont
  {L\'opez~Nacir}}, \bibinfo {author} {\bibfnamefont {M.}~\bibnamefont
  {Simonovi\'c}}, \bibinfo {author} {\bibfnamefont {G.}~\bibnamefont
  {Trevisan}}, \ and\ \bibinfo {author} {\bibfnamefont {M.}~\bibnamefont
  {Zaldarriaga}},\ }\href {\doibase 10.1103/PhysRevLett.112.241303} {\bibfield
  {journal} {\bibinfo  {journal} {Phys. Rev. Lett.}\ }\textbf {\bibinfo
  {volume} {112}},\ \bibinfo {pages} {241303} (\bibinfo {year} {2014})},\
  \Eprint {http://arxiv.org/abs/1404.1065} {arXiv:1404.1065 [astro-ph.CO]}
  \BibitemShut {NoStop}%
\bibitem [{\citenamefont {Dai}\ \emph {et~al.}(2014)\citenamefont {Dai},
  \citenamefont {Kamionkowski},\ and\ \citenamefont {Wang}}]{Dai:2014jja}%
  \BibitemOpen
  \bibfield  {author} {\bibinfo {author} {\bibfnamefont {L.}~\bibnamefont
  {Dai}}, \bibinfo {author} {\bibfnamefont {M.}~\bibnamefont {Kamionkowski}}, \
  and\ \bibinfo {author} {\bibfnamefont {J.}~\bibnamefont {Wang}},\ }\href
  {\doibase 10.1103/PhysRevLett.113.041302} {\bibfield  {journal} {\bibinfo
  {journal} {Phys. Rev. Lett.}\ }\textbf {\bibinfo {volume} {113}},\ \bibinfo
  {pages} {041302} (\bibinfo {year} {2014})},\ \Eprint
  {http://arxiv.org/abs/1404.6704} {arXiv:1404.6704 [astro-ph.CO]} \BibitemShut
  {NoStop}%
\bibitem [{\citenamefont {Rinaldi}\ \emph {et~al.}(2014)\citenamefont
  {Rinaldi}, \citenamefont {Cognola}, \citenamefont {Vanzo},\ and\
  \citenamefont {Zerbini}}]{Rinaldi:2014gua}%
  \BibitemOpen
  \bibfield  {author} {\bibinfo {author} {\bibfnamefont {M.}~\bibnamefont
  {Rinaldi}}, \bibinfo {author} {\bibfnamefont {G.}~\bibnamefont {Cognola}},
  \bibinfo {author} {\bibfnamefont {L.}~\bibnamefont {Vanzo}}, \ and\ \bibinfo
  {author} {\bibfnamefont {S.}~\bibnamefont {Zerbini}},\ }\href {\doibase
  10.1088/1475-7516/2014/08/015} {\bibfield  {journal} {\bibinfo  {journal}
  {JCAP}\ }\textbf {\bibinfo {volume} {08}},\ \bibinfo {pages} {015} (\bibinfo
  {year} {2014})},\ \Eprint {http://arxiv.org/abs/1406.1096} {arXiv:1406.1096
  [gr-qc]} \BibitemShut {NoStop}%
\bibitem [{\citenamefont {Rinaldi}\ \emph {et~al.}(2015)\citenamefont
  {Rinaldi}, \citenamefont {Cognola}, \citenamefont {Vanzo},\ and\
  \citenamefont {Zerbini}}]{Rinaldi:2014gha}%
  \BibitemOpen
  \bibfield  {author} {\bibinfo {author} {\bibfnamefont {M.}~\bibnamefont
  {Rinaldi}}, \bibinfo {author} {\bibfnamefont {G.}~\bibnamefont {Cognola}},
  \bibinfo {author} {\bibfnamefont {L.}~\bibnamefont {Vanzo}}, \ and\ \bibinfo
  {author} {\bibfnamefont {S.}~\bibnamefont {Zerbini}},\ }\href {\doibase
  10.1103/PhysRevD.91.123527} {\bibfield  {journal} {\bibinfo  {journal} {Phys.
  Rev. D}\ }\textbf {\bibinfo {volume} {91}},\ \bibinfo {pages} {123527}
  (\bibinfo {year} {2015})},\ \Eprint {http://arxiv.org/abs/1410.0631}
  {arXiv:1410.0631 [gr-qc]} \BibitemShut {NoStop}%
\bibitem [{\citenamefont {Myrzakulov}\ \emph
  {et~al.}(2015{\natexlab{a}})\citenamefont {Myrzakulov}, \citenamefont
  {Sebastiani},\ and\ \citenamefont {Zerbini}}]{Myrzakulov:2015fra}%
  \BibitemOpen
  \bibfield  {author} {\bibinfo {author} {\bibfnamefont {R.}~\bibnamefont
  {Myrzakulov}}, \bibinfo {author} {\bibfnamefont {L.}~\bibnamefont
  {Sebastiani}}, \ and\ \bibinfo {author} {\bibfnamefont {S.}~\bibnamefont
  {Zerbini}},\ }\href {\doibase 10.1140/epjc/s10052-015-3443-4} {\bibfield
  {journal} {\bibinfo  {journal} {Eur. Phys. J. C}\ }\textbf {\bibinfo {volume}
  {75}},\ \bibinfo {pages} {215} (\bibinfo {year} {2015}{\natexlab{a}})},\
  \Eprint {http://arxiv.org/abs/1502.04432} {arXiv:1502.04432 [gr-qc]}
  \BibitemShut {NoStop}%
\bibitem [{\citenamefont {Rinaldi}\ \emph {et~al.}(2016)\citenamefont
  {Rinaldi}, \citenamefont {Vanzo}, \citenamefont {Zerbini},\ and\
  \citenamefont {Venturi}}]{Rinaldi:2015yoa}%
  \BibitemOpen
  \bibfield  {author} {\bibinfo {author} {\bibfnamefont {M.}~\bibnamefont
  {Rinaldi}}, \bibinfo {author} {\bibfnamefont {L.}~\bibnamefont {Vanzo}},
  \bibinfo {author} {\bibfnamefont {S.}~\bibnamefont {Zerbini}}, \ and\
  \bibinfo {author} {\bibfnamefont {G.}~\bibnamefont {Venturi}},\ }\href
  {\doibase 10.1103/PhysRevD.93.024040} {\bibfield  {journal} {\bibinfo
  {journal} {Phys. Rev. D}\ }\textbf {\bibinfo {volume} {93}},\ \bibinfo
  {pages} {024040} (\bibinfo {year} {2016})},\ \Eprint
  {http://arxiv.org/abs/1505.03386} {arXiv:1505.03386 [hep-th]} \BibitemShut
  {NoStop}%
\bibitem [{\citenamefont {Escudero}\ \emph {et~al.}(2016)\citenamefont
  {Escudero}, \citenamefont {Ram\'\i{}rez}, \citenamefont {Boubekeur},
  \citenamefont {Giusarma},\ and\ \citenamefont {Mena}}]{Escudero:2015wba}%
  \BibitemOpen
  \bibfield  {author} {\bibinfo {author} {\bibfnamefont {M.}~\bibnamefont
  {Escudero}}, \bibinfo {author} {\bibfnamefont {H.}~\bibnamefont
  {Ram\'\i{}rez}}, \bibinfo {author} {\bibfnamefont {L.}~\bibnamefont
  {Boubekeur}}, \bibinfo {author} {\bibfnamefont {E.}~\bibnamefont {Giusarma}},
  \ and\ \bibinfo {author} {\bibfnamefont {O.}~\bibnamefont {Mena}},\ }\href
  {\doibase 10.1088/1475-7516/2016/02/020} {\bibfield  {journal} {\bibinfo
  {journal} {JCAP}\ }\textbf {\bibinfo {volume} {02}},\ \bibinfo {pages} {020}
  (\bibinfo {year} {2016})},\ \Eprint {http://arxiv.org/abs/1509.05419}
  {arXiv:1509.05419 [astro-ph.CO]} \BibitemShut {NoStop}%
\bibitem [{\citenamefont {Benetti}\ and\ \citenamefont
  {Alcaniz}(2016)}]{Benetti:2016tvm}%
  \BibitemOpen
  \bibfield  {author} {\bibinfo {author} {\bibfnamefont {M.}~\bibnamefont
  {Benetti}}\ and\ \bibinfo {author} {\bibfnamefont {J.~S.}\ \bibnamefont
  {Alcaniz}},\ }\href {\doibase 10.1103/PhysRevD.94.023526} {\bibfield
  {journal} {\bibinfo  {journal} {Phys. Rev. D}\ }\textbf {\bibinfo {volume}
  {94}},\ \bibinfo {pages} {023526} (\bibinfo {year} {2016})},\ \Eprint
  {http://arxiv.org/abs/1604.08156} {arXiv:1604.08156 [astro-ph.CO]}
  \BibitemShut {NoStop}%
\bibitem [{\citenamefont {Benetti}\ \emph {et~al.}(2016)\citenamefont
  {Benetti}, \citenamefont {Landau},\ and\ \citenamefont
  {Alcaniz}}]{Benetti:2016ycg}%
  \BibitemOpen
  \bibfield  {author} {\bibinfo {author} {\bibfnamefont {M.}~\bibnamefont
  {Benetti}}, \bibinfo {author} {\bibfnamefont {S.~J.}\ \bibnamefont {Landau}},
  \ and\ \bibinfo {author} {\bibfnamefont {J.~S.}\ \bibnamefont {Alcaniz}},\
  }\href {\doibase 10.1088/1475-7516/2016/12/035} {\bibfield  {journal}
  {\bibinfo  {journal} {JCAP}\ }\textbf {\bibinfo {volume} {12}},\ \bibinfo
  {pages} {035} (\bibinfo {year} {2016})},\ \Eprint
  {http://arxiv.org/abs/1610.03091} {arXiv:1610.03091 [astro-ph.CO]}
  \BibitemShut {NoStop}%
\bibitem [{\citenamefont {Benetti}\ and\ \citenamefont
  {Ramos}(2017)}]{Benetti:2016jhf}%
  \BibitemOpen
  \bibfield  {author} {\bibinfo {author} {\bibfnamefont {M.}~\bibnamefont
  {Benetti}}\ and\ \bibinfo {author} {\bibfnamefont {R.~O.}\ \bibnamefont
  {Ramos}},\ }\href {\doibase 10.1103/PhysRevD.95.023517} {\bibfield  {journal}
  {\bibinfo  {journal} {Phys. Rev. D}\ }\textbf {\bibinfo {volume} {95}},\
  \bibinfo {pages} {023517} (\bibinfo {year} {2017})},\ \Eprint
  {http://arxiv.org/abs/1610.08758} {arXiv:1610.08758 [astro-ph.CO]}
  \BibitemShut {NoStop}%
\bibitem [{\citenamefont {Guo}\ and\ \citenamefont
  {Zhang}(2017)}]{Guo:2017qjt}%
  \BibitemOpen
  \bibfield  {author} {\bibinfo {author} {\bibfnamefont {R.-Y.}\ \bibnamefont
  {Guo}}\ and\ \bibinfo {author} {\bibfnamefont {X.}~\bibnamefont {Zhang}},\
  }\href {\doibase 10.1140/epjc/s10052-017-5454-9} {\bibfield  {journal}
  {\bibinfo  {journal} {Eur. Phys. J. C}\ }\textbf {\bibinfo {volume} {77}},\
  \bibinfo {pages} {882} (\bibinfo {year} {2017})},\ \Eprint
  {http://arxiv.org/abs/1704.04784} {arXiv:1704.04784 [astro-ph.CO]}
  \BibitemShut {NoStop}%
\bibitem [{\citenamefont {Campista}\ \emph {et~al.}(2017)\citenamefont
  {Campista}, \citenamefont {Benetti},\ and\ \citenamefont
  {Alcaniz}}]{Campista:2017ovq}%
  \BibitemOpen
  \bibfield  {author} {\bibinfo {author} {\bibfnamefont {M.}~\bibnamefont
  {Campista}}, \bibinfo {author} {\bibfnamefont {M.}~\bibnamefont {Benetti}}, \
  and\ \bibinfo {author} {\bibfnamefont {J.}~\bibnamefont {Alcaniz}},\ }\href
  {\doibase 10.1088/1475-7516/2017/09/010} {\bibfield  {journal} {\bibinfo
  {journal} {JCAP}\ }\textbf {\bibinfo {volume} {09}},\ \bibinfo {pages} {010}
  (\bibinfo {year} {2017})},\ \Eprint {http://arxiv.org/abs/1705.08877}
  {arXiv:1705.08877 [astro-ph.CO]} \BibitemShut {NoStop}%
\bibitem [{\citenamefont {Ni}\ \emph {et~al.}(2018)\citenamefont {Ni},
  \citenamefont {Li}, \citenamefont {Qiu}, \citenamefont {Zheng},\ and\
  \citenamefont {Zhang}}]{Ni:2017jxw}%
  \BibitemOpen
  \bibfield  {author} {\bibinfo {author} {\bibfnamefont {S.}~\bibnamefont
  {Ni}}, \bibinfo {author} {\bibfnamefont {H.}~\bibnamefont {Li}}, \bibinfo
  {author} {\bibfnamefont {T.}~\bibnamefont {Qiu}}, \bibinfo {author}
  {\bibfnamefont {W.}~\bibnamefont {Zheng}}, \ and\ \bibinfo {author}
  {\bibfnamefont {X.}~\bibnamefont {Zhang}},\ }\href {\doibase
  10.1140/epjc/s10052-018-6085-5} {\bibfield  {journal} {\bibinfo  {journal}
  {Eur. Phys. J. C}\ }\textbf {\bibinfo {volume} {78}},\ \bibinfo {pages} {608}
  (\bibinfo {year} {2018})},\ \Eprint {http://arxiv.org/abs/1707.05570}
  {arXiv:1707.05570 [astro-ph.CO]} \BibitemShut {NoStop}%
\bibitem [{\citenamefont {Santos~da Costa}\ \emph {et~al.}(2018)\citenamefont
  {Santos~da Costa}, \citenamefont {Benetti},\ and\ \citenamefont
  {Alcaniz}}]{SantosdaCosta:2017ctv}%
  \BibitemOpen
  \bibfield  {author} {\bibinfo {author} {\bibfnamefont {S.}~\bibnamefont
  {Santos~da Costa}}, \bibinfo {author} {\bibfnamefont {M.}~\bibnamefont
  {Benetti}}, \ and\ \bibinfo {author} {\bibfnamefont {J.}~\bibnamefont
  {Alcaniz}},\ }\href {\doibase 10.1088/1475-7516/2018/03/004} {\bibfield
  {journal} {\bibinfo  {journal} {JCAP}\ }\textbf {\bibinfo {volume} {03}},\
  \bibinfo {pages} {004} (\bibinfo {year} {2018})},\ \Eprint
  {http://arxiv.org/abs/1710.01613} {arXiv:1710.01613 [astro-ph.CO]}
  \BibitemShut {NoStop}%
\bibitem [{\citenamefont {Park}\ and\ \citenamefont
  {Ratra}(2019)}]{Park:2017xbl}%
  \BibitemOpen
  \bibfield  {author} {\bibinfo {author} {\bibfnamefont {C.-G.}\ \bibnamefont
  {Park}}\ and\ \bibinfo {author} {\bibfnamefont {B.}~\bibnamefont {Ratra}},\
  }\href {\doibase 10.3847/1538-4357/ab3641} {\bibfield  {journal} {\bibinfo
  {journal} {Astrophys. J.}\ }\textbf {\bibinfo {volume} {882}},\ \bibinfo
  {pages} {158} (\bibinfo {year} {2019})},\ \Eprint
  {http://arxiv.org/abs/1801.00213} {arXiv:1801.00213 [astro-ph.CO]}
  \BibitemShut {NoStop}%
\bibitem [{\citenamefont {Guo}\ \emph {et~al.}(2019)\citenamefont {Guo},
  \citenamefont {Zhang}, \citenamefont {Zhang},\ and\ \citenamefont
  {Zhang}}]{Guo:2018uic}%
  \BibitemOpen
  \bibfield  {author} {\bibinfo {author} {\bibfnamefont {R.-Y.}\ \bibnamefont
  {Guo}}, \bibinfo {author} {\bibfnamefont {L.}~\bibnamefont {Zhang}}, \bibinfo
  {author} {\bibfnamefont {J.-F.}\ \bibnamefont {Zhang}}, \ and\ \bibinfo
  {author} {\bibfnamefont {X.}~\bibnamefont {Zhang}},\ }\href {\doibase
  10.1007/s11433-018-9278-1} {\bibfield  {journal} {\bibinfo  {journal} {Sci.
  China Phys. Mech. Astron.}\ }\textbf {\bibinfo {volume} {62}},\ \bibinfo
  {pages} {30411} (\bibinfo {year} {2019})},\ \Eprint
  {http://arxiv.org/abs/1801.02187} {arXiv:1801.02187 [astro-ph.CO]}
  \BibitemShut {NoStop}%
\bibitem [{\citenamefont {Di~Valentino}\ and\ \citenamefont
  {Mersini-Houghton}(2019)}]{DiValentino:2018wum}%
  \BibitemOpen
  \bibfield  {author} {\bibinfo {author} {\bibfnamefont {E.}~\bibnamefont
  {Di~Valentino}}\ and\ \bibinfo {author} {\bibfnamefont {L.}~\bibnamefont
  {Mersini-Houghton}},\ }\href {\doibase 10.3390/sym11040520} {\bibfield
  {journal} {\bibinfo  {journal} {Symmetry}\ }\textbf {\bibinfo {volume}
  {11}},\ \bibinfo {pages} {520} (\bibinfo {year} {2019})},\ \Eprint
  {http://arxiv.org/abs/1807.10833} {arXiv:1807.10833 [astro-ph.CO]}
  \BibitemShut {NoStop}%
\bibitem [{\citenamefont {Chowdhury}\ \emph {et~al.}(2019)\citenamefont
  {Chowdhury}, \citenamefont {Martin}, \citenamefont {Ringeval},\ and\
  \citenamefont {Vennin}}]{Chowdhury:2019otk}%
  \BibitemOpen
  \bibfield  {author} {\bibinfo {author} {\bibfnamefont {D.}~\bibnamefont
  {Chowdhury}}, \bibinfo {author} {\bibfnamefont {J.}~\bibnamefont {Martin}},
  \bibinfo {author} {\bibfnamefont {C.}~\bibnamefont {Ringeval}}, \ and\
  \bibinfo {author} {\bibfnamefont {V.}~\bibnamefont {Vennin}},\ }\href
  {\doibase 10.1103/PhysRevD.100.083537} {\bibfield  {journal} {\bibinfo
  {journal} {Phys. Rev. D}\ }\textbf {\bibinfo {volume} {100}},\ \bibinfo
  {pages} {083537} (\bibinfo {year} {2019})},\ \Eprint
  {http://arxiv.org/abs/1902.03951} {arXiv:1902.03951 [astro-ph.CO]}
  \BibitemShut {NoStop}%
\bibitem [{\citenamefont {Benetti}\ \emph {et~al.}(2019)\citenamefont
  {Benetti}, \citenamefont {Graef},\ and\ \citenamefont
  {Ramos}}]{Benetti:2019kgw}%
  \BibitemOpen
  \bibfield  {author} {\bibinfo {author} {\bibfnamefont {M.}~\bibnamefont
  {Benetti}}, \bibinfo {author} {\bibfnamefont {L.}~\bibnamefont {Graef}}, \
  and\ \bibinfo {author} {\bibfnamefont {R.~O.}\ \bibnamefont {Ramos}},\ }\href
  {\doibase 10.1088/1475-7516/2019/10/066} {\bibfield  {journal} {\bibinfo
  {journal} {JCAP}\ }\textbf {\bibinfo {volume} {10}},\ \bibinfo {pages} {066}
  (\bibinfo {year} {2019})},\ \Eprint {http://arxiv.org/abs/1907.03633}
  {arXiv:1907.03633 [astro-ph.CO]} \BibitemShut {NoStop}%
\bibitem [{\citenamefont {Haro}\ \emph {et~al.}(2020)\citenamefont {Haro},
  \citenamefont {Amor\'os},\ and\ \citenamefont {Pan}}]{Haro:2019peq}%
  \BibitemOpen
  \bibfield  {author} {\bibinfo {author} {\bibfnamefont {J.}~\bibnamefont
  {Haro}}, \bibinfo {author} {\bibfnamefont {J.}~\bibnamefont {Amor\'os}}, \
  and\ \bibinfo {author} {\bibfnamefont {S.}~\bibnamefont {Pan}},\ }\href
  {\doibase 10.1140/epjc/s10052-020-7950-6} {\bibfield  {journal} {\bibinfo
  {journal} {Eur. Phys. J. C}\ }\textbf {\bibinfo {volume} {80}},\ \bibinfo
  {pages} {404} (\bibinfo {year} {2020})},\ \Eprint
  {http://arxiv.org/abs/1908.01516} {arXiv:1908.01516 [gr-qc]} \BibitemShut
  {NoStop}%
\bibitem [{\citenamefont {Guo}\ \emph {et~al.}(2020)\citenamefont {Guo},
  \citenamefont {Zhang},\ and\ \citenamefont {Zhang}}]{Guo:2019dui}%
  \BibitemOpen
  \bibfield  {author} {\bibinfo {author} {\bibfnamefont {R.-Y.}\ \bibnamefont
  {Guo}}, \bibinfo {author} {\bibfnamefont {J.-F.}\ \bibnamefont {Zhang}}, \
  and\ \bibinfo {author} {\bibfnamefont {X.}~\bibnamefont {Zhang}},\ }\href
  {\doibase 10.1007/s11433-019-1514-0} {\bibfield  {journal} {\bibinfo
  {journal} {Sci. China Phys. Mech. Astron.}\ }\textbf {\bibinfo {volume}
  {63}},\ \bibinfo {pages} {290406} (\bibinfo {year} {2020})},\ \Eprint
  {http://arxiv.org/abs/1910.13944} {arXiv:1910.13944 [astro-ph.CO]}
  \BibitemShut {NoStop}%
\bibitem [{\citenamefont {Li}\ \emph {et~al.}(2020)\citenamefont {Li},
  \citenamefont {Ye}, \citenamefont {Cai},\ and\ \citenamefont
  {Piao}}]{Li:2019ipk}%
  \BibitemOpen
  \bibfield  {author} {\bibinfo {author} {\bibfnamefont {H.-H.}\ \bibnamefont
  {Li}}, \bibinfo {author} {\bibfnamefont {G.}~\bibnamefont {Ye}}, \bibinfo
  {author} {\bibfnamefont {Y.}~\bibnamefont {Cai}}, \ and\ \bibinfo {author}
  {\bibfnamefont {Y.-S.}\ \bibnamefont {Piao}},\ }\href {\doibase
  10.1103/PhysRevD.101.063527} {\bibfield  {journal} {\bibinfo  {journal}
  {Phys. Rev. D}\ }\textbf {\bibinfo {volume} {101}},\ \bibinfo {pages}
  {063527} (\bibinfo {year} {2020})},\ \Eprint
  {http://arxiv.org/abs/1911.06148} {arXiv:1911.06148 [gr-qc]} \BibitemShut
  {NoStop}%
\bibitem [{\citenamefont {Aich}\ \emph {et~al.}(2020)\citenamefont {Aich},
  \citenamefont {Ma}, \citenamefont {Dai},\ and\ \citenamefont
  {Xia}}]{Aich:2019obd}%
  \BibitemOpen
  \bibfield  {author} {\bibinfo {author} {\bibfnamefont {M.}~\bibnamefont
  {Aich}}, \bibinfo {author} {\bibfnamefont {Y.-Z.}\ \bibnamefont {Ma}},
  \bibinfo {author} {\bibfnamefont {W.-M.}\ \bibnamefont {Dai}}, \ and\
  \bibinfo {author} {\bibfnamefont {J.-Q.}\ \bibnamefont {Xia}},\ }\href
  {\doibase 10.1103/PhysRevD.101.063536} {\bibfield  {journal} {\bibinfo
  {journal} {Phys. Rev. D}\ }\textbf {\bibinfo {volume} {101}},\ \bibinfo
  {pages} {063536} (\bibinfo {year} {2020})},\ \Eprint
  {http://arxiv.org/abs/1912.00995} {arXiv:1912.00995 [astro-ph.CO]}
  \BibitemShut {NoStop}%
\bibitem [{\citenamefont {Braglia}\ \emph {et~al.}(2020)\citenamefont
  {Braglia}, \citenamefont {Hazra}, \citenamefont {Sriramkumar},\ and\
  \citenamefont {Finelli}}]{Braglia:2020fms}%
  \BibitemOpen
  \bibfield  {author} {\bibinfo {author} {\bibfnamefont {M.}~\bibnamefont
  {Braglia}}, \bibinfo {author} {\bibfnamefont {D.~K.}\ \bibnamefont {Hazra}},
  \bibinfo {author} {\bibfnamefont {L.}~\bibnamefont {Sriramkumar}}, \ and\
  \bibinfo {author} {\bibfnamefont {F.}~\bibnamefont {Finelli}},\ }\href
  {\doibase 10.1088/1475-7516/2020/08/025} {\bibfield  {journal} {\bibinfo
  {journal} {JCAP}\ }\textbf {\bibinfo {volume} {08}},\ \bibinfo {pages} {025}
  (\bibinfo {year} {2020})},\ \Eprint {http://arxiv.org/abs/2004.00672}
  {arXiv:2004.00672 [astro-ph.CO]} \BibitemShut {NoStop}%
\bibitem [{\citenamefont {Cicoli}\ and\ \citenamefont
  {Di~Valentino}(2020)}]{Cicoli:2020bao}%
  \BibitemOpen
  \bibfield  {author} {\bibinfo {author} {\bibfnamefont {M.}~\bibnamefont
  {Cicoli}}\ and\ \bibinfo {author} {\bibfnamefont {E.}~\bibnamefont
  {Di~Valentino}},\ }\href {\doibase 10.1103/PhysRevD.102.043521} {\bibfield
  {journal} {\bibinfo  {journal} {Phys. Rev. D}\ }\textbf {\bibinfo {volume}
  {102}},\ \bibinfo {pages} {043521} (\bibinfo {year} {2020})},\ \Eprint
  {http://arxiv.org/abs/2004.01210} {arXiv:2004.01210 [astro-ph.CO]}
  \BibitemShut {NoStop}%
\bibitem [{\citenamefont {Keeley}\ \emph {et~al.}(2020)\citenamefont {Keeley},
  \citenamefont {Shafieloo}, \citenamefont {Hazra},\ and\ \citenamefont
  {Souradeep}}]{Keeley:2020rmo}%
  \BibitemOpen
  \bibfield  {author} {\bibinfo {author} {\bibfnamefont {R.~E.}\ \bibnamefont
  {Keeley}}, \bibinfo {author} {\bibfnamefont {A.}~\bibnamefont {Shafieloo}},
  \bibinfo {author} {\bibfnamefont {D.~K.}\ \bibnamefont {Hazra}}, \ and\
  \bibinfo {author} {\bibfnamefont {T.}~\bibnamefont {Souradeep}},\ }\href
  {\doibase 10.1088/1475-7516/2020/09/055} {\bibfield  {journal} {\bibinfo
  {journal} {JCAP}\ }\textbf {\bibinfo {volume} {09}},\ \bibinfo {pages} {055}
  (\bibinfo {year} {2020})},\ \Eprint {http://arxiv.org/abs/2006.12710}
  {arXiv:2006.12710 [astro-ph.CO]} \BibitemShut {NoStop}%
\bibitem [{\citenamefont {Santos~da Costa}\ \emph {et~al.}(2021)\citenamefont
  {Santos~da Costa}, \citenamefont {Benetti}, \citenamefont {Neves},
  \citenamefont {Brito}, \citenamefont {Silva},\ and\ \citenamefont
  {Alcaniz}}]{SantosdaCosta:2020dyl}%
  \BibitemOpen
  \bibfield  {author} {\bibinfo {author} {\bibfnamefont {S.}~\bibnamefont
  {Santos~da Costa}}, \bibinfo {author} {\bibfnamefont {M.}~\bibnamefont
  {Benetti}}, \bibinfo {author} {\bibfnamefont {R.~M.~P.}\ \bibnamefont
  {Neves}}, \bibinfo {author} {\bibfnamefont {F.~A.}\ \bibnamefont {Brito}},
  \bibinfo {author} {\bibfnamefont {R.}~\bibnamefont {Silva}}, \ and\ \bibinfo
  {author} {\bibfnamefont {J.~S.}\ \bibnamefont {Alcaniz}},\ }\href {\doibase
  10.1140/epjp/s13360-020-01015-1} {\bibfield  {journal} {\bibinfo  {journal}
  {Eur. Phys. J. Plus}\ }\textbf {\bibinfo {volume} {136}},\ \bibinfo {pages}
  {84} (\bibinfo {year} {2021})},\ \Eprint {http://arxiv.org/abs/2007.09211}
  {arXiv:2007.09211 [astro-ph.CO]} \BibitemShut {NoStop}%
\bibitem [{\citenamefont {Rodrigues}\ \emph {et~al.}(2021)\citenamefont
  {Rodrigues}, \citenamefont {Santos~da Costa},\ and\ \citenamefont
  {Alcaniz}}]{Rodrigues:2020fle}%
  \BibitemOpen
  \bibfield  {author} {\bibinfo {author} {\bibfnamefont {J.~G.}\ \bibnamefont
  {Rodrigues}}, \bibinfo {author} {\bibfnamefont {S.}~\bibnamefont {Santos~da
  Costa}}, \ and\ \bibinfo {author} {\bibfnamefont {J.~S.}\ \bibnamefont
  {Alcaniz}},\ }\href {\doibase 10.1016/j.physletb.2021.136156} {\bibfield
  {journal} {\bibinfo  {journal} {Phys. Lett. B}\ }\textbf {\bibinfo {volume}
  {815}},\ \bibinfo {pages} {136156} (\bibinfo {year} {2021})},\ \Eprint
  {http://arxiv.org/abs/2007.10763} {arXiv:2007.10763 [astro-ph.CO]}
  \BibitemShut {NoStop}%
\bibitem [{\citenamefont {Vagnozzi}\ \emph
  {et~al.}(2021{\natexlab{a}})\citenamefont {Vagnozzi}, \citenamefont
  {Di~Valentino}, \citenamefont {Gariazzo}, \citenamefont {Melchiorri},
  \citenamefont {Mena},\ and\ \citenamefont {Silk}}]{Vagnozzi:2020rcz}%
  \BibitemOpen
  \bibfield  {author} {\bibinfo {author} {\bibfnamefont {S.}~\bibnamefont
  {Vagnozzi}}, \bibinfo {author} {\bibfnamefont {E.}~\bibnamefont
  {Di~Valentino}}, \bibinfo {author} {\bibfnamefont {S.}~\bibnamefont
  {Gariazzo}}, \bibinfo {author} {\bibfnamefont {A.}~\bibnamefont
  {Melchiorri}}, \bibinfo {author} {\bibfnamefont {O.}~\bibnamefont {Mena}}, \
  and\ \bibinfo {author} {\bibfnamefont {J.}~\bibnamefont {Silk}},\ }\href
  {\doibase 10.1016/j.dark.2021.100851} {\bibfield  {journal} {\bibinfo
  {journal} {Phys. Dark Univ.}\ }\textbf {\bibinfo {volume} {33}},\ \bibinfo
  {pages} {100851} (\bibinfo {year} {2021}{\natexlab{a}})},\ \Eprint
  {http://arxiv.org/abs/2010.02230} {arXiv:2010.02230 [astro-ph.CO]}
  \BibitemShut {NoStop}%
\bibitem [{\citenamefont {Neves}\ \emph {et~al.}(2022)\citenamefont {Neves},
  \citenamefont {Santos Da~Costa}, \citenamefont {Brito},\ and\ \citenamefont
  {Alcaniz}}]{Neves:2020anh}%
  \BibitemOpen
  \bibfield  {author} {\bibinfo {author} {\bibfnamefont {R.~M.~P.}\
  \bibnamefont {Neves}}, \bibinfo {author} {\bibfnamefont {S.}~\bibnamefont
  {Santos Da~Costa}}, \bibinfo {author} {\bibfnamefont {F.~A.}\ \bibnamefont
  {Brito}}, \ and\ \bibinfo {author} {\bibfnamefont {J.~S.}\ \bibnamefont
  {Alcaniz}},\ }\href {\doibase 10.1088/1475-7516/2022/07/024} {\bibfield
  {journal} {\bibinfo  {journal} {JCAP}\ }\textbf {\bibinfo {volume} {07}},\
  \bibinfo {pages} {024} (\bibinfo {year} {2022})},\ \Eprint
  {http://arxiv.org/abs/2011.05264} {arXiv:2011.05264 [hep-th]} \BibitemShut
  {NoStop}%
\bibitem [{\citenamefont {Vagnozzi}\ \emph
  {et~al.}(2021{\natexlab{b}})\citenamefont {Vagnozzi}, \citenamefont {Loeb},\
  and\ \citenamefont {Moresco}}]{Vagnozzi:2020dfn}%
  \BibitemOpen
  \bibfield  {author} {\bibinfo {author} {\bibfnamefont {S.}~\bibnamefont
  {Vagnozzi}}, \bibinfo {author} {\bibfnamefont {A.}~\bibnamefont {Loeb}}, \
  and\ \bibinfo {author} {\bibfnamefont {M.}~\bibnamefont {Moresco}},\ }\href
  {\doibase 10.3847/1538-4357/abd4df} {\bibfield  {journal} {\bibinfo
  {journal} {Astrophys. J.}\ }\textbf {\bibinfo {volume} {908}},\ \bibinfo
  {pages} {84} (\bibinfo {year} {2021}{\natexlab{b}})},\ \Eprint
  {http://arxiv.org/abs/2011.11645} {arXiv:2011.11645 [astro-ph.CO]}
  \BibitemShut {NoStop}%
\bibitem [{\citenamefont {Ye}\ \emph {et~al.}(2021)\citenamefont {Ye},
  \citenamefont {Hu},\ and\ \citenamefont {Piao}}]{Ye:2021nej}%
  \BibitemOpen
  \bibfield  {author} {\bibinfo {author} {\bibfnamefont {G.}~\bibnamefont
  {Ye}}, \bibinfo {author} {\bibfnamefont {B.}~\bibnamefont {Hu}}, \ and\
  \bibinfo {author} {\bibfnamefont {Y.-S.}\ \bibnamefont {Piao}},\ }\href
  {\doibase 10.1103/PhysRevD.104.063510} {\bibfield  {journal} {\bibinfo
  {journal} {Phys. Rev. D}\ }\textbf {\bibinfo {volume} {104}},\ \bibinfo
  {pages} {063510} (\bibinfo {year} {2021})},\ \Eprint
  {http://arxiv.org/abs/2103.09729} {arXiv:2103.09729 [astro-ph.CO]}
  \BibitemShut {NoStop}%
\bibitem [{\citenamefont {Dhawan}\ \emph {et~al.}(2021)\citenamefont {Dhawan},
  \citenamefont {Alsing},\ and\ \citenamefont {Vagnozzi}}]{Dhawan:2021mel}%
  \BibitemOpen
  \bibfield  {author} {\bibinfo {author} {\bibfnamefont {S.}~\bibnamefont
  {Dhawan}}, \bibinfo {author} {\bibfnamefont {J.}~\bibnamefont {Alsing}}, \
  and\ \bibinfo {author} {\bibfnamefont {S.}~\bibnamefont {Vagnozzi}},\ }\href
  {\doibase 10.1093/mnrasl/slab058} {\bibfield  {journal} {\bibinfo  {journal}
  {Mon. Not. Roy. Astron. Soc.}\ }\textbf {\bibinfo {volume} {506}},\ \bibinfo
  {pages} {L1} (\bibinfo {year} {2021})},\ \Eprint
  {http://arxiv.org/abs/2104.02485} {arXiv:2104.02485 [astro-ph.CO]}
  \BibitemShut {NoStop}%
\bibitem [{\citenamefont {Stein}\ and\ \citenamefont
  {Kinney}(2022)}]{Stein:2021uge}%
  \BibitemOpen
  \bibfield  {author} {\bibinfo {author} {\bibfnamefont {N.~K.}\ \bibnamefont
  {Stein}}\ and\ \bibinfo {author} {\bibfnamefont {W.~H.}\ \bibnamefont
  {Kinney}},\ }\href {\doibase 10.1088/1475-7516/2022/01/022} {\bibfield
  {journal} {\bibinfo  {journal} {JCAP}\ }\textbf {\bibinfo {volume} {01}},\
  \bibinfo {pages} {022} (\bibinfo {year} {2022})},\ \Eprint
  {http://arxiv.org/abs/2106.02089} {arXiv:2106.02089 [astro-ph.CO]}
  \BibitemShut {NoStop}%
\bibitem [{\citenamefont {Forconi}\ \emph {et~al.}(2021)\citenamefont
  {Forconi}, \citenamefont {Giar\`e}, \citenamefont {Di~Valentino},\ and\
  \citenamefont {Melchiorri}}]{Forconi:2021que}%
  \BibitemOpen
  \bibfield  {author} {\bibinfo {author} {\bibfnamefont {M.}~\bibnamefont
  {Forconi}}, \bibinfo {author} {\bibfnamefont {W.}~\bibnamefont {Giar\`e}},
  \bibinfo {author} {\bibfnamefont {E.}~\bibnamefont {Di~Valentino}}, \ and\
  \bibinfo {author} {\bibfnamefont {A.}~\bibnamefont {Melchiorri}},\ }\href
  {\doibase 10.1103/PhysRevD.104.103528} {\bibfield  {journal} {\bibinfo
  {journal} {Phys. Rev. D}\ }\textbf {\bibinfo {volume} {104}},\ \bibinfo
  {pages} {103528} (\bibinfo {year} {2021})},\ \Eprint
  {http://arxiv.org/abs/2110.01695} {arXiv:2110.01695 [astro-ph.CO]}
  \BibitemShut {NoStop}%
\bibitem [{\citenamefont {dos Santos}\ \emph {et~al.}(2022)\citenamefont {dos
  Santos}, \citenamefont {Santos~da Costa}, \citenamefont {Silva},
  \citenamefont {Benetti},\ and\ \citenamefont {Alcaniz}}]{dosSantos:2021vis}%
  \BibitemOpen
  \bibfield  {author} {\bibinfo {author} {\bibfnamefont {F.~B.~M.}\
  \bibnamefont {dos Santos}}, \bibinfo {author} {\bibfnamefont
  {S.}~\bibnamefont {Santos~da Costa}}, \bibinfo {author} {\bibfnamefont
  {R.}~\bibnamefont {Silva}}, \bibinfo {author} {\bibfnamefont
  {M.}~\bibnamefont {Benetti}}, \ and\ \bibinfo {author} {\bibfnamefont
  {J.}~\bibnamefont {Alcaniz}},\ }\href {\doibase
  10.1088/1475-7516/2022/06/001} {\bibfield  {journal} {\bibinfo  {journal}
  {JCAP}\ }\textbf {\bibinfo {volume} {06}},\ \bibinfo {pages} {001} (\bibinfo
  {year} {2022})},\ \Eprint {http://arxiv.org/abs/2110.14758} {arXiv:2110.14758
  [astro-ph.CO]} \BibitemShut {NoStop}%
\bibitem [{\citenamefont {Cabass}\ \emph
  {et~al.}(2022{\natexlab{a}})\citenamefont {Cabass}, \citenamefont {Ivanov},
  \citenamefont {Philcox}, \citenamefont {Simonovi\'c},\ and\ \citenamefont
  {Zaldarriaga}}]{Cabass:2022wjy}%
  \BibitemOpen
  \bibfield  {author} {\bibinfo {author} {\bibfnamefont {G.}~\bibnamefont
  {Cabass}}, \bibinfo {author} {\bibfnamefont {M.~M.}\ \bibnamefont {Ivanov}},
  \bibinfo {author} {\bibfnamefont {O.~H.~E.}\ \bibnamefont {Philcox}},
  \bibinfo {author} {\bibfnamefont {M.}~\bibnamefont {Simonovi\'c}}, \ and\
  \bibinfo {author} {\bibfnamefont {M.}~\bibnamefont {Zaldarriaga}},\ }\href
  {\doibase 10.1103/PhysRevLett.129.021301} {\bibfield  {journal} {\bibinfo
  {journal} {Phys. Rev. Lett.}\ }\textbf {\bibinfo {volume} {129}},\ \bibinfo
  {pages} {021301} (\bibinfo {year} {2022}{\natexlab{a}})},\ \Eprint
  {http://arxiv.org/abs/2201.07238} {arXiv:2201.07238 [astro-ph.CO]}
  \BibitemShut {NoStop}%
\bibitem [{\citenamefont {Ye}\ and\ \citenamefont {Piao}(2022)}]{Ye:2022afu}%
  \BibitemOpen
  \bibfield  {author} {\bibinfo {author} {\bibfnamefont {G.}~\bibnamefont
  {Ye}}\ and\ \bibinfo {author} {\bibfnamefont {Y.-S.}\ \bibnamefont {Piao}},\
  }\href {\doibase 10.1103/PhysRevD.106.043536} {\bibfield  {journal} {\bibinfo
   {journal} {Phys. Rev. D}\ }\textbf {\bibinfo {volume} {106}},\ \bibinfo
  {pages} {043536} (\bibinfo {year} {2022})},\ \Eprint
  {http://arxiv.org/abs/2202.10055} {arXiv:2202.10055 [astro-ph.CO]}
  \BibitemShut {NoStop}%
\bibitem [{\citenamefont {Antony}\ \emph {et~al.}(2023)\citenamefont {Antony},
  \citenamefont {Finelli}, \citenamefont {Hazra},\ and\ \citenamefont
  {Shafieloo}}]{Antony:2022ert}%
  \BibitemOpen
  \bibfield  {author} {\bibinfo {author} {\bibfnamefont {A.}~\bibnamefont
  {Antony}}, \bibinfo {author} {\bibfnamefont {F.}~\bibnamefont {Finelli}},
  \bibinfo {author} {\bibfnamefont {D.~K.}\ \bibnamefont {Hazra}}, \ and\
  \bibinfo {author} {\bibfnamefont {A.}~\bibnamefont {Shafieloo}},\ }\href
  {\doibase 10.1103/PhysRevLett.130.111001} {\bibfield  {journal} {\bibinfo
  {journal} {Phys. Rev. Lett.}\ }\textbf {\bibinfo {volume} {130}},\ \bibinfo
  {pages} {111001} (\bibinfo {year} {2023})},\ \Eprint
  {http://arxiv.org/abs/2202.14028} {arXiv:2202.14028 [astro-ph.CO]}
  \BibitemShut {NoStop}%
\bibitem [{\citenamefont {Cabass}\ \emph
  {et~al.}(2022{\natexlab{b}})\citenamefont {Cabass}, \citenamefont {Ivanov},
  \citenamefont {Philcox}, \citenamefont {Simonovi\'c},\ and\ \citenamefont
  {Zaldarriaga}}]{Cabass:2022ymb}%
  \BibitemOpen
  \bibfield  {author} {\bibinfo {author} {\bibfnamefont {G.}~\bibnamefont
  {Cabass}}, \bibinfo {author} {\bibfnamefont {M.~M.}\ \bibnamefont {Ivanov}},
  \bibinfo {author} {\bibfnamefont {O.~H.~E.}\ \bibnamefont {Philcox}},
  \bibinfo {author} {\bibfnamefont {M.}~\bibnamefont {Simonovi\'c}}, \ and\
  \bibinfo {author} {\bibfnamefont {M.}~\bibnamefont {Zaldarriaga}},\ }\href
  {\doibase 10.1103/PhysRevD.106.043506} {\bibfield  {journal} {\bibinfo
  {journal} {Phys. Rev. D}\ }\textbf {\bibinfo {volume} {106}},\ \bibinfo
  {pages} {043506} (\bibinfo {year} {2022}{\natexlab{b}})},\ \Eprint
  {http://arxiv.org/abs/2204.01781} {arXiv:2204.01781 [astro-ph.CO]}
  \BibitemShut {NoStop}%
\bibitem [{\citenamefont {Ye}\ \emph {et~al.}(2022)\citenamefont {Ye},
  \citenamefont {Jiang},\ and\ \citenamefont {Piao}}]{Ye:2022efx}%
  \BibitemOpen
  \bibfield  {author} {\bibinfo {author} {\bibfnamefont {G.}~\bibnamefont
  {Ye}}, \bibinfo {author} {\bibfnamefont {J.-Q.}\ \bibnamefont {Jiang}}, \
  and\ \bibinfo {author} {\bibfnamefont {Y.-S.}\ \bibnamefont {Piao}},\ }\href
  {\doibase 10.1103/PhysRevD.106.103528} {\bibfield  {journal} {\bibinfo
  {journal} {Phys. Rev. D}\ }\textbf {\bibinfo {volume} {106}},\ \bibinfo
  {pages} {103528} (\bibinfo {year} {2022})},\ \Eprint
  {http://arxiv.org/abs/2205.02478} {arXiv:2205.02478 [astro-ph.CO]}
  \BibitemShut {NoStop}%
\bibitem [{\citenamefont {Ghoshal}\ \emph
  {et~al.}(2023{\natexlab{a}})\citenamefont {Ghoshal}, \citenamefont
  {Mukherjee},\ and\ \citenamefont {Rinaldi}}]{Ghoshal:2022qxk}%
  \BibitemOpen
  \bibfield  {author} {\bibinfo {author} {\bibfnamefont {A.}~\bibnamefont
  {Ghoshal}}, \bibinfo {author} {\bibfnamefont {D.}~\bibnamefont {Mukherjee}},
  \ and\ \bibinfo {author} {\bibfnamefont {M.}~\bibnamefont {Rinaldi}},\ }\href
  {\doibase 10.1007/JHEP05(2023)023} {\bibfield  {journal} {\bibinfo  {journal}
  {JHEP}\ }\textbf {\bibinfo {volume} {05}},\ \bibinfo {pages} {023} (\bibinfo
  {year} {2023}{\natexlab{a}})},\ \Eprint {http://arxiv.org/abs/2205.06475}
  {arXiv:2205.06475 [gr-qc]} \BibitemShut {NoStop}%
\bibitem [{\citenamefont {Gangopadhyay}\ \emph {et~al.}(2023)\citenamefont
  {Gangopadhyay}, \citenamefont {Khan},\ and\ \citenamefont
  {Yogesh}}]{Gangopadhyay:2022vgh}%
  \BibitemOpen
  \bibfield  {author} {\bibinfo {author} {\bibfnamefont {M.~R.}\ \bibnamefont
  {Gangopadhyay}}, \bibinfo {author} {\bibfnamefont {H.~A.}\ \bibnamefont
  {Khan}}, \ and\ \bibinfo {author} {\bibnamefont {Yogesh}},\ }\href {\doibase
  10.1016/j.dark.2023.101177} {\bibfield  {journal} {\bibinfo  {journal} {Phys.
  Dark Univ.}\ }\textbf {\bibinfo {volume} {40}},\ \bibinfo {pages} {101177}
  (\bibinfo {year} {2023})},\ \Eprint {http://arxiv.org/abs/2205.15261}
  {arXiv:2205.15261 [astro-ph.CO]} \BibitemShut {NoStop}%
\bibitem [{\citenamefont {Montefalcone}\ \emph
  {et~al.}(2023{\natexlab{a}})\citenamefont {Montefalcone}, \citenamefont
  {Aragam}, \citenamefont {Visinelli},\ and\ \citenamefont
  {Freese}}]{Montefalcone:2022owy}%
  \BibitemOpen
  \bibfield  {author} {\bibinfo {author} {\bibfnamefont {G.}~\bibnamefont
  {Montefalcone}}, \bibinfo {author} {\bibfnamefont {V.}~\bibnamefont
  {Aragam}}, \bibinfo {author} {\bibfnamefont {L.}~\bibnamefont {Visinelli}}, \
  and\ \bibinfo {author} {\bibfnamefont {K.}~\bibnamefont {Freese}},\ }\href
  {\doibase 10.1103/PhysRevD.107.063543} {\bibfield  {journal} {\bibinfo
  {journal} {Phys. Rev. D}\ }\textbf {\bibinfo {volume} {107}},\ \bibinfo
  {pages} {063543} (\bibinfo {year} {2023}{\natexlab{a}})},\ \Eprint
  {http://arxiv.org/abs/2209.14908} {arXiv:2209.14908 [gr-qc]} \BibitemShut
  {NoStop}%
\bibitem [{\citenamefont {Stein}\ and\ \citenamefont
  {Kinney}(2023)}]{Stein:2022cpk}%
  \BibitemOpen
  \bibfield  {author} {\bibinfo {author} {\bibfnamefont {N.~K.}\ \bibnamefont
  {Stein}}\ and\ \bibinfo {author} {\bibfnamefont {W.~H.}\ \bibnamefont
  {Kinney}},\ }\href {\doibase 10.1088/1475-7516/2023/03/027} {\bibfield
  {journal} {\bibinfo  {journal} {JCAP}\ }\textbf {\bibinfo {volume} {03}},\
  \bibinfo {pages} {027} (\bibinfo {year} {2023})},\ \Eprint
  {http://arxiv.org/abs/2210.05757} {arXiv:2210.05757 [astro-ph.CO]}
  \BibitemShut {NoStop}%
\bibitem [{\citenamefont {Cabass}\ \emph {et~al.}(2023)\citenamefont {Cabass},
  \citenamefont {Ivanov},\ and\ \citenamefont {Philcox}}]{Cabass:2022oap}%
  \BibitemOpen
  \bibfield  {author} {\bibinfo {author} {\bibfnamefont {G.}~\bibnamefont
  {Cabass}}, \bibinfo {author} {\bibfnamefont {M.~M.}\ \bibnamefont {Ivanov}},
  \ and\ \bibinfo {author} {\bibfnamefont {O.~H.~E.}\ \bibnamefont {Philcox}},\
  }\href {\doibase 10.1103/PhysRevD.107.023523} {\bibfield  {journal} {\bibinfo
   {journal} {Phys. Rev. D}\ }\textbf {\bibinfo {volume} {107}},\ \bibinfo
  {pages} {023523} (\bibinfo {year} {2023})},\ \Eprint
  {http://arxiv.org/abs/2210.16320} {arXiv:2210.16320 [astro-ph.CO]}
  \BibitemShut {NoStop}%
\bibitem [{\citenamefont {Montefalcone}\ \emph
  {et~al.}(2023{\natexlab{b}})\citenamefont {Montefalcone}, \citenamefont
  {Aragam}, \citenamefont {Visinelli},\ and\ \citenamefont
  {Freese}}]{Montefalcone:2022jfw}%
  \BibitemOpen
  \bibfield  {author} {\bibinfo {author} {\bibfnamefont {G.}~\bibnamefont
  {Montefalcone}}, \bibinfo {author} {\bibfnamefont {V.}~\bibnamefont
  {Aragam}}, \bibinfo {author} {\bibfnamefont {L.}~\bibnamefont {Visinelli}}, \
  and\ \bibinfo {author} {\bibfnamefont {K.}~\bibnamefont {Freese}},\ }\href
  {\doibase 10.1088/1475-7516/2023/03/002} {\bibfield  {journal} {\bibinfo
  {journal} {JCAP}\ }\textbf {\bibinfo {volume} {03}},\ \bibinfo {pages} {002}
  (\bibinfo {year} {2023}{\natexlab{b}})},\ \Eprint
  {http://arxiv.org/abs/2212.04482} {arXiv:2212.04482 [gr-qc]} \BibitemShut
  {NoStop}%
\bibitem [{\citenamefont {Abazajian}\ \emph {et~al.}(2016)\citenamefont
  {Abazajian} \emph {et~al.}}]{CMB-S4:2016ple}%
  \BibitemOpen
  \bibfield  {author} {\bibinfo {author} {\bibfnamefont {K.~N.}\ \bibnamefont
  {Abazajian}} \emph {et~al.} (\bibinfo {collaboration} {CMB-S4}),\ }\href@noop
  {} {\  (\bibinfo {year} {2016})},\ \Eprint {http://arxiv.org/abs/1610.02743}
  {arXiv:1610.02743 [astro-ph.CO]} \BibitemShut {NoStop}%
\bibitem [{\citenamefont {Ade}\ \emph {et~al.}(2019)\citenamefont {Ade} \emph
  {et~al.}}]{SimonsObservatory:2018koc}%
  \BibitemOpen
  \bibfield  {author} {\bibinfo {author} {\bibfnamefont {P.}~\bibnamefont
  {Ade}} \emph {et~al.} (\bibinfo {collaboration} {Simons Observatory}),\
  }\href {\doibase 10.1088/1475-7516/2019/02/056} {\bibfield  {journal}
  {\bibinfo  {journal} {JCAP}\ }\textbf {\bibinfo {volume} {02}},\ \bibinfo
  {pages} {056} (\bibinfo {year} {2019})},\ \Eprint
  {http://arxiv.org/abs/1808.07445} {arXiv:1808.07445 [astro-ph.CO]}
  \BibitemShut {NoStop}%
\bibitem [{\citenamefont {Abitbol}\ \emph {et~al.}(2019)\citenamefont {Abitbol}
  \emph {et~al.}}]{SimonsObservatory:2019qwx}%
  \BibitemOpen
  \bibfield  {author} {\bibinfo {author} {\bibfnamefont {M.~H.}\ \bibnamefont
  {Abitbol}} \emph {et~al.} (\bibinfo {collaboration} {Simons Observatory}),\
  }\href@noop {} {\bibfield  {journal} {\bibinfo  {journal} {Bull. Am. Astron.
  Soc.}\ }\textbf {\bibinfo {volume} {51}},\ \bibinfo {pages} {147} (\bibinfo
  {year} {2019})},\ \Eprint {http://arxiv.org/abs/1907.08284} {arXiv:1907.08284
  [astro-ph.IM]} \BibitemShut {NoStop}%
\bibitem [{\citenamefont {Kamionkowski}\ and\ \citenamefont
  {Kovetz}(2016)}]{Kamionkowski:2015yta}%
  \BibitemOpen
  \bibfield  {author} {\bibinfo {author} {\bibfnamefont {M.}~\bibnamefont
  {Kamionkowski}}\ and\ \bibinfo {author} {\bibfnamefont {E.~D.}\ \bibnamefont
  {Kovetz}},\ }\href {\doibase 10.1146/annurev-astro-081915-023433} {\bibfield
  {journal} {\bibinfo  {journal} {Ann. Rev. Astron. Astrophys.}\ }\textbf
  {\bibinfo {volume} {54}},\ \bibinfo {pages} {227} (\bibinfo {year} {2016})},\
  \Eprint {http://arxiv.org/abs/1510.06042} {arXiv:1510.06042 [astro-ph.CO]}
  \BibitemShut {NoStop}%
\bibitem [{\citenamefont {Vagnozzi}(2021)}]{Vagnozzi:2020gtf}%
  \BibitemOpen
  \bibfield  {author} {\bibinfo {author} {\bibfnamefont {S.}~\bibnamefont
  {Vagnozzi}},\ }\href {\doibase 10.1093/mnrasl/slaa203} {\bibfield  {journal}
  {\bibinfo  {journal} {Mon. Not. Roy. Astron. Soc.}\ }\textbf {\bibinfo
  {volume} {502}},\ \bibinfo {pages} {L11} (\bibinfo {year} {2021})},\ \Eprint
  {http://arxiv.org/abs/2009.13432} {arXiv:2009.13432 [astro-ph.CO]}
  \BibitemShut {NoStop}%
\bibitem [{\citenamefont {Zhao}\ \emph {et~al.}(2013)\citenamefont {Zhao},
  \citenamefont {Zhang}, \citenamefont {You},\ and\ \citenamefont
  {Zhu}}]{Zhao:2013bba}%
  \BibitemOpen
  \bibfield  {author} {\bibinfo {author} {\bibfnamefont {W.}~\bibnamefont
  {Zhao}}, \bibinfo {author} {\bibfnamefont {Y.}~\bibnamefont {Zhang}},
  \bibinfo {author} {\bibfnamefont {X.-P.}\ \bibnamefont {You}}, \ and\
  \bibinfo {author} {\bibfnamefont {Z.-H.}\ \bibnamefont {Zhu}},\ }\href
  {\doibase 10.1103/PhysRevD.87.124012} {\bibfield  {journal} {\bibinfo
  {journal} {Phys. Rev. D}\ }\textbf {\bibinfo {volume} {87}},\ \bibinfo
  {pages} {124012} (\bibinfo {year} {2013})},\ \Eprint
  {http://arxiv.org/abs/1303.6718} {arXiv:1303.6718 [astro-ph.CO]} \BibitemShut
  {NoStop}%
\bibitem [{\citenamefont {Liu}\ \emph {et~al.}(2016)\citenamefont {Liu},
  \citenamefont {Zhao}, \citenamefont {Zhang},\ and\ \citenamefont
  {Zhu}}]{Liu:2015psa}%
  \BibitemOpen
  \bibfield  {author} {\bibinfo {author} {\bibfnamefont {X.-J.}\ \bibnamefont
  {Liu}}, \bibinfo {author} {\bibfnamefont {W.}~\bibnamefont {Zhao}}, \bibinfo
  {author} {\bibfnamefont {Y.}~\bibnamefont {Zhang}}, \ and\ \bibinfo {author}
  {\bibfnamefont {Z.-H.}\ \bibnamefont {Zhu}},\ }\href {\doibase
  10.1103/PhysRevD.93.024031} {\bibfield  {journal} {\bibinfo  {journal} {Phys.
  Rev. D}\ }\textbf {\bibinfo {volume} {93}},\ \bibinfo {pages} {024031}
  (\bibinfo {year} {2016})},\ \Eprint {http://arxiv.org/abs/1509.03524}
  {arXiv:1509.03524 [astro-ph.CO]} \BibitemShut {NoStop}%
\bibitem [{\citenamefont {Kuroyanagi}\ \emph {et~al.}(2009)\citenamefont
  {Kuroyanagi}, \citenamefont {Chiba},\ and\ \citenamefont
  {Sugiyama}}]{Kuroyanagi:2008ye}%
  \BibitemOpen
  \bibfield  {author} {\bibinfo {author} {\bibfnamefont {S.}~\bibnamefont
  {Kuroyanagi}}, \bibinfo {author} {\bibfnamefont {T.}~\bibnamefont {Chiba}}, \
  and\ \bibinfo {author} {\bibfnamefont {N.}~\bibnamefont {Sugiyama}},\ }\href
  {\doibase 10.1103/PhysRevD.79.103501} {\bibfield  {journal} {\bibinfo
  {journal} {Phys. Rev. D}\ }\textbf {\bibinfo {volume} {79}},\ \bibinfo
  {pages} {103501} (\bibinfo {year} {2009})},\ \Eprint
  {http://arxiv.org/abs/0804.3249} {arXiv:0804.3249 [astro-ph]} \BibitemShut
  {NoStop}%
\bibitem [{\citenamefont {Kuroyanagi}\ \emph
  {et~al.}(2011{\natexlab{a}})\citenamefont {Kuroyanagi}, \citenamefont
  {Chiba},\ and\ \citenamefont {Sugiyama}}]{Kuroyanagi:2010mm}%
  \BibitemOpen
  \bibfield  {author} {\bibinfo {author} {\bibfnamefont {S.}~\bibnamefont
  {Kuroyanagi}}, \bibinfo {author} {\bibfnamefont {T.}~\bibnamefont {Chiba}}, \
  and\ \bibinfo {author} {\bibfnamefont {N.}~\bibnamefont {Sugiyama}},\ }\href
  {\doibase 10.1103/PhysRevD.83.043514} {\bibfield  {journal} {\bibinfo
  {journal} {Phys. Rev. D}\ }\textbf {\bibinfo {volume} {83}},\ \bibinfo
  {pages} {043514} (\bibinfo {year} {2011}{\natexlab{a}})},\ \Eprint
  {http://arxiv.org/abs/1010.5246} {arXiv:1010.5246 [astro-ph.CO]} \BibitemShut
  {NoStop}%
\bibitem [{\citenamefont {Kuroyanagi}\ \emph
  {et~al.}(2011{\natexlab{b}})\citenamefont {Kuroyanagi}, \citenamefont
  {Nakayama},\ and\ \citenamefont {Saito}}]{Kuroyanagi:2011fy}%
  \BibitemOpen
  \bibfield  {author} {\bibinfo {author} {\bibfnamefont {S.}~\bibnamefont
  {Kuroyanagi}}, \bibinfo {author} {\bibfnamefont {K.}~\bibnamefont
  {Nakayama}}, \ and\ \bibinfo {author} {\bibfnamefont {S.}~\bibnamefont
  {Saito}},\ }\href {\doibase 10.1103/PhysRevD.84.123513} {\bibfield  {journal}
  {\bibinfo  {journal} {Phys. Rev. D}\ }\textbf {\bibinfo {volume} {84}},\
  \bibinfo {pages} {123513} (\bibinfo {year} {2011}{\natexlab{b}})},\ \Eprint
  {http://arxiv.org/abs/1110.4169} {arXiv:1110.4169 [astro-ph.CO]} \BibitemShut
  {NoStop}%
\bibitem [{\citenamefont {Giovannini}(2020)}]{Giovannini:2019oii}%
  \BibitemOpen
  \bibfield  {author} {\bibinfo {author} {\bibfnamefont {M.}~\bibnamefont
  {Giovannini}},\ }\href {\doibase 10.1016/j.ppnp.2020.103774} {\bibfield
  {journal} {\bibinfo  {journal} {Prog. Part. Nucl. Phys.}\ }\textbf {\bibinfo
  {volume} {112}},\ \bibinfo {pages} {103774} (\bibinfo {year} {2020})},\
  \Eprint {http://arxiv.org/abs/1912.07065} {arXiv:1912.07065 [astro-ph.CO]}
  \BibitemShut {NoStop}%
\bibitem [{\citenamefont {Odintsov}\ \emph
  {et~al.}(2022{\natexlab{a}})\citenamefont {Odintsov}, \citenamefont
  {Oikonomou},\ and\ \citenamefont {Myrzakulov}}]{Odintsov:2022cbm}%
  \BibitemOpen
  \bibfield  {author} {\bibinfo {author} {\bibfnamefont {S.~D.}\ \bibnamefont
  {Odintsov}}, \bibinfo {author} {\bibfnamefont {V.~K.}\ \bibnamefont
  {Oikonomou}}, \ and\ \bibinfo {author} {\bibfnamefont {R.}~\bibnamefont
  {Myrzakulov}},\ }\href {\doibase 10.3390/sym14040729} {\bibfield  {journal}
  {\bibinfo  {journal} {Symmetry}\ }\textbf {\bibinfo {volume} {14}},\ \bibinfo
  {pages} {729} (\bibinfo {year} {2022}{\natexlab{a}})},\ \Eprint
  {http://arxiv.org/abs/2204.00876} {arXiv:2204.00876 [gr-qc]} \BibitemShut
  {NoStop}%
\bibitem [{\citenamefont {Copeland}\ \emph {et~al.}(1993)\citenamefont
  {Copeland}, \citenamefont {Kolb}, \citenamefont {Liddle},\ and\ \citenamefont
  {Lidsey}}]{Copeland:1993ie}%
  \BibitemOpen
  \bibfield  {author} {\bibinfo {author} {\bibfnamefont {E.~J.}\ \bibnamefont
  {Copeland}}, \bibinfo {author} {\bibfnamefont {E.~W.}\ \bibnamefont {Kolb}},
  \bibinfo {author} {\bibfnamefont {A.~R.}\ \bibnamefont {Liddle}}, \ and\
  \bibinfo {author} {\bibfnamefont {J.~E.}\ \bibnamefont {Lidsey}},\ }\href
  {\doibase 10.1103/PhysRevLett.71.219} {\bibfield  {journal} {\bibinfo
  {journal} {Phys. Rev. Lett.}\ }\textbf {\bibinfo {volume} {71}},\ \bibinfo
  {pages} {219} (\bibinfo {year} {1993})},\ \Eprint
  {http://arxiv.org/abs/hep-ph/9304228} {arXiv:hep-ph/9304228} \BibitemShut
  {NoStop}%
\bibitem [{\citenamefont {Turner}\ \emph {et~al.}(1993)\citenamefont {Turner},
  \citenamefont {White},\ and\ \citenamefont {Lidsey}}]{Turner:1993vb}%
  \BibitemOpen
  \bibfield  {author} {\bibinfo {author} {\bibfnamefont {M.~S.}\ \bibnamefont
  {Turner}}, \bibinfo {author} {\bibfnamefont {M.~J.}\ \bibnamefont {White}}, \
  and\ \bibinfo {author} {\bibfnamefont {J.~E.}\ \bibnamefont {Lidsey}},\
  }\href {\doibase 10.1103/PhysRevD.48.4613} {\bibfield  {journal} {\bibinfo
  {journal} {Phys. Rev. D}\ }\textbf {\bibinfo {volume} {48}},\ \bibinfo
  {pages} {4613} (\bibinfo {year} {1993})},\ \Eprint
  {http://arxiv.org/abs/astro-ph/9306029} {arXiv:astro-ph/9306029} \BibitemShut
  {NoStop}%
\bibitem [{\citenamefont {Chongchitnan}\ and\ \citenamefont
  {Efstathiou}(2006)}]{Chongchitnan:2006pe}%
  \BibitemOpen
  \bibfield  {author} {\bibinfo {author} {\bibfnamefont {S.}~\bibnamefont
  {Chongchitnan}}\ and\ \bibinfo {author} {\bibfnamefont {G.}~\bibnamefont
  {Efstathiou}},\ }\href {\doibase 10.1103/PhysRevD.73.083511} {\bibfield
  {journal} {\bibinfo  {journal} {Phys. Rev. D}\ }\textbf {\bibinfo {volume}
  {73}},\ \bibinfo {pages} {083511} (\bibinfo {year} {2006})},\ \Eprint
  {http://arxiv.org/abs/astro-ph/0602594} {arXiv:astro-ph/0602594} \BibitemShut
  {NoStop}%
\bibitem [{\citenamefont {Watanabe}\ and\ \citenamefont
  {Komatsu}(2006)}]{Watanabe:2006qe}%
  \BibitemOpen
  \bibfield  {author} {\bibinfo {author} {\bibfnamefont {Y.}~\bibnamefont
  {Watanabe}}\ and\ \bibinfo {author} {\bibfnamefont {E.}~\bibnamefont
  {Komatsu}},\ }\href {\doibase 10.1103/PhysRevD.73.123515} {\bibfield
  {journal} {\bibinfo  {journal} {Phys. Rev. D}\ }\textbf {\bibinfo {volume}
  {73}},\ \bibinfo {pages} {123515} (\bibinfo {year} {2006})},\ \Eprint
  {http://arxiv.org/abs/astro-ph/0604176} {arXiv:astro-ph/0604176} \BibitemShut
  {NoStop}%
\bibitem [{\citenamefont {Zhao}\ and\ \citenamefont
  {Zhang}(2006)}]{Zhao:2006mm}%
  \BibitemOpen
  \bibfield  {author} {\bibinfo {author} {\bibfnamefont {W.}~\bibnamefont
  {Zhao}}\ and\ \bibinfo {author} {\bibfnamefont {Y.}~\bibnamefont {Zhang}},\
  }\href {\doibase 10.1103/PhysRevD.74.043503} {\bibfield  {journal} {\bibinfo
  {journal} {Phys. Rev. D}\ }\textbf {\bibinfo {volume} {74}},\ \bibinfo
  {pages} {043503} (\bibinfo {year} {2006})},\ \Eprint
  {http://arxiv.org/abs/astro-ph/0604458} {arXiv:astro-ph/0604458} \BibitemShut
  {NoStop}%
\bibitem [{\citenamefont {Giovannini}(2010)}]{Giovannini:2009kg}%
  \BibitemOpen
  \bibfield  {author} {\bibinfo {author} {\bibfnamefont {M.}~\bibnamefont
  {Giovannini}},\ }\href {\doibase 10.1186/1754-0410-4-1} {\bibfield  {journal}
  {\bibinfo  {journal} {PMC Phys. A}\ }\textbf {\bibinfo {volume} {4}},\
  \bibinfo {pages} {1} (\bibinfo {year} {2010})},\ \Eprint
  {http://arxiv.org/abs/0901.3026} {arXiv:0901.3026 [astro-ph.CO]} \BibitemShut
  {NoStop}%
\bibitem [{\citenamefont {Kite}\ \emph {et~al.}(2021)\citenamefont {Kite},
  \citenamefont {Chluba}, \citenamefont {Ravenni},\ and\ \citenamefont
  {Patil}}]{Kite:2021yoe}%
  \BibitemOpen
  \bibfield  {author} {\bibinfo {author} {\bibfnamefont {T.}~\bibnamefont
  {Kite}}, \bibinfo {author} {\bibfnamefont {J.}~\bibnamefont {Chluba}},
  \bibinfo {author} {\bibfnamefont {A.}~\bibnamefont {Ravenni}}, \ and\
  \bibinfo {author} {\bibfnamefont {S.~P.}\ \bibnamefont {Patil}},\ }\href
  {\doibase 10.1093/mnras/stab3125} {\bibfield  {journal} {\bibinfo  {journal}
  {Mon. Not. Roy. Astron. Soc.}\ }\textbf {\bibinfo {volume} {509}},\ \bibinfo
  {pages} {1366} (\bibinfo {year} {2021})},\ \Eprint
  {http://arxiv.org/abs/2107.13351} {arXiv:2107.13351 [astro-ph.CO]}
  \BibitemShut {NoStop}%
\bibitem [{\citenamefont {{Phinney}}(2001)}]{Phinney:2001di}%
  \BibitemOpen
  \bibfield  {author} {\bibinfo {author} {\bibfnamefont {E.~S.}\ \bibnamefont
  {{Phinney}}},\ }\href {\doibase 10.48550/arXiv.astro-ph/0108028} {\bibfield
  {journal} {\bibinfo  {journal} {arXiv e-prints}\ ,\ \bibinfo {eid}
  {astro-ph/0108028}} (\bibinfo {year} {2001})},\ \Eprint
  {http://arxiv.org/abs/astro-ph/0108028} {arXiv:astro-ph/0108028 [astro-ph]}
  \BibitemShut {NoStop}%
\bibitem [{\citenamefont {Aghanim}\ \emph {et~al.}(2020)\citenamefont {Aghanim}
  \emph {et~al.}}]{Planck:2018vyg}%
  \BibitemOpen
  \bibfield  {author} {\bibinfo {author} {\bibfnamefont {N.}~\bibnamefont
  {Aghanim}} \emph {et~al.} (\bibinfo {collaboration} {Planck}),\ }\href
  {\doibase 10.1051/0004-6361/201833910} {\bibfield  {journal} {\bibinfo
  {journal} {Astron. Astrophys.}\ }\textbf {\bibinfo {volume} {641}},\ \bibinfo
  {pages} {A6} (\bibinfo {year} {2020})},\ \bibinfo {note} {[Erratum:
  Astron.Astrophys. 652, C4 (2021)]},\ \Eprint
  {http://arxiv.org/abs/1807.06209} {arXiv:1807.06209 [astro-ph.CO]}
  \BibitemShut {NoStop}%
\bibitem [{\citenamefont {Di~Valentino}\ \emph {et~al.}(2021)\citenamefont
  {Di~Valentino}, \citenamefont {Mena}, \citenamefont {Pan}, \citenamefont
  {Visinelli}, \citenamefont {Yang}, \citenamefont {Melchiorri}, \citenamefont
  {Mota}, \citenamefont {Riess},\ and\ \citenamefont
  {Silk}}]{DiValentino:2021izs}%
  \BibitemOpen
  \bibfield  {author} {\bibinfo {author} {\bibfnamefont {E.}~\bibnamefont
  {Di~Valentino}}, \bibinfo {author} {\bibfnamefont {O.}~\bibnamefont {Mena}},
  \bibinfo {author} {\bibfnamefont {S.}~\bibnamefont {Pan}}, \bibinfo {author}
  {\bibfnamefont {L.}~\bibnamefont {Visinelli}}, \bibinfo {author}
  {\bibfnamefont {W.}~\bibnamefont {Yang}}, \bibinfo {author} {\bibfnamefont
  {A.}~\bibnamefont {Melchiorri}}, \bibinfo {author} {\bibfnamefont {D.~F.}\
  \bibnamefont {Mota}}, \bibinfo {author} {\bibfnamefont {A.~G.}\ \bibnamefont
  {Riess}}, \ and\ \bibinfo {author} {\bibfnamefont {J.}~\bibnamefont {Silk}},\
  }\href {\doibase 10.1088/1361-6382/ac086d} {\bibfield  {journal} {\bibinfo
  {journal} {Class. Quant. Grav.}\ }\textbf {\bibinfo {volume} {38}},\ \bibinfo
  {pages} {153001} (\bibinfo {year} {2021})},\ \Eprint
  {http://arxiv.org/abs/2103.01183} {arXiv:2103.01183 [astro-ph.CO]}
  \BibitemShut {NoStop}%
\bibitem [{\citenamefont {Mangano}\ \emph {et~al.}(2002)\citenamefont
  {Mangano}, \citenamefont {Miele}, \citenamefont {Pastor},\ and\ \citenamefont
  {Peloso}}]{Mangano:2001iu}%
  \BibitemOpen
  \bibfield  {author} {\bibinfo {author} {\bibfnamefont {G.}~\bibnamefont
  {Mangano}}, \bibinfo {author} {\bibfnamefont {G.}~\bibnamefont {Miele}},
  \bibinfo {author} {\bibfnamefont {S.}~\bibnamefont {Pastor}}, \ and\ \bibinfo
  {author} {\bibfnamefont {M.}~\bibnamefont {Peloso}},\ }\href {\doibase
  10.1016/S0370-2693(02)01622-2} {\bibfield  {journal} {\bibinfo  {journal}
  {Phys. Lett. B}\ }\textbf {\bibinfo {volume} {534}},\ \bibinfo {pages} {8}
  (\bibinfo {year} {2002})},\ \Eprint {http://arxiv.org/abs/astro-ph/0111408}
  {arXiv:astro-ph/0111408} \BibitemShut {NoStop}%
\bibitem [{\citenamefont {Mangano}\ \emph {et~al.}(2005)\citenamefont
  {Mangano}, \citenamefont {Miele}, \citenamefont {Pastor}, \citenamefont
  {Pinto}, \citenamefont {Pisanti},\ and\ \citenamefont
  {Serpico}}]{Mangano:2005cc}%
  \BibitemOpen
  \bibfield  {author} {\bibinfo {author} {\bibfnamefont {G.}~\bibnamefont
  {Mangano}}, \bibinfo {author} {\bibfnamefont {G.}~\bibnamefont {Miele}},
  \bibinfo {author} {\bibfnamefont {S.}~\bibnamefont {Pastor}}, \bibinfo
  {author} {\bibfnamefont {T.}~\bibnamefont {Pinto}}, \bibinfo {author}
  {\bibfnamefont {O.}~\bibnamefont {Pisanti}}, \ and\ \bibinfo {author}
  {\bibfnamefont {P.~D.}\ \bibnamefont {Serpico}},\ }\href {\doibase
  10.1016/j.nuclphysb.2005.09.041} {\bibfield  {journal} {\bibinfo  {journal}
  {Nucl. Phys. B}\ }\textbf {\bibinfo {volume} {729}},\ \bibinfo {pages} {221}
  (\bibinfo {year} {2005})},\ \Eprint {http://arxiv.org/abs/hep-ph/0506164}
  {arXiv:hep-ph/0506164} \BibitemShut {NoStop}%
\bibitem [{\citenamefont {Bennett}\ \emph {et~al.}(2020)\citenamefont
  {Bennett}, \citenamefont {Buldgen}, \citenamefont {Drewes},\ and\
  \citenamefont {Wong}}]{Bennett:2019ewm}%
  \BibitemOpen
  \bibfield  {author} {\bibinfo {author} {\bibfnamefont {J.~J.}\ \bibnamefont
  {Bennett}}, \bibinfo {author} {\bibfnamefont {G.}~\bibnamefont {Buldgen}},
  \bibinfo {author} {\bibfnamefont {M.}~\bibnamefont {Drewes}}, \ and\ \bibinfo
  {author} {\bibfnamefont {Y.~Y.~Y.}\ \bibnamefont {Wong}},\ }\href {\doibase
  10.1088/1475-7516/2020/03/003} {\bibfield  {journal} {\bibinfo  {journal}
  {JCAP}\ }\textbf {\bibinfo {volume} {03}},\ \bibinfo {pages} {003} (\bibinfo
  {year} {2020})},\ \bibinfo {note} {[Addendum: JCAP 03, A01 (2021)]},\ \Eprint
  {http://arxiv.org/abs/1911.04504} {arXiv:1911.04504 [hep-ph]} \BibitemShut
  {NoStop}%
\bibitem [{\citenamefont {Akita}\ and\ \citenamefont
  {Yamaguchi}(2020)}]{Akita:2020szl}%
  \BibitemOpen
  \bibfield  {author} {\bibinfo {author} {\bibfnamefont {K.}~\bibnamefont
  {Akita}}\ and\ \bibinfo {author} {\bibfnamefont {M.}~\bibnamefont
  {Yamaguchi}},\ }\href {\doibase 10.1088/1475-7516/2020/08/012} {\bibfield
  {journal} {\bibinfo  {journal} {JCAP}\ }\textbf {\bibinfo {volume} {08}},\
  \bibinfo {pages} {012} (\bibinfo {year} {2020})},\ \Eprint
  {http://arxiv.org/abs/2005.07047} {arXiv:2005.07047 [hep-ph]} \BibitemShut
  {NoStop}%
\bibitem [{\citenamefont {Froustey}\ \emph {et~al.}(2020)\citenamefont
  {Froustey}, \citenamefont {Pitrou},\ and\ \citenamefont
  {Volpe}}]{Froustey:2020mcq}%
  \BibitemOpen
  \bibfield  {author} {\bibinfo {author} {\bibfnamefont {J.}~\bibnamefont
  {Froustey}}, \bibinfo {author} {\bibfnamefont {C.}~\bibnamefont {Pitrou}}, \
  and\ \bibinfo {author} {\bibfnamefont {M.~C.}\ \bibnamefont {Volpe}},\ }\href
  {\doibase 10.1088/1475-7516/2020/12/015} {\bibfield  {journal} {\bibinfo
  {journal} {JCAP}\ }\textbf {\bibinfo {volume} {12}},\ \bibinfo {pages} {015}
  (\bibinfo {year} {2020})},\ \Eprint {http://arxiv.org/abs/2008.01074}
  {arXiv:2008.01074 [hep-ph]} \BibitemShut {NoStop}%
\bibitem [{\citenamefont {Bennett}\ \emph {et~al.}(2021)\citenamefont
  {Bennett}, \citenamefont {Buldgen}, \citenamefont {De~Salas}, \citenamefont
  {Drewes}, \citenamefont {Gariazzo}, \citenamefont {Pastor},\ and\
  \citenamefont {Wong}}]{Bennett:2020zkv}%
  \BibitemOpen
  \bibfield  {author} {\bibinfo {author} {\bibfnamefont {J.~J.}\ \bibnamefont
  {Bennett}}, \bibinfo {author} {\bibfnamefont {G.}~\bibnamefont {Buldgen}},
  \bibinfo {author} {\bibfnamefont {P.~F.}\ \bibnamefont {De~Salas}}, \bibinfo
  {author} {\bibfnamefont {M.}~\bibnamefont {Drewes}}, \bibinfo {author}
  {\bibfnamefont {S.}~\bibnamefont {Gariazzo}}, \bibinfo {author}
  {\bibfnamefont {S.}~\bibnamefont {Pastor}}, \ and\ \bibinfo {author}
  {\bibfnamefont {Y.~Y.~Y.}\ \bibnamefont {Wong}},\ }\href {\doibase
  10.1088/1475-7516/2021/04/073} {\bibfield  {journal} {\bibinfo  {journal}
  {JCAP}\ }\textbf {\bibinfo {volume} {04}},\ \bibinfo {pages} {073} (\bibinfo
  {year} {2021})},\ \Eprint {http://arxiv.org/abs/2012.02726} {arXiv:2012.02726
  [hep-ph]} \BibitemShut {NoStop}%
\bibitem [{\citenamefont {Cielo}\ \emph {et~al.}(2023)\citenamefont {Cielo},
  \citenamefont {Escudero}, \citenamefont {Mangano},\ and\ \citenamefont
  {Pisanti}}]{Cielo:2023bqp}%
  \BibitemOpen
  \bibfield  {author} {\bibinfo {author} {\bibfnamefont {M.}~\bibnamefont
  {Cielo}}, \bibinfo {author} {\bibfnamefont {M.}~\bibnamefont {Escudero}},
  \bibinfo {author} {\bibfnamefont {G.}~\bibnamefont {Mangano}}, \ and\
  \bibinfo {author} {\bibfnamefont {O.}~\bibnamefont {Pisanti}},\ }\href@noop
  {} {\  (\bibinfo {year} {2023})},\ \Eprint {http://arxiv.org/abs/2306.05460}
  {arXiv:2306.05460 [hep-ph]} \BibitemShut {NoStop}%
\bibitem [{\citenamefont {Allen}\ and\ \citenamefont
  {Romano}(1999)}]{Allen:1997ad}%
  \BibitemOpen
  \bibfield  {author} {\bibinfo {author} {\bibfnamefont {B.}~\bibnamefont
  {Allen}}\ and\ \bibinfo {author} {\bibfnamefont {J.~D.}\ \bibnamefont
  {Romano}},\ }\href {\doibase 10.1103/PhysRevD.59.102001} {\bibfield
  {journal} {\bibinfo  {journal} {Phys. Rev. D}\ }\textbf {\bibinfo {volume}
  {59}},\ \bibinfo {pages} {102001} (\bibinfo {year} {1999})},\ \Eprint
  {http://arxiv.org/abs/gr-qc/9710117} {arXiv:gr-qc/9710117} \BibitemShut
  {NoStop}%
\bibitem [{\citenamefont {Smith}\ \emph {et~al.}(2006)\citenamefont {Smith},
  \citenamefont {Pierpaoli},\ and\ \citenamefont
  {Kamionkowski}}]{Smith:2006nka}%
  \BibitemOpen
  \bibfield  {author} {\bibinfo {author} {\bibfnamefont {T.~L.}\ \bibnamefont
  {Smith}}, \bibinfo {author} {\bibfnamefont {E.}~\bibnamefont {Pierpaoli}}, \
  and\ \bibinfo {author} {\bibfnamefont {M.}~\bibnamefont {Kamionkowski}},\
  }\href {\doibase 10.1103/PhysRevLett.97.021301} {\bibfield  {journal}
  {\bibinfo  {journal} {Phys. Rev. Lett.}\ }\textbf {\bibinfo {volume} {97}},\
  \bibinfo {pages} {021301} (\bibinfo {year} {2006})},\ \Eprint
  {http://arxiv.org/abs/astro-ph/0603144} {arXiv:astro-ph/0603144} \BibitemShut
  {NoStop}%
\bibitem [{\citenamefont {Boyle}\ and\ \citenamefont
  {Buonanno}(2008)}]{Boyle:2007zx}%
  \BibitemOpen
  \bibfield  {author} {\bibinfo {author} {\bibfnamefont {L.~A.}\ \bibnamefont
  {Boyle}}\ and\ \bibinfo {author} {\bibfnamefont {A.}~\bibnamefont
  {Buonanno}},\ }\href {\doibase 10.1103/PhysRevD.78.043531} {\bibfield
  {journal} {\bibinfo  {journal} {Phys. Rev. D}\ }\textbf {\bibinfo {volume}
  {78}},\ \bibinfo {pages} {043531} (\bibinfo {year} {2008})},\ \Eprint
  {http://arxiv.org/abs/0708.2279} {arXiv:0708.2279 [astro-ph]} \BibitemShut
  {NoStop}%
\bibitem [{\citenamefont {Kuroyanagi}\ \emph {et~al.}(2015)\citenamefont
  {Kuroyanagi}, \citenamefont {Takahashi},\ and\ \citenamefont
  {Yokoyama}}]{Kuroyanagi:2014nba}%
  \BibitemOpen
  \bibfield  {author} {\bibinfo {author} {\bibfnamefont {S.}~\bibnamefont
  {Kuroyanagi}}, \bibinfo {author} {\bibfnamefont {T.}~\bibnamefont
  {Takahashi}}, \ and\ \bibinfo {author} {\bibfnamefont {S.}~\bibnamefont
  {Yokoyama}},\ }\href {\doibase 10.1088/1475-7516/2015/02/003} {\bibfield
  {journal} {\bibinfo  {journal} {JCAP}\ }\textbf {\bibinfo {volume} {02}},\
  \bibinfo {pages} {003} (\bibinfo {year} {2015})},\ \Eprint
  {http://arxiv.org/abs/1407.4785} {arXiv:1407.4785 [astro-ph.CO]} \BibitemShut
  {NoStop}%
\bibitem [{\citenamefont {Vagnozzi}\ and\ \citenamefont
  {Loeb}(2022)}]{Vagnozzi:2022qmc}%
  \BibitemOpen
  \bibfield  {author} {\bibinfo {author} {\bibfnamefont {S.}~\bibnamefont
  {Vagnozzi}}\ and\ \bibinfo {author} {\bibfnamefont {A.}~\bibnamefont
  {Loeb}},\ }\href {\doibase 10.3847/2041-8213/ac9b0e} {\bibfield  {journal}
  {\bibinfo  {journal} {Astrophys. J. Lett.}\ }\textbf {\bibinfo {volume}
  {939}},\ \bibinfo {pages} {L22} (\bibinfo {year} {2022})},\ \Eprint
  {http://arxiv.org/abs/2208.14088} {arXiv:2208.14088 [astro-ph.CO]}
  \BibitemShut {NoStop}%
\bibitem [{\citenamefont {Giar\`e}\ \emph
  {et~al.}(2023{\natexlab{a}})\citenamefont {Giar\`e}, \citenamefont {Forconi},
  \citenamefont {Di~Valentino},\ and\ \citenamefont
  {Melchiorri}}]{Giare:2022wxq}%
  \BibitemOpen
  \bibfield  {author} {\bibinfo {author} {\bibfnamefont {W.}~\bibnamefont
  {Giar\`e}}, \bibinfo {author} {\bibfnamefont {M.}~\bibnamefont {Forconi}},
  \bibinfo {author} {\bibfnamefont {E.}~\bibnamefont {Di~Valentino}}, \ and\
  \bibinfo {author} {\bibfnamefont {A.}~\bibnamefont {Melchiorri}},\ }\href
  {\doibase 10.1093/mnras/stad258} {\bibfield  {journal} {\bibinfo  {journal}
  {Mon. Not. Roy. Astron. Soc.}\ }\textbf {\bibinfo {volume} {520}},\ \bibinfo
  {pages} {2} (\bibinfo {year} {2023}{\natexlab{a}})},\ \Eprint
  {http://arxiv.org/abs/2210.14159} {arXiv:2210.14159 [astro-ph.CO]}
  \BibitemShut {NoStop}%
\bibitem [{\citenamefont {de~Salas}\ \emph {et~al.}(2015)\citenamefont
  {de~Salas}, \citenamefont {Lattanzi}, \citenamefont {Mangano}, \citenamefont
  {Miele}, \citenamefont {Pastor},\ and\ \citenamefont
  {Pisanti}}]{deSalas:2015glj}%
  \BibitemOpen
  \bibfield  {author} {\bibinfo {author} {\bibfnamefont {P.~F.}\ \bibnamefont
  {de~Salas}}, \bibinfo {author} {\bibfnamefont {M.}~\bibnamefont {Lattanzi}},
  \bibinfo {author} {\bibfnamefont {G.}~\bibnamefont {Mangano}}, \bibinfo
  {author} {\bibfnamefont {G.}~\bibnamefont {Miele}}, \bibinfo {author}
  {\bibfnamefont {S.}~\bibnamefont {Pastor}}, \ and\ \bibinfo {author}
  {\bibfnamefont {O.}~\bibnamefont {Pisanti}},\ }\href {\doibase
  10.1103/PhysRevD.92.123534} {\bibfield  {journal} {\bibinfo  {journal} {Phys.
  Rev. D}\ }\textbf {\bibinfo {volume} {92}},\ \bibinfo {pages} {123534}
  (\bibinfo {year} {2015})},\ \Eprint {http://arxiv.org/abs/1511.00672}
  {arXiv:1511.00672 [astro-ph.CO]} \BibitemShut {NoStop}%
\bibitem [{\citenamefont {Aver}\ \emph {et~al.}(2015)\citenamefont {Aver},
  \citenamefont {Olive},\ and\ \citenamefont {Skillman}}]{Aver:2015iza}%
  \BibitemOpen
  \bibfield  {author} {\bibinfo {author} {\bibfnamefont {E.}~\bibnamefont
  {Aver}}, \bibinfo {author} {\bibfnamefont {K.~A.}\ \bibnamefont {Olive}}, \
  and\ \bibinfo {author} {\bibfnamefont {E.~D.}\ \bibnamefont {Skillman}},\
  }\href {\doibase 10.1088/1475-7516/2015/07/011} {\bibfield  {journal}
  {\bibinfo  {journal} {JCAP}\ }\textbf {\bibinfo {volume} {07}},\ \bibinfo
  {pages} {011} (\bibinfo {year} {2015})},\ \Eprint
  {http://arxiv.org/abs/1503.08146} {arXiv:1503.08146 [astro-ph.CO]}
  \BibitemShut {NoStop}%
\bibitem [{\citenamefont {Cooke}\ \emph {et~al.}(2018)\citenamefont {Cooke},
  \citenamefont {Pettini},\ and\ \citenamefont {Steidel}}]{Cooke:2017cwo}%
  \BibitemOpen
  \bibfield  {author} {\bibinfo {author} {\bibfnamefont {R.~J.}\ \bibnamefont
  {Cooke}}, \bibinfo {author} {\bibfnamefont {M.}~\bibnamefont {Pettini}}, \
  and\ \bibinfo {author} {\bibfnamefont {C.~C.}\ \bibnamefont {Steidel}},\
  }\href {\doibase 10.3847/1538-4357/aaab53} {\bibfield  {journal} {\bibinfo
  {journal} {Astrophys. J.}\ }\textbf {\bibinfo {volume} {855}},\ \bibinfo
  {pages} {102} (\bibinfo {year} {2018})},\ \Eprint
  {http://arxiv.org/abs/1710.11129} {arXiv:1710.11129 [astro-ph.CO]}
  \BibitemShut {NoStop}%
\bibitem [{\citenamefont {Vagnozzi}(2020)}]{Vagnozzi:2019ezj}%
  \BibitemOpen
  \bibfield  {author} {\bibinfo {author} {\bibfnamefont {S.}~\bibnamefont
  {Vagnozzi}},\ }\href {\doibase 10.1103/PhysRevD.102.023518} {\bibfield
  {journal} {\bibinfo  {journal} {Phys. Rev. D}\ }\textbf {\bibinfo {volume}
  {102}},\ \bibinfo {pages} {023518} (\bibinfo {year} {2020})},\ \Eprint
  {http://arxiv.org/abs/1907.07569} {arXiv:1907.07569 [astro-ph.CO]}
  \BibitemShut {NoStop}%
\bibitem [{\citenamefont {Hsyu}\ \emph {et~al.}(2020)\citenamefont {Hsyu},
  \citenamefont {Cooke}, \citenamefont {Prochaska},\ and\ \citenamefont
  {Bolte}}]{Hsyu:2020uqb}%
  \BibitemOpen
  \bibfield  {author} {\bibinfo {author} {\bibfnamefont {T.}~\bibnamefont
  {Hsyu}}, \bibinfo {author} {\bibfnamefont {R.~J.}\ \bibnamefont {Cooke}},
  \bibinfo {author} {\bibfnamefont {J.~X.}\ \bibnamefont {Prochaska}}, \ and\
  \bibinfo {author} {\bibfnamefont {M.}~\bibnamefont {Bolte}},\ }\href
  {\doibase 10.3847/1538-4357/ab91af} {\bibfield  {journal} {\bibinfo
  {journal} {Astrophys. J.}\ }\textbf {\bibinfo {volume} {896}},\ \bibinfo
  {pages} {77} (\bibinfo {year} {2020})},\ \Eprint
  {http://arxiv.org/abs/2005.12290} {arXiv:2005.12290 [astro-ph.GA]}
  \BibitemShut {NoStop}%
\bibitem [{\citenamefont {Aiola}\ \emph {et~al.}(2020)\citenamefont {Aiola}
  \emph {et~al.}}]{ACT:2020gnv}%
  \BibitemOpen
  \bibfield  {author} {\bibinfo {author} {\bibfnamefont {S.}~\bibnamefont
  {Aiola}} \emph {et~al.} (\bibinfo {collaboration} {ACT}),\ }\href {\doibase
  10.1088/1475-7516/2020/12/047} {\bibfield  {journal} {\bibinfo  {journal}
  {JCAP}\ }\textbf {\bibinfo {volume} {12}},\ \bibinfo {pages} {047} (\bibinfo
  {year} {2020})},\ \Eprint {http://arxiv.org/abs/2007.07288} {arXiv:2007.07288
  [astro-ph.CO]} \BibitemShut {NoStop}%
\bibitem [{\citenamefont {Mossa}\ \emph {et~al.}(2020)\citenamefont {Mossa}
  \emph {et~al.}}]{Mossa:2020gjc}%
  \BibitemOpen
  \bibfield  {author} {\bibinfo {author} {\bibfnamefont {V.}~\bibnamefont
  {Mossa}} \emph {et~al.},\ }\href {\doibase 10.1038/s41586-020-2878-4}
  {\bibfield  {journal} {\bibinfo  {journal} {Nature}\ }\textbf {\bibinfo
  {volume} {587}},\ \bibinfo {pages} {210} (\bibinfo {year}
  {2020})}\BibitemShut {NoStop}%
\bibitem [{\citenamefont {Kuroyanagi}\ \emph {et~al.}(2021)\citenamefont
  {Kuroyanagi}, \citenamefont {Takahashi},\ and\ \citenamefont
  {Yokoyama}}]{Kuroyanagi:2020sfw}%
  \BibitemOpen
  \bibfield  {author} {\bibinfo {author} {\bibfnamefont {S.}~\bibnamefont
  {Kuroyanagi}}, \bibinfo {author} {\bibfnamefont {T.}~\bibnamefont
  {Takahashi}}, \ and\ \bibinfo {author} {\bibfnamefont {S.}~\bibnamefont
  {Yokoyama}},\ }\href {\doibase 10.1088/1475-7516/2021/01/071} {\bibfield
  {journal} {\bibinfo  {journal} {JCAP}\ }\textbf {\bibinfo {volume} {01}},\
  \bibinfo {pages} {071} (\bibinfo {year} {2021})},\ \Eprint
  {http://arxiv.org/abs/2011.03323} {arXiv:2011.03323 [astro-ph.CO]}
  \BibitemShut {NoStop}%
\bibitem [{\citenamefont {Vagnozzi}\ \emph {et~al.}(2017)\citenamefont
  {Vagnozzi}, \citenamefont {Giusarma}, \citenamefont {Mena}, \citenamefont
  {Freese}, \citenamefont {Gerbino}, \citenamefont {Ho},\ and\ \citenamefont
  {Lattanzi}}]{Vagnozzi:2017ovm}%
  \BibitemOpen
  \bibfield  {author} {\bibinfo {author} {\bibfnamefont {S.}~\bibnamefont
  {Vagnozzi}}, \bibinfo {author} {\bibfnamefont {E.}~\bibnamefont {Giusarma}},
  \bibinfo {author} {\bibfnamefont {O.}~\bibnamefont {Mena}}, \bibinfo {author}
  {\bibfnamefont {K.}~\bibnamefont {Freese}}, \bibinfo {author} {\bibfnamefont
  {M.}~\bibnamefont {Gerbino}}, \bibinfo {author} {\bibfnamefont
  {S.}~\bibnamefont {Ho}}, \ and\ \bibinfo {author} {\bibfnamefont
  {M.}~\bibnamefont {Lattanzi}},\ }\href {\doibase 10.1103/PhysRevD.96.123503}
  {\bibfield  {journal} {\bibinfo  {journal} {Phys. Rev. D}\ }\textbf {\bibinfo
  {volume} {96}},\ \bibinfo {pages} {123503} (\bibinfo {year} {2017})},\
  \Eprint {http://arxiv.org/abs/1701.08172} {arXiv:1701.08172 [astro-ph.CO]}
  \BibitemShut {NoStop}%
\bibitem [{\citenamefont {Vagnozzi}\ \emph {et~al.}(2018)\citenamefont
  {Vagnozzi}, \citenamefont {Dhawan}, \citenamefont {Gerbino}, \citenamefont
  {Freese}, \citenamefont {Goobar},\ and\ \citenamefont
  {Mena}}]{Vagnozzi:2018jhn}%
  \BibitemOpen
  \bibfield  {author} {\bibinfo {author} {\bibfnamefont {S.}~\bibnamefont
  {Vagnozzi}}, \bibinfo {author} {\bibfnamefont {S.}~\bibnamefont {Dhawan}},
  \bibinfo {author} {\bibfnamefont {M.}~\bibnamefont {Gerbino}}, \bibinfo
  {author} {\bibfnamefont {K.}~\bibnamefont {Freese}}, \bibinfo {author}
  {\bibfnamefont {A.}~\bibnamefont {Goobar}}, \ and\ \bibinfo {author}
  {\bibfnamefont {O.}~\bibnamefont {Mena}},\ }\href {\doibase
  10.1103/PhysRevD.98.083501} {\bibfield  {journal} {\bibinfo  {journal} {Phys.
  Rev. D}\ }\textbf {\bibinfo {volume} {98}},\ \bibinfo {pages} {083501}
  (\bibinfo {year} {2018})},\ \Eprint {http://arxiv.org/abs/1801.08553}
  {arXiv:1801.08553 [astro-ph.CO]} \BibitemShut {NoStop}%
\bibitem [{\citenamefont {Roy~Choudhury}\ and\ \citenamefont
  {Hannestad}(2020)}]{RoyChoudhury:2019hls}%
  \BibitemOpen
  \bibfield  {author} {\bibinfo {author} {\bibfnamefont {S.}~\bibnamefont
  {Roy~Choudhury}}\ and\ \bibinfo {author} {\bibfnamefont {S.}~\bibnamefont
  {Hannestad}},\ }\href {\doibase 10.1088/1475-7516/2020/07/037} {\bibfield
  {journal} {\bibinfo  {journal} {JCAP}\ }\textbf {\bibinfo {volume} {07}},\
  \bibinfo {pages} {037} (\bibinfo {year} {2020})},\ \Eprint
  {http://arxiv.org/abs/1907.12598} {arXiv:1907.12598 [astro-ph.CO]}
  \BibitemShut {NoStop}%
\bibitem [{\citenamefont {Giar\`e}\ \emph
  {et~al.}(2021{\natexlab{a}})\citenamefont {Giar\`e}, \citenamefont
  {Di~Valentino}, \citenamefont {Melchiorri},\ and\ \citenamefont
  {Mena}}]{Giare:2020vzo}%
  \BibitemOpen
  \bibfield  {author} {\bibinfo {author} {\bibfnamefont {W.}~\bibnamefont
  {Giar\`e}}, \bibinfo {author} {\bibfnamefont {E.}~\bibnamefont
  {Di~Valentino}}, \bibinfo {author} {\bibfnamefont {A.}~\bibnamefont
  {Melchiorri}}, \ and\ \bibinfo {author} {\bibfnamefont {O.}~\bibnamefont
  {Mena}},\ }\href {\doibase 10.1093/mnras/stab1442} {\bibfield  {journal}
  {\bibinfo  {journal} {Mon. Not. Roy. Astron. Soc.}\ }\textbf {\bibinfo
  {volume} {505}},\ \bibinfo {pages} {2703} (\bibinfo {year}
  {2021}{\natexlab{a}})},\ \Eprint {http://arxiv.org/abs/2011.14704}
  {arXiv:2011.14704 [astro-ph.CO]} \BibitemShut {NoStop}%
\bibitem [{\citenamefont {Gariazzo}\ \emph {et~al.}(2022)\citenamefont
  {Gariazzo} \emph {et~al.}}]{Gariazzo:2022ahe}%
  \BibitemOpen
  \bibfield  {author} {\bibinfo {author} {\bibfnamefont {S.}~\bibnamefont
  {Gariazzo}} \emph {et~al.},\ }\href {\doibase 10.1088/1475-7516/2022/10/010}
  {\bibfield  {journal} {\bibinfo  {journal} {JCAP}\ }\textbf {\bibinfo
  {volume} {10}},\ \bibinfo {pages} {010} (\bibinfo {year} {2022})},\ \Eprint
  {http://arxiv.org/abs/2205.02195} {arXiv:2205.02195 [hep-ph]} \BibitemShut
  {NoStop}%
\bibitem [{\citenamefont {Alam}\ \emph
  {et~al.}(2021{\natexlab{a}})\citenamefont {Alam} \emph
  {et~al.}}]{NANOGrav:2020gpb}%
  \BibitemOpen
  \bibfield  {author} {\bibinfo {author} {\bibfnamefont {M.~F.}\ \bibnamefont
  {Alam}} \emph {et~al.} (\bibinfo {collaboration} {NANOGrav}),\ }\href
  {\doibase 10.3847/1538-4365/abc6a0} {\bibfield  {journal} {\bibinfo
  {journal} {Astrophys. J. Suppl.}\ }\textbf {\bibinfo {volume} {252}},\
  \bibinfo {pages} {4} (\bibinfo {year} {2021}{\natexlab{a}})},\ \Eprint
  {http://arxiv.org/abs/2005.06490} {arXiv:2005.06490 [astro-ph.HE]}
  \BibitemShut {NoStop}%
\bibitem [{\citenamefont {Alam}\ \emph
  {et~al.}(2021{\natexlab{b}})\citenamefont {Alam} \emph
  {et~al.}}]{NANOGrav:2020qll}%
  \BibitemOpen
  \bibfield  {author} {\bibinfo {author} {\bibfnamefont {M.~F.}\ \bibnamefont
  {Alam}} \emph {et~al.} (\bibinfo {collaboration} {NANOGrav}),\ }\href
  {\doibase 10.3847/1538-4365/abc6a1} {\bibfield  {journal} {\bibinfo
  {journal} {Astrophys. J. Suppl.}\ }\textbf {\bibinfo {volume} {252}},\
  \bibinfo {pages} {5} (\bibinfo {year} {2021}{\natexlab{b}})},\ \Eprint
  {http://arxiv.org/abs/2005.06495} {arXiv:2005.06495 [astro-ph.HE]}
  \BibitemShut {NoStop}%
\bibitem [{\citenamefont {{Park}}\ \emph {et~al.}(2021)\citenamefont {{Park}},
  \citenamefont {{Folkner}}, \citenamefont {{Williams}},\ and\ \citenamefont
  {{Boggs}}}]{Park:2021ghw}%
  \BibitemOpen
  \bibfield  {author} {\bibinfo {author} {\bibfnamefont {R.~S.}\ \bibnamefont
  {{Park}}}, \bibinfo {author} {\bibfnamefont {W.~M.}\ \bibnamefont
  {{Folkner}}}, \bibinfo {author} {\bibfnamefont {J.~G.}\ \bibnamefont
  {{Williams}}}, \ and\ \bibinfo {author} {\bibfnamefont {D.~H.}\ \bibnamefont
  {{Boggs}}},\ }\href {\doibase 10.3847/1538-3881/abd414} {\bibfield  {journal}
  {\bibinfo  {journal} {Astron. J.}\ }\textbf {\bibinfo {volume} {161}},\
  \bibinfo {eid} {105} (\bibinfo {year} {2021})}\BibitemShut {NoStop}%
\bibitem [{\citenamefont {Visinelli}\ and\ \citenamefont
  {Vagnozzi}(2019)}]{Visinelli:2018utg}%
  \BibitemOpen
  \bibfield  {author} {\bibinfo {author} {\bibfnamefont {L.}~\bibnamefont
  {Visinelli}}\ and\ \bibinfo {author} {\bibfnamefont {S.}~\bibnamefont
  {Vagnozzi}},\ }\href {\doibase 10.1103/PhysRevD.99.063517} {\bibfield
  {journal} {\bibinfo  {journal} {Phys. Rev. D}\ }\textbf {\bibinfo {volume}
  {99}},\ \bibinfo {pages} {063517} (\bibinfo {year} {2019})},\ \Eprint
  {http://arxiv.org/abs/1809.06382} {arXiv:1809.06382 [hep-ph]} \BibitemShut
  {NoStop}%
\bibitem [{\citenamefont {Giar\`e}\ \emph
  {et~al.}(2023{\natexlab{b}})\citenamefont {Giar\`e}, \citenamefont
  {De~Angelis}, \citenamefont {van~de Bruck},\ and\ \citenamefont
  {Di~Valentino}}]{Giare:2023kiv}%
  \BibitemOpen
  \bibfield  {author} {\bibinfo {author} {\bibfnamefont {W.}~\bibnamefont
  {Giar\`e}}, \bibinfo {author} {\bibfnamefont {M.}~\bibnamefont {De~Angelis}},
  \bibinfo {author} {\bibfnamefont {C.}~\bibnamefont {van~de Bruck}}, \ and\
  \bibinfo {author} {\bibfnamefont {E.}~\bibnamefont {Di~Valentino}},\
  }\href@noop {} {\  (\bibinfo {year} {2023}{\natexlab{b}})},\ \Eprint
  {http://arxiv.org/abs/2306.12414} {arXiv:2306.12414 [astro-ph.CO]}
  \BibitemShut {NoStop}%
\bibitem [{\citenamefont {Audren}\ \emph {et~al.}(2013)\citenamefont {Audren},
  \citenamefont {Lesgourgues}, \citenamefont {Benabed},\ and\ \citenamefont
  {Prunet}}]{Audren:2012wb}%
  \BibitemOpen
  \bibfield  {author} {\bibinfo {author} {\bibfnamefont {B.}~\bibnamefont
  {Audren}}, \bibinfo {author} {\bibfnamefont {J.}~\bibnamefont {Lesgourgues}},
  \bibinfo {author} {\bibfnamefont {K.}~\bibnamefont {Benabed}}, \ and\
  \bibinfo {author} {\bibfnamefont {S.}~\bibnamefont {Prunet}},\ }\href
  {\doibase 10.1088/1475-7516/2013/02/001} {\bibfield  {journal} {\bibinfo
  {journal} {JCAP}\ }\textbf {\bibinfo {volume} {02}},\ \bibinfo {pages} {001}
  (\bibinfo {year} {2013})},\ \Eprint {http://arxiv.org/abs/1210.7183}
  {arXiv:1210.7183 [astro-ph.CO]} \BibitemShut {NoStop}%
\bibitem [{\citenamefont {Brinckmann}\ and\ \citenamefont
  {Lesgourgues}(2019)}]{Brinckmann:2018cvx}%
  \BibitemOpen
  \bibfield  {author} {\bibinfo {author} {\bibfnamefont {T.}~\bibnamefont
  {Brinckmann}}\ and\ \bibinfo {author} {\bibfnamefont {J.}~\bibnamefont
  {Lesgourgues}},\ }\href {\doibase 10.1016/j.dark.2018.100260} {\bibfield
  {journal} {\bibinfo  {journal} {Phys. Dark Univ.}\ }\textbf {\bibinfo
  {volume} {24}},\ \bibinfo {pages} {100260} (\bibinfo {year} {2019})},\
  \Eprint {http://arxiv.org/abs/1804.07261} {arXiv:1804.07261 [astro-ph.CO]}
  \BibitemShut {NoStop}%
\bibitem [{\citenamefont {Ade}\ \emph {et~al.}(2021)\citenamefont {Ade} \emph
  {et~al.}}]{BICEP:2021xfz}%
  \BibitemOpen
  \bibfield  {author} {\bibinfo {author} {\bibfnamefont {P.~A.~R.}\
  \bibnamefont {Ade}} \emph {et~al.} (\bibinfo {collaboration} {BICEP, Keck}),\
  }\href {\doibase 10.1103/PhysRevLett.127.151301} {\bibfield  {journal}
  {\bibinfo  {journal} {Phys. Rev. Lett.}\ }\textbf {\bibinfo {volume} {127}},\
  \bibinfo {pages} {151301} (\bibinfo {year} {2021})},\ \Eprint
  {http://arxiv.org/abs/2110.00483} {arXiv:2110.00483 [astro-ph.CO]}
  \BibitemShut {NoStop}%
\bibitem [{\citenamefont {Gelman}\ and\ \citenamefont
  {Rubin}(1992)}]{Gelman:1992zz}%
  \BibitemOpen
  \bibfield  {author} {\bibinfo {author} {\bibfnamefont {A.}~\bibnamefont
  {Gelman}}\ and\ \bibinfo {author} {\bibfnamefont {D.~B.}\ \bibnamefont
  {Rubin}},\ }\href {\doibase 10.1214/ss/1177011136} {\bibfield  {journal}
  {\bibinfo  {journal} {Statist. Sci.}\ }\textbf {\bibinfo {volume} {7}},\
  \bibinfo {pages} {457} (\bibinfo {year} {1992})}\BibitemShut {NoStop}%
\bibitem [{\citenamefont {Ellis}\ and\ \citenamefont
  {Lewicki}(2021)}]{Ellis:2020ena}%
  \BibitemOpen
  \bibfield  {author} {\bibinfo {author} {\bibfnamefont {J.}~\bibnamefont
  {Ellis}}\ and\ \bibinfo {author} {\bibfnamefont {M.}~\bibnamefont
  {Lewicki}},\ }\href {\doibase 10.1103/PhysRevLett.126.041304} {\bibfield
  {journal} {\bibinfo  {journal} {Phys. Rev. Lett.}\ }\textbf {\bibinfo
  {volume} {126}},\ \bibinfo {pages} {041304} (\bibinfo {year} {2021})},\
  \Eprint {http://arxiv.org/abs/2009.06555} {arXiv:2009.06555 [astro-ph.CO]}
  \BibitemShut {NoStop}%
\bibitem [{\citenamefont {Blasi}\ \emph {et~al.}(2021)\citenamefont {Blasi},
  \citenamefont {Brdar},\ and\ \citenamefont {Schmitz}}]{Blasi:2020mfx}%
  \BibitemOpen
  \bibfield  {author} {\bibinfo {author} {\bibfnamefont {S.}~\bibnamefont
  {Blasi}}, \bibinfo {author} {\bibfnamefont {V.}~\bibnamefont {Brdar}}, \ and\
  \bibinfo {author} {\bibfnamefont {K.}~\bibnamefont {Schmitz}},\ }\href
  {\doibase 10.1103/PhysRevLett.126.041305} {\bibfield  {journal} {\bibinfo
  {journal} {Phys. Rev. Lett.}\ }\textbf {\bibinfo {volume} {126}},\ \bibinfo
  {pages} {041305} (\bibinfo {year} {2021})},\ \Eprint
  {http://arxiv.org/abs/2009.06607} {arXiv:2009.06607 [astro-ph.CO]}
  \BibitemShut {NoStop}%
\bibitem [{\citenamefont {Vaskonen}\ and\ \citenamefont
  {Veerm\"ae}(2021)}]{Vaskonen:2020lbd}%
  \BibitemOpen
  \bibfield  {author} {\bibinfo {author} {\bibfnamefont {V.}~\bibnamefont
  {Vaskonen}}\ and\ \bibinfo {author} {\bibfnamefont {H.}~\bibnamefont
  {Veerm\"ae}},\ }\href {\doibase 10.1103/PhysRevLett.126.051303} {\bibfield
  {journal} {\bibinfo  {journal} {Phys. Rev. Lett.}\ }\textbf {\bibinfo
  {volume} {126}},\ \bibinfo {pages} {051303} (\bibinfo {year} {2021})},\
  \Eprint {http://arxiv.org/abs/2009.07832} {arXiv:2009.07832 [astro-ph.CO]}
  \BibitemShut {NoStop}%
\bibitem [{\citenamefont {De~Luca}\ \emph {et~al.}(2021)\citenamefont
  {De~Luca}, \citenamefont {Franciolini},\ and\ \citenamefont
  {Riotto}}]{DeLuca:2020agl}%
  \BibitemOpen
  \bibfield  {author} {\bibinfo {author} {\bibfnamefont {V.}~\bibnamefont
  {De~Luca}}, \bibinfo {author} {\bibfnamefont {G.}~\bibnamefont
  {Franciolini}}, \ and\ \bibinfo {author} {\bibfnamefont {A.}~\bibnamefont
  {Riotto}},\ }\href {\doibase 10.1103/PhysRevLett.126.041303} {\bibfield
  {journal} {\bibinfo  {journal} {Phys. Rev. Lett.}\ }\textbf {\bibinfo
  {volume} {126}},\ \bibinfo {pages} {041303} (\bibinfo {year} {2021})},\
  \Eprint {http://arxiv.org/abs/2009.08268} {arXiv:2009.08268 [astro-ph.CO]}
  \BibitemShut {NoStop}%
\bibitem [{\citenamefont {Buchmuller}\ \emph {et~al.}(2020)\citenamefont
  {Buchmuller}, \citenamefont {Domcke},\ and\ \citenamefont
  {Schmitz}}]{Buchmuller:2020lbh}%
  \BibitemOpen
  \bibfield  {author} {\bibinfo {author} {\bibfnamefont {W.}~\bibnamefont
  {Buchmuller}}, \bibinfo {author} {\bibfnamefont {V.}~\bibnamefont {Domcke}},
  \ and\ \bibinfo {author} {\bibfnamefont {K.}~\bibnamefont {Schmitz}},\ }\href
  {\doibase 10.1016/j.physletb.2020.135914} {\bibfield  {journal} {\bibinfo
  {journal} {Phys. Lett. B}\ }\textbf {\bibinfo {volume} {811}},\ \bibinfo
  {pages} {135914} (\bibinfo {year} {2020})},\ \Eprint
  {http://arxiv.org/abs/2009.10649} {arXiv:2009.10649 [astro-ph.CO]}
  \BibitemShut {NoStop}%
\bibitem [{\citenamefont {Nakai}\ \emph {et~al.}(2021)\citenamefont {Nakai},
  \citenamefont {Suzuki}, \citenamefont {Takahashi},\ and\ \citenamefont
  {Yamada}}]{Nakai:2020oit}%
  \BibitemOpen
  \bibfield  {author} {\bibinfo {author} {\bibfnamefont {Y.}~\bibnamefont
  {Nakai}}, \bibinfo {author} {\bibfnamefont {M.}~\bibnamefont {Suzuki}},
  \bibinfo {author} {\bibfnamefont {F.}~\bibnamefont {Takahashi}}, \ and\
  \bibinfo {author} {\bibfnamefont {M.}~\bibnamefont {Yamada}},\ }\href
  {\doibase 10.1016/j.physletb.2021.136238} {\bibfield  {journal} {\bibinfo
  {journal} {Phys. Lett. B}\ }\textbf {\bibinfo {volume} {816}},\ \bibinfo
  {pages} {136238} (\bibinfo {year} {2021})},\ \Eprint
  {http://arxiv.org/abs/2009.09754} {arXiv:2009.09754 [astro-ph.CO]}
  \BibitemShut {NoStop}%
\bibitem [{\citenamefont {Addazi}\ \emph {et~al.}(2021)\citenamefont {Addazi},
  \citenamefont {Cai}, \citenamefont {Gan}, \citenamefont {Marciano},\ and\
  \citenamefont {Zeng}}]{Addazi:2020zcj}%
  \BibitemOpen
  \bibfield  {author} {\bibinfo {author} {\bibfnamefont {A.}~\bibnamefont
  {Addazi}}, \bibinfo {author} {\bibfnamefont {Y.-F.}\ \bibnamefont {Cai}},
  \bibinfo {author} {\bibfnamefont {Q.}~\bibnamefont {Gan}}, \bibinfo {author}
  {\bibfnamefont {A.}~\bibnamefont {Marciano}}, \ and\ \bibinfo {author}
  {\bibfnamefont {K.}~\bibnamefont {Zeng}},\ }\href {\doibase
  10.1007/s11433-021-1724-6} {\bibfield  {journal} {\bibinfo  {journal} {Sci.
  China Phys. Mech. Astron.}\ }\textbf {\bibinfo {volume} {64}},\ \bibinfo
  {pages} {290411} (\bibinfo {year} {2021})},\ \Eprint
  {http://arxiv.org/abs/2009.10327} {arXiv:2009.10327 [hep-ph]} \BibitemShut
  {NoStop}%
\bibitem [{\citenamefont {Kohri}\ and\ \citenamefont
  {Terada}(2021)}]{Kohri:2020qqd}%
  \BibitemOpen
  \bibfield  {author} {\bibinfo {author} {\bibfnamefont {K.}~\bibnamefont
  {Kohri}}\ and\ \bibinfo {author} {\bibfnamefont {T.}~\bibnamefont {Terada}},\
  }\href {\doibase 10.1016/j.physletb.2020.136040} {\bibfield  {journal}
  {\bibinfo  {journal} {Phys. Lett. B}\ }\textbf {\bibinfo {volume} {813}},\
  \bibinfo {pages} {136040} (\bibinfo {year} {2021})},\ \Eprint
  {http://arxiv.org/abs/2009.11853} {arXiv:2009.11853 [astro-ph.CO]}
  \BibitemShut {NoStop}%
\bibitem [{\citenamefont {Ratzinger}\ and\ \citenamefont
  {Schwaller}(2021)}]{Ratzinger:2020koh}%
  \BibitemOpen
  \bibfield  {author} {\bibinfo {author} {\bibfnamefont {W.}~\bibnamefont
  {Ratzinger}}\ and\ \bibinfo {author} {\bibfnamefont {P.}~\bibnamefont
  {Schwaller}},\ }\href {\doibase 10.21468/SciPostPhys.10.2.047} {\bibfield
  {journal} {\bibinfo  {journal} {SciPost Phys.}\ }\textbf {\bibinfo {volume}
  {10}},\ \bibinfo {pages} {047} (\bibinfo {year} {2021})},\ \Eprint
  {http://arxiv.org/abs/2009.11875} {arXiv:2009.11875 [astro-ph.CO]}
  \BibitemShut {NoStop}%
\bibitem [{\citenamefont {Samanta}\ and\ \citenamefont
  {Datta}(2021)}]{Samanta:2020cdk}%
  \BibitemOpen
  \bibfield  {author} {\bibinfo {author} {\bibfnamefont {R.}~\bibnamefont
  {Samanta}}\ and\ \bibinfo {author} {\bibfnamefont {S.}~\bibnamefont
  {Datta}},\ }\href {\doibase 10.1007/JHEP05(2021)211} {\bibfield  {journal}
  {\bibinfo  {journal} {JHEP}\ }\textbf {\bibinfo {volume} {05}},\ \bibinfo
  {pages} {211} (\bibinfo {year} {2021})},\ \Eprint
  {http://arxiv.org/abs/2009.13452} {arXiv:2009.13452 [hep-ph]} \BibitemShut
  {NoStop}%
\bibitem [{\citenamefont {Bian}\ \emph {et~al.}(2021)\citenamefont {Bian},
  \citenamefont {Cai}, \citenamefont {Liu}, \citenamefont {Yang},\ and\
  \citenamefont {Zhou}}]{Bian:2020urb}%
  \BibitemOpen
  \bibfield  {author} {\bibinfo {author} {\bibfnamefont {L.}~\bibnamefont
  {Bian}}, \bibinfo {author} {\bibfnamefont {R.-G.}\ \bibnamefont {Cai}},
  \bibinfo {author} {\bibfnamefont {J.}~\bibnamefont {Liu}}, \bibinfo {author}
  {\bibfnamefont {X.-Y.}\ \bibnamefont {Yang}}, \ and\ \bibinfo {author}
  {\bibfnamefont {R.}~\bibnamefont {Zhou}},\ }\href {\doibase
  10.1103/PhysRevD.103.L081301} {\bibfield  {journal} {\bibinfo  {journal}
  {Phys. Rev. D}\ }\textbf {\bibinfo {volume} {103}},\ \bibinfo {pages}
  {L081301} (\bibinfo {year} {2021})},\ \Eprint
  {http://arxiv.org/abs/2009.13893} {arXiv:2009.13893 [astro-ph.CO]}
  \BibitemShut {NoStop}%
\bibitem [{\citenamefont {Namba}\ and\ \citenamefont
  {Suzuki}(2020)}]{Namba:2020kij}%
  \BibitemOpen
  \bibfield  {author} {\bibinfo {author} {\bibfnamefont {R.}~\bibnamefont
  {Namba}}\ and\ \bibinfo {author} {\bibfnamefont {M.}~\bibnamefont {Suzuki}},\
  }\href {\doibase 10.1103/PhysRevD.102.123527} {\bibfield  {journal} {\bibinfo
   {journal} {Phys. Rev. D}\ }\textbf {\bibinfo {volume} {102}},\ \bibinfo
  {pages} {123527} (\bibinfo {year} {2020})},\ \Eprint
  {http://arxiv.org/abs/2009.13909} {arXiv:2009.13909 [astro-ph.CO]}
  \BibitemShut {NoStop}%
\bibitem [{\citenamefont {Neronov}\ \emph {et~al.}(2021)\citenamefont
  {Neronov}, \citenamefont {Roper~Pol}, \citenamefont {Caprini},\ and\
  \citenamefont {Semikoz}}]{Neronov:2020qrl}%
  \BibitemOpen
  \bibfield  {author} {\bibinfo {author} {\bibfnamefont {A.}~\bibnamefont
  {Neronov}}, \bibinfo {author} {\bibfnamefont {A.}~\bibnamefont {Roper~Pol}},
  \bibinfo {author} {\bibfnamefont {C.}~\bibnamefont {Caprini}}, \ and\
  \bibinfo {author} {\bibfnamefont {D.}~\bibnamefont {Semikoz}},\ }\href
  {\doibase 10.1103/PhysRevD.103.L041302} {\bibfield  {journal} {\bibinfo
  {journal} {Phys. Rev. D}\ }\textbf {\bibinfo {volume} {103}},\ \bibinfo
  {pages} {041302} (\bibinfo {year} {2021})},\ \Eprint
  {http://arxiv.org/abs/2009.14174} {arXiv:2009.14174 [astro-ph.CO]}
  \BibitemShut {NoStop}%
\bibitem [{\citenamefont {Li}\ \emph {et~al.}(2021{\natexlab{a}})\citenamefont
  {Li}, \citenamefont {Ye},\ and\ \citenamefont {Piao}}]{Li:2020cjj}%
  \BibitemOpen
  \bibfield  {author} {\bibinfo {author} {\bibfnamefont {H.-H.}\ \bibnamefont
  {Li}}, \bibinfo {author} {\bibfnamefont {G.}~\bibnamefont {Ye}}, \ and\
  \bibinfo {author} {\bibfnamefont {Y.-S.}\ \bibnamefont {Piao}},\ }\href
  {\doibase 10.1016/j.physletb.2021.136211} {\bibfield  {journal} {\bibinfo
  {journal} {Phys. Lett. B}\ }\textbf {\bibinfo {volume} {816}},\ \bibinfo
  {pages} {136211} (\bibinfo {year} {2021}{\natexlab{a}})},\ \Eprint
  {http://arxiv.org/abs/2009.14663} {arXiv:2009.14663 [astro-ph.CO]}
  \BibitemShut {NoStop}%
\bibitem [{\citenamefont {Sugiyama}\ \emph {et~al.}(2021)\citenamefont
  {Sugiyama}, \citenamefont {Takhistov}, \citenamefont {Vitagliano},
  \citenamefont {Kusenko}, \citenamefont {Sasaki},\ and\ \citenamefont
  {Takada}}]{Sugiyama:2020roc}%
  \BibitemOpen
  \bibfield  {author} {\bibinfo {author} {\bibfnamefont {S.}~\bibnamefont
  {Sugiyama}}, \bibinfo {author} {\bibfnamefont {V.}~\bibnamefont {Takhistov}},
  \bibinfo {author} {\bibfnamefont {E.}~\bibnamefont {Vitagliano}}, \bibinfo
  {author} {\bibfnamefont {A.}~\bibnamefont {Kusenko}}, \bibinfo {author}
  {\bibfnamefont {M.}~\bibnamefont {Sasaki}}, \ and\ \bibinfo {author}
  {\bibfnamefont {M.}~\bibnamefont {Takada}},\ }\href {\doibase
  10.1016/j.physletb.2021.136097} {\bibfield  {journal} {\bibinfo  {journal}
  {Phys. Lett. B}\ }\textbf {\bibinfo {volume} {814}},\ \bibinfo {pages}
  {136097} (\bibinfo {year} {2021})},\ \Eprint
  {http://arxiv.org/abs/2010.02189} {arXiv:2010.02189 [astro-ph.CO]}
  \BibitemShut {NoStop}%
\bibitem [{\citenamefont {Liu}\ \emph {et~al.}(2021)\citenamefont {Liu},
  \citenamefont {Cai},\ and\ \citenamefont {Guo}}]{Liu:2020mru}%
  \BibitemOpen
  \bibfield  {author} {\bibinfo {author} {\bibfnamefont {J.}~\bibnamefont
  {Liu}}, \bibinfo {author} {\bibfnamefont {R.-G.}\ \bibnamefont {Cai}}, \ and\
  \bibinfo {author} {\bibfnamefont {Z.-K.}\ \bibnamefont {Guo}},\ }\href
  {\doibase 10.1103/PhysRevLett.126.141303} {\bibfield  {journal} {\bibinfo
  {journal} {Phys. Rev. Lett.}\ }\textbf {\bibinfo {volume} {126}},\ \bibinfo
  {pages} {141303} (\bibinfo {year} {2021})},\ \Eprint
  {http://arxiv.org/abs/2010.03225} {arXiv:2010.03225 [astro-ph.CO]}
  \BibitemShut {NoStop}%
\bibitem [{\citenamefont {Paul}\ \emph {et~al.}(2021)\citenamefont {Paul},
  \citenamefont {Mukhopadhyay},\ and\ \citenamefont {Majumdar}}]{Paul:2020wbz}%
  \BibitemOpen
  \bibfield  {author} {\bibinfo {author} {\bibfnamefont {A.}~\bibnamefont
  {Paul}}, \bibinfo {author} {\bibfnamefont {U.}~\bibnamefont {Mukhopadhyay}},
  \ and\ \bibinfo {author} {\bibfnamefont {D.}~\bibnamefont {Majumdar}},\
  }\href {\doibase 10.1007/JHEP05(2021)223} {\bibfield  {journal} {\bibinfo
  {journal} {JHEP}\ }\textbf {\bibinfo {volume} {05}},\ \bibinfo {pages} {223}
  (\bibinfo {year} {2021})},\ \Eprint {http://arxiv.org/abs/2010.03439}
  {arXiv:2010.03439 [hep-ph]} \BibitemShut {NoStop}%
\bibitem [{\citenamefont {Zhou}\ \emph {et~al.}(2020)\citenamefont {Zhou},
  \citenamefont {Jiang}, \citenamefont {Cai}, \citenamefont {Sasaki},\ and\
  \citenamefont {Pi}}]{Zhou:2020kkf}%
  \BibitemOpen
  \bibfield  {author} {\bibinfo {author} {\bibfnamefont {Z.}~\bibnamefont
  {Zhou}}, \bibinfo {author} {\bibfnamefont {J.}~\bibnamefont {Jiang}},
  \bibinfo {author} {\bibfnamefont {Y.-F.}\ \bibnamefont {Cai}}, \bibinfo
  {author} {\bibfnamefont {M.}~\bibnamefont {Sasaki}}, \ and\ \bibinfo {author}
  {\bibfnamefont {S.}~\bibnamefont {Pi}},\ }\href {\doibase
  10.1103/PhysRevD.102.103527} {\bibfield  {journal} {\bibinfo  {journal}
  {Phys. Rev. D}\ }\textbf {\bibinfo {volume} {102}},\ \bibinfo {pages}
  {103527} (\bibinfo {year} {2020})},\ \Eprint
  {http://arxiv.org/abs/2010.03537} {arXiv:2010.03537 [astro-ph.CO]}
  \BibitemShut {NoStop}%
\bibitem [{\citenamefont {Dom\`enech}\ and\ \citenamefont
  {Pi}(2022)}]{Domenech:2020ers}%
  \BibitemOpen
  \bibfield  {author} {\bibinfo {author} {\bibfnamefont {G.}~\bibnamefont
  {Dom\`enech}}\ and\ \bibinfo {author} {\bibfnamefont {S.}~\bibnamefont
  {Pi}},\ }\href {\doibase 10.1007/s11433-021-1839-6} {\bibfield  {journal}
  {\bibinfo  {journal} {Sci. China Phys. Mech. Astron.}\ }\textbf {\bibinfo
  {volume} {65}},\ \bibinfo {pages} {230411} (\bibinfo {year} {2022})},\
  \Eprint {http://arxiv.org/abs/2010.03976} {arXiv:2010.03976 [astro-ph.CO]}
  \BibitemShut {NoStop}%
\bibitem [{\citenamefont {Bhattacharya}\ \emph {et~al.}(2021)\citenamefont
  {Bhattacharya}, \citenamefont {Mohanty},\ and\ \citenamefont
  {Parashari}}]{Bhattacharya:2020lhc}%
  \BibitemOpen
  \bibfield  {author} {\bibinfo {author} {\bibfnamefont {S.}~\bibnamefont
  {Bhattacharya}}, \bibinfo {author} {\bibfnamefont {S.}~\bibnamefont
  {Mohanty}}, \ and\ \bibinfo {author} {\bibfnamefont {P.}~\bibnamefont
  {Parashari}},\ }\href {\doibase 10.1103/PhysRevD.103.063532} {\bibfield
  {journal} {\bibinfo  {journal} {Phys. Rev. D}\ }\textbf {\bibinfo {volume}
  {103}},\ \bibinfo {pages} {063532} (\bibinfo {year} {2021})},\ \Eprint
  {http://arxiv.org/abs/2010.05071} {arXiv:2010.05071 [astro-ph.CO]}
  \BibitemShut {NoStop}%
\bibitem [{\citenamefont {Abe}\ \emph {et~al.}(2021)\citenamefont {Abe},
  \citenamefont {Tada},\ and\ \citenamefont {Ueda}}]{Abe:2020sqb}%
  \BibitemOpen
  \bibfield  {author} {\bibinfo {author} {\bibfnamefont {K.~T.}\ \bibnamefont
  {Abe}}, \bibinfo {author} {\bibfnamefont {Y.}~\bibnamefont {Tada}}, \ and\
  \bibinfo {author} {\bibfnamefont {I.}~\bibnamefont {Ueda}},\ }\href {\doibase
  10.1088/1475-7516/2021/06/048} {\bibfield  {journal} {\bibinfo  {journal}
  {JCAP}\ }\textbf {\bibinfo {volume} {06}},\ \bibinfo {pages} {048} (\bibinfo
  {year} {2021})},\ \Eprint {http://arxiv.org/abs/2010.06193} {arXiv:2010.06193
  [astro-ph.CO]} \BibitemShut {NoStop}%
\bibitem [{\citenamefont {Kitajima}\ \emph {et~al.}(2021)\citenamefont
  {Kitajima}, \citenamefont {Soda},\ and\ \citenamefont
  {Urakawa}}]{Kitajima:2020rpm}%
  \BibitemOpen
  \bibfield  {author} {\bibinfo {author} {\bibfnamefont {N.}~\bibnamefont
  {Kitajima}}, \bibinfo {author} {\bibfnamefont {J.}~\bibnamefont {Soda}}, \
  and\ \bibinfo {author} {\bibfnamefont {Y.}~\bibnamefont {Urakawa}},\ }\href
  {\doibase 10.1103/PhysRevLett.126.121301} {\bibfield  {journal} {\bibinfo
  {journal} {Phys. Rev. Lett.}\ }\textbf {\bibinfo {volume} {126}},\ \bibinfo
  {pages} {121301} (\bibinfo {year} {2021})},\ \Eprint
  {http://arxiv.org/abs/2010.10990} {arXiv:2010.10990 [astro-ph.CO]}
  \BibitemShut {NoStop}%
\bibitem [{\citenamefont {Middleton}\ \emph {et~al.}(2021)\citenamefont
  {Middleton}, \citenamefont {Sesana}, \citenamefont {Chen}, \citenamefont
  {Vecchio}, \citenamefont {Del~Pozzo},\ and\ \citenamefont
  {Rosado}}]{Middleton:2020asl}%
  \BibitemOpen
  \bibfield  {author} {\bibinfo {author} {\bibfnamefont {H.}~\bibnamefont
  {Middleton}}, \bibinfo {author} {\bibfnamefont {A.}~\bibnamefont {Sesana}},
  \bibinfo {author} {\bibfnamefont {S.}~\bibnamefont {Chen}}, \bibinfo {author}
  {\bibfnamefont {A.}~\bibnamefont {Vecchio}}, \bibinfo {author} {\bibfnamefont
  {W.}~\bibnamefont {Del~Pozzo}}, \ and\ \bibinfo {author} {\bibfnamefont
  {P.~A.}\ \bibnamefont {Rosado}},\ }\href {\doibase 10.1093/mnrasl/slab008}
  {\bibfield  {journal} {\bibinfo  {journal} {Mon. Not. Roy. Astron. Soc.}\
  }\textbf {\bibinfo {volume} {502}},\ \bibinfo {pages} {L99} (\bibinfo {year}
  {2021})},\ \Eprint {http://arxiv.org/abs/2011.01246} {arXiv:2011.01246
  [astro-ph.HE]} \BibitemShut {NoStop}%
\bibitem [{\citenamefont {Inomata}\ \emph {et~al.}(2021)\citenamefont
  {Inomata}, \citenamefont {Kawasaki}, \citenamefont {Mukaida},\ and\
  \citenamefont {Yanagida}}]{Inomata:2020xad}%
  \BibitemOpen
  \bibfield  {author} {\bibinfo {author} {\bibfnamefont {K.}~\bibnamefont
  {Inomata}}, \bibinfo {author} {\bibfnamefont {M.}~\bibnamefont {Kawasaki}},
  \bibinfo {author} {\bibfnamefont {K.}~\bibnamefont {Mukaida}}, \ and\
  \bibinfo {author} {\bibfnamefont {T.~T.}\ \bibnamefont {Yanagida}},\ }\href
  {\doibase 10.1103/PhysRevLett.126.131301} {\bibfield  {journal} {\bibinfo
  {journal} {Phys. Rev. Lett.}\ }\textbf {\bibinfo {volume} {126}},\ \bibinfo
  {pages} {131301} (\bibinfo {year} {2021})},\ \Eprint
  {http://arxiv.org/abs/2011.01270} {arXiv:2011.01270 [astro-ph.CO]}
  \BibitemShut {NoStop}%
\bibitem [{\citenamefont {Tahara}\ and\ \citenamefont
  {Kobayashi}(2020)}]{Tahara:2020fmn}%
  \BibitemOpen
  \bibfield  {author} {\bibinfo {author} {\bibfnamefont {H.~W.~H.}\
  \bibnamefont {Tahara}}\ and\ \bibinfo {author} {\bibfnamefont
  {T.}~\bibnamefont {Kobayashi}},\ }\href {\doibase
  10.1103/PhysRevD.102.123533} {\bibfield  {journal} {\bibinfo  {journal}
  {Phys. Rev. D}\ }\textbf {\bibinfo {volume} {102}},\ \bibinfo {pages}
  {123533} (\bibinfo {year} {2020})},\ \Eprint
  {http://arxiv.org/abs/2011.01605} {arXiv:2011.01605 [gr-qc]} \BibitemShut
  {NoStop}%
\bibitem [{\citenamefont {Chigusa}\ \emph {et~al.}(2021)\citenamefont
  {Chigusa}, \citenamefont {Nakai},\ and\ \citenamefont
  {Zheng}}]{Chigusa:2020rks}%
  \BibitemOpen
  \bibfield  {author} {\bibinfo {author} {\bibfnamefont {S.}~\bibnamefont
  {Chigusa}}, \bibinfo {author} {\bibfnamefont {Y.}~\bibnamefont {Nakai}}, \
  and\ \bibinfo {author} {\bibfnamefont {J.}~\bibnamefont {Zheng}},\ }\href
  {\doibase 10.1103/PhysRevD.104.035031} {\bibfield  {journal} {\bibinfo
  {journal} {Phys. Rev. D}\ }\textbf {\bibinfo {volume} {104}},\ \bibinfo
  {pages} {035031} (\bibinfo {year} {2021})},\ \Eprint
  {http://arxiv.org/abs/2011.04090} {arXiv:2011.04090 [hep-ph]} \BibitemShut
  {NoStop}%
\bibitem [{\citenamefont {Pandey}(2021)}]{Pandey:2020gjy}%
  \BibitemOpen
  \bibfield  {author} {\bibinfo {author} {\bibfnamefont {A.~K.}\ \bibnamefont
  {Pandey}},\ }\href {\doibase 10.1140/epjc/s10052-021-09190-w} {\bibfield
  {journal} {\bibinfo  {journal} {Eur. Phys. J. C}\ }\textbf {\bibinfo {volume}
  {81}},\ \bibinfo {pages} {399} (\bibinfo {year} {2021})},\ \Eprint
  {http://arxiv.org/abs/2011.05821} {arXiv:2011.05821 [astro-ph.CO]}
  \BibitemShut {NoStop}%
\bibitem [{\citenamefont {Bigazzi}\ \emph {et~al.}(2021)\citenamefont
  {Bigazzi}, \citenamefont {Caddeo}, \citenamefont {Cotrone},\ and\
  \citenamefont {Paredes}}]{Bigazzi:2020avc}%
  \BibitemOpen
  \bibfield  {author} {\bibinfo {author} {\bibfnamefont {F.}~\bibnamefont
  {Bigazzi}}, \bibinfo {author} {\bibfnamefont {A.}~\bibnamefont {Caddeo}},
  \bibinfo {author} {\bibfnamefont {A.~L.}\ \bibnamefont {Cotrone}}, \ and\
  \bibinfo {author} {\bibfnamefont {A.}~\bibnamefont {Paredes}},\ }\href
  {\doibase 10.1007/JHEP04(2021)094} {\bibfield  {journal} {\bibinfo  {journal}
  {JHEP}\ }\textbf {\bibinfo {volume} {04}},\ \bibinfo {pages} {094} (\bibinfo
  {year} {2021})},\ \Eprint {http://arxiv.org/abs/2011.08757} {arXiv:2011.08757
  [hep-ph]} \BibitemShut {NoStop}%
\bibitem [{\citenamefont {Ramberg}\ and\ \citenamefont
  {Visinelli}(2021)}]{Ramberg:2020oct}%
  \BibitemOpen
  \bibfield  {author} {\bibinfo {author} {\bibfnamefont {N.}~\bibnamefont
  {Ramberg}}\ and\ \bibinfo {author} {\bibfnamefont {L.}~\bibnamefont
  {Visinelli}},\ }\href {\doibase 10.1103/PhysRevD.103.063031} {\bibfield
  {journal} {\bibinfo  {journal} {Phys. Rev. D}\ }\textbf {\bibinfo {volume}
  {103}},\ \bibinfo {pages} {063031} (\bibinfo {year} {2021})},\ \Eprint
  {http://arxiv.org/abs/2012.06882} {arXiv:2012.06882 [astro-ph.CO]}
  \BibitemShut {NoStop}%
\bibitem [{\citenamefont {Cai}\ and\ \citenamefont {Piao}(2021)}]{Cai:2020qpu}%
  \BibitemOpen
  \bibfield  {author} {\bibinfo {author} {\bibfnamefont {Y.}~\bibnamefont
  {Cai}}\ and\ \bibinfo {author} {\bibfnamefont {Y.-S.}\ \bibnamefont {Piao}},\
  }\href {\doibase 10.1103/PhysRevD.103.083521} {\bibfield  {journal} {\bibinfo
   {journal} {Phys. Rev. D}\ }\textbf {\bibinfo {volume} {103}},\ \bibinfo
  {pages} {083521} (\bibinfo {year} {2021})},\ \Eprint
  {http://arxiv.org/abs/2012.11304} {arXiv:2012.11304 [gr-qc]} \BibitemShut
  {NoStop}%
\bibitem [{\citenamefont {Barman}\ \emph {et~al.}(2020)\citenamefont {Barman},
  \citenamefont {Dutta~Banik},\ and\ \citenamefont {Paul}}]{Barman:2020jrf}%
  \BibitemOpen
  \bibfield  {author} {\bibinfo {author} {\bibfnamefont {B.}~\bibnamefont
  {Barman}}, \bibinfo {author} {\bibfnamefont {A.}~\bibnamefont {Dutta~Banik}},
  \ and\ \bibinfo {author} {\bibfnamefont {A.}~\bibnamefont {Paul}},\
  }\href@noop {} {\  (\bibinfo {year} {2020})},\ \Eprint
  {http://arxiv.org/abs/2012.11969} {arXiv:2012.11969 [astro-ph.CO]}
  \BibitemShut {NoStop}%
\bibitem [{\citenamefont {Chiang}\ and\ \citenamefont
  {Lu}(2021)}]{Chiang:2020aui}%
  \BibitemOpen
  \bibfield  {author} {\bibinfo {author} {\bibfnamefont {C.-W.}\ \bibnamefont
  {Chiang}}\ and\ \bibinfo {author} {\bibfnamefont {B.-Q.}\ \bibnamefont
  {Lu}},\ }\href {\doibase 10.1088/1475-7516/2021/05/049} {\bibfield  {journal}
  {\bibinfo  {journal} {JCAP}\ }\textbf {\bibinfo {volume} {05}},\ \bibinfo
  {pages} {049} (\bibinfo {year} {2021})},\ \Eprint
  {http://arxiv.org/abs/2012.14071} {arXiv:2012.14071 [hep-ph]} \BibitemShut
  {NoStop}%
\bibitem [{\citenamefont {Atal}\ \emph {et~al.}(2021)\citenamefont {Atal},
  \citenamefont {Sanglas},\ and\ \citenamefont {Triantafyllou}}]{Atal:2020yic}%
  \BibitemOpen
  \bibfield  {author} {\bibinfo {author} {\bibfnamefont {V.}~\bibnamefont
  {Atal}}, \bibinfo {author} {\bibfnamefont {A.}~\bibnamefont {Sanglas}}, \
  and\ \bibinfo {author} {\bibfnamefont {N.}~\bibnamefont {Triantafyllou}},\
  }\href {\doibase 10.1088/1475-7516/2021/06/022} {\bibfield  {journal}
  {\bibinfo  {journal} {JCAP}\ }\textbf {\bibinfo {volume} {06}},\ \bibinfo
  {pages} {022} (\bibinfo {year} {2021})},\ \Eprint
  {http://arxiv.org/abs/2012.14721} {arXiv:2012.14721 [astro-ph.CO]}
  \BibitemShut {NoStop}%
\bibitem [{\citenamefont {Datta}\ \emph {et~al.}(2021)\citenamefont {Datta},
  \citenamefont {Ghosal},\ and\ \citenamefont {Samanta}}]{Datta:2020bht}%
  \BibitemOpen
  \bibfield  {author} {\bibinfo {author} {\bibfnamefont {S.}~\bibnamefont
  {Datta}}, \bibinfo {author} {\bibfnamefont {A.}~\bibnamefont {Ghosal}}, \
  and\ \bibinfo {author} {\bibfnamefont {R.}~\bibnamefont {Samanta}},\ }\href
  {\doibase 10.1088/1475-7516/2021/08/021} {\bibfield  {journal} {\bibinfo
  {journal} {JCAP}\ }\textbf {\bibinfo {volume} {08}},\ \bibinfo {pages} {021}
  (\bibinfo {year} {2021})},\ \Eprint {http://arxiv.org/abs/2012.14981}
  {arXiv:2012.14981 [hep-ph]} \BibitemShut {NoStop}%
\bibitem [{\citenamefont {Chen}\ \emph
  {et~al.}(2021{\natexlab{b}})\citenamefont {Chen}, \citenamefont {Yuan},\ and\
  \citenamefont {Huang}}]{Chen:2021wdo}%
  \BibitemOpen
  \bibfield  {author} {\bibinfo {author} {\bibfnamefont {Z.-C.}\ \bibnamefont
  {Chen}}, \bibinfo {author} {\bibfnamefont {C.}~\bibnamefont {Yuan}}, \ and\
  \bibinfo {author} {\bibfnamefont {Q.-G.}\ \bibnamefont {Huang}},\ }\href
  {\doibase 10.1007/s11433-021-1797-y} {\bibfield  {journal} {\bibinfo
  {journal} {Sci. China Phys. Mech. Astron.}\ }\textbf {\bibinfo {volume}
  {64}},\ \bibinfo {pages} {120412} (\bibinfo {year} {2021}{\natexlab{b}})},\
  \Eprint {http://arxiv.org/abs/2101.06869} {arXiv:2101.06869 [astro-ph.CO]}
  \BibitemShut {NoStop}%
\bibitem [{\citenamefont {Gao}\ and\ \citenamefont {Yang}(2021)}]{Gao:2021dfi}%
  \BibitemOpen
  \bibfield  {author} {\bibinfo {author} {\bibfnamefont {T.-J.}\ \bibnamefont
  {Gao}}\ and\ \bibinfo {author} {\bibfnamefont {X.-Y.}\ \bibnamefont {Yang}},\
  }\href {\doibase 10.1140/epjc/s10052-021-09269-4} {\bibfield  {journal}
  {\bibinfo  {journal} {Eur. Phys. J. C}\ }\textbf {\bibinfo {volume} {81}},\
  \bibinfo {pages} {494} (\bibinfo {year} {2021})},\ \Eprint
  {http://arxiv.org/abs/2101.07616} {arXiv:2101.07616 [astro-ph.CO]}
  \BibitemShut {NoStop}%
\bibitem [{\citenamefont {Li}\ \emph {et~al.}(2021{\natexlab{b}})\citenamefont
  {Li}, \citenamefont {Shao}, \citenamefont {Wu},\ and\ \citenamefont
  {Yu}}]{Li:2021qer}%
  \BibitemOpen
  \bibfield  {author} {\bibinfo {author} {\bibfnamefont {S.-L.}\ \bibnamefont
  {Li}}, \bibinfo {author} {\bibfnamefont {L.}~\bibnamefont {Shao}}, \bibinfo
  {author} {\bibfnamefont {P.}~\bibnamefont {Wu}}, \ and\ \bibinfo {author}
  {\bibfnamefont {H.}~\bibnamefont {Yu}},\ }\href {\doibase
  10.1103/PhysRevD.104.043510} {\bibfield  {journal} {\bibinfo  {journal}
  {Phys. Rev. D}\ }\textbf {\bibinfo {volume} {104}},\ \bibinfo {pages}
  {043510} (\bibinfo {year} {2021}{\natexlab{b}})},\ \Eprint
  {http://arxiv.org/abs/2101.08012} {arXiv:2101.08012 [astro-ph.CO]}
  \BibitemShut {NoStop}%
\bibitem [{\citenamefont {Gorghetto}\ \emph {et~al.}(2021)\citenamefont
  {Gorghetto}, \citenamefont {Hardy},\ and\ \citenamefont
  {Nicolaescu}}]{Gorghetto:2021fsn}%
  \BibitemOpen
  \bibfield  {author} {\bibinfo {author} {\bibfnamefont {M.}~\bibnamefont
  {Gorghetto}}, \bibinfo {author} {\bibfnamefont {E.}~\bibnamefont {Hardy}}, \
  and\ \bibinfo {author} {\bibfnamefont {H.}~\bibnamefont {Nicolaescu}},\
  }\href {\doibase 10.1088/1475-7516/2021/06/034} {\bibfield  {journal}
  {\bibinfo  {journal} {JCAP}\ }\textbf {\bibinfo {volume} {06}},\ \bibinfo
  {pages} {034} (\bibinfo {year} {2021})},\ \Eprint
  {http://arxiv.org/abs/2101.11007} {arXiv:2101.11007 [hep-ph]} \BibitemShut
  {NoStop}%
\bibitem [{\citenamefont {Kawasaki}\ and\ \citenamefont
  {Nakatsuka}(2021)}]{Kawasaki:2021ycf}%
  \BibitemOpen
  \bibfield  {author} {\bibinfo {author} {\bibfnamefont {M.}~\bibnamefont
  {Kawasaki}}\ and\ \bibinfo {author} {\bibfnamefont {H.}~\bibnamefont
  {Nakatsuka}},\ }\href {\doibase 10.1088/1475-7516/2021/05/023} {\bibfield
  {journal} {\bibinfo  {journal} {JCAP}\ }\textbf {\bibinfo {volume} {05}},\
  \bibinfo {pages} {023} (\bibinfo {year} {2021})},\ \Eprint
  {http://arxiv.org/abs/2101.11244} {arXiv:2101.11244 [astro-ph.CO]}
  \BibitemShut {NoStop}%
\bibitem [{\citenamefont {Blanco-Pillado}\ \emph {et~al.}(2021)\citenamefont
  {Blanco-Pillado}, \citenamefont {Olum},\ and\ \citenamefont
  {Wachter}}]{Blanco-Pillado:2021ygr}%
  \BibitemOpen
  \bibfield  {author} {\bibinfo {author} {\bibfnamefont {J.~J.}\ \bibnamefont
  {Blanco-Pillado}}, \bibinfo {author} {\bibfnamefont {K.~D.}\ \bibnamefont
  {Olum}}, \ and\ \bibinfo {author} {\bibfnamefont {J.~M.}\ \bibnamefont
  {Wachter}},\ }\href {\doibase 10.1103/PhysRevD.103.103512} {\bibfield
  {journal} {\bibinfo  {journal} {Phys. Rev. D}\ }\textbf {\bibinfo {volume}
  {103}},\ \bibinfo {pages} {103512} (\bibinfo {year} {2021})},\ \Eprint
  {http://arxiv.org/abs/2102.08194} {arXiv:2102.08194 [astro-ph.CO]}
  \BibitemShut {NoStop}%
\bibitem [{\citenamefont {Sharma}(2022)}]{Sharma:2021rot}%
  \BibitemOpen
  \bibfield  {author} {\bibinfo {author} {\bibfnamefont {R.}~\bibnamefont
  {Sharma}},\ }\href {\doibase 10.1103/PhysRevD.105.L041302} {\bibfield
  {journal} {\bibinfo  {journal} {Phys. Rev. D}\ }\textbf {\bibinfo {volume}
  {105}},\ \bibinfo {pages} {L041302} (\bibinfo {year} {2022})},\ \Eprint
  {http://arxiv.org/abs/2102.09358} {arXiv:2102.09358 [astro-ph.CO]}
  \BibitemShut {NoStop}%
\bibitem [{\citenamefont {Brandenburg}\ \emph {et~al.}(2021)\citenamefont
  {Brandenburg}, \citenamefont {Clarke}, \citenamefont {He},\ and\
  \citenamefont {Kahniashvili}}]{Brandenburg:2021tmp}%
  \BibitemOpen
  \bibfield  {author} {\bibinfo {author} {\bibfnamefont {A.}~\bibnamefont
  {Brandenburg}}, \bibinfo {author} {\bibfnamefont {E.}~\bibnamefont {Clarke}},
  \bibinfo {author} {\bibfnamefont {Y.}~\bibnamefont {He}}, \ and\ \bibinfo
  {author} {\bibfnamefont {T.}~\bibnamefont {Kahniashvili}},\ }\href {\doibase
  10.1103/PhysRevD.104.043513} {\bibfield  {journal} {\bibinfo  {journal}
  {Phys. Rev. D}\ }\textbf {\bibinfo {volume} {104}},\ \bibinfo {pages}
  {043513} (\bibinfo {year} {2021})},\ \Eprint
  {http://arxiv.org/abs/2102.12428} {arXiv:2102.12428 [astro-ph.CO]}
  \BibitemShut {NoStop}%
\bibitem [{\citenamefont {Hindmarsh}\ \emph {et~al.}(2021)\citenamefont
  {Hindmarsh}, \citenamefont {Lizarraga}, \citenamefont {Urio},\ and\
  \citenamefont {Urrestilla}}]{Hindmarsh:2021mnl}%
  \BibitemOpen
  \bibfield  {author} {\bibinfo {author} {\bibfnamefont {M.}~\bibnamefont
  {Hindmarsh}}, \bibinfo {author} {\bibfnamefont {J.}~\bibnamefont
  {Lizarraga}}, \bibinfo {author} {\bibfnamefont {A.}~\bibnamefont {Urio}}, \
  and\ \bibinfo {author} {\bibfnamefont {J.}~\bibnamefont {Urrestilla}},\
  }\href {\doibase 10.1103/PhysRevD.104.043519} {\bibfield  {journal} {\bibinfo
   {journal} {Phys. Rev. D}\ }\textbf {\bibinfo {volume} {104}},\ \bibinfo
  {pages} {043519} (\bibinfo {year} {2021})},\ \Eprint
  {http://arxiv.org/abs/2103.16248} {arXiv:2103.16248 [astro-ph.CO]}
  \BibitemShut {NoStop}%
\bibitem [{\citenamefont {Lazarides}\ \emph {et~al.}(2021)\citenamefont
  {Lazarides}, \citenamefont {Maji},\ and\ \citenamefont
  {Shafi}}]{Lazarides:2021uxv}%
  \BibitemOpen
  \bibfield  {author} {\bibinfo {author} {\bibfnamefont {G.}~\bibnamefont
  {Lazarides}}, \bibinfo {author} {\bibfnamefont {R.}~\bibnamefont {Maji}}, \
  and\ \bibinfo {author} {\bibfnamefont {Q.}~\bibnamefont {Shafi}},\ }\href
  {\doibase 10.1103/PhysRevD.104.095004} {\bibfield  {journal} {\bibinfo
  {journal} {Phys. Rev. D}\ }\textbf {\bibinfo {volume} {104}},\ \bibinfo
  {pages} {095004} (\bibinfo {year} {2021})},\ \Eprint
  {http://arxiv.org/abs/2104.02016} {arXiv:2104.02016 [hep-ph]} \BibitemShut
  {NoStop}%
\bibitem [{\citenamefont {Zhou}\ \emph {et~al.}(2021)\citenamefont {Zhou},
  \citenamefont {Bian},\ and\ \citenamefont {Shu}}]{Zhou:2021cfu}%
  \BibitemOpen
  \bibfield  {author} {\bibinfo {author} {\bibfnamefont {R.}~\bibnamefont
  {Zhou}}, \bibinfo {author} {\bibfnamefont {L.}~\bibnamefont {Bian}}, \ and\
  \bibinfo {author} {\bibfnamefont {J.}~\bibnamefont {Shu}},\ }\href@noop {} {\
   (\bibinfo {year} {2021})},\ \Eprint {http://arxiv.org/abs/2104.03519}
  {arXiv:2104.03519 [hep-ph]} \BibitemShut {NoStop}%
\bibitem [{\citenamefont {Sakharov}\ \emph {et~al.}(2021)\citenamefont
  {Sakharov}, \citenamefont {Eroshenko},\ and\ \citenamefont
  {Rubin}}]{Sakharov:2021dim}%
  \BibitemOpen
  \bibfield  {author} {\bibinfo {author} {\bibfnamefont {A.~S.}\ \bibnamefont
  {Sakharov}}, \bibinfo {author} {\bibfnamefont {Y.~N.}\ \bibnamefont
  {Eroshenko}}, \ and\ \bibinfo {author} {\bibfnamefont {S.~G.}\ \bibnamefont
  {Rubin}},\ }\href {\doibase 10.1103/PhysRevD.104.043005} {\bibfield
  {journal} {\bibinfo  {journal} {Phys. Rev. D}\ }\textbf {\bibinfo {volume}
  {104}},\ \bibinfo {pages} {043005} (\bibinfo {year} {2021})},\ \Eprint
  {http://arxiv.org/abs/2104.08750} {arXiv:2104.08750 [hep-ph]} \BibitemShut
  {NoStop}%
\bibitem [{\citenamefont {Arzoumanian}\ \emph
  {et~al.}(2021{\natexlab{a}})\citenamefont {Arzoumanian} \emph
  {et~al.}}]{NANOGrav:2021flc}%
  \BibitemOpen
  \bibfield  {author} {\bibinfo {author} {\bibfnamefont {Z.}~\bibnamefont
  {Arzoumanian}} \emph {et~al.} (\bibinfo {collaboration} {NANOGrav}),\ }\href
  {\doibase 10.1103/PhysRevLett.127.251302} {\bibfield  {journal} {\bibinfo
  {journal} {Phys. Rev. Lett.}\ }\textbf {\bibinfo {volume} {127}},\ \bibinfo
  {pages} {251302} (\bibinfo {year} {2021}{\natexlab{a}})},\ \Eprint
  {http://arxiv.org/abs/2104.13930} {arXiv:2104.13930 [astro-ph.CO]}
  \BibitemShut {NoStop}%
\bibitem [{\citenamefont {Moore}\ and\ \citenamefont
  {Vecchio}(2021)}]{Moore:2021ibq}%
  \BibitemOpen
  \bibfield  {author} {\bibinfo {author} {\bibfnamefont {C.~J.}\ \bibnamefont
  {Moore}}\ and\ \bibinfo {author} {\bibfnamefont {A.}~\bibnamefont
  {Vecchio}},\ }\href {\doibase 10.1038/s41550-021-01489-8} {\bibfield
  {journal} {\bibinfo  {journal} {Nature Astron.}\ }\textbf {\bibinfo {volume}
  {5}},\ \bibinfo {pages} {1268} (\bibinfo {year} {2021})},\ \Eprint
  {http://arxiv.org/abs/2104.15130} {arXiv:2104.15130 [astro-ph.CO]}
  \BibitemShut {NoStop}%
\bibitem [{\citenamefont {Borah}\ \emph
  {et~al.}(2021{\natexlab{a}})\citenamefont {Borah}, \citenamefont {Dasgupta},\
  and\ \citenamefont {Kang}}]{Borah:2021ocu}%
  \BibitemOpen
  \bibfield  {author} {\bibinfo {author} {\bibfnamefont {D.}~\bibnamefont
  {Borah}}, \bibinfo {author} {\bibfnamefont {A.}~\bibnamefont {Dasgupta}}, \
  and\ \bibinfo {author} {\bibfnamefont {S.~K.}\ \bibnamefont {Kang}},\ }\href
  {\doibase 10.1103/PhysRevD.104.063501} {\bibfield  {journal} {\bibinfo
  {journal} {Phys. Rev. D}\ }\textbf {\bibinfo {volume} {104}},\ \bibinfo
  {pages} {063501} (\bibinfo {year} {2021}{\natexlab{a}})},\ \Eprint
  {http://arxiv.org/abs/2105.01007} {arXiv:2105.01007 [hep-ph]} \BibitemShut
  {NoStop}%
\bibitem [{\citenamefont {Yi}\ and\ \citenamefont {Zhu}(2022)}]{Yi:2021lxc}%
  \BibitemOpen
  \bibfield  {author} {\bibinfo {author} {\bibfnamefont {Z.}~\bibnamefont
  {Yi}}\ and\ \bibinfo {author} {\bibfnamefont {Z.-H.}\ \bibnamefont {Zhu}},\
  }\href {\doibase 10.1088/1475-7516/2022/05/046} {\bibfield  {journal}
  {\bibinfo  {journal} {JCAP}\ }\textbf {\bibinfo {volume} {05}},\ \bibinfo
  {pages} {046} (\bibinfo {year} {2022})},\ \Eprint
  {http://arxiv.org/abs/2105.01943} {arXiv:2105.01943 [gr-qc]} \BibitemShut
  {NoStop}%
\bibitem [{\citenamefont {Wu}\ \emph {et~al.}(2021)\citenamefont {Wu},
  \citenamefont {Gong},\ and\ \citenamefont {Li}}]{Wu:2021zta}%
  \BibitemOpen
  \bibfield  {author} {\bibinfo {author} {\bibfnamefont {L.}~\bibnamefont
  {Wu}}, \bibinfo {author} {\bibfnamefont {Y.}~\bibnamefont {Gong}}, \ and\
  \bibinfo {author} {\bibfnamefont {T.}~\bibnamefont {Li}},\ }\href {\doibase
  10.1103/PhysRevD.104.123544} {\bibfield  {journal} {\bibinfo  {journal}
  {Phys. Rev. D}\ }\textbf {\bibinfo {volume} {104}},\ \bibinfo {pages}
  {123544} (\bibinfo {year} {2021})},\ \Eprint
  {http://arxiv.org/abs/2105.07694} {arXiv:2105.07694 [gr-qc]} \BibitemShut
  {NoStop}%
\bibitem [{\citenamefont {Haque}\ \emph {et~al.}(2021)\citenamefont {Haque},
  \citenamefont {Maity}, \citenamefont {Paul},\ and\ \citenamefont
  {Sriramkumar}}]{Haque:2021dha}%
  \BibitemOpen
  \bibfield  {author} {\bibinfo {author} {\bibfnamefont {M.~R.}\ \bibnamefont
  {Haque}}, \bibinfo {author} {\bibfnamefont {D.}~\bibnamefont {Maity}},
  \bibinfo {author} {\bibfnamefont {T.}~\bibnamefont {Paul}}, \ and\ \bibinfo
  {author} {\bibfnamefont {L.}~\bibnamefont {Sriramkumar}},\ }\href {\doibase
  10.1103/PhysRevD.104.063513} {\bibfield  {journal} {\bibinfo  {journal}
  {Phys. Rev. D}\ }\textbf {\bibinfo {volume} {104}},\ \bibinfo {pages}
  {063513} (\bibinfo {year} {2021})},\ \Eprint
  {http://arxiv.org/abs/2105.09242} {arXiv:2105.09242 [astro-ph.CO]}
  \BibitemShut {NoStop}%
\bibitem [{\citenamefont {Liu}\ \emph {et~al.}(2022)\citenamefont {Liu},
  \citenamefont {Bian}, \citenamefont {Cai}, \citenamefont {Guo},\ and\
  \citenamefont {Wang}}]{Liu:2021svg}%
  \BibitemOpen
  \bibfield  {author} {\bibinfo {author} {\bibfnamefont {J.}~\bibnamefont
  {Liu}}, \bibinfo {author} {\bibfnamefont {L.}~\bibnamefont {Bian}}, \bibinfo
  {author} {\bibfnamefont {R.-G.}\ \bibnamefont {Cai}}, \bibinfo {author}
  {\bibfnamefont {Z.-K.}\ \bibnamefont {Guo}}, \ and\ \bibinfo {author}
  {\bibfnamefont {S.-J.}\ \bibnamefont {Wang}},\ }\href {\doibase
  10.1103/PhysRevD.105.L021303} {\bibfield  {journal} {\bibinfo  {journal}
  {Phys. Rev. D}\ }\textbf {\bibinfo {volume} {105}},\ \bibinfo {pages}
  {L021303} (\bibinfo {year} {2022})},\ \Eprint
  {http://arxiv.org/abs/2106.05637} {arXiv:2106.05637 [astro-ph.CO]}
  \BibitemShut {NoStop}%
\bibitem [{\citenamefont {Lewicki}\ \emph {et~al.}(2021)\citenamefont
  {Lewicki}, \citenamefont {Pujol\`as},\ and\ \citenamefont
  {Vaskonen}}]{Lewicki:2021xku}%
  \BibitemOpen
  \bibfield  {author} {\bibinfo {author} {\bibfnamefont {M.}~\bibnamefont
  {Lewicki}}, \bibinfo {author} {\bibfnamefont {O.}~\bibnamefont {Pujol\`as}},
  \ and\ \bibinfo {author} {\bibfnamefont {V.}~\bibnamefont {Vaskonen}},\
  }\href {\doibase 10.1140/epjc/s10052-021-09669-6} {\bibfield  {journal}
  {\bibinfo  {journal} {Eur. Phys. J. C}\ }\textbf {\bibinfo {volume} {81}},\
  \bibinfo {pages} {857} (\bibinfo {year} {2021})},\ \Eprint
  {http://arxiv.org/abs/2106.09706} {arXiv:2106.09706 [astro-ph.CO]}
  \BibitemShut {NoStop}%
\bibitem [{\citenamefont {Chang}\ and\ \citenamefont
  {Cui}(2022)}]{Chang:2021afa}%
  \BibitemOpen
  \bibfield  {author} {\bibinfo {author} {\bibfnamefont {C.-F.}\ \bibnamefont
  {Chang}}\ and\ \bibinfo {author} {\bibfnamefont {Y.}~\bibnamefont {Cui}},\
  }\href {\doibase 10.1007/JHEP03(2022)114} {\bibfield  {journal} {\bibinfo
  {journal} {JHEP}\ }\textbf {\bibinfo {volume} {03}},\ \bibinfo {pages} {114}
  (\bibinfo {year} {2022})},\ \Eprint {http://arxiv.org/abs/2106.09746}
  {arXiv:2106.09746 [hep-ph]} \BibitemShut {NoStop}%
\bibitem [{\citenamefont {Buchmuller}\ \emph {et~al.}(2021)\citenamefont
  {Buchmuller}, \citenamefont {Domcke},\ and\ \citenamefont
  {Schmitz}}]{Buchmuller:2021mbb}%
  \BibitemOpen
  \bibfield  {author} {\bibinfo {author} {\bibfnamefont {W.}~\bibnamefont
  {Buchmuller}}, \bibinfo {author} {\bibfnamefont {V.}~\bibnamefont {Domcke}},
  \ and\ \bibinfo {author} {\bibfnamefont {K.}~\bibnamefont {Schmitz}},\ }\href
  {\doibase 10.1088/1475-7516/2021/12/006} {\bibfield  {journal} {\bibinfo
  {journal} {JCAP}\ }\textbf {\bibinfo {volume} {12}},\ \bibinfo {pages} {006}
  (\bibinfo {year} {2021})},\ \Eprint {http://arxiv.org/abs/2107.04578}
  {arXiv:2107.04578 [hep-ph]} \BibitemShut {NoStop}%
\bibitem [{\citenamefont {Masoud}\ \emph {et~al.}(2021)\citenamefont {Masoud},
  \citenamefont {Rehman},\ and\ \citenamefont {Shafi}}]{Masoud:2021prr}%
  \BibitemOpen
  \bibfield  {author} {\bibinfo {author} {\bibfnamefont {M.~A.}\ \bibnamefont
  {Masoud}}, \bibinfo {author} {\bibfnamefont {M.~U.}\ \bibnamefont {Rehman}},
  \ and\ \bibinfo {author} {\bibfnamefont {Q.}~\bibnamefont {Shafi}},\ }\href
  {\doibase 10.1088/1475-7516/2021/11/022} {\bibfield  {journal} {\bibinfo
  {journal} {JCAP}\ }\textbf {\bibinfo {volume} {11}},\ \bibinfo {pages} {022}
  (\bibinfo {year} {2021})},\ \Eprint {http://arxiv.org/abs/2107.09689}
  {arXiv:2107.09689 [hep-ph]} \BibitemShut {NoStop}%
\bibitem [{\citenamefont {Casey-Clyde}\ \emph {et~al.}(2022)\citenamefont
  {Casey-Clyde}, \citenamefont {Mingarelli}, \citenamefont {Greene},
  \citenamefont {Pardo}, \citenamefont {Na\~nez},\ and\ \citenamefont
  {Goulding}}]{Casey-Clyde:2021xro}%
  \BibitemOpen
  \bibfield  {author} {\bibinfo {author} {\bibfnamefont {J.~A.}\ \bibnamefont
  {Casey-Clyde}}, \bibinfo {author} {\bibfnamefont {C.~M.~F.}\ \bibnamefont
  {Mingarelli}}, \bibinfo {author} {\bibfnamefont {J.~E.}\ \bibnamefont
  {Greene}}, \bibinfo {author} {\bibfnamefont {K.}~\bibnamefont {Pardo}},
  \bibinfo {author} {\bibfnamefont {M.}~\bibnamefont {Na\~nez}}, \ and\
  \bibinfo {author} {\bibfnamefont {A.~D.}\ \bibnamefont {Goulding}},\ }\href
  {\doibase 10.3847/1538-4357/ac32de} {\bibfield  {journal} {\bibinfo
  {journal} {Astrophys. J.}\ }\textbf {\bibinfo {volume} {924}},\ \bibinfo
  {pages} {93} (\bibinfo {year} {2022})},\ \Eprint
  {http://arxiv.org/abs/2107.11390} {arXiv:2107.11390 [astro-ph.HE]}
  \BibitemShut {NoStop}%
\bibitem [{\citenamefont {Li}\ and\ \citenamefont
  {Shapiro}(2021)}]{Li:2021htg}%
  \BibitemOpen
  \bibfield  {author} {\bibinfo {author} {\bibfnamefont {B.}~\bibnamefont
  {Li}}\ and\ \bibinfo {author} {\bibfnamefont {P.~R.}\ \bibnamefont
  {Shapiro}},\ }\href {\doibase 10.1088/1475-7516/2021/10/024} {\bibfield
  {journal} {\bibinfo  {journal} {JCAP}\ }\textbf {\bibinfo {volume} {10}},\
  \bibinfo {pages} {024} (\bibinfo {year} {2021})},\ \Eprint
  {http://arxiv.org/abs/2107.12229} {arXiv:2107.12229 [astro-ph.CO]}
  \BibitemShut {NoStop}%
\bibitem [{\citenamefont {Cai}\ \emph {et~al.}(2021{\natexlab{a}})\citenamefont
  {Cai}, \citenamefont {Chen},\ and\ \citenamefont {Fu}}]{Cai:2021wzd}%
  \BibitemOpen
  \bibfield  {author} {\bibinfo {author} {\bibfnamefont {R.-G.}\ \bibnamefont
  {Cai}}, \bibinfo {author} {\bibfnamefont {C.}~\bibnamefont {Chen}}, \ and\
  \bibinfo {author} {\bibfnamefont {C.}~\bibnamefont {Fu}},\ }\href {\doibase
  10.1103/PhysRevD.104.083537} {\bibfield  {journal} {\bibinfo  {journal}
  {Phys. Rev. D}\ }\textbf {\bibinfo {volume} {104}},\ \bibinfo {pages}
  {083537} (\bibinfo {year} {2021}{\natexlab{a}})},\ \Eprint
  {http://arxiv.org/abs/2108.03422} {arXiv:2108.03422 [astro-ph.CO]}
  \BibitemShut {NoStop}%
\bibitem [{\citenamefont {Spanos}\ and\ \citenamefont
  {Stamou}(2021)}]{Spanos:2021hpk}%
  \BibitemOpen
  \bibfield  {author} {\bibinfo {author} {\bibfnamefont {V.~C.}\ \bibnamefont
  {Spanos}}\ and\ \bibinfo {author} {\bibfnamefont {I.~D.}\ \bibnamefont
  {Stamou}},\ }\href {\doibase 10.1103/PhysRevD.104.123537} {\bibfield
  {journal} {\bibinfo  {journal} {Phys. Rev. D}\ }\textbf {\bibinfo {volume}
  {104}},\ \bibinfo {pages} {123537} (\bibinfo {year} {2021})},\ \Eprint
  {http://arxiv.org/abs/2108.05671} {arXiv:2108.05671 [astro-ph.CO]}
  \BibitemShut {NoStop}%
\bibitem [{\citenamefont {Khodadi}\ \emph {et~al.}(2021)\citenamefont
  {Khodadi}, \citenamefont {Dey},\ and\ \citenamefont
  {Lambiase}}]{Khodadi:2021ees}%
  \BibitemOpen
  \bibfield  {author} {\bibinfo {author} {\bibfnamefont {M.}~\bibnamefont
  {Khodadi}}, \bibinfo {author} {\bibfnamefont {U.~K.}\ \bibnamefont {Dey}}, \
  and\ \bibinfo {author} {\bibfnamefont {G.}~\bibnamefont {Lambiase}},\ }\href
  {\doibase 10.1103/PhysRevD.104.063039} {\bibfield  {journal} {\bibinfo
  {journal} {Phys. Rev. D}\ }\textbf {\bibinfo {volume} {104}},\ \bibinfo
  {pages} {063039} (\bibinfo {year} {2021})},\ \Eprint
  {http://arxiv.org/abs/2108.09320} {arXiv:2108.09320 [gr-qc]} \BibitemShut
  {NoStop}%
\bibitem [{\citenamefont {Wu}\ \emph {et~al.}(2022)\citenamefont {Wu},
  \citenamefont {Chen},\ and\ \citenamefont {Huang}}]{Wu:2021kmd}%
  \BibitemOpen
  \bibfield  {author} {\bibinfo {author} {\bibfnamefont {Y.-M.}\ \bibnamefont
  {Wu}}, \bibinfo {author} {\bibfnamefont {Z.-C.}\ \bibnamefont {Chen}}, \ and\
  \bibinfo {author} {\bibfnamefont {Q.-G.}\ \bibnamefont {Huang}},\ }\href
  {\doibase 10.3847/1538-4357/ac35cc} {\bibfield  {journal} {\bibinfo
  {journal} {Astrophys. J.}\ }\textbf {\bibinfo {volume} {925}},\ \bibinfo
  {pages} {37} (\bibinfo {year} {2022})},\ \Eprint
  {http://arxiv.org/abs/2108.10518} {arXiv:2108.10518 [astro-ph.CO]}
  \BibitemShut {NoStop}%
\bibitem [{\citenamefont {Izquierdo-Villalba}\ \emph
  {et~al.}(2021)\citenamefont {Izquierdo-Villalba}, \citenamefont {Sesana},
  \citenamefont {Bonoli},\ and\ \citenamefont
  {Colpi}}]{Izquierdo-Villalba:2021prf}%
  \BibitemOpen
  \bibfield  {author} {\bibinfo {author} {\bibfnamefont {D.}~\bibnamefont
  {Izquierdo-Villalba}}, \bibinfo {author} {\bibfnamefont {A.}~\bibnamefont
  {Sesana}}, \bibinfo {author} {\bibfnamefont {S.}~\bibnamefont {Bonoli}}, \
  and\ \bibinfo {author} {\bibfnamefont {M.}~\bibnamefont {Colpi}},\ }\href
  {\doibase 10.1093/mnras/stab3239} {\bibfield  {journal} {\bibinfo  {journal}
  {Mon. Not. Roy. Astron. Soc.}\ }\textbf {\bibinfo {volume} {509}},\ \bibinfo
  {pages} {3488} (\bibinfo {year} {2021})},\ \Eprint
  {http://arxiv.org/abs/2108.11671} {arXiv:2108.11671 [astro-ph.GA]}
  \BibitemShut {NoStop}%
\bibitem [{\citenamefont {Chen}\ \emph {et~al.}(2022)\citenamefont {Chen},
  \citenamefont {Wu},\ and\ \citenamefont {Huang}}]{Chen:2021ncc}%
  \BibitemOpen
  \bibfield  {author} {\bibinfo {author} {\bibfnamefont {Z.-C.}\ \bibnamefont
  {Chen}}, \bibinfo {author} {\bibfnamefont {Y.-M.}\ \bibnamefont {Wu}}, \ and\
  \bibinfo {author} {\bibfnamefont {Q.-G.}\ \bibnamefont {Huang}},\ }\href
  {\doibase 10.1088/1572-9494/ac7cdf} {\bibfield  {journal} {\bibinfo
  {journal} {Commun. Theor. Phys.}\ }\textbf {\bibinfo {volume} {74}},\
  \bibinfo {pages} {105402} (\bibinfo {year} {2022})},\ \Eprint
  {http://arxiv.org/abs/2109.00296} {arXiv:2109.00296 [astro-ph.CO]}
  \BibitemShut {NoStop}%
\bibitem [{\citenamefont {Borah}\ \emph
  {et~al.}(2021{\natexlab{b}})\citenamefont {Borah}, \citenamefont {Dasgupta},\
  and\ \citenamefont {Kang}}]{Borah:2021ftr}%
  \BibitemOpen
  \bibfield  {author} {\bibinfo {author} {\bibfnamefont {D.}~\bibnamefont
  {Borah}}, \bibinfo {author} {\bibfnamefont {A.}~\bibnamefont {Dasgupta}}, \
  and\ \bibinfo {author} {\bibfnamefont {S.~K.}\ \bibnamefont {Kang}},\ }\href
  {\doibase 10.1088/1475-7516/2021/12/039} {\bibfield  {journal} {\bibinfo
  {journal} {JCAP}\ }\textbf {\bibinfo {volume} {12}},\ \bibinfo {pages} {039}
  (\bibinfo {year} {2021}{\natexlab{b}})},\ \Eprint
  {http://arxiv.org/abs/2109.11558} {arXiv:2109.11558 [hep-ph]} \BibitemShut
  {NoStop}%
\bibitem [{\citenamefont {Arzoumanian}\ \emph
  {et~al.}(2021{\natexlab{b}})\citenamefont {Arzoumanian} \emph
  {et~al.}}]{NANOGrav:2021ini}%
  \BibitemOpen
  \bibfield  {author} {\bibinfo {author} {\bibfnamefont {Z.}~\bibnamefont
  {Arzoumanian}} \emph {et~al.} (\bibinfo {collaboration} {NANOGrav}),\ }\href
  {\doibase 10.3847/2041-8213/ac401c} {\bibfield  {journal} {\bibinfo
  {journal} {Astrophys. J. Lett.}\ }\textbf {\bibinfo {volume} {923}},\
  \bibinfo {pages} {L22} (\bibinfo {year} {2021}{\natexlab{b}})},\ \Eprint
  {http://arxiv.org/abs/2109.14706} {arXiv:2109.14706 [gr-qc]} \BibitemShut
  {NoStop}%
\bibitem [{\citenamefont {Gao}(2021)}]{Gao:2021lno}%
  \BibitemOpen
  \bibfield  {author} {\bibinfo {author} {\bibfnamefont {T.-J.}\ \bibnamefont
  {Gao}},\ }\href@noop {} {\  (\bibinfo {year} {2021})},\ \Eprint
  {http://arxiv.org/abs/2110.00205} {arXiv:2110.00205 [gr-qc]} \BibitemShut
  {NoStop}%
\bibitem [{\citenamefont {Lin}\ \emph {et~al.}(2023)\citenamefont {Lin},
  \citenamefont {Gao}, \citenamefont {Gong}, \citenamefont {Lu}, \citenamefont
  {Wang},\ and\ \citenamefont {Zhang}}]{Lin:2021vwc}%
  \BibitemOpen
  \bibfield  {author} {\bibinfo {author} {\bibfnamefont {J.}~\bibnamefont
  {Lin}}, \bibinfo {author} {\bibfnamefont {S.}~\bibnamefont {Gao}}, \bibinfo
  {author} {\bibfnamefont {Y.}~\bibnamefont {Gong}}, \bibinfo {author}
  {\bibfnamefont {Y.}~\bibnamefont {Lu}}, \bibinfo {author} {\bibfnamefont
  {Z.}~\bibnamefont {Wang}}, \ and\ \bibinfo {author} {\bibfnamefont
  {F.}~\bibnamefont {Zhang}},\ }\href {\doibase 10.1103/PhysRevD.107.043517}
  {\bibfield  {journal} {\bibinfo  {journal} {Phys. Rev. D}\ }\textbf {\bibinfo
  {volume} {107}},\ \bibinfo {pages} {043517} (\bibinfo {year} {2023})},\
  \Eprint {http://arxiv.org/abs/2111.01362} {arXiv:2111.01362 [gr-qc]}
  \BibitemShut {NoStop}%
\bibitem [{\citenamefont {Benetti}\ \emph {et~al.}(2022)\citenamefont
  {Benetti}, \citenamefont {Graef},\ and\ \citenamefont
  {Vagnozzi}}]{Benetti:2021uea}%
  \BibitemOpen
  \bibfield  {author} {\bibinfo {author} {\bibfnamefont {M.}~\bibnamefont
  {Benetti}}, \bibinfo {author} {\bibfnamefont {L.~L.}\ \bibnamefont {Graef}},
  \ and\ \bibinfo {author} {\bibfnamefont {S.}~\bibnamefont {Vagnozzi}},\
  }\href {\doibase 10.1103/PhysRevD.105.043520} {\bibfield  {journal} {\bibinfo
   {journal} {Phys. Rev. D}\ }\textbf {\bibinfo {volume} {105}},\ \bibinfo
  {pages} {043520} (\bibinfo {year} {2022})},\ \Eprint
  {http://arxiv.org/abs/2111.04758} {arXiv:2111.04758 [astro-ph.CO]}
  \BibitemShut {NoStop}%
\bibitem [{\citenamefont {Dunsky}\ \emph {et~al.}(2022)\citenamefont {Dunsky},
  \citenamefont {Ghoshal}, \citenamefont {Murayama}, \citenamefont
  {Sakakihara},\ and\ \citenamefont {White}}]{Dunsky:2021tih}%
  \BibitemOpen
  \bibfield  {author} {\bibinfo {author} {\bibfnamefont {D.~I.}\ \bibnamefont
  {Dunsky}}, \bibinfo {author} {\bibfnamefont {A.}~\bibnamefont {Ghoshal}},
  \bibinfo {author} {\bibfnamefont {H.}~\bibnamefont {Murayama}}, \bibinfo
  {author} {\bibfnamefont {Y.}~\bibnamefont {Sakakihara}}, \ and\ \bibinfo
  {author} {\bibfnamefont {G.}~\bibnamefont {White}},\ }\href {\doibase
  10.1103/PhysRevD.106.075030} {\bibfield  {journal} {\bibinfo  {journal}
  {Phys. Rev. D}\ }\textbf {\bibinfo {volume} {106}},\ \bibinfo {pages}
  {075030} (\bibinfo {year} {2022})},\ \Eprint
  {http://arxiv.org/abs/2111.08750} {arXiv:2111.08750 [hep-ph]} \BibitemShut
  {NoStop}%
\bibitem [{\citenamefont {Zhang}(2022)}]{Zhang:2021rqs}%
  \BibitemOpen
  \bibfield  {author} {\bibinfo {author} {\bibfnamefont {F.}~\bibnamefont
  {Zhang}},\ }\href {\doibase 10.1103/PhysRevD.105.063539} {\bibfield
  {journal} {\bibinfo  {journal} {Phys. Rev. D}\ }\textbf {\bibinfo {volume}
  {105}},\ \bibinfo {pages} {063539} (\bibinfo {year} {2022})},\ \Eprint
  {http://arxiv.org/abs/2112.10516} {arXiv:2112.10516 [gr-qc]} \BibitemShut
  {NoStop}%
\bibitem [{\citenamefont {Cai}\ and\ \citenamefont {Piao}(2022)}]{Cai:2022nqv}%
  \BibitemOpen
  \bibfield  {author} {\bibinfo {author} {\bibfnamefont {Y.}~\bibnamefont
  {Cai}}\ and\ \bibinfo {author} {\bibfnamefont {Y.-S.}\ \bibnamefont {Piao}},\
  }\href {\doibase 10.1007/JHEP06(2022)067} {\bibfield  {journal} {\bibinfo
  {journal} {JHEP}\ }\textbf {\bibinfo {volume} {06}},\ \bibinfo {pages} {067}
  (\bibinfo {year} {2022})},\ \Eprint {http://arxiv.org/abs/2201.04552}
  {arXiv:2201.04552 [gr-qc]} \BibitemShut {NoStop}%
\bibitem [{\citenamefont {Roper~Pol}\ \emph {et~al.}(2022)\citenamefont
  {Roper~Pol}, \citenamefont {Caprini}, \citenamefont {Neronov},\ and\
  \citenamefont {Semikoz}}]{RoperPol:2022iel}%
  \BibitemOpen
  \bibfield  {author} {\bibinfo {author} {\bibfnamefont {A.}~\bibnamefont
  {Roper~Pol}}, \bibinfo {author} {\bibfnamefont {C.}~\bibnamefont {Caprini}},
  \bibinfo {author} {\bibfnamefont {A.}~\bibnamefont {Neronov}}, \ and\
  \bibinfo {author} {\bibfnamefont {D.}~\bibnamefont {Semikoz}},\ }\href
  {\doibase 10.1103/PhysRevD.105.123502} {\bibfield  {journal} {\bibinfo
  {journal} {Phys. Rev. D}\ }\textbf {\bibinfo {volume} {105}},\ \bibinfo
  {pages} {123502} (\bibinfo {year} {2022})},\ \Eprint
  {http://arxiv.org/abs/2201.05630} {arXiv:2201.05630 [astro-ph.CO]}
  \BibitemShut {NoStop}%
\bibitem [{\citenamefont {Wang}(2022{\natexlab{a}})}]{Wang:2022wwj}%
  \BibitemOpen
  \bibfield  {author} {\bibinfo {author} {\bibfnamefont {D.}~\bibnamefont
  {Wang}},\ }\href@noop {} {\  (\bibinfo {year} {2022}{\natexlab{a}})},\
  \Eprint {http://arxiv.org/abs/2201.09295} {arXiv:2201.09295 [astro-ph.CO]}
  \BibitemShut {NoStop}%
\bibitem [{\citenamefont {Ashoorioon}\ \emph {et~al.}(2022)\citenamefont
  {Ashoorioon}, \citenamefont {Rezazadeh},\ and\ \citenamefont
  {Rostami}}]{Ashoorioon:2022raz}%
  \BibitemOpen
  \bibfield  {author} {\bibinfo {author} {\bibfnamefont {A.}~\bibnamefont
  {Ashoorioon}}, \bibinfo {author} {\bibfnamefont {K.}~\bibnamefont
  {Rezazadeh}}, \ and\ \bibinfo {author} {\bibfnamefont {A.}~\bibnamefont
  {Rostami}},\ }\href {\doibase 10.1016/j.physletb.2022.137542} {\bibfield
  {journal} {\bibinfo  {journal} {Phys. Lett. B}\ }\textbf {\bibinfo {volume}
  {835}},\ \bibinfo {pages} {137542} (\bibinfo {year} {2022})},\ \Eprint
  {http://arxiv.org/abs/2202.01131} {arXiv:2202.01131 [astro-ph.CO]}
  \BibitemShut {NoStop}%
\bibitem [{\citenamefont {Ahmed}\ \emph {et~al.}(2022)\citenamefont {Ahmed},
  \citenamefont {Junaid}, \citenamefont {Nasri},\ and\ \citenamefont
  {Zubair}}]{Ahmed:2022rwy}%
  \BibitemOpen
  \bibfield  {author} {\bibinfo {author} {\bibfnamefont {W.}~\bibnamefont
  {Ahmed}}, \bibinfo {author} {\bibfnamefont {M.}~\bibnamefont {Junaid}},
  \bibinfo {author} {\bibfnamefont {S.}~\bibnamefont {Nasri}}, \ and\ \bibinfo
  {author} {\bibfnamefont {U.}~\bibnamefont {Zubair}},\ }\href {\doibase
  10.1103/PhysRevD.105.115008} {\bibfield  {journal} {\bibinfo  {journal}
  {Phys. Rev. D}\ }\textbf {\bibinfo {volume} {105}},\ \bibinfo {pages}
  {115008} (\bibinfo {year} {2022})},\ \Eprint
  {http://arxiv.org/abs/2202.06216} {arXiv:2202.06216 [hep-ph]} \BibitemShut
  {NoStop}%
\bibitem [{\citenamefont {Afzal}\ \emph {et~al.}(2022)\citenamefont {Afzal},
  \citenamefont {Ahmed}, \citenamefont {Rehman},\ and\ \citenamefont
  {Shafi}}]{Afzal:2022vjx}%
  \BibitemOpen
  \bibfield  {author} {\bibinfo {author} {\bibfnamefont {A.}~\bibnamefont
  {Afzal}}, \bibinfo {author} {\bibfnamefont {W.}~\bibnamefont {Ahmed}},
  \bibinfo {author} {\bibfnamefont {M.~U.}\ \bibnamefont {Rehman}}, \ and\
  \bibinfo {author} {\bibfnamefont {Q.}~\bibnamefont {Shafi}},\ }\href
  {\doibase 10.1103/PhysRevD.105.103539} {\bibfield  {journal} {\bibinfo
  {journal} {Phys. Rev. D}\ }\textbf {\bibinfo {volume} {105}},\ \bibinfo
  {pages} {103539} (\bibinfo {year} {2022})},\ \Eprint
  {http://arxiv.org/abs/2202.07386} {arXiv:2202.07386 [hep-ph]} \BibitemShut
  {NoStop}%
\bibitem [{\citenamefont {Wang}(2022{\natexlab{b}})}]{Wang:2022rjz}%
  \BibitemOpen
  \bibfield  {author} {\bibinfo {author} {\bibfnamefont {D.}~\bibnamefont
  {Wang}},\ }\href@noop {} {\  (\bibinfo {year} {2022}{\natexlab{b}})},\
  \Eprint {http://arxiv.org/abs/2203.10959} {arXiv:2203.10959 [astro-ph.CO]}
  \BibitemShut {NoStop}%
\bibitem [{\citenamefont {Cheong}\ \emph {et~al.}(2022)\citenamefont {Cheong},
  \citenamefont {Kohri},\ and\ \citenamefont {Park}}]{Cheong:2022gfc}%
  \BibitemOpen
  \bibfield  {author} {\bibinfo {author} {\bibfnamefont {D.~Y.}\ \bibnamefont
  {Cheong}}, \bibinfo {author} {\bibfnamefont {K.}~\bibnamefont {Kohri}}, \
  and\ \bibinfo {author} {\bibfnamefont {S.~C.}\ \bibnamefont {Park}},\ }\href
  {\doibase 10.1088/1475-7516/2022/10/015} {\bibfield  {journal} {\bibinfo
  {journal} {JCAP}\ }\textbf {\bibinfo {volume} {10}},\ \bibinfo {pages} {015}
  (\bibinfo {year} {2022})},\ \Eprint {http://arxiv.org/abs/2205.14813}
  {arXiv:2205.14813 [hep-ph]} \BibitemShut {NoStop}%
\bibitem [{\citenamefont {Bernardo}\ and\ \citenamefont
  {Ng}(2023{\natexlab{a}})}]{Bernardo:2022vlj}%
  \BibitemOpen
  \bibfield  {author} {\bibinfo {author} {\bibfnamefont {R.~C.}\ \bibnamefont
  {Bernardo}}\ and\ \bibinfo {author} {\bibfnamefont {K.-W.}\ \bibnamefont
  {Ng}},\ }\href {\doibase 10.1016/j.physletb.2023.137939} {\bibfield
  {journal} {\bibinfo  {journal} {Phys. Lett. B}\ }\textbf {\bibinfo {volume}
  {841}},\ \bibinfo {pages} {137939} (\bibinfo {year} {2023}{\natexlab{a}})},\
  \Eprint {http://arxiv.org/abs/2206.01056} {arXiv:2206.01056 [astro-ph.CO]}
  \BibitemShut {NoStop}%
\bibitem [{\citenamefont {Ghoshal}\ \emph {et~al.}(2022)\citenamefont
  {Ghoshal}, \citenamefont {Heurtier},\ and\ \citenamefont
  {Paul}}]{Ghoshal:2022ruy}%
  \BibitemOpen
  \bibfield  {author} {\bibinfo {author} {\bibfnamefont {A.}~\bibnamefont
  {Ghoshal}}, \bibinfo {author} {\bibfnamefont {L.}~\bibnamefont {Heurtier}}, \
  and\ \bibinfo {author} {\bibfnamefont {A.}~\bibnamefont {Paul}},\ }\href
  {\doibase 10.1007/JHEP12(2022)105} {\bibfield  {journal} {\bibinfo  {journal}
  {JHEP}\ }\textbf {\bibinfo {volume} {12}},\ \bibinfo {pages} {105} (\bibinfo
  {year} {2022})},\ \Eprint {http://arxiv.org/abs/2208.01670} {arXiv:2208.01670
  [hep-ph]} \BibitemShut {NoStop}%
\bibitem [{\citenamefont {Datta}\ and\ \citenamefont
  {Samanta}(2022)}]{Datta:2022tab}%
  \BibitemOpen
  \bibfield  {author} {\bibinfo {author} {\bibfnamefont {S.}~\bibnamefont
  {Datta}}\ and\ \bibinfo {author} {\bibfnamefont {R.}~\bibnamefont
  {Samanta}},\ }\href {\doibase 10.1007/JHEP11(2022)159} {\bibfield  {journal}
  {\bibinfo  {journal} {JHEP}\ }\textbf {\bibinfo {volume} {11}},\ \bibinfo
  {pages} {159} (\bibinfo {year} {2022})},\ \Eprint
  {http://arxiv.org/abs/2208.09949} {arXiv:2208.09949 [hep-ph]} \BibitemShut
  {NoStop}%
\bibitem [{\citenamefont {El~Bourakadi}\ \emph {et~al.}(2022)\citenamefont
  {El~Bourakadi}, \citenamefont {Asfour}, \citenamefont {Sakhi}, \citenamefont
  {Bennai},\ and\ \citenamefont {Ouali}}]{ElBourakadi:2022anr}%
  \BibitemOpen
  \bibfield  {author} {\bibinfo {author} {\bibfnamefont {K.}~\bibnamefont
  {El~Bourakadi}}, \bibinfo {author} {\bibfnamefont {B.}~\bibnamefont
  {Asfour}}, \bibinfo {author} {\bibfnamefont {Z.}~\bibnamefont {Sakhi}},
  \bibinfo {author} {\bibfnamefont {Z.~M.}\ \bibnamefont {Bennai}}, \ and\
  \bibinfo {author} {\bibfnamefont {T.}~\bibnamefont {Ouali}},\ }\href
  {\doibase 10.1140/epjc/s10052-022-10762-7} {\bibfield  {journal} {\bibinfo
  {journal} {Eur. Phys. J. C}\ }\textbf {\bibinfo {volume} {82}},\ \bibinfo
  {pages} {792} (\bibinfo {year} {2022})},\ \Eprint
  {http://arxiv.org/abs/2209.08585} {arXiv:2209.08585 [gr-qc]} \BibitemShut
  {NoStop}%
\bibitem [{\citenamefont {Blasi}\ \emph
  {et~al.}(2023{\natexlab{a}})\citenamefont {Blasi}, \citenamefont {Mariotti},
  \citenamefont {Rase}, \citenamefont {Sevrin},\ and\ \citenamefont
  {Turbang}}]{Blasi:2022ayo}%
  \BibitemOpen
  \bibfield  {author} {\bibinfo {author} {\bibfnamefont {S.}~\bibnamefont
  {Blasi}}, \bibinfo {author} {\bibfnamefont {A.}~\bibnamefont {Mariotti}},
  \bibinfo {author} {\bibfnamefont {A.}~\bibnamefont {Rase}}, \bibinfo {author}
  {\bibfnamefont {A.}~\bibnamefont {Sevrin}}, \ and\ \bibinfo {author}
  {\bibfnamefont {K.}~\bibnamefont {Turbang}},\ }\href {\doibase
  10.1088/1475-7516/2023/04/008} {\bibfield  {journal} {\bibinfo  {journal}
  {JCAP}\ }\textbf {\bibinfo {volume} {04}},\ \bibinfo {pages} {008} (\bibinfo
  {year} {2023}{\natexlab{a}})},\ \Eprint {http://arxiv.org/abs/2210.14246}
  {arXiv:2210.14246 [hep-ph]} \BibitemShut {NoStop}%
\bibitem [{\citenamefont {Borah}\ \emph
  {et~al.}(2023{\natexlab{a}})\citenamefont {Borah}, \citenamefont {Jyoti~Das},
  \citenamefont {Samanta},\ and\ \citenamefont {Urban}}]{Borah:2022iym}%
  \BibitemOpen
  \bibfield  {author} {\bibinfo {author} {\bibfnamefont {D.}~\bibnamefont
  {Borah}}, \bibinfo {author} {\bibfnamefont {S.}~\bibnamefont {Jyoti~Das}},
  \bibinfo {author} {\bibfnamefont {R.}~\bibnamefont {Samanta}}, \ and\
  \bibinfo {author} {\bibfnamefont {F.~R.}\ \bibnamefont {Urban}},\ }\href
  {\doibase 10.1007/JHEP03(2023)127} {\bibfield  {journal} {\bibinfo  {journal}
  {JHEP}\ }\textbf {\bibinfo {volume} {03}},\ \bibinfo {pages} {127} (\bibinfo
  {year} {2023}{\natexlab{a}})},\ \Eprint {http://arxiv.org/abs/2211.15726}
  {arXiv:2211.15726 [hep-ph]} \BibitemShut {NoStop}%
\bibitem [{\citenamefont {Saad}(2023)}]{Saad:2022mzu}%
  \BibitemOpen
  \bibfield  {author} {\bibinfo {author} {\bibfnamefont {S.}~\bibnamefont
  {Saad}},\ }\href {\doibase 10.1007/JHEP04(2023)058} {\bibfield  {journal}
  {\bibinfo  {journal} {JHEP}\ }\textbf {\bibinfo {volume} {04}},\ \bibinfo
  {pages} {058} (\bibinfo {year} {2023})},\ \Eprint
  {http://arxiv.org/abs/2212.05291} {arXiv:2212.05291 [hep-ph]} \BibitemShut
  {NoStop}%
\bibitem [{\citenamefont {Bian}\ \emph {et~al.}(2022)\citenamefont {Bian},
  \citenamefont {Ge}, \citenamefont {Li}, \citenamefont {Shu},\ and\
  \citenamefont {Zong}}]{Bian:2022qbh}%
  \BibitemOpen
  \bibfield  {author} {\bibinfo {author} {\bibfnamefont {L.}~\bibnamefont
  {Bian}}, \bibinfo {author} {\bibfnamefont {S.}~\bibnamefont {Ge}}, \bibinfo
  {author} {\bibfnamefont {C.}~\bibnamefont {Li}}, \bibinfo {author}
  {\bibfnamefont {J.}~\bibnamefont {Shu}}, \ and\ \bibinfo {author}
  {\bibfnamefont {J.}~\bibnamefont {Zong}},\ }\href@noop {} {\  (\bibinfo
  {year} {2022})},\ \Eprint {http://arxiv.org/abs/2212.07871} {arXiv:2212.07871
  [hep-ph]} \BibitemShut {NoStop}%
\bibitem [{\citenamefont {Cai}(2023)}]{Cai:2022lec}%
  \BibitemOpen
  \bibfield  {author} {\bibinfo {author} {\bibfnamefont {Y.}~\bibnamefont
  {Cai}},\ }\href {\doibase 10.1103/PhysRevD.107.063512} {\bibfield  {journal}
  {\bibinfo  {journal} {Phys. Rev. D}\ }\textbf {\bibinfo {volume} {107}},\
  \bibinfo {pages} {063512} (\bibinfo {year} {2023})},\ \Eprint
  {http://arxiv.org/abs/2212.10893} {arXiv:2212.10893 [gr-qc]} \BibitemShut
  {NoStop}%
\bibitem [{\citenamefont {Arzoumanian}\ \emph {et~al.}(2023)\citenamefont
  {Arzoumanian} \emph {et~al.}}]{NANOGrav:2023bts}%
  \BibitemOpen
  \bibfield  {author} {\bibinfo {author} {\bibfnamefont {Z.}~\bibnamefont
  {Arzoumanian}} \emph {et~al.} (\bibinfo {collaboration} {NANOGrav}),\
  }\href@noop {} {\  (\bibinfo {year} {2023})},\ \Eprint
  {http://arxiv.org/abs/2301.03608} {arXiv:2301.03608 [astro-ph.GA]}
  \BibitemShut {NoStop}%
\bibitem [{\citenamefont {Berbig}\ and\ \citenamefont
  {Ghoshal}(2023)}]{Berbig:2023yyy}%
  \BibitemOpen
  \bibfield  {author} {\bibinfo {author} {\bibfnamefont {M.}~\bibnamefont
  {Berbig}}\ and\ \bibinfo {author} {\bibfnamefont {A.}~\bibnamefont
  {Ghoshal}},\ }\href {\doibase 10.1007/JHEP05(2023)172} {\bibfield  {journal}
  {\bibinfo  {journal} {JHEP}\ }\textbf {\bibinfo {volume} {05}},\ \bibinfo
  {pages} {172} (\bibinfo {year} {2023})},\ \Eprint
  {http://arxiv.org/abs/2301.05672} {arXiv:2301.05672 [hep-ph]} \BibitemShut
  {NoStop}%
\bibitem [{\citenamefont {Wu}\ \emph {et~al.}(2023)\citenamefont {Wu},
  \citenamefont {Chen},\ and\ \citenamefont {Huang}}]{Wu:2023pbt}%
  \BibitemOpen
  \bibfield  {author} {\bibinfo {author} {\bibfnamefont {Y.-M.}\ \bibnamefont
  {Wu}}, \bibinfo {author} {\bibfnamefont {Z.-C.}\ \bibnamefont {Chen}}, \ and\
  \bibinfo {author} {\bibfnamefont {Q.-G.}\ \bibnamefont {Huang}},\ }\href
  {\doibase 10.1103/PhysRevD.107.042003} {\bibfield  {journal} {\bibinfo
  {journal} {Phys. Rev. D}\ }\textbf {\bibinfo {volume} {107}},\ \bibinfo
  {pages} {042003} (\bibinfo {year} {2023})},\ \Eprint
  {http://arxiv.org/abs/2302.00229} {arXiv:2302.00229 [gr-qc]} \BibitemShut
  {NoStop}%
\bibitem [{\citenamefont {Dandoy}\ \emph {et~al.}(2023)\citenamefont {Dandoy},
  \citenamefont {Domcke},\ and\ \citenamefont {Rompineve}}]{Dandoy:2023jot}%
  \BibitemOpen
  \bibfield  {author} {\bibinfo {author} {\bibfnamefont {V.}~\bibnamefont
  {Dandoy}}, \bibinfo {author} {\bibfnamefont {V.}~\bibnamefont {Domcke}}, \
  and\ \bibinfo {author} {\bibfnamefont {F.}~\bibnamefont {Rompineve}},\
  }\href@noop {} {\  (\bibinfo {year} {2023})},\ \Eprint
  {http://arxiv.org/abs/2302.07901} {arXiv:2302.07901 [astro-ph.CO]}
  \BibitemShut {NoStop}%
\bibitem [{\citenamefont {Bernardo}\ and\ \citenamefont
  {Ng}(2023{\natexlab{b}})}]{Bernardo:2023mxc}%
  \BibitemOpen
  \bibfield  {author} {\bibinfo {author} {\bibfnamefont {R.~C.}\ \bibnamefont
  {Bernardo}}\ and\ \bibinfo {author} {\bibfnamefont {K.-W.}\ \bibnamefont
  {Ng}},\ }\href {\doibase 10.1103/PhysRevD.107.L101502} {\bibfield  {journal}
  {\bibinfo  {journal} {Phys. Rev. D}\ }\textbf {\bibinfo {volume} {107}},\
  \bibinfo {pages} {L101502} (\bibinfo {year} {2023}{\natexlab{b}})},\ \Eprint
  {http://arxiv.org/abs/2302.11796} {arXiv:2302.11796 [gr-qc]} \BibitemShut
  {NoStop}%
\bibitem [{\citenamefont {An}\ and\ \citenamefont {Yang}(2023)}]{An:2023idh}%
  \BibitemOpen
  \bibfield  {author} {\bibinfo {author} {\bibfnamefont {H.}~\bibnamefont
  {An}}\ and\ \bibinfo {author} {\bibfnamefont {C.}~\bibnamefont {Yang}},\
  }\href@noop {} {\  (\bibinfo {year} {2023})},\ \Eprint
  {http://arxiv.org/abs/2304.02361} {arXiv:2304.02361 [hep-ph]} \BibitemShut
  {NoStop}%
\bibitem [{\citenamefont {Ferrer}\ \emph {et~al.}(2023)\citenamefont {Ferrer},
  \citenamefont {Ghoshal},\ and\ \citenamefont {Lewicki}}]{Ferrer:2023uwz}%
  \BibitemOpen
  \bibfield  {author} {\bibinfo {author} {\bibfnamefont {F.}~\bibnamefont
  {Ferrer}}, \bibinfo {author} {\bibfnamefont {A.}~\bibnamefont {Ghoshal}}, \
  and\ \bibinfo {author} {\bibfnamefont {M.}~\bibnamefont {Lewicki}},\
  }\href@noop {} {\  (\bibinfo {year} {2023})},\ \Eprint
  {http://arxiv.org/abs/2304.02636} {arXiv:2304.02636 [astro-ph.CO]}
  \BibitemShut {NoStop}%
\bibitem [{\citenamefont {Ghoshal}\ \emph
  {et~al.}(2023{\natexlab{b}})\citenamefont {Ghoshal}, \citenamefont
  {Gouttenoire}, \citenamefont {Heurtier},\ and\ \citenamefont
  {Simakachorn}}]{Ghoshal:2023sfa}%
  \BibitemOpen
  \bibfield  {author} {\bibinfo {author} {\bibfnamefont {A.}~\bibnamefont
  {Ghoshal}}, \bibinfo {author} {\bibfnamefont {Y.}~\bibnamefont
  {Gouttenoire}}, \bibinfo {author} {\bibfnamefont {L.}~\bibnamefont
  {Heurtier}}, \ and\ \bibinfo {author} {\bibfnamefont {P.}~\bibnamefont
  {Simakachorn}},\ }\href@noop {} {\  (\bibinfo {year} {2023}{\natexlab{b}})},\
  \Eprint {http://arxiv.org/abs/2304.04793} {arXiv:2304.04793 [hep-ph]}
  \BibitemShut {NoStop}%
\bibitem [{\citenamefont {Ferrante}\ \emph {et~al.}(2023)\citenamefont
  {Ferrante}, \citenamefont {Franciolini}, \citenamefont {Iovino},\ and\
  \citenamefont {Urbano}}]{Ferrante:2023bgz}%
  \BibitemOpen
  \bibfield  {author} {\bibinfo {author} {\bibfnamefont {G.}~\bibnamefont
  {Ferrante}}, \bibinfo {author} {\bibfnamefont {G.}~\bibnamefont
  {Franciolini}}, \bibinfo {author} {\bibfnamefont {A.}~\bibnamefont {Iovino},
  \bibfnamefont {Junior.}}, \ and\ \bibinfo {author} {\bibfnamefont
  {A.}~\bibnamefont {Urbano}},\ }\href {\doibase 10.1088/1475-7516/2023/06/057}
  {\bibfield  {journal} {\bibinfo  {journal} {JCAP}\ }\textbf {\bibinfo
  {volume} {06}},\ \bibinfo {pages} {057} (\bibinfo {year} {2023})},\ \Eprint
  {http://arxiv.org/abs/2305.13382} {arXiv:2305.13382 [astro-ph.CO]}
  \BibitemShut {NoStop}%
\bibitem [{\citenamefont {Bringmann}\ \emph {et~al.}(2023)\citenamefont
  {Bringmann}, \citenamefont {Depta}, \citenamefont {Konstandin}, \citenamefont
  {Schmidt-Hoberg},\ and\ \citenamefont {Tasillo}}]{Bringmann:2023opz}%
  \BibitemOpen
  \bibfield  {author} {\bibinfo {author} {\bibfnamefont {T.}~\bibnamefont
  {Bringmann}}, \bibinfo {author} {\bibfnamefont {P.~F.}\ \bibnamefont
  {Depta}}, \bibinfo {author} {\bibfnamefont {T.}~\bibnamefont {Konstandin}},
  \bibinfo {author} {\bibfnamefont {K.}~\bibnamefont {Schmidt-Hoberg}}, \ and\
  \bibinfo {author} {\bibfnamefont {C.}~\bibnamefont {Tasillo}},\ }\href@noop
  {} {\  (\bibinfo {year} {2023})},\ \Eprint {http://arxiv.org/abs/2306.09411}
  {arXiv:2306.09411 [astro-ph.CO]} \BibitemShut {NoStop}%
\bibitem [{\citenamefont {Madge}\ \emph {et~al.}(2023)\citenamefont {Madge},
  \citenamefont {Morgante}, \citenamefont {Ib\'a\~nez}, \citenamefont
  {Ramberg}, \citenamefont {Ratzinger}, \citenamefont {Schenk},\ and\
  \citenamefont {Schwaller}}]{Madge:2023cak}%
  \BibitemOpen
  \bibfield  {author} {\bibinfo {author} {\bibfnamefont {E.}~\bibnamefont
  {Madge}}, \bibinfo {author} {\bibfnamefont {E.}~\bibnamefont {Morgante}},
  \bibinfo {author} {\bibfnamefont {C.~P.}\ \bibnamefont {Ib\'a\~nez}},
  \bibinfo {author} {\bibfnamefont {N.}~\bibnamefont {Ramberg}}, \bibinfo
  {author} {\bibfnamefont {W.}~\bibnamefont {Ratzinger}}, \bibinfo {author}
  {\bibfnamefont {S.}~\bibnamefont {Schenk}}, \ and\ \bibinfo {author}
  {\bibfnamefont {P.}~\bibnamefont {Schwaller}},\ }\href@noop {} {\  (\bibinfo
  {year} {2023})},\ \Eprint {http://arxiv.org/abs/2306.14856} {arXiv:2306.14856
  [hep-ph]} \BibitemShut {NoStop}%
\bibitem [{\citenamefont {Baldes}\ \emph {et~al.}(2023)\citenamefont {Baldes},
  \citenamefont {Dichtl}, \citenamefont {Gouttenoire},\ and\ \citenamefont
  {Sala}}]{Baldes:2023fsp}%
  \BibitemOpen
  \bibfield  {author} {\bibinfo {author} {\bibfnamefont {I.}~\bibnamefont
  {Baldes}}, \bibinfo {author} {\bibfnamefont {M.}~\bibnamefont {Dichtl}},
  \bibinfo {author} {\bibfnamefont {Y.}~\bibnamefont {Gouttenoire}}, \ and\
  \bibinfo {author} {\bibfnamefont {F.}~\bibnamefont {Sala}},\ }\href@noop {}
  {\  (\bibinfo {year} {2023})},\ \Eprint {http://arxiv.org/abs/2306.15555}
  {arXiv:2306.15555 [hep-ph]} \BibitemShut {NoStop}%
\bibitem [{\citenamefont {Piao}\ and\ \citenamefont
  {Zhang}(2004)}]{Piao:2004tq}%
  \BibitemOpen
  \bibfield  {author} {\bibinfo {author} {\bibfnamefont {Y.-S.}\ \bibnamefont
  {Piao}}\ and\ \bibinfo {author} {\bibfnamefont {Y.-Z.}\ \bibnamefont
  {Zhang}},\ }\href {\doibase 10.1103/PhysRevD.70.063513} {\bibfield  {journal}
  {\bibinfo  {journal} {Phys. Rev. D}\ }\textbf {\bibinfo {volume} {70}},\
  \bibinfo {pages} {063513} (\bibinfo {year} {2004})},\ \Eprint
  {http://arxiv.org/abs/astro-ph/0401231} {arXiv:astro-ph/0401231} \BibitemShut
  {NoStop}%
\bibitem [{\citenamefont {Liu}\ \emph {et~al.}(2011)\citenamefont {Liu},
  \citenamefont {Zhang},\ and\ \citenamefont {Piao}}]{Liu:2010dh}%
  \BibitemOpen
  \bibfield  {author} {\bibinfo {author} {\bibfnamefont {Z.-G.}\ \bibnamefont
  {Liu}}, \bibinfo {author} {\bibfnamefont {J.}~\bibnamefont {Zhang}}, \ and\
  \bibinfo {author} {\bibfnamefont {Y.-S.}\ \bibnamefont {Piao}},\ }\href
  {\doibase 10.1016/j.physletb.2010.12.055} {\bibfield  {journal} {\bibinfo
  {journal} {Phys. Lett. B}\ }\textbf {\bibinfo {volume} {697}},\ \bibinfo
  {pages} {407} (\bibinfo {year} {2011})},\ \Eprint
  {http://arxiv.org/abs/1012.0673} {arXiv:1012.0673 [gr-qc]} \BibitemShut
  {NoStop}%
\bibitem [{\citenamefont {Liu}\ and\ \citenamefont {Piao}(2012)}]{Liu:2012iba}%
  \BibitemOpen
  \bibfield  {author} {\bibinfo {author} {\bibfnamefont {Z.-G.}\ \bibnamefont
  {Liu}}\ and\ \bibinfo {author} {\bibfnamefont {Y.-S.}\ \bibnamefont {Piao}},\
  }\href {\doibase 10.1016/j.physletb.2012.05.027} {\bibfield  {journal}
  {\bibinfo  {journal} {Phys. Lett. B}\ }\textbf {\bibinfo {volume} {713}},\
  \bibinfo {pages} {53} (\bibinfo {year} {2012})},\ \Eprint
  {http://arxiv.org/abs/1203.4901} {arXiv:1203.4901 [gr-qc]} \BibitemShut
  {NoStop}%
\bibitem [{\citenamefont {Dinda}\ \emph {et~al.}(2014)\citenamefont {Dinda},
  \citenamefont {Kumar},\ and\ \citenamefont {Sen}}]{Dinda:2014zta}%
  \BibitemOpen
  \bibfield  {author} {\bibinfo {author} {\bibfnamefont {B.~R.}\ \bibnamefont
  {Dinda}}, \bibinfo {author} {\bibfnamefont {S.}~\bibnamefont {Kumar}}, \ and\
  \bibinfo {author} {\bibfnamefont {A.~A.}\ \bibnamefont {Sen}},\ }\href
  {\doibase 10.1103/PhysRevD.90.083515} {\bibfield  {journal} {\bibinfo
  {journal} {Phys. Rev. D}\ }\textbf {\bibinfo {volume} {90}},\ \bibinfo
  {pages} {083515} (\bibinfo {year} {2014})},\ \Eprint
  {http://arxiv.org/abs/1404.3683} {arXiv:1404.3683 [astro-ph.CO]} \BibitemShut
  {NoStop}%
\bibitem [{\citenamefont {Richarte}\ and\ \citenamefont
  {Kremer}(2017)}]{Richarte:2016qqm}%
  \BibitemOpen
  \bibfield  {author} {\bibinfo {author} {\bibfnamefont {M.~G.}\ \bibnamefont
  {Richarte}}\ and\ \bibinfo {author} {\bibfnamefont {G.~M.}\ \bibnamefont
  {Kremer}},\ }\href {\doibase 10.1140/epjc/s10052-017-4629-8} {\bibfield
  {journal} {\bibinfo  {journal} {Eur. Phys. J. C}\ }\textbf {\bibinfo {volume}
  {77}},\ \bibinfo {pages} {51} (\bibinfo {year} {2017})},\ \Eprint
  {http://arxiv.org/abs/1612.03822} {arXiv:1612.03822 [gr-qc]} \BibitemShut
  {NoStop}%
\bibitem [{\citenamefont {Kobayashi}\ \emph {et~al.}(2010)\citenamefont
  {Kobayashi}, \citenamefont {Yamaguchi},\ and\ \citenamefont
  {Yokoyama}}]{Kobayashi:2010cm}%
  \BibitemOpen
  \bibfield  {author} {\bibinfo {author} {\bibfnamefont {T.}~\bibnamefont
  {Kobayashi}}, \bibinfo {author} {\bibfnamefont {M.}~\bibnamefont
  {Yamaguchi}}, \ and\ \bibinfo {author} {\bibfnamefont {J.}~\bibnamefont
  {Yokoyama}},\ }\href {\doibase 10.1103/PhysRevLett.105.231302} {\bibfield
  {journal} {\bibinfo  {journal} {Phys. Rev. Lett.}\ }\textbf {\bibinfo
  {volume} {105}},\ \bibinfo {pages} {231302} (\bibinfo {year} {2010})},\
  \Eprint {http://arxiv.org/abs/1008.0603} {arXiv:1008.0603 [hep-th]}
  \BibitemShut {NoStop}%
\bibitem [{\citenamefont {Myrzakulov}\ \emph
  {et~al.}(2015{\natexlab{b}})\citenamefont {Myrzakulov}, \citenamefont
  {Sebastiani},\ and\ \citenamefont {Vagnozzi}}]{Myrzakulov:2015qaa}%
  \BibitemOpen
  \bibfield  {author} {\bibinfo {author} {\bibfnamefont {R.}~\bibnamefont
  {Myrzakulov}}, \bibinfo {author} {\bibfnamefont {L.}~\bibnamefont
  {Sebastiani}}, \ and\ \bibinfo {author} {\bibfnamefont {S.}~\bibnamefont
  {Vagnozzi}},\ }\href {\doibase 10.1140/epjc/s10052-015-3672-6} {\bibfield
  {journal} {\bibinfo  {journal} {Eur. Phys. J. C}\ }\textbf {\bibinfo {volume}
  {75}},\ \bibinfo {pages} {444} (\bibinfo {year} {2015}{\natexlab{b}})},\
  \Eprint {http://arxiv.org/abs/1504.07984} {arXiv:1504.07984 [gr-qc]}
  \BibitemShut {NoStop}%
\bibitem [{\citenamefont {Fujita}\ \emph {et~al.}(2019)\citenamefont {Fujita},
  \citenamefont {Kuroyanagi}, \citenamefont {Mizuno},\ and\ \citenamefont
  {Mukohyama}}]{Fujita:2018ehq}%
  \BibitemOpen
  \bibfield  {author} {\bibinfo {author} {\bibfnamefont {T.}~\bibnamefont
  {Fujita}}, \bibinfo {author} {\bibfnamefont {S.}~\bibnamefont {Kuroyanagi}},
  \bibinfo {author} {\bibfnamefont {S.}~\bibnamefont {Mizuno}}, \ and\ \bibinfo
  {author} {\bibfnamefont {S.}~\bibnamefont {Mukohyama}},\ }\href {\doibase
  10.1016/j.physletb.2018.12.025} {\bibfield  {journal} {\bibinfo  {journal}
  {Phys. Lett. B}\ }\textbf {\bibinfo {volume} {789}},\ \bibinfo {pages} {215}
  (\bibinfo {year} {2019})},\ \Eprint {http://arxiv.org/abs/1808.02381}
  {arXiv:1808.02381 [gr-qc]} \BibitemShut {NoStop}%
\bibitem [{\citenamefont {Odintsov}\ and\ \citenamefont
  {Oikonomou}(2019)}]{Odintsov:2019evb}%
  \BibitemOpen
  \bibfield  {author} {\bibinfo {author} {\bibfnamefont {S.~D.}\ \bibnamefont
  {Odintsov}}\ and\ \bibinfo {author} {\bibfnamefont {V.~K.}\ \bibnamefont
  {Oikonomou}},\ }\href {\doibase 10.1103/PhysRevD.99.104070} {\bibfield
  {journal} {\bibinfo  {journal} {Phys. Rev. D}\ }\textbf {\bibinfo {volume}
  {99}},\ \bibinfo {pages} {104070} (\bibinfo {year} {2019})},\ \Eprint
  {http://arxiv.org/abs/1905.03496} {arXiv:1905.03496 [gr-qc]} \BibitemShut
  {NoStop}%
\bibitem [{\citenamefont {Nojiri}\ \emph
  {et~al.}(2020{\natexlab{a}})\citenamefont {Nojiri}, \citenamefont
  {Odintsov},\ and\ \citenamefont {Oikonomou}}]{Nojiri:2019riz}%
  \BibitemOpen
  \bibfield  {author} {\bibinfo {author} {\bibfnamefont {S.}~\bibnamefont
  {Nojiri}}, \bibinfo {author} {\bibfnamefont {S.~D.}\ \bibnamefont
  {Odintsov}}, \ and\ \bibinfo {author} {\bibfnamefont {V.~K.}\ \bibnamefont
  {Oikonomou}},\ }\href {\doibase 10.1016/j.aop.2020.168186} {\bibfield
  {journal} {\bibinfo  {journal} {Annals Phys.}\ }\textbf {\bibinfo {volume}
  {418}},\ \bibinfo {pages} {168186} (\bibinfo {year} {2020}{\natexlab{a}})},\
  \Eprint {http://arxiv.org/abs/1907.01625} {arXiv:1907.01625 [gr-qc]}
  \BibitemShut {NoStop}%
\bibitem [{\citenamefont {Nojiri}\ \emph
  {et~al.}(2020{\natexlab{b}})\citenamefont {Nojiri}, \citenamefont
  {Odintsov},\ and\ \citenamefont {Oikonomou}}]{Nojiri:2019fft}%
  \BibitemOpen
  \bibfield  {author} {\bibinfo {author} {\bibfnamefont {S.}~\bibnamefont
  {Nojiri}}, \bibinfo {author} {\bibfnamefont {S.~D.}\ \bibnamefont
  {Odintsov}}, \ and\ \bibinfo {author} {\bibfnamefont {V.~K.}\ \bibnamefont
  {Oikonomou}},\ }\href {\doibase 10.1016/j.dark.2020.100602} {\bibfield
  {journal} {\bibinfo  {journal} {Phys. Dark Univ.}\ }\textbf {\bibinfo
  {volume} {29}},\ \bibinfo {pages} {100602} (\bibinfo {year}
  {2020}{\natexlab{b}})},\ \Eprint {http://arxiv.org/abs/1912.13128}
  {arXiv:1912.13128 [gr-qc]} \BibitemShut {NoStop}%
\bibitem [{\citenamefont {Odintsov}\ and\ \citenamefont
  {Oikonomou}(2020)}]{Odintsov:2020nwm}%
  \BibitemOpen
  \bibfield  {author} {\bibinfo {author} {\bibfnamefont {S.~D.}\ \bibnamefont
  {Odintsov}}\ and\ \bibinfo {author} {\bibfnamefont {V.~K.}\ \bibnamefont
  {Oikonomou}},\ }\href {\doibase 10.1103/PhysRevD.101.044009} {\bibfield
  {journal} {\bibinfo  {journal} {Phys. Rev. D}\ }\textbf {\bibinfo {volume}
  {101}},\ \bibinfo {pages} {044009} (\bibinfo {year} {2020})},\ \Eprint
  {http://arxiv.org/abs/2001.06830} {arXiv:2001.06830 [gr-qc]} \BibitemShut
  {NoStop}%
\bibitem [{\citenamefont {Nojiri}\ \emph
  {et~al.}(2020{\natexlab{c}})\citenamefont {Nojiri}, \citenamefont {Odintsov},
  \citenamefont {Oikonomou},\ and\ \citenamefont {Popov}}]{Nojiri:2020pqr}%
  \BibitemOpen
  \bibfield  {author} {\bibinfo {author} {\bibfnamefont {S.}~\bibnamefont
  {Nojiri}}, \bibinfo {author} {\bibfnamefont {S.~D.}\ \bibnamefont
  {Odintsov}}, \bibinfo {author} {\bibfnamefont {V.~K.}\ \bibnamefont
  {Oikonomou}}, \ and\ \bibinfo {author} {\bibfnamefont {A.~A.}\ \bibnamefont
  {Popov}},\ }\href {\doibase 10.1016/j.dark.2020.100514} {\bibfield  {journal}
  {\bibinfo  {journal} {Phys. Dark Univ.}\ }\textbf {\bibinfo {volume} {28}},\
  \bibinfo {pages} {100514} (\bibinfo {year} {2020}{\natexlab{c}})},\ \Eprint
  {http://arxiv.org/abs/2002.10402} {arXiv:2002.10402 [gr-qc]} \BibitemShut
  {NoStop}%
\bibitem [{\citenamefont {Odintsov}\ \emph {et~al.}(2020)\citenamefont
  {Odintsov}, \citenamefont {Oikonomou}, \citenamefont {Fronimos},\ and\
  \citenamefont {Fasoulakos}}]{Odintsov:2020vjb}%
  \BibitemOpen
  \bibfield  {author} {\bibinfo {author} {\bibfnamefont {S.~D.}\ \bibnamefont
  {Odintsov}}, \bibinfo {author} {\bibfnamefont {V.~K.}\ \bibnamefont
  {Oikonomou}}, \bibinfo {author} {\bibfnamefont {F.~P.}\ \bibnamefont
  {Fronimos}}, \ and\ \bibinfo {author} {\bibfnamefont {K.~V.}\ \bibnamefont
  {Fasoulakos}},\ }\href {\doibase 10.1103/PhysRevD.102.104042} {\bibfield
  {journal} {\bibinfo  {journal} {Phys. Rev. D}\ }\textbf {\bibinfo {volume}
  {102}},\ \bibinfo {pages} {104042} (\bibinfo {year} {2020})},\ \Eprint
  {http://arxiv.org/abs/2010.13580} {arXiv:2010.13580 [gr-qc]} \BibitemShut
  {NoStop}%
\bibitem [{\citenamefont {Kawai}\ and\ \citenamefont
  {Kim}(2021)}]{Kawai:2021edk}%
  \BibitemOpen
  \bibfield  {author} {\bibinfo {author} {\bibfnamefont {S.}~\bibnamefont
  {Kawai}}\ and\ \bibinfo {author} {\bibfnamefont {J.}~\bibnamefont {Kim}},\
  }\href {\doibase 10.1103/PhysRevD.104.083545} {\bibfield  {journal} {\bibinfo
   {journal} {Phys. Rev. D}\ }\textbf {\bibinfo {volume} {104}},\ \bibinfo
  {pages} {083545} (\bibinfo {year} {2021})},\ \Eprint
  {http://arxiv.org/abs/2108.01340} {arXiv:2108.01340 [astro-ph.CO]}
  \BibitemShut {NoStop}%
\bibitem [{\citenamefont {Oikonomou}(2021)}]{Oikonomou:2021kql}%
  \BibitemOpen
  \bibfield  {author} {\bibinfo {author} {\bibfnamefont {V.~K.}\ \bibnamefont
  {Oikonomou}},\ }\href {\doibase 10.1088/1361-6382/ac2168} {\bibfield
  {journal} {\bibinfo  {journal} {Class. Quant. Grav.}\ }\textbf {\bibinfo
  {volume} {38}},\ \bibinfo {pages} {195025} (\bibinfo {year} {2021})},\
  \Eprint {http://arxiv.org/abs/2108.10460} {arXiv:2108.10460 [gr-qc]}
  \BibitemShut {NoStop}%
\bibitem [{\citenamefont {Odintsov}\ \emph
  {et~al.}(2022{\natexlab{b}})\citenamefont {Odintsov}, \citenamefont
  {Oikonomou},\ and\ \citenamefont {Fronimos}}]{Odintsov:2021kup}%
  \BibitemOpen
  \bibfield  {author} {\bibinfo {author} {\bibfnamefont {S.~D.}\ \bibnamefont
  {Odintsov}}, \bibinfo {author} {\bibfnamefont {V.~K.}\ \bibnamefont
  {Oikonomou}}, \ and\ \bibinfo {author} {\bibfnamefont {F.~P.}\ \bibnamefont
  {Fronimos}},\ }\href {\doibase 10.1016/j.dark.2022.100950} {\bibfield
  {journal} {\bibinfo  {journal} {Phys. Dark Univ.}\ }\textbf {\bibinfo
  {volume} {35}},\ \bibinfo {pages} {100950} (\bibinfo {year}
  {2022}{\natexlab{b}})},\ \Eprint {http://arxiv.org/abs/2108.11231}
  {arXiv:2108.11231 [gr-qc]} \BibitemShut {NoStop}%
\bibitem [{\citenamefont {Odintsov}\ and\ \citenamefont
  {Oikonomou}(2022{\natexlab{a}})}]{Odintsov:2021urx}%
  \BibitemOpen
  \bibfield  {author} {\bibinfo {author} {\bibfnamefont {S.~D.}\ \bibnamefont
  {Odintsov}}\ and\ \bibinfo {author} {\bibfnamefont {V.~K.}\ \bibnamefont
  {Oikonomou}},\ }\href {\doibase 10.1016/j.physletb.2021.136817} {\bibfield
  {journal} {\bibinfo  {journal} {Phys. Lett. B}\ }\textbf {\bibinfo {volume}
  {824}},\ \bibinfo {pages} {136817} (\bibinfo {year} {2022}{\natexlab{a}})},\
  \Eprint {http://arxiv.org/abs/2112.02584} {arXiv:2112.02584 [gr-qc]}
  \BibitemShut {NoStop}%
\bibitem [{\citenamefont {Odintsov}\ and\ \citenamefont
  {Oikonomou}(2022{\natexlab{b}})}]{Odintsov:2022sdk}%
  \BibitemOpen
  \bibfield  {author} {\bibinfo {author} {\bibfnamefont {S.~D.}\ \bibnamefont
  {Odintsov}}\ and\ \bibinfo {author} {\bibfnamefont {V.~K.}\ \bibnamefont
  {Oikonomou}},\ }\href {\doibase 10.1002/prop.202100167} {\bibfield  {journal}
  {\bibinfo  {journal} {Fortsch. Phys.}\ }\textbf {\bibinfo {volume} {70}},\
  \bibinfo {pages} {2100167} (\bibinfo {year} {2022}{\natexlab{b}})},\ \Eprint
  {http://arxiv.org/abs/2203.10599} {arXiv:2203.10599 [gr-qc]} \BibitemShut
  {NoStop}%
\bibitem [{\citenamefont {Oikonomou}(2022{\natexlab{a}})}]{Oikonomou:2022xoq}%
  \BibitemOpen
  \bibfield  {author} {\bibinfo {author} {\bibfnamefont {V.~K.}\ \bibnamefont
  {Oikonomou}},\ }\href {\doibase 10.1016/j.astropartphys.2022.102718}
  {\bibfield  {journal} {\bibinfo  {journal} {Astropart. Phys.}\ }\textbf
  {\bibinfo {volume} {141}},\ \bibinfo {pages} {102718} (\bibinfo {year}
  {2022}{\natexlab{a}})},\ \Eprint {http://arxiv.org/abs/2204.06304}
  {arXiv:2204.06304 [gr-qc]} \BibitemShut {NoStop}%
\bibitem [{\citenamefont {Odintsov}\ and\ \citenamefont
  {Oikonomou}(2022{\natexlab{c}})}]{Odintsov:2022hxu}%
  \BibitemOpen
  \bibfield  {author} {\bibinfo {author} {\bibfnamefont {S.~D.}\ \bibnamefont
  {Odintsov}}\ and\ \bibinfo {author} {\bibfnamefont {V.~K.}\ \bibnamefont
  {Oikonomou}},\ }\href {\doibase 10.1103/PhysRevD.105.104054} {\bibfield
  {journal} {\bibinfo  {journal} {Phys. Rev. D}\ }\textbf {\bibinfo {volume}
  {105}},\ \bibinfo {pages} {104054} (\bibinfo {year} {2022}{\natexlab{c}})},\
  \Eprint {http://arxiv.org/abs/2205.07304} {arXiv:2205.07304 [gr-qc]}
  \BibitemShut {NoStop}%
\bibitem [{\citenamefont {Oikonomou}(2022{\natexlab{b}})}]{Oikonomou:2022pdf}%
  \BibitemOpen
  \bibfield  {author} {\bibinfo {author} {\bibfnamefont {V.~K.}\ \bibnamefont
  {Oikonomou}},\ }\href {\doibase 10.1002/andp.202200134} {\bibfield  {journal}
  {\bibinfo  {journal} {Annalen Phys.}\ }\textbf {\bibinfo {volume} {534}},\
  \bibinfo {pages} {2200134} (\bibinfo {year} {2022}{\natexlab{b}})},\ \Eprint
  {http://arxiv.org/abs/2205.15405} {arXiv:2205.15405 [gr-qc]} \BibitemShut
  {NoStop}%
\bibitem [{\citenamefont {Oikonomou}\ and\ \citenamefont
  {Lymperiadou}(2022)}]{Oikonomou:2022yle}%
  \BibitemOpen
  \bibfield  {author} {\bibinfo {author} {\bibfnamefont {V.~K.}\ \bibnamefont
  {Oikonomou}}\ and\ \bibinfo {author} {\bibfnamefont {E.~C.}\ \bibnamefont
  {Lymperiadou}},\ }\href {\doibase 10.3390/sym14061143} {\bibfield  {journal}
  {\bibinfo  {journal} {Symmetry}\ }\textbf {\bibinfo {volume} {14}},\ \bibinfo
  {pages} {1143} (\bibinfo {year} {2022})},\ \Eprint
  {http://arxiv.org/abs/2206.00721} {arXiv:2206.00721 [gr-qc]} \BibitemShut
  {NoStop}%
\bibitem [{\citenamefont {Oikonomou}(2023{\natexlab{a}})}]{Oikonomou:2022ijs}%
  \BibitemOpen
  \bibfield  {author} {\bibinfo {author} {\bibfnamefont {V.~K.}\ \bibnamefont
  {Oikonomou}},\ }\href {\doibase 10.1016/j.astropartphys.2022.102777}
  {\bibfield  {journal} {\bibinfo  {journal} {Astropart. Phys.}\ }\textbf
  {\bibinfo {volume} {144}},\ \bibinfo {pages} {102777} (\bibinfo {year}
  {2023}{\natexlab{a}})},\ \Eprint {http://arxiv.org/abs/2209.09781}
  {arXiv:2209.09781 [gr-qc]} \BibitemShut {NoStop}%
\bibitem [{\citenamefont {Oikonomou}(2022{\natexlab{c}})}]{Oikonomou:2022irx}%
  \BibitemOpen
  \bibfield  {author} {\bibinfo {author} {\bibfnamefont {V.~K.}\ \bibnamefont
  {Oikonomou}},\ }\href {\doibase 10.1016/j.nuclphysb.2022.115985} {\bibfield
  {journal} {\bibinfo  {journal} {Nucl. Phys. B}\ }\textbf {\bibinfo {volume}
  {984}},\ \bibinfo {pages} {115985} (\bibinfo {year} {2022}{\natexlab{c}})},\
  \Eprint {http://arxiv.org/abs/2210.02861} {arXiv:2210.02861 [gr-qc]}
  \BibitemShut {NoStop}%
\bibitem [{\citenamefont {Calcagni}\ and\ \citenamefont
  {Tsujikawa}(2004)}]{Calcagni:2004as}%
  \BibitemOpen
  \bibfield  {author} {\bibinfo {author} {\bibfnamefont {G.}~\bibnamefont
  {Calcagni}}\ and\ \bibinfo {author} {\bibfnamefont {S.}~\bibnamefont
  {Tsujikawa}},\ }\href {\doibase 10.1103/PhysRevD.70.103514} {\bibfield
  {journal} {\bibinfo  {journal} {Phys. Rev. D}\ }\textbf {\bibinfo {volume}
  {70}},\ \bibinfo {pages} {103514} (\bibinfo {year} {2004})},\ \Eprint
  {http://arxiv.org/abs/astro-ph/0407543} {arXiv:astro-ph/0407543} \BibitemShut
  {NoStop}%
\bibitem [{\citenamefont {Calcagni}\ \emph {et~al.}(2014)\citenamefont
  {Calcagni}, \citenamefont {Kuroyanagi}, \citenamefont {Ohashi},\ and\
  \citenamefont {Tsujikawa}}]{Calcagni:2013lya}%
  \BibitemOpen
  \bibfield  {author} {\bibinfo {author} {\bibfnamefont {G.}~\bibnamefont
  {Calcagni}}, \bibinfo {author} {\bibfnamefont {S.}~\bibnamefont
  {Kuroyanagi}}, \bibinfo {author} {\bibfnamefont {J.}~\bibnamefont {Ohashi}},
  \ and\ \bibinfo {author} {\bibfnamefont {S.}~\bibnamefont {Tsujikawa}},\
  }\href {\doibase 10.1088/1475-7516/2014/03/052} {\bibfield  {journal}
  {\bibinfo  {journal} {JCAP}\ }\textbf {\bibinfo {volume} {03}},\ \bibinfo
  {pages} {052} (\bibinfo {year} {2014})},\ \Eprint
  {http://arxiv.org/abs/1310.5186} {arXiv:1310.5186 [astro-ph.CO]} \BibitemShut
  {NoStop}%
\bibitem [{\citenamefont {Endlich}\ \emph {et~al.}(2013)\citenamefont
  {Endlich}, \citenamefont {Nicolis},\ and\ \citenamefont
  {Wang}}]{Endlich:2012pz}%
  \BibitemOpen
  \bibfield  {author} {\bibinfo {author} {\bibfnamefont {S.}~\bibnamefont
  {Endlich}}, \bibinfo {author} {\bibfnamefont {A.}~\bibnamefont {Nicolis}}, \
  and\ \bibinfo {author} {\bibfnamefont {J.}~\bibnamefont {Wang}},\ }\href
  {\doibase 10.1088/1475-7516/2013/10/011} {\bibfield  {journal} {\bibinfo
  {journal} {JCAP}\ }\textbf {\bibinfo {volume} {10}},\ \bibinfo {pages} {011}
  (\bibinfo {year} {2013})},\ \Eprint {http://arxiv.org/abs/1210.0569}
  {arXiv:1210.0569 [hep-th]} \BibitemShut {NoStop}%
\bibitem [{\citenamefont {Cannone}\ \emph {et~al.}(2015)\citenamefont
  {Cannone}, \citenamefont {Tasinato},\ and\ \citenamefont
  {Wands}}]{Cannone:2014uqa}%
  \BibitemOpen
  \bibfield  {author} {\bibinfo {author} {\bibfnamefont {D.}~\bibnamefont
  {Cannone}}, \bibinfo {author} {\bibfnamefont {G.}~\bibnamefont {Tasinato}}, \
  and\ \bibinfo {author} {\bibfnamefont {D.}~\bibnamefont {Wands}},\ }\href
  {\doibase 10.1088/1475-7516/2015/01/029} {\bibfield  {journal} {\bibinfo
  {journal} {JCAP}\ }\textbf {\bibinfo {volume} {01}},\ \bibinfo {pages} {029}
  (\bibinfo {year} {2015})},\ \Eprint {http://arxiv.org/abs/1409.6568}
  {arXiv:1409.6568 [astro-ph.CO]} \BibitemShut {NoStop}%
\bibitem [{\citenamefont {Graef}\ and\ \citenamefont
  {Brandenberger}(2015)}]{Graef:2015ova}%
  \BibitemOpen
  \bibfield  {author} {\bibinfo {author} {\bibfnamefont {L.}~\bibnamefont
  {Graef}}\ and\ \bibinfo {author} {\bibfnamefont {R.}~\bibnamefont
  {Brandenberger}},\ }\href {\doibase 10.1088/1475-7516/2015/10/009} {\bibfield
   {journal} {\bibinfo  {journal} {JCAP}\ }\textbf {\bibinfo {volume} {10}},\
  \bibinfo {pages} {009} (\bibinfo {year} {2015})},\ \Eprint
  {http://arxiv.org/abs/1506.00896} {arXiv:1506.00896 [astro-ph.CO]}
  \BibitemShut {NoStop}%
\bibitem [{\citenamefont {Ricciardone}\ and\ \citenamefont
  {Tasinato}(2017)}]{Ricciardone:2016lym}%
  \BibitemOpen
  \bibfield  {author} {\bibinfo {author} {\bibfnamefont {A.}~\bibnamefont
  {Ricciardone}}\ and\ \bibinfo {author} {\bibfnamefont {G.}~\bibnamefont
  {Tasinato}},\ }\href {\doibase 10.1103/PhysRevD.96.023508} {\bibfield
  {journal} {\bibinfo  {journal} {Phys. Rev. D}\ }\textbf {\bibinfo {volume}
  {96}},\ \bibinfo {pages} {023508} (\bibinfo {year} {2017})},\ \Eprint
  {http://arxiv.org/abs/1611.04516} {arXiv:1611.04516 [astro-ph.CO]}
  \BibitemShut {NoStop}%
\bibitem [{\citenamefont {Graef}\ \emph {et~al.}(2017)\citenamefont {Graef},
  \citenamefont {Benetti},\ and\ \citenamefont {Alcaniz}}]{Graef:2017cfy}%
  \BibitemOpen
  \bibfield  {author} {\bibinfo {author} {\bibfnamefont {L.~L.}\ \bibnamefont
  {Graef}}, \bibinfo {author} {\bibfnamefont {M.}~\bibnamefont {Benetti}}, \
  and\ \bibinfo {author} {\bibfnamefont {J.~S.}\ \bibnamefont {Alcaniz}},\
  }\href {\doibase 10.1088/1475-7516/2017/07/013} {\bibfield  {journal}
  {\bibinfo  {journal} {JCAP}\ }\textbf {\bibinfo {volume} {07}},\ \bibinfo
  {pages} {013} (\bibinfo {year} {2017})},\ \Eprint
  {http://arxiv.org/abs/1705.01961} {arXiv:1705.01961 [astro-ph.CO]}
  \BibitemShut {NoStop}%
\bibitem [{\citenamefont {Baldi}\ \emph {et~al.}(2005)\citenamefont {Baldi},
  \citenamefont {Finelli},\ and\ \citenamefont {Matarrese}}]{Baldi:2005gk}%
  \BibitemOpen
  \bibfield  {author} {\bibinfo {author} {\bibfnamefont {M.}~\bibnamefont
  {Baldi}}, \bibinfo {author} {\bibfnamefont {F.}~\bibnamefont {Finelli}}, \
  and\ \bibinfo {author} {\bibfnamefont {S.}~\bibnamefont {Matarrese}},\ }\href
  {\doibase 10.1103/PhysRevD.72.083504} {\bibfield  {journal} {\bibinfo
  {journal} {Phys. Rev. D}\ }\textbf {\bibinfo {volume} {72}},\ \bibinfo
  {pages} {083504} (\bibinfo {year} {2005})},\ \Eprint
  {http://arxiv.org/abs/astro-ph/0505552} {arXiv:astro-ph/0505552} \BibitemShut
  {NoStop}%
\bibitem [{\citenamefont {Maleknejad}\ and\ \citenamefont
  {Sheikh-Jabbari}(2011)}]{Maleknejad:2011sq}%
  \BibitemOpen
  \bibfield  {author} {\bibinfo {author} {\bibfnamefont {A.}~\bibnamefont
  {Maleknejad}}\ and\ \bibinfo {author} {\bibfnamefont {M.~M.}\ \bibnamefont
  {Sheikh-Jabbari}},\ }\href {\doibase 10.1103/PhysRevD.84.043515} {\bibfield
  {journal} {\bibinfo  {journal} {Phys. Rev. D}\ }\textbf {\bibinfo {volume}
  {84}},\ \bibinfo {pages} {043515} (\bibinfo {year} {2011})},\ \Eprint
  {http://arxiv.org/abs/1102.1932} {arXiv:1102.1932 [hep-ph]} \BibitemShut
  {NoStop}%
\bibitem [{\citenamefont {Adshead}\ and\ \citenamefont
  {Wyman}(2012)}]{Adshead:2012kp}%
  \BibitemOpen
  \bibfield  {author} {\bibinfo {author} {\bibfnamefont {P.}~\bibnamefont
  {Adshead}}\ and\ \bibinfo {author} {\bibfnamefont {M.}~\bibnamefont
  {Wyman}},\ }\href {\doibase 10.1103/PhysRevLett.108.261302} {\bibfield
  {journal} {\bibinfo  {journal} {Phys. Rev. Lett.}\ }\textbf {\bibinfo
  {volume} {108}},\ \bibinfo {pages} {261302} (\bibinfo {year} {2012})},\
  \Eprint {http://arxiv.org/abs/1202.2366} {arXiv:1202.2366 [hep-th]}
  \BibitemShut {NoStop}%
\bibitem [{\citenamefont {Maleknejad}(2016)}]{Maleknejad:2016qjz}%
  \BibitemOpen
  \bibfield  {author} {\bibinfo {author} {\bibfnamefont {A.}~\bibnamefont
  {Maleknejad}},\ }\href {\doibase 10.1007/JHEP07(2016)104} {\bibfield
  {journal} {\bibinfo  {journal} {JHEP}\ }\textbf {\bibinfo {volume} {07}},\
  \bibinfo {pages} {104} (\bibinfo {year} {2016})},\ \Eprint
  {http://arxiv.org/abs/1604.03327} {arXiv:1604.03327 [hep-ph]} \BibitemShut
  {NoStop}%
\bibitem [{\citenamefont {Dimastrogiovanni}\ \emph {et~al.}(2017)\citenamefont
  {Dimastrogiovanni}, \citenamefont {Fasiello},\ and\ \citenamefont
  {Fujita}}]{Dimastrogiovanni:2016fuu}%
  \BibitemOpen
  \bibfield  {author} {\bibinfo {author} {\bibfnamefont {E.}~\bibnamefont
  {Dimastrogiovanni}}, \bibinfo {author} {\bibfnamefont {M.}~\bibnamefont
  {Fasiello}}, \ and\ \bibinfo {author} {\bibfnamefont {T.}~\bibnamefont
  {Fujita}},\ }\href {\doibase 10.1088/1475-7516/2017/01/019} {\bibfield
  {journal} {\bibinfo  {journal} {JCAP}\ }\textbf {\bibinfo {volume} {01}},\
  \bibinfo {pages} {019} (\bibinfo {year} {2017})},\ \Eprint
  {http://arxiv.org/abs/1608.04216} {arXiv:1608.04216 [astro-ph.CO]}
  \BibitemShut {NoStop}%
\bibitem [{\citenamefont {Adshead}\ \emph {et~al.}(2016)\citenamefont
  {Adshead}, \citenamefont {Martinec}, \citenamefont {Sfakianakis},\ and\
  \citenamefont {Wyman}}]{Adshead:2016omu}%
  \BibitemOpen
  \bibfield  {author} {\bibinfo {author} {\bibfnamefont {P.}~\bibnamefont
  {Adshead}}, \bibinfo {author} {\bibfnamefont {E.}~\bibnamefont {Martinec}},
  \bibinfo {author} {\bibfnamefont {E.~I.}\ \bibnamefont {Sfakianakis}}, \ and\
  \bibinfo {author} {\bibfnamefont {M.}~\bibnamefont {Wyman}},\ }\href
  {\doibase 10.1007/JHEP12(2016)137} {\bibfield  {journal} {\bibinfo  {journal}
  {JHEP}\ }\textbf {\bibinfo {volume} {12}},\ \bibinfo {pages} {137} (\bibinfo
  {year} {2016})},\ \Eprint {http://arxiv.org/abs/1609.04025} {arXiv:1609.04025
  [hep-th]} \BibitemShut {NoStop}%
\bibitem [{\citenamefont {Obata}(2017)}]{Obata:2016oym}%
  \BibitemOpen
  \bibfield  {author} {\bibinfo {author} {\bibfnamefont {I.}~\bibnamefont
  {Obata}},\ }\href {\doibase 10.1088/1475-7516/2017/06/050} {\bibfield
  {journal} {\bibinfo  {journal} {JCAP}\ }\textbf {\bibinfo {volume} {06}},\
  \bibinfo {pages} {050} (\bibinfo {year} {2017})},\ \Eprint
  {http://arxiv.org/abs/1612.08817} {arXiv:1612.08817 [astro-ph.CO]}
  \BibitemShut {NoStop}%
\bibitem [{\citenamefont {Nojiri}\ \emph {et~al.}(2019)\citenamefont {Nojiri},
  \citenamefont {Odintsov}, \citenamefont {Oikonomou},\ and\ \citenamefont
  {Popov}}]{Nojiri:2019nar}%
  \BibitemOpen
  \bibfield  {author} {\bibinfo {author} {\bibfnamefont {S.}~\bibnamefont
  {Nojiri}}, \bibinfo {author} {\bibfnamefont {S.~D.}\ \bibnamefont
  {Odintsov}}, \bibinfo {author} {\bibfnamefont {V.~K.}\ \bibnamefont
  {Oikonomou}}, \ and\ \bibinfo {author} {\bibfnamefont {A.~A.}\ \bibnamefont
  {Popov}},\ }\href {\doibase 10.1103/PhysRevD.100.084009} {\bibfield
  {journal} {\bibinfo  {journal} {Phys. Rev. D}\ }\textbf {\bibinfo {volume}
  {100}},\ \bibinfo {pages} {084009} (\bibinfo {year} {2019})},\ \Eprint
  {http://arxiv.org/abs/1909.01324} {arXiv:1909.01324 [gr-qc]} \BibitemShut
  {NoStop}%
\bibitem [{\citenamefont {Iacconi}\ \emph {et~al.}(2020)\citenamefont
  {Iacconi}, \citenamefont {Fasiello}, \citenamefont {Assadullahi},\ and\
  \citenamefont {Wands}}]{Iacconi:2020yxn}%
  \BibitemOpen
  \bibfield  {author} {\bibinfo {author} {\bibfnamefont {L.}~\bibnamefont
  {Iacconi}}, \bibinfo {author} {\bibfnamefont {M.}~\bibnamefont {Fasiello}},
  \bibinfo {author} {\bibfnamefont {H.}~\bibnamefont {Assadullahi}}, \ and\
  \bibinfo {author} {\bibfnamefont {D.}~\bibnamefont {Wands}},\ }\href
  {\doibase 10.1088/1475-7516/2020/12/005} {\bibfield  {journal} {\bibinfo
  {journal} {JCAP}\ }\textbf {\bibinfo {volume} {12}},\ \bibinfo {pages} {005}
  (\bibinfo {year} {2020})},\ \Eprint {http://arxiv.org/abs/2008.00452}
  {arXiv:2008.00452 [astro-ph.CO]} \BibitemShut {NoStop}%
\bibitem [{\citenamefont {Oikonomou}(2023{\natexlab{b}})}]{Oikonomou:2023bah}%
  \BibitemOpen
  \bibfield  {author} {\bibinfo {author} {\bibfnamefont {V.~K.}\ \bibnamefont
  {Oikonomou}},\ }\href {\doibase 10.1103/PhysRevD.107.064071} {\bibfield
  {journal} {\bibinfo  {journal} {Phys. Rev. D}\ }\textbf {\bibinfo {volume}
  {107}},\ \bibinfo {pages} {064071} (\bibinfo {year} {2023}{\natexlab{b}})},\
  \Eprint {http://arxiv.org/abs/2303.05889} {arXiv:2303.05889 [hep-ph]}
  \BibitemShut {NoStop}%
\bibitem [{\citenamefont {Cook}\ and\ \citenamefont
  {Sorbo}(2012)}]{Cook:2011hg}%
  \BibitemOpen
  \bibfield  {author} {\bibinfo {author} {\bibfnamefont {J.~L.}\ \bibnamefont
  {Cook}}\ and\ \bibinfo {author} {\bibfnamefont {L.}~\bibnamefont {Sorbo}},\
  }\href {\doibase 10.1103/PhysRevD.85.023534} {\bibfield  {journal} {\bibinfo
  {journal} {Phys. Rev. D}\ }\textbf {\bibinfo {volume} {85}},\ \bibinfo
  {pages} {023534} (\bibinfo {year} {2012})},\ \bibinfo {note} {[Erratum:
  Phys.Rev.D 86, 069901 (2012)]},\ \Eprint {http://arxiv.org/abs/1109.0022}
  {arXiv:1109.0022 [astro-ph.CO]} \BibitemShut {NoStop}%
\bibitem [{\citenamefont {Pajer}\ and\ \citenamefont
  {Peloso}(2013)}]{Pajer:2013fsa}%
  \BibitemOpen
  \bibfield  {author} {\bibinfo {author} {\bibfnamefont {E.}~\bibnamefont
  {Pajer}}\ and\ \bibinfo {author} {\bibfnamefont {M.}~\bibnamefont {Peloso}},\
  }\href {\doibase 10.1088/0264-9381/30/21/214002} {\bibfield  {journal}
  {\bibinfo  {journal} {Class. Quant. Grav.}\ }\textbf {\bibinfo {volume}
  {30}},\ \bibinfo {pages} {214002} (\bibinfo {year} {2013})},\ \Eprint
  {http://arxiv.org/abs/1305.3557} {arXiv:1305.3557 [hep-th]} \BibitemShut
  {NoStop}%
\bibitem [{\citenamefont {Mukohyama}\ \emph {et~al.}(2014)\citenamefont
  {Mukohyama}, \citenamefont {Namba}, \citenamefont {Peloso},\ and\
  \citenamefont {Shiu}}]{Mukohyama:2014gba}%
  \BibitemOpen
  \bibfield  {author} {\bibinfo {author} {\bibfnamefont {S.}~\bibnamefont
  {Mukohyama}}, \bibinfo {author} {\bibfnamefont {R.}~\bibnamefont {Namba}},
  \bibinfo {author} {\bibfnamefont {M.}~\bibnamefont {Peloso}}, \ and\ \bibinfo
  {author} {\bibfnamefont {G.}~\bibnamefont {Shiu}},\ }\href {\doibase
  10.1088/1475-7516/2014/08/036} {\bibfield  {journal} {\bibinfo  {journal}
  {JCAP}\ }\textbf {\bibinfo {volume} {08}},\ \bibinfo {pages} {036} (\bibinfo
  {year} {2014})},\ \Eprint {http://arxiv.org/abs/1405.0346} {arXiv:1405.0346
  [astro-ph.CO]} \BibitemShut {NoStop}%
\bibitem [{\citenamefont {Ashoorioon}\ \emph {et~al.}(2014)\citenamefont
  {Ashoorioon}, \citenamefont {Dimopoulos}, \citenamefont {Sheikh-Jabbari},\
  and\ \citenamefont {Shiu}}]{Ashoorioon:2014nta}%
  \BibitemOpen
  \bibfield  {author} {\bibinfo {author} {\bibfnamefont {A.}~\bibnamefont
  {Ashoorioon}}, \bibinfo {author} {\bibfnamefont {K.}~\bibnamefont
  {Dimopoulos}}, \bibinfo {author} {\bibfnamefont {M.~M.}\ \bibnamefont
  {Sheikh-Jabbari}}, \ and\ \bibinfo {author} {\bibfnamefont {G.}~\bibnamefont
  {Shiu}},\ }\href {\doibase 10.1016/j.physletb.2014.08.038} {\bibfield
  {journal} {\bibinfo  {journal} {Phys. Lett. B}\ }\textbf {\bibinfo {volume}
  {737}},\ \bibinfo {pages} {98} (\bibinfo {year} {2014})},\ \Eprint
  {http://arxiv.org/abs/1403.6099} {arXiv:1403.6099 [hep-th]} \BibitemShut
  {NoStop}%
\bibitem [{\citenamefont {Gruzinov}(2004)}]{Gruzinov:2004ty}%
  \BibitemOpen
  \bibfield  {author} {\bibinfo {author} {\bibfnamefont {A.}~\bibnamefont
  {Gruzinov}},\ }\href {\doibase 10.1103/PhysRevD.70.063518} {\bibfield
  {journal} {\bibinfo  {journal} {Phys. Rev. D}\ }\textbf {\bibinfo {volume}
  {70}},\ \bibinfo {pages} {063518} (\bibinfo {year} {2004})},\ \Eprint
  {http://arxiv.org/abs/astro-ph/0404548} {arXiv:astro-ph/0404548} \BibitemShut
  {NoStop}%
\bibitem [{\citenamefont {Giar\`e}\ \emph
  {et~al.}(2021{\natexlab{b}})\citenamefont {Giar\`e}, \citenamefont {Renzi},\
  and\ \citenamefont {Melchiorri}}]{Giare:2020plo}%
  \BibitemOpen
  \bibfield  {author} {\bibinfo {author} {\bibfnamefont {W.}~\bibnamefont
  {Giar\`e}}, \bibinfo {author} {\bibfnamefont {F.}~\bibnamefont {Renzi}}, \
  and\ \bibinfo {author} {\bibfnamefont {A.}~\bibnamefont {Melchiorri}},\
  }\href {\doibase 10.1103/PhysRevD.103.043515} {\bibfield  {journal} {\bibinfo
   {journal} {Phys. Rev. D}\ }\textbf {\bibinfo {volume} {103}},\ \bibinfo
  {pages} {043515} (\bibinfo {year} {2021}{\natexlab{b}})},\ \Eprint
  {http://arxiv.org/abs/2012.00527} {arXiv:2012.00527 [astro-ph.CO]}
  \BibitemShut {NoStop}%
\bibitem [{\citenamefont {Biagetti}\ \emph {et~al.}(2013)\citenamefont
  {Biagetti}, \citenamefont {Fasiello},\ and\ \citenamefont
  {Riotto}}]{Biagetti:2013kwa}%
  \BibitemOpen
  \bibfield  {author} {\bibinfo {author} {\bibfnamefont {M.}~\bibnamefont
  {Biagetti}}, \bibinfo {author} {\bibfnamefont {M.}~\bibnamefont {Fasiello}},
  \ and\ \bibinfo {author} {\bibfnamefont {A.}~\bibnamefont {Riotto}},\ }\href
  {\doibase 10.1103/PhysRevD.88.103518} {\bibfield  {journal} {\bibinfo
  {journal} {Phys. Rev. D}\ }\textbf {\bibinfo {volume} {88}},\ \bibinfo
  {pages} {103518} (\bibinfo {year} {2013})},\ \Eprint
  {http://arxiv.org/abs/1305.7241} {arXiv:1305.7241 [astro-ph.CO]} \BibitemShut
  {NoStop}%
\bibitem [{\citenamefont {Cai}\ \emph {et~al.}(2016)\citenamefont {Cai},
  \citenamefont {Wang},\ and\ \citenamefont {Piao}}]{Cai:2016ldn}%
  \BibitemOpen
  \bibfield  {author} {\bibinfo {author} {\bibfnamefont {Y.}~\bibnamefont
  {Cai}}, \bibinfo {author} {\bibfnamefont {Y.-T.}\ \bibnamefont {Wang}}, \
  and\ \bibinfo {author} {\bibfnamefont {Y.-S.}\ \bibnamefont {Piao}},\ }\href
  {\doibase 10.1103/PhysRevD.94.043002} {\bibfield  {journal} {\bibinfo
  {journal} {Phys. Rev. D}\ }\textbf {\bibinfo {volume} {94}},\ \bibinfo
  {pages} {043002} (\bibinfo {year} {2016})},\ \Eprint
  {http://arxiv.org/abs/1602.05431} {arXiv:1602.05431 [astro-ph.CO]}
  \BibitemShut {NoStop}%
\bibitem [{\citenamefont {Cai}\ \emph {et~al.}(2021{\natexlab{b}})\citenamefont
  {Cai}, \citenamefont {Lin}, \citenamefont {Wang},\ and\ \citenamefont
  {Yan}}]{Cai:2020ovp}%
  \BibitemOpen
  \bibfield  {author} {\bibinfo {author} {\bibfnamefont {Y.-F.}\ \bibnamefont
  {Cai}}, \bibinfo {author} {\bibfnamefont {C.}~\bibnamefont {Lin}}, \bibinfo
  {author} {\bibfnamefont {B.}~\bibnamefont {Wang}}, \ and\ \bibinfo {author}
  {\bibfnamefont {S.-F.}\ \bibnamefont {Yan}},\ }\href {\doibase
  10.1103/PhysRevLett.126.071303} {\bibfield  {journal} {\bibinfo  {journal}
  {Phys. Rev. Lett.}\ }\textbf {\bibinfo {volume} {126}},\ \bibinfo {pages}
  {071303} (\bibinfo {year} {2021}{\natexlab{b}})},\ \Eprint
  {http://arxiv.org/abs/2009.09833} {arXiv:2009.09833 [gr-qc]} \BibitemShut
  {NoStop}%
\bibitem [{\citenamefont {Kleidis}\ and\ \citenamefont
  {Oikonomou}(2016)}]{Kleidis:2016vmd}%
  \BibitemOpen
  \bibfield  {author} {\bibinfo {author} {\bibfnamefont {K.}~\bibnamefont
  {Kleidis}}\ and\ \bibinfo {author} {\bibfnamefont {V.~K.}\ \bibnamefont
  {Oikonomou}},\ }\href {\doibase 10.1007/s10509-016-2914-x} {\bibfield
  {journal} {\bibinfo  {journal} {Astrophys. Space Sci.}\ }\textbf {\bibinfo
  {volume} {361}},\ \bibinfo {pages} {326} (\bibinfo {year} {2016})},\ \Eprint
  {http://arxiv.org/abs/1609.00848} {arXiv:1609.00848 [gr-qc]} \BibitemShut
  {NoStop}%
\bibitem [{\citenamefont {Brandenberger}\ and\ \citenamefont
  {Vafa}(1989)}]{Brandenberger:1988aj}%
  \BibitemOpen
  \bibfield  {author} {\bibinfo {author} {\bibfnamefont {R.~H.}\ \bibnamefont
  {Brandenberger}}\ and\ \bibinfo {author} {\bibfnamefont {C.}~\bibnamefont
  {Vafa}},\ }\href {\doibase 10.1016/0550-3213(89)90037-0} {\bibfield
  {journal} {\bibinfo  {journal} {Nucl. Phys. B}\ }\textbf {\bibinfo {volume}
  {316}},\ \bibinfo {pages} {391} (\bibinfo {year} {1989})}\BibitemShut
  {NoStop}%
\bibitem [{\citenamefont {Brandenberger}\ \emph
  {et~al.}(2007{\natexlab{a}})\citenamefont {Brandenberger}, \citenamefont
  {Nayeri}, \citenamefont {Patil},\ and\ \citenamefont
  {Vafa}}]{Brandenberger:2006vv}%
  \BibitemOpen
  \bibfield  {author} {\bibinfo {author} {\bibfnamefont {R.~H.}\ \bibnamefont
  {Brandenberger}}, \bibinfo {author} {\bibfnamefont {A.}~\bibnamefont
  {Nayeri}}, \bibinfo {author} {\bibfnamefont {S.~P.}\ \bibnamefont {Patil}}, \
  and\ \bibinfo {author} {\bibfnamefont {C.}~\bibnamefont {Vafa}},\ }\href
  {\doibase 10.1142/S0217751X07037159} {\bibfield  {journal} {\bibinfo
  {journal} {Int. J. Mod. Phys. A}\ }\textbf {\bibinfo {volume} {22}},\
  \bibinfo {pages} {3621} (\bibinfo {year} {2007}{\natexlab{a}})},\ \Eprint
  {http://arxiv.org/abs/hep-th/0608121} {arXiv:hep-th/0608121} \BibitemShut
  {NoStop}%
\bibitem [{\citenamefont {Brandenberger}\ \emph
  {et~al.}(2007{\natexlab{b}})\citenamefont {Brandenberger}, \citenamefont
  {Nayeri}, \citenamefont {Patil},\ and\ \citenamefont
  {Vafa}}]{Brandenberger:2006xi}%
  \BibitemOpen
  \bibfield  {author} {\bibinfo {author} {\bibfnamefont {R.~H.}\ \bibnamefont
  {Brandenberger}}, \bibinfo {author} {\bibfnamefont {A.}~\bibnamefont
  {Nayeri}}, \bibinfo {author} {\bibfnamefont {S.~P.}\ \bibnamefont {Patil}}, \
  and\ \bibinfo {author} {\bibfnamefont {C.}~\bibnamefont {Vafa}},\ }\href
  {\doibase 10.1103/PhysRevLett.98.231302} {\bibfield  {journal} {\bibinfo
  {journal} {Phys. Rev. Lett.}\ }\textbf {\bibinfo {volume} {98}},\ \bibinfo
  {pages} {231302} (\bibinfo {year} {2007}{\natexlab{b}})},\ \Eprint
  {http://arxiv.org/abs/hep-th/0604126} {arXiv:hep-th/0604126} \BibitemShut
  {NoStop}%
\bibitem [{\citenamefont {Stewart}\ and\ \citenamefont
  {Brandenberger}(2008)}]{Stewart:2007fu}%
  \BibitemOpen
  \bibfield  {author} {\bibinfo {author} {\bibfnamefont {A.}~\bibnamefont
  {Stewart}}\ and\ \bibinfo {author} {\bibfnamefont {R.}~\bibnamefont
  {Brandenberger}},\ }\href {\doibase 10.1088/1475-7516/2008/08/012} {\bibfield
   {journal} {\bibinfo  {journal} {JCAP}\ }\textbf {\bibinfo {volume} {08}},\
  \bibinfo {pages} {012} (\bibinfo {year} {2008})},\ \Eprint
  {http://arxiv.org/abs/0711.4602} {arXiv:0711.4602 [astro-ph]} \BibitemShut
  {NoStop}%
\bibitem [{\citenamefont {Brandenberger}\ \emph {et~al.}(2014)\citenamefont
  {Brandenberger}, \citenamefont {Nayeri},\ and\ \citenamefont
  {Patil}}]{Brandenberger:2014faa}%
  \BibitemOpen
  \bibfield  {author} {\bibinfo {author} {\bibfnamefont {R.~H.}\ \bibnamefont
  {Brandenberger}}, \bibinfo {author} {\bibfnamefont {A.}~\bibnamefont
  {Nayeri}}, \ and\ \bibinfo {author} {\bibfnamefont {S.~P.}\ \bibnamefont
  {Patil}},\ }\href {\doibase 10.1103/PhysRevD.90.067301} {\bibfield  {journal}
  {\bibinfo  {journal} {Phys. Rev. D}\ }\textbf {\bibinfo {volume} {90}},\
  \bibinfo {pages} {067301} (\bibinfo {year} {2014})},\ \Eprint
  {http://arxiv.org/abs/1403.4927} {arXiv:1403.4927 [astro-ph.CO]} \BibitemShut
  {NoStop}%
\bibitem [{\citenamefont {Khoury}\ \emph {et~al.}(2001)\citenamefont {Khoury},
  \citenamefont {Ovrut}, \citenamefont {Steinhardt},\ and\ \citenamefont
  {Turok}}]{Khoury:2001wf}%
  \BibitemOpen
  \bibfield  {author} {\bibinfo {author} {\bibfnamefont {J.}~\bibnamefont
  {Khoury}}, \bibinfo {author} {\bibfnamefont {B.~A.}\ \bibnamefont {Ovrut}},
  \bibinfo {author} {\bibfnamefont {P.~J.}\ \bibnamefont {Steinhardt}}, \ and\
  \bibinfo {author} {\bibfnamefont {N.}~\bibnamefont {Turok}},\ }\href
  {\doibase 10.1103/PhysRevD.64.123522} {\bibfield  {journal} {\bibinfo
  {journal} {Phys. Rev. D}\ }\textbf {\bibinfo {volume} {64}},\ \bibinfo
  {pages} {123522} (\bibinfo {year} {2001})},\ \Eprint
  {http://arxiv.org/abs/hep-th/0103239} {arXiv:hep-th/0103239} \BibitemShut
  {NoStop}%
\bibitem [{\citenamefont {Hipolito-Ricaldi}\ \emph {et~al.}(2016)\citenamefont
  {Hipolito-Ricaldi}, \citenamefont {Brandenberger}, \citenamefont {Ferreira},\
  and\ \citenamefont {Graef}}]{Hipolito-Ricaldi:2016kqq}%
  \BibitemOpen
  \bibfield  {author} {\bibinfo {author} {\bibfnamefont {W.~S.}\ \bibnamefont
  {Hipolito-Ricaldi}}, \bibinfo {author} {\bibfnamefont {R.}~\bibnamefont
  {Brandenberger}}, \bibinfo {author} {\bibfnamefont {E.~G.~M.}\ \bibnamefont
  {Ferreira}}, \ and\ \bibinfo {author} {\bibfnamefont {L.~L.}\ \bibnamefont
  {Graef}},\ }\href {\doibase 10.1088/1475-7516/2016/11/024} {\bibfield
  {journal} {\bibinfo  {journal} {JCAP}\ }\textbf {\bibinfo {volume} {11}},\
  \bibinfo {pages} {024} (\bibinfo {year} {2016})},\ \Eprint
  {http://arxiv.org/abs/1605.04670} {arXiv:1605.04670 [hep-th]} \BibitemShut
  {NoStop}%
\bibitem [{\citenamefont {Brandenberger}(2011)}]{Brandenberger:2009jq}%
  \BibitemOpen
  \bibfield  {author} {\bibinfo {author} {\bibfnamefont {R.~H.}\ \bibnamefont
  {Brandenberger}},\ }\href {\doibase 10.1142/S2010194511000109} {\bibfield
  {journal} {\bibinfo  {journal} {Int. J. Mod. Phys. Conf. Ser.}\ }\textbf
  {\bibinfo {volume} {01}},\ \bibinfo {pages} {67} (\bibinfo {year} {2011})},\
  \Eprint {http://arxiv.org/abs/0902.4731} {arXiv:0902.4731 [hep-th]}
  \BibitemShut {NoStop}%
\bibitem [{\citenamefont {Wang}\ and\ \citenamefont
  {Xue}(2014)}]{Wang:2014kqa}%
  \BibitemOpen
  \bibfield  {author} {\bibinfo {author} {\bibfnamefont {Y.}~\bibnamefont
  {Wang}}\ and\ \bibinfo {author} {\bibfnamefont {W.}~\bibnamefont {Xue}},\
  }\href {\doibase 10.1088/1475-7516/2014/10/075} {\bibfield  {journal}
  {\bibinfo  {journal} {JCAP}\ }\textbf {\bibinfo {volume} {10}},\ \bibinfo
  {pages} {075} (\bibinfo {year} {2014})},\ \Eprint
  {http://arxiv.org/abs/1403.5817} {arXiv:1403.5817 [astro-ph.CO]} \BibitemShut
  {NoStop}%
\bibitem [{\citenamefont {Meerburg}\ \emph {et~al.}(2015)\citenamefont
  {Meerburg}, \citenamefont {Hlo\v{z}ek}, \citenamefont {Hadzhiyska},\ and\
  \citenamefont {Meyers}}]{Meerburg:2015zua}%
  \BibitemOpen
  \bibfield  {author} {\bibinfo {author} {\bibfnamefont {P.~D.}\ \bibnamefont
  {Meerburg}}, \bibinfo {author} {\bibfnamefont {R.}~\bibnamefont
  {Hlo\v{z}ek}}, \bibinfo {author} {\bibfnamefont {B.}~\bibnamefont
  {Hadzhiyska}}, \ and\ \bibinfo {author} {\bibfnamefont {J.}~\bibnamefont
  {Meyers}},\ }\href {\doibase 10.1103/PhysRevD.91.103505} {\bibfield
  {journal} {\bibinfo  {journal} {Phys. Rev. D}\ }\textbf {\bibinfo {volume}
  {91}},\ \bibinfo {pages} {103505} (\bibinfo {year} {2015})},\ \Eprint
  {http://arxiv.org/abs/1502.00302} {arXiv:1502.00302 [astro-ph.CO]}
  \BibitemShut {NoStop}%
\bibitem [{\citenamefont {Cabass}\ \emph {et~al.}(2016)\citenamefont {Cabass},
  \citenamefont {Pagano}, \citenamefont {Salvati}, \citenamefont {Gerbino},
  \citenamefont {Giusarma},\ and\ \citenamefont {Melchiorri}}]{Cabass:2015jwe}%
  \BibitemOpen
  \bibfield  {author} {\bibinfo {author} {\bibfnamefont {G.}~\bibnamefont
  {Cabass}}, \bibinfo {author} {\bibfnamefont {L.}~\bibnamefont {Pagano}},
  \bibinfo {author} {\bibfnamefont {L.}~\bibnamefont {Salvati}}, \bibinfo
  {author} {\bibfnamefont {M.}~\bibnamefont {Gerbino}}, \bibinfo {author}
  {\bibfnamefont {E.}~\bibnamefont {Giusarma}}, \ and\ \bibinfo {author}
  {\bibfnamefont {A.}~\bibnamefont {Melchiorri}},\ }\href {\doibase
  10.1103/PhysRevD.93.063508} {\bibfield  {journal} {\bibinfo  {journal} {Phys.
  Rev. D}\ }\textbf {\bibinfo {volume} {93}},\ \bibinfo {pages} {063508}
  (\bibinfo {year} {2016})},\ \Eprint {http://arxiv.org/abs/1511.05146}
  {arXiv:1511.05146 [astro-ph.CO]} \BibitemShut {NoStop}%
\bibitem [{\citenamefont {Wang}\ \emph {et~al.}(2017)\citenamefont {Wang},
  \citenamefont {Cai}, \citenamefont {Liu},\ and\ \citenamefont
  {Piao}}]{Wang:2016tbj}%
  \BibitemOpen
  \bibfield  {author} {\bibinfo {author} {\bibfnamefont {Y.-T.}\ \bibnamefont
  {Wang}}, \bibinfo {author} {\bibfnamefont {Y.}~\bibnamefont {Cai}}, \bibinfo
  {author} {\bibfnamefont {Z.-G.}\ \bibnamefont {Liu}}, \ and\ \bibinfo
  {author} {\bibfnamefont {Y.-S.}\ \bibnamefont {Piao}},\ }\href {\doibase
  10.1088/1475-7516/2017/01/010} {\bibfield  {journal} {\bibinfo  {journal}
  {JCAP}\ }\textbf {\bibinfo {volume} {01}},\ \bibinfo {pages} {010} (\bibinfo
  {year} {2017})},\ \Eprint {http://arxiv.org/abs/1612.05088} {arXiv:1612.05088
  [astro-ph.CO]} \BibitemShut {NoStop}%
\bibitem [{\citenamefont {Graef}\ \emph {et~al.}(2019)\citenamefont {Graef},
  \citenamefont {Benetti},\ and\ \citenamefont {Alcaniz}}]{Graef:2018fzu}%
  \BibitemOpen
  \bibfield  {author} {\bibinfo {author} {\bibfnamefont {L.~L.}\ \bibnamefont
  {Graef}}, \bibinfo {author} {\bibfnamefont {M.}~\bibnamefont {Benetti}}, \
  and\ \bibinfo {author} {\bibfnamefont {J.~S.}\ \bibnamefont {Alcaniz}},\
  }\href {\doibase 10.1103/PhysRevD.99.043519} {\bibfield  {journal} {\bibinfo
  {journal} {Phys. Rev. D}\ }\textbf {\bibinfo {volume} {99}},\ \bibinfo
  {pages} {043519} (\bibinfo {year} {2019})},\ \Eprint
  {http://arxiv.org/abs/1809.04501} {arXiv:1809.04501 [astro-ph.CO]}
  \BibitemShut {NoStop}%
\bibitem [{\citenamefont {Giar\`e}\ \emph {et~al.}(2019)\citenamefont
  {Giar\`e}, \citenamefont {Di~Valentino},\ and\ \citenamefont
  {Melchiorri}}]{Giare:2019snj}%
  \BibitemOpen
  \bibfield  {author} {\bibinfo {author} {\bibfnamefont {W.}~\bibnamefont
  {Giar\`e}}, \bibinfo {author} {\bibfnamefont {E.}~\bibnamefont
  {Di~Valentino}}, \ and\ \bibinfo {author} {\bibfnamefont {A.}~\bibnamefont
  {Melchiorri}},\ }\href {\doibase 10.1103/PhysRevD.99.123522} {\bibfield
  {journal} {\bibinfo  {journal} {Phys. Rev. D}\ }\textbf {\bibinfo {volume}
  {99}},\ \bibinfo {pages} {123522} (\bibinfo {year} {2019})}\BibitemShut
  {NoStop}%
\bibitem [{\citenamefont {Galloni}\ \emph {et~al.}(2023)\citenamefont
  {Galloni}, \citenamefont {Bartolo}, \citenamefont {Matarrese}, \citenamefont
  {Migliaccio}, \citenamefont {Ricciardone},\ and\ \citenamefont
  {Vittorio}}]{Galloni:2022mok}%
  \BibitemOpen
  \bibfield  {author} {\bibinfo {author} {\bibfnamefont {G.}~\bibnamefont
  {Galloni}}, \bibinfo {author} {\bibfnamefont {N.}~\bibnamefont {Bartolo}},
  \bibinfo {author} {\bibfnamefont {S.}~\bibnamefont {Matarrese}}, \bibinfo
  {author} {\bibfnamefont {M.}~\bibnamefont {Migliaccio}}, \bibinfo {author}
  {\bibfnamefont {A.}~\bibnamefont {Ricciardone}}, \ and\ \bibinfo {author}
  {\bibfnamefont {N.}~\bibnamefont {Vittorio}},\ }\href {\doibase
  10.1088/1475-7516/2023/04/062} {\bibfield  {journal} {\bibinfo  {journal}
  {JCAP}\ }\textbf {\bibinfo {volume} {04}},\ \bibinfo {pages} {062} (\bibinfo
  {year} {2023})},\ \Eprint {http://arxiv.org/abs/2208.00188} {arXiv:2208.00188
  [astro-ph.CO]} \BibitemShut {NoStop}%
\bibitem [{\citenamefont {Giar\`e}\ and\ \citenamefont
  {Renzi}(2020)}]{Giare:2020vss}%
  \BibitemOpen
  \bibfield  {author} {\bibinfo {author} {\bibfnamefont {W.}~\bibnamefont
  {Giar\`e}}\ and\ \bibinfo {author} {\bibfnamefont {F.}~\bibnamefont
  {Renzi}},\ }\href {\doibase 10.1103/PhysRevD.102.083530} {\bibfield
  {journal} {\bibinfo  {journal} {Phys. Rev. D}\ }\textbf {\bibinfo {volume}
  {102}},\ \bibinfo {pages} {083530} (\bibinfo {year} {2020})},\ \Eprint
  {http://arxiv.org/abs/2007.04256} {arXiv:2007.04256 [astro-ph.CO]}
  \BibitemShut {NoStop}%
\bibitem [{\citenamefont {Kawasaki}\ \emph {et~al.}(2000)\citenamefont
  {Kawasaki}, \citenamefont {Kohri},\ and\ \citenamefont
  {Sugiyama}}]{Kawasaki:2000en}%
  \BibitemOpen
  \bibfield  {author} {\bibinfo {author} {\bibfnamefont {M.}~\bibnamefont
  {Kawasaki}}, \bibinfo {author} {\bibfnamefont {K.}~\bibnamefont {Kohri}}, \
  and\ \bibinfo {author} {\bibfnamefont {N.}~\bibnamefont {Sugiyama}},\ }\href
  {\doibase 10.1103/PhysRevD.62.023506} {\bibfield  {journal} {\bibinfo
  {journal} {Phys. Rev. D}\ }\textbf {\bibinfo {volume} {62}},\ \bibinfo
  {pages} {023506} (\bibinfo {year} {2000})},\ \Eprint
  {http://arxiv.org/abs/astro-ph/0002127} {arXiv:astro-ph/0002127} \BibitemShut
  {NoStop}%
\bibitem [{\citenamefont {Giudice}\ \emph {et~al.}(2001)\citenamefont
  {Giudice}, \citenamefont {Kolb},\ and\ \citenamefont
  {Riotto}}]{Giudice:2000ex}%
  \BibitemOpen
  \bibfield  {author} {\bibinfo {author} {\bibfnamefont {G.~F.}\ \bibnamefont
  {Giudice}}, \bibinfo {author} {\bibfnamefont {E.~W.}\ \bibnamefont {Kolb}}, \
  and\ \bibinfo {author} {\bibfnamefont {A.}~\bibnamefont {Riotto}},\ }\href
  {\doibase 10.1103/PhysRevD.64.023508} {\bibfield  {journal} {\bibinfo
  {journal} {Phys. Rev. D}\ }\textbf {\bibinfo {volume} {64}},\ \bibinfo
  {pages} {023508} (\bibinfo {year} {2001})},\ \Eprint
  {http://arxiv.org/abs/hep-ph/0005123} {arXiv:hep-ph/0005123} \BibitemShut
  {NoStop}%
\bibitem [{\citenamefont {Hannestad}(2004)}]{Hannestad:2004px}%
  \BibitemOpen
  \bibfield  {author} {\bibinfo {author} {\bibfnamefont {S.}~\bibnamefont
  {Hannestad}},\ }\href {\doibase 10.1103/PhysRevD.70.043506} {\bibfield
  {journal} {\bibinfo  {journal} {Phys. Rev. D}\ }\textbf {\bibinfo {volume}
  {70}},\ \bibinfo {pages} {043506} (\bibinfo {year} {2004})},\ \Eprint
  {http://arxiv.org/abs/astro-ph/0403291} {arXiv:astro-ph/0403291} \BibitemShut
  {NoStop}%
\bibitem [{\citenamefont {Gerbino}\ \emph {et~al.}(2017)\citenamefont
  {Gerbino}, \citenamefont {Freese}, \citenamefont {Vagnozzi}, \citenamefont
  {Lattanzi}, \citenamefont {Mena}, \citenamefont {Giusarma},\ and\
  \citenamefont {Ho}}]{Gerbino:2016sgw}%
  \BibitemOpen
  \bibfield  {author} {\bibinfo {author} {\bibfnamefont {M.}~\bibnamefont
  {Gerbino}}, \bibinfo {author} {\bibfnamefont {K.}~\bibnamefont {Freese}},
  \bibinfo {author} {\bibfnamefont {S.}~\bibnamefont {Vagnozzi}}, \bibinfo
  {author} {\bibfnamefont {M.}~\bibnamefont {Lattanzi}}, \bibinfo {author}
  {\bibfnamefont {O.}~\bibnamefont {Mena}}, \bibinfo {author} {\bibfnamefont
  {E.}~\bibnamefont {Giusarma}}, \ and\ \bibinfo {author} {\bibfnamefont
  {S.}~\bibnamefont {Ho}},\ }\href {\doibase 10.1103/PhysRevD.95.043512}
  {\bibfield  {journal} {\bibinfo  {journal} {Phys. Rev. D}\ }\textbf {\bibinfo
  {volume} {95}},\ \bibinfo {pages} {043512} (\bibinfo {year} {2017})},\
  \Eprint {http://arxiv.org/abs/1610.08830} {arXiv:1610.08830 [astro-ph.CO]}
  \BibitemShut {NoStop}%
\bibitem [{\citenamefont {Handley}\ and\ \citenamefont
  {Lemos}(2021)}]{Handley:2020hdp}%
  \BibitemOpen
  \bibfield  {author} {\bibinfo {author} {\bibfnamefont {W.}~\bibnamefont
  {Handley}}\ and\ \bibinfo {author} {\bibfnamefont {P.}~\bibnamefont
  {Lemos}},\ }\href {\doibase 10.1103/PhysRevD.103.063529} {\bibfield
  {journal} {\bibinfo  {journal} {Phys. Rev. D}\ }\textbf {\bibinfo {volume}
  {103}},\ \bibinfo {pages} {063529} (\bibinfo {year} {2021})},\ \Eprint
  {http://arxiv.org/abs/2007.08496} {arXiv:2007.08496 [astro-ph.CO]}
  \BibitemShut {NoStop}%
\bibitem [{\citenamefont {Giar\`e}\ \emph
  {et~al.}(2023{\natexlab{c}})\citenamefont {Giar\`e}, \citenamefont {Renzi},
  \citenamefont {Mena}, \citenamefont {Di~Valentino},\ and\ \citenamefont
  {Melchiorri}}]{Giare:2022rvg}%
  \BibitemOpen
  \bibfield  {author} {\bibinfo {author} {\bibfnamefont {W.}~\bibnamefont
  {Giar\`e}}, \bibinfo {author} {\bibfnamefont {F.}~\bibnamefont {Renzi}},
  \bibinfo {author} {\bibfnamefont {O.}~\bibnamefont {Mena}}, \bibinfo {author}
  {\bibfnamefont {E.}~\bibnamefont {Di~Valentino}}, \ and\ \bibinfo {author}
  {\bibfnamefont {A.}~\bibnamefont {Melchiorri}},\ }\href {\doibase
  10.1093/mnras/stad724} {\bibfield  {journal} {\bibinfo  {journal} {Mon. Not.
  Roy. Astron. Soc.}\ }\textbf {\bibinfo {volume} {521}},\ \bibinfo {pages}
  {2911} (\bibinfo {year} {2023}{\natexlab{c}})},\ \Eprint
  {http://arxiv.org/abs/2210.09018} {arXiv:2210.09018 [astro-ph.CO]}
  \BibitemShut {NoStop}%
\bibitem [{\citenamefont {Zhai}\ \emph {et~al.}(2023)\citenamefont {Zhai},
  \citenamefont {Giar\`e}, \citenamefont {van~de Bruck}, \citenamefont
  {Di~Valentino}, \citenamefont {Mena},\ and\ \citenamefont
  {Nunes}}]{Zhai:2023yny}%
  \BibitemOpen
  \bibfield  {author} {\bibinfo {author} {\bibfnamefont {Y.}~\bibnamefont
  {Zhai}}, \bibinfo {author} {\bibfnamefont {W.}~\bibnamefont {Giar\`e}},
  \bibinfo {author} {\bibfnamefont {C.}~\bibnamefont {van~de Bruck}}, \bibinfo
  {author} {\bibfnamefont {E.}~\bibnamefont {Di~Valentino}}, \bibinfo {author}
  {\bibfnamefont {O.}~\bibnamefont {Mena}}, \ and\ \bibinfo {author}
  {\bibfnamefont {R.~C.}\ \bibnamefont {Nunes}},\ }\href@noop {} {\  (\bibinfo
  {year} {2023})},\ \Eprint {http://arxiv.org/abs/2303.08201} {arXiv:2303.08201
  [astro-ph.CO]} \BibitemShut {NoStop}%
\bibitem [{\citenamefont {Di~Valentino}\ \emph {et~al.}(2023)\citenamefont
  {Di~Valentino}, \citenamefont {Gariazzo}, \citenamefont {Giar\`e},\ and\
  \citenamefont {Mena}}]{DiValentino:2023fei}%
  \BibitemOpen
  \bibfield  {author} {\bibinfo {author} {\bibfnamefont {E.}~\bibnamefont
  {Di~Valentino}}, \bibinfo {author} {\bibfnamefont {S.}~\bibnamefont
  {Gariazzo}}, \bibinfo {author} {\bibfnamefont {W.}~\bibnamefont {Giar\`e}}, \
  and\ \bibinfo {author} {\bibfnamefont {O.}~\bibnamefont {Mena}},\ }\href@noop
  {} {\  (\bibinfo {year} {2023})},\ \Eprint {http://arxiv.org/abs/2305.12989}
  {arXiv:2305.12989 [astro-ph.CO]} \BibitemShut {NoStop}%
\bibitem [{\citenamefont {Giar\`e}\ \emph
  {et~al.}(2023{\natexlab{d}})\citenamefont {Giar\`e}, \citenamefont {Pan},
  \citenamefont {Di~Valentino}, \citenamefont {Yang}, \citenamefont {de~Haro},\
  and\ \citenamefont {Melchiorri}}]{Giare:2023wzl}%
  \BibitemOpen
  \bibfield  {author} {\bibinfo {author} {\bibfnamefont {W.}~\bibnamefont
  {Giar\`e}}, \bibinfo {author} {\bibfnamefont {S.}~\bibnamefont {Pan}},
  \bibinfo {author} {\bibfnamefont {E.}~\bibnamefont {Di~Valentino}}, \bibinfo
  {author} {\bibfnamefont {W.}~\bibnamefont {Yang}}, \bibinfo {author}
  {\bibfnamefont {J.}~\bibnamefont {de~Haro}}, \ and\ \bibinfo {author}
  {\bibfnamefont {A.}~\bibnamefont {Melchiorri}},\ }\href@noop {} {\  (\bibinfo
  {year} {2023}{\natexlab{d}})},\ \Eprint {http://arxiv.org/abs/2305.15378}
  {arXiv:2305.15378 [astro-ph.CO]} \BibitemShut {NoStop}%
\bibitem [{\citenamefont {Akrami}\ \emph {et~al.}(2020)\citenamefont {Akrami}
  \emph {et~al.}}]{Planck:2018jri}%
  \BibitemOpen
  \bibfield  {author} {\bibinfo {author} {\bibfnamefont {Y.}~\bibnamefont
  {Akrami}} \emph {et~al.} (\bibinfo {collaboration} {Planck}),\ }\href
  {\doibase 10.1051/0004-6361/201833887} {\bibfield  {journal} {\bibinfo
  {journal} {Astron. Astrophys.}\ }\textbf {\bibinfo {volume} {641}},\ \bibinfo
  {pages} {A10} (\bibinfo {year} {2020})},\ \Eprint
  {http://arxiv.org/abs/1807.06211} {arXiv:1807.06211 [astro-ph.CO]}
  \BibitemShut {NoStop}%
\bibitem [{\citenamefont {Campeti}\ \emph {et~al.}(2021)\citenamefont
  {Campeti}, \citenamefont {Komatsu}, \citenamefont {Poletti},\ and\
  \citenamefont {Baccigalupi}}]{Campeti:2020xwn}%
  \BibitemOpen
  \bibfield  {author} {\bibinfo {author} {\bibfnamefont {P.}~\bibnamefont
  {Campeti}}, \bibinfo {author} {\bibfnamefont {E.}~\bibnamefont {Komatsu}},
  \bibinfo {author} {\bibfnamefont {D.}~\bibnamefont {Poletti}}, \ and\
  \bibinfo {author} {\bibfnamefont {C.}~\bibnamefont {Baccigalupi}},\ }\href
  {\doibase 10.1088/1475-7516/2021/01/012} {\bibfield  {journal} {\bibinfo
  {journal} {JCAP}\ }\textbf {\bibinfo {volume} {01}},\ \bibinfo {pages} {012}
  (\bibinfo {year} {2021})},\ \Eprint {http://arxiv.org/abs/2007.04241}
  {arXiv:2007.04241 [astro-ph.CO]} \BibitemShut {NoStop}%
\bibitem [{\citenamefont {Giar\`e}\ and\ \citenamefont
  {Melchiorri}(2021)}]{Giare:2020vhn}%
  \BibitemOpen
  \bibfield  {author} {\bibinfo {author} {\bibfnamefont {W.}~\bibnamefont
  {Giar\`e}}\ and\ \bibinfo {author} {\bibfnamefont {A.}~\bibnamefont
  {Melchiorri}},\ }\href {\doibase 10.1016/j.physletb.2021.136137} {\bibfield
  {journal} {\bibinfo  {journal} {Phys. Lett. B}\ }\textbf {\bibinfo {volume}
  {815}},\ \bibinfo {pages} {136137} (\bibinfo {year} {2021})},\ \Eprint
  {http://arxiv.org/abs/2003.04783} {arXiv:2003.04783 [astro-ph.CO]}
  \BibitemShut {NoStop}%
\bibitem [{\citenamefont {Kinney}(2021)}]{Kinney:2021nje}%
  \BibitemOpen
  \bibfield  {author} {\bibinfo {author} {\bibfnamefont {W.~H.}\ \bibnamefont
  {Kinney}},\ }\href {\doibase 10.21105/astro.2103.00281} {\  (\bibinfo {year}
  {2021}),\ 10.21105/astro.2103.00281},\ \Eprint
  {http://arxiv.org/abs/2103.00281} {arXiv:2103.00281 [astro-ph.CO]}
  \BibitemShut {NoStop}%
\bibitem [{\citenamefont {Kinney}\ and\ \citenamefont
  {Riotto}(2006)}]{Kinney:2005in}%
  \BibitemOpen
  \bibfield  {author} {\bibinfo {author} {\bibfnamefont {W.~H.}\ \bibnamefont
  {Kinney}}\ and\ \bibinfo {author} {\bibfnamefont {A.}~\bibnamefont
  {Riotto}},\ }\href {\doibase 10.1088/1475-7516/2006/03/011} {\bibfield
  {journal} {\bibinfo  {journal} {JCAP}\ }\textbf {\bibinfo {volume} {03}},\
  \bibinfo {pages} {011} (\bibinfo {year} {2006})},\ \Eprint
  {http://arxiv.org/abs/astro-ph/0511127} {arXiv:astro-ph/0511127} \BibitemShut
  {NoStop}%
\bibitem [{\citenamefont {Caligiuri}\ \emph {et~al.}(2015)\citenamefont
  {Caligiuri}, \citenamefont {Kosowsky}, \citenamefont {Kinney},\ and\
  \citenamefont {Seto}}]{Caligiuri:2014ola}%
  \BibitemOpen
  \bibfield  {author} {\bibinfo {author} {\bibfnamefont {J.}~\bibnamefont
  {Caligiuri}}, \bibinfo {author} {\bibfnamefont {A.}~\bibnamefont {Kosowsky}},
  \bibinfo {author} {\bibfnamefont {W.~H.}\ \bibnamefont {Kinney}}, \ and\
  \bibinfo {author} {\bibfnamefont {N.}~\bibnamefont {Seto}},\ }\href {\doibase
  10.1103/PhysRevD.91.103529} {\bibfield  {journal} {\bibinfo  {journal} {Phys.
  Rev. D}\ }\textbf {\bibinfo {volume} {91}},\ \bibinfo {pages} {103529}
  (\bibinfo {year} {2015})},\ \Eprint {http://arxiv.org/abs/1409.3195}
  {arXiv:1409.3195 [astro-ph.CO]} \BibitemShut {NoStop}%
\bibitem [{\citenamefont {Abbott}\ \emph {et~al.}(2021)\citenamefont {Abbott}
  \emph {et~al.}}]{KAGRA:2021kbb}%
  \BibitemOpen
  \bibfield  {author} {\bibinfo {author} {\bibfnamefont {R.}~\bibnamefont
  {Abbott}} \emph {et~al.} (\bibinfo {collaboration} {KAGRA, Virgo, LIGO
  Scientific}),\ }\href {\doibase 10.1103/PhysRevD.104.022004} {\bibfield
  {journal} {\bibinfo  {journal} {Phys. Rev. D}\ }\textbf {\bibinfo {volume}
  {104}},\ \bibinfo {pages} {022004} (\bibinfo {year} {2021})},\ \Eprint
  {http://arxiv.org/abs/2101.12130} {arXiv:2101.12130 [gr-qc]} \BibitemShut
  {NoStop}%
\bibitem [{\citenamefont {Freese}\ \emph {et~al.}(2018)\citenamefont {Freese},
  \citenamefont {Sfakianakis}, \citenamefont {Stengel},\ and\ \citenamefont
  {Visinelli}}]{Freese:2017ace}%
  \BibitemOpen
  \bibfield  {author} {\bibinfo {author} {\bibfnamefont {K.}~\bibnamefont
  {Freese}}, \bibinfo {author} {\bibfnamefont {E.~I.}\ \bibnamefont
  {Sfakianakis}}, \bibinfo {author} {\bibfnamefont {P.}~\bibnamefont
  {Stengel}}, \ and\ \bibinfo {author} {\bibfnamefont {L.}~\bibnamefont
  {Visinelli}},\ }\href {\doibase 10.1088/1475-7516/2018/05/067} {\bibfield
  {journal} {\bibinfo  {journal} {JCAP}\ }\textbf {\bibinfo {volume} {05}},\
  \bibinfo {pages} {067} (\bibinfo {year} {2018})},\ \Eprint
  {http://arxiv.org/abs/1712.03791} {arXiv:1712.03791 [hep-ph]} \BibitemShut
  {NoStop}%
\bibitem [{\citenamefont {Litsa}\ \emph {et~al.}(2021)\citenamefont {Litsa},
  \citenamefont {Freese}, \citenamefont {Sfakianakis}, \citenamefont
  {Stengel},\ and\ \citenamefont {Visinelli}}]{Litsa:2020rsm}%
  \BibitemOpen
  \bibfield  {author} {\bibinfo {author} {\bibfnamefont {A.}~\bibnamefont
  {Litsa}}, \bibinfo {author} {\bibfnamefont {K.}~\bibnamefont {Freese}},
  \bibinfo {author} {\bibfnamefont {E.~I.}\ \bibnamefont {Sfakianakis}},
  \bibinfo {author} {\bibfnamefont {P.}~\bibnamefont {Stengel}}, \ and\
  \bibinfo {author} {\bibfnamefont {L.}~\bibnamefont {Visinelli}},\ }\href
  {\doibase 10.1103/PhysRevD.104.123546} {\bibfield  {journal} {\bibinfo
  {journal} {Phys. Rev. D}\ }\textbf {\bibinfo {volume} {104}},\ \bibinfo
  {pages} {123546} (\bibinfo {year} {2021})},\ \Eprint
  {http://arxiv.org/abs/2009.14218} {arXiv:2009.14218 [hep-ph]} \BibitemShut
  {NoStop}%
\bibitem [{\citenamefont {Litsa}\ \emph {et~al.}(2023)\citenamefont {Litsa},
  \citenamefont {Freese}, \citenamefont {Sfakianakis}, \citenamefont
  {Stengel},\ and\ \citenamefont {Visinelli}}]{Litsa:2020mvj}%
  \BibitemOpen
  \bibfield  {author} {\bibinfo {author} {\bibfnamefont {A.}~\bibnamefont
  {Litsa}}, \bibinfo {author} {\bibfnamefont {K.}~\bibnamefont {Freese}},
  \bibinfo {author} {\bibfnamefont {E.~I.}\ \bibnamefont {Sfakianakis}},
  \bibinfo {author} {\bibfnamefont {P.}~\bibnamefont {Stengel}}, \ and\
  \bibinfo {author} {\bibfnamefont {L.}~\bibnamefont {Visinelli}},\ }\href
  {\doibase 10.1088/1475-7516/2023/03/033} {\bibfield  {journal} {\bibinfo
  {journal} {JCAP}\ }\textbf {\bibinfo {volume} {03}},\ \bibinfo {pages} {033}
  (\bibinfo {year} {2023})},\ \Eprint {http://arxiv.org/abs/2011.11649}
  {arXiv:2011.11649 [hep-ph]} \BibitemShut {NoStop}%
\bibitem [{\citenamefont {Deffayet}\ \emph {et~al.}(2009)\citenamefont
  {Deffayet}, \citenamefont {Deser},\ and\ \citenamefont
  {Esposito-Farese}}]{Deffayet:2009mn}%
  \BibitemOpen
  \bibfield  {author} {\bibinfo {author} {\bibfnamefont {C.}~\bibnamefont
  {Deffayet}}, \bibinfo {author} {\bibfnamefont {S.}~\bibnamefont {Deser}}, \
  and\ \bibinfo {author} {\bibfnamefont {G.}~\bibnamefont {Esposito-Farese}},\
  }\href {\doibase 10.1103/PhysRevD.80.064015} {\bibfield  {journal} {\bibinfo
  {journal} {Phys. Rev. D}\ }\textbf {\bibinfo {volume} {80}},\ \bibinfo
  {pages} {064015} (\bibinfo {year} {2009})},\ \Eprint
  {http://arxiv.org/abs/0906.1967} {arXiv:0906.1967 [gr-qc]} \BibitemShut
  {NoStop}%
\bibitem [{\citenamefont {He}\ \emph {et~al.}(2016)\citenamefont {He},
  \citenamefont {Liu}, \citenamefont {Lu}, \citenamefont {Zhou}, \citenamefont
  {Cai}, \citenamefont {Wang},\ and\ \citenamefont
  {Brandenberger}}]{He:2016uiy}%
  \BibitemOpen
  \bibfield  {author} {\bibinfo {author} {\bibfnamefont {M.}~\bibnamefont
  {He}}, \bibinfo {author} {\bibfnamefont {J.}~\bibnamefont {Liu}}, \bibinfo
  {author} {\bibfnamefont {S.}~\bibnamefont {Lu}}, \bibinfo {author}
  {\bibfnamefont {S.}~\bibnamefont {Zhou}}, \bibinfo {author} {\bibfnamefont
  {Y.-F.}\ \bibnamefont {Cai}}, \bibinfo {author} {\bibfnamefont
  {Y.}~\bibnamefont {Wang}}, \ and\ \bibinfo {author} {\bibfnamefont
  {R.}~\bibnamefont {Brandenberger}},\ }\href {\doibase
  10.1088/1475-7516/2016/12/040} {\bibfield  {journal} {\bibinfo  {journal}
  {JCAP}\ }\textbf {\bibinfo {volume} {12}},\ \bibinfo {pages} {040} (\bibinfo
  {year} {2016})},\ \Eprint {http://arxiv.org/abs/1608.05079} {arXiv:1608.05079
  [astro-ph.CO]} \BibitemShut {NoStop}%
\bibitem [{\citenamefont {Koshelev}\ \emph {et~al.}(2016)\citenamefont
  {Koshelev}, \citenamefont {Modesto}, \citenamefont {Rachwal},\ and\
  \citenamefont {Starobinsky}}]{Koshelev:2016xqb}%
  \BibitemOpen
  \bibfield  {author} {\bibinfo {author} {\bibfnamefont {A.~S.}\ \bibnamefont
  {Koshelev}}, \bibinfo {author} {\bibfnamefont {L.}~\bibnamefont {Modesto}},
  \bibinfo {author} {\bibfnamefont {L.}~\bibnamefont {Rachwal}}, \ and\
  \bibinfo {author} {\bibfnamefont {A.~A.}\ \bibnamefont {Starobinsky}},\
  }\href {\doibase 10.1007/JHEP11(2016)067} {\bibfield  {journal} {\bibinfo
  {journal} {JHEP}\ }\textbf {\bibinfo {volume} {11}},\ \bibinfo {pages} {067}
  (\bibinfo {year} {2016})},\ \Eprint {http://arxiv.org/abs/1604.03127}
  {arXiv:1604.03127 [hep-th]} \BibitemShut {NoStop}%
\bibitem [{\citenamefont {Koshelev}\ \emph {et~al.}(2018)\citenamefont
  {Koshelev}, \citenamefont {Sravan~Kumar},\ and\ \citenamefont
  {Starobinsky}}]{Koshelev:2017tvv}%
  \BibitemOpen
  \bibfield  {author} {\bibinfo {author} {\bibfnamefont {A.~S.}\ \bibnamefont
  {Koshelev}}, \bibinfo {author} {\bibfnamefont {K.}~\bibnamefont
  {Sravan~Kumar}}, \ and\ \bibinfo {author} {\bibfnamefont {A.~A.}\
  \bibnamefont {Starobinsky}},\ }\href {\doibase 10.1007/JHEP03(2018)071}
  {\bibfield  {journal} {\bibinfo  {journal} {JHEP}\ }\textbf {\bibinfo
  {volume} {03}},\ \bibinfo {pages} {071} (\bibinfo {year} {2018})},\ \Eprint
  {http://arxiv.org/abs/1711.08864} {arXiv:1711.08864 [hep-th]} \BibitemShut
  {NoStop}%
\bibitem [{\citenamefont {Koshelev}\ \emph
  {et~al.}(2020{\natexlab{a}})\citenamefont {Koshelev}, \citenamefont
  {Sravan~Kumar}, \citenamefont {Mazumdar},\ and\ \citenamefont
  {Starobinsky}}]{Koshelev:2020foq}%
  \BibitemOpen
  \bibfield  {author} {\bibinfo {author} {\bibfnamefont {A.~S.}\ \bibnamefont
  {Koshelev}}, \bibinfo {author} {\bibfnamefont {K.}~\bibnamefont
  {Sravan~Kumar}}, \bibinfo {author} {\bibfnamefont {A.}~\bibnamefont
  {Mazumdar}}, \ and\ \bibinfo {author} {\bibfnamefont {A.~A.}\ \bibnamefont
  {Starobinsky}},\ }\href {\doibase 10.1007/JHEP06(2020)152} {\bibfield
  {journal} {\bibinfo  {journal} {JHEP}\ }\textbf {\bibinfo {volume} {06}},\
  \bibinfo {pages} {152} (\bibinfo {year} {2020}{\natexlab{a}})},\ \Eprint
  {http://arxiv.org/abs/2003.00629} {arXiv:2003.00629 [hep-th]} \BibitemShut
  {NoStop}%
\bibitem [{\citenamefont {Koshelev}\ \emph
  {et~al.}(2020{\natexlab{b}})\citenamefont {Koshelev}, \citenamefont {Kumar},\
  and\ \citenamefont {Starobinsky}}]{Koshelev:2020xby}%
  \BibitemOpen
  \bibfield  {author} {\bibinfo {author} {\bibfnamefont {A.~S.}\ \bibnamefont
  {Koshelev}}, \bibinfo {author} {\bibfnamefont {K.~S.}\ \bibnamefont {Kumar}},
  \ and\ \bibinfo {author} {\bibfnamefont {A.~A.}\ \bibnamefont
  {Starobinsky}},\ }\href {\doibase 10.1142/S021827182043018X} {\bibfield
  {journal} {\bibinfo  {journal} {Int. J. Mod. Phys. D}\ }\textbf {\bibinfo
  {volume} {29}},\ \bibinfo {pages} {2043018} (\bibinfo {year}
  {2020}{\natexlab{b}})},\ \Eprint {http://arxiv.org/abs/2005.09550}
  {arXiv:2005.09550 [hep-th]} \BibitemShut {NoStop}%
\bibitem [{\citenamefont {Koshelev}\ \emph
  {et~al.}(2022{\natexlab{a}})\citenamefont {Koshelev}, \citenamefont {Kumar},\
  and\ \citenamefont {Starobinsky}}]{Koshelev:2022olc}%
  \BibitemOpen
  \bibfield  {author} {\bibinfo {author} {\bibfnamefont {A.~S.}\ \bibnamefont
  {Koshelev}}, \bibinfo {author} {\bibfnamefont {K.~S.}\ \bibnamefont {Kumar}},
  \ and\ \bibinfo {author} {\bibfnamefont {A.~A.}\ \bibnamefont
  {Starobinsky}},\ }\href@noop {} {\  (\bibinfo {year} {2022}{\natexlab{a}})},\
  \Eprint {http://arxiv.org/abs/2209.02515} {arXiv:2209.02515 [hep-th]}
  \BibitemShut {NoStop}%
\bibitem [{\citenamefont {Koshelev}\ \emph
  {et~al.}(2022{\natexlab{b}})\citenamefont {Koshelev}, \citenamefont {Kumar},\
  and\ \citenamefont {Starobinsky}}]{Koshelev:2022bvg}%
  \BibitemOpen
  \bibfield  {author} {\bibinfo {author} {\bibfnamefont {A.~S.}\ \bibnamefont
  {Koshelev}}, \bibinfo {author} {\bibfnamefont {K.~S.}\ \bibnamefont {Kumar}},
  \ and\ \bibinfo {author} {\bibfnamefont {A.~A.}\ \bibnamefont
  {Starobinsky}},\ }\href@noop {} {\  (\bibinfo {year} {2022}{\natexlab{b}})},\
  \Eprint {http://arxiv.org/abs/2210.16459} {arXiv:2210.16459 [hep-th]}
  \BibitemShut {NoStop}%
\bibitem [{\citenamefont {Koshelev}\ \emph {et~al.}(2023)\citenamefont
  {Koshelev}, \citenamefont {Kumar},\ and\ \citenamefont
  {Starobinsky}}]{Koshelev:2023elc}%
  \BibitemOpen
  \bibfield  {author} {\bibinfo {author} {\bibfnamefont {A.~S.}\ \bibnamefont
  {Koshelev}}, \bibinfo {author} {\bibfnamefont {K.~S.}\ \bibnamefont {Kumar}},
  \ and\ \bibinfo {author} {\bibfnamefont {A.~A.}\ \bibnamefont
  {Starobinsky}},\ }\href@noop {} {\  (\bibinfo {year} {2023})},\ \Eprint
  {http://arxiv.org/abs/2305.18716} {arXiv:2305.18716 [hep-th]} \BibitemShut
  {NoStop}%
\bibitem [{\citenamefont {Biswas}\ \emph {et~al.}(2014)\citenamefont {Biswas},
  \citenamefont {Koivisto},\ and\ \citenamefont {Mazumdar}}]{Biswas:2014kva}%
  \BibitemOpen
  \bibfield  {author} {\bibinfo {author} {\bibfnamefont {T.}~\bibnamefont
  {Biswas}}, \bibinfo {author} {\bibfnamefont {T.}~\bibnamefont {Koivisto}}, \
  and\ \bibinfo {author} {\bibfnamefont {A.}~\bibnamefont {Mazumdar}},\ }\href
  {\doibase 10.1007/JHEP08(2014)116} {\bibfield  {journal} {\bibinfo  {journal}
  {JHEP}\ }\textbf {\bibinfo {volume} {08}},\ \bibinfo {pages} {116} (\bibinfo
  {year} {2014})},\ \Eprint {http://arxiv.org/abs/1403.7163} {arXiv:1403.7163
  [hep-th]} \BibitemShut {NoStop}%
\bibitem [{\citenamefont {Khoury}\ \emph {et~al.}(2002)\citenamefont {Khoury},
  \citenamefont {Ovrut}, \citenamefont {Steinhardt},\ and\ \citenamefont
  {Turok}}]{Khoury:2001zk}%
  \BibitemOpen
  \bibfield  {author} {\bibinfo {author} {\bibfnamefont {J.}~\bibnamefont
  {Khoury}}, \bibinfo {author} {\bibfnamefont {B.~A.}\ \bibnamefont {Ovrut}},
  \bibinfo {author} {\bibfnamefont {P.~J.}\ \bibnamefont {Steinhardt}}, \ and\
  \bibinfo {author} {\bibfnamefont {N.}~\bibnamefont {Turok}},\ }\href
  {\doibase 10.1103/PhysRevD.66.046005} {\bibfield  {journal} {\bibinfo
  {journal} {Phys. Rev. D}\ }\textbf {\bibinfo {volume} {66}},\ \bibinfo
  {pages} {046005} (\bibinfo {year} {2002})},\ \Eprint
  {http://arxiv.org/abs/hep-th/0109050} {arXiv:hep-th/0109050} \BibitemShut
  {NoStop}%
\bibitem [{\citenamefont {Buchbinder}\ \emph {et~al.}(2007)\citenamefont
  {Buchbinder}, \citenamefont {Khoury},\ and\ \citenamefont
  {Ovrut}}]{Buchbinder:2007ad}%
  \BibitemOpen
  \bibfield  {author} {\bibinfo {author} {\bibfnamefont {E.~I.}\ \bibnamefont
  {Buchbinder}}, \bibinfo {author} {\bibfnamefont {J.}~\bibnamefont {Khoury}},
  \ and\ \bibinfo {author} {\bibfnamefont {B.~A.}\ \bibnamefont {Ovrut}},\
  }\href {\doibase 10.1103/PhysRevD.76.123503} {\bibfield  {journal} {\bibinfo
  {journal} {Phys. Rev. D}\ }\textbf {\bibinfo {volume} {76}},\ \bibinfo
  {pages} {123503} (\bibinfo {year} {2007})},\ \Eprint
  {http://arxiv.org/abs/hep-th/0702154} {arXiv:hep-th/0702154} \BibitemShut
  {NoStop}%
\bibitem [{\citenamefont {Lehners}(2008)}]{Lehners:2008vx}%
  \BibitemOpen
  \bibfield  {author} {\bibinfo {author} {\bibfnamefont {J.-L.}\ \bibnamefont
  {Lehners}},\ }\href {\doibase 10.1016/j.physrep.2008.06.001} {\bibfield
  {journal} {\bibinfo  {journal} {Phys. Rept.}\ }\textbf {\bibinfo {volume}
  {465}},\ \bibinfo {pages} {223} (\bibinfo {year} {2008})},\ \Eprint
  {http://arxiv.org/abs/0806.1245} {arXiv:0806.1245 [astro-ph]} \BibitemShut
  {NoStop}%
\bibitem [{\citenamefont {Oikonomou}(2015)}]{Oikonomou:2014yua}%
  \BibitemOpen
  \bibfield  {author} {\bibinfo {author} {\bibfnamefont {V.~K.}\ \bibnamefont
  {Oikonomou}},\ }\href {\doibase 10.1007/s10509-015-2478-1} {\bibfield
  {journal} {\bibinfo  {journal} {Astrophys. Space Sci.}\ }\textbf {\bibinfo
  {volume} {359}},\ \bibinfo {pages} {30} (\bibinfo {year} {2015})},\ \Eprint
  {http://arxiv.org/abs/1412.4343} {arXiv:1412.4343 [gr-qc]} \BibitemShut
  {NoStop}%
\bibitem [{\citenamefont {Odintsov}\ \emph {et~al.}(2015)\citenamefont
  {Odintsov}, \citenamefont {Oikonomou},\ and\ \citenamefont
  {Saridakis}}]{Odintsov:2015uca}%
  \BibitemOpen
  \bibfield  {author} {\bibinfo {author} {\bibfnamefont {S.~D.}\ \bibnamefont
  {Odintsov}}, \bibinfo {author} {\bibfnamefont {V.~K.}\ \bibnamefont
  {Oikonomou}}, \ and\ \bibinfo {author} {\bibfnamefont {E.~N.}\ \bibnamefont
  {Saridakis}},\ }\href {\doibase 10.1016/j.aop.2015.08.021} {\bibfield
  {journal} {\bibinfo  {journal} {Annals Phys.}\ }\textbf {\bibinfo {volume}
  {363}},\ \bibinfo {pages} {141} (\bibinfo {year} {2015})},\ \Eprint
  {http://arxiv.org/abs/1501.06591} {arXiv:1501.06591 [gr-qc]} \BibitemShut
  {NoStop}%
\bibitem [{\citenamefont {Boyle}\ \emph {et~al.}(2004)\citenamefont {Boyle},
  \citenamefont {Steinhardt},\ and\ \citenamefont {Turok}}]{Boyle:2003km}%
  \BibitemOpen
  \bibfield  {author} {\bibinfo {author} {\bibfnamefont {L.~A.}\ \bibnamefont
  {Boyle}}, \bibinfo {author} {\bibfnamefont {P.~J.}\ \bibnamefont
  {Steinhardt}}, \ and\ \bibinfo {author} {\bibfnamefont {N.}~\bibnamefont
  {Turok}},\ }\href {\doibase 10.1103/PhysRevD.69.127302} {\bibfield  {journal}
  {\bibinfo  {journal} {Phys. Rev. D}\ }\textbf {\bibinfo {volume} {69}},\
  \bibinfo {pages} {127302} (\bibinfo {year} {2004})},\ \Eprint
  {http://arxiv.org/abs/hep-th/0307170} {arXiv:hep-th/0307170} \BibitemShut
  {NoStop}%
\bibitem [{\citenamefont {Calcagni}\ and\ \citenamefont
  {Kuroyanagi}(2021)}]{Calcagni:2020tvw}%
  \BibitemOpen
  \bibfield  {author} {\bibinfo {author} {\bibfnamefont {G.}~\bibnamefont
  {Calcagni}}\ and\ \bibinfo {author} {\bibfnamefont {S.}~\bibnamefont
  {Kuroyanagi}},\ }\href {\doibase 10.1088/1475-7516/2021/03/019} {\bibfield
  {journal} {\bibinfo  {journal} {JCAP}\ }\textbf {\bibinfo {volume} {03}},\
  \bibinfo {pages} {019} (\bibinfo {year} {2021})},\ \Eprint
  {http://arxiv.org/abs/2012.00170} {arXiv:2012.00170 [gr-qc]} \BibitemShut
  {NoStop}%
\bibitem [{\citenamefont {Brandenberger}\ and\ \citenamefont
  {Wang}(2020{\natexlab{a}})}]{Brandenberger:2020tcr}%
  \BibitemOpen
  \bibfield  {author} {\bibinfo {author} {\bibfnamefont {R.}~\bibnamefont
  {Brandenberger}}\ and\ \bibinfo {author} {\bibfnamefont {Z.}~\bibnamefont
  {Wang}},\ }\href {\doibase 10.1103/PhysRevD.101.063522} {\bibfield  {journal}
  {\bibinfo  {journal} {Phys. Rev. D}\ }\textbf {\bibinfo {volume} {101}},\
  \bibinfo {pages} {063522} (\bibinfo {year} {2020}{\natexlab{a}})},\ \Eprint
  {http://arxiv.org/abs/2001.00638} {arXiv:2001.00638 [hep-th]} \BibitemShut
  {NoStop}%
\bibitem [{\citenamefont {Brandenberger}\ and\ \citenamefont
  {Wang}(2020{\natexlab{b}})}]{Brandenberger:2020eyf}%
  \BibitemOpen
  \bibfield  {author} {\bibinfo {author} {\bibfnamefont {R.}~\bibnamefont
  {Brandenberger}}\ and\ \bibinfo {author} {\bibfnamefont {Z.}~\bibnamefont
  {Wang}},\ }\href {\doibase 10.1103/PhysRevD.102.023516} {\bibfield  {journal}
  {\bibinfo  {journal} {Phys. Rev. D}\ }\textbf {\bibinfo {volume} {102}},\
  \bibinfo {pages} {023516} (\bibinfo {year} {2020}{\natexlab{b}})},\ \Eprint
  {http://arxiv.org/abs/2004.06437} {arXiv:2004.06437 [hep-th]} \BibitemShut
  {NoStop}%
\bibitem [{\citenamefont {Brandenberger}\ \emph {et~al.}(2020)\citenamefont
  {Brandenberger}, \citenamefont {Dasgupta},\ and\ \citenamefont
  {Wang}}]{Brandenberger:2020wha}%
  \BibitemOpen
  \bibfield  {author} {\bibinfo {author} {\bibfnamefont {R.}~\bibnamefont
  {Brandenberger}}, \bibinfo {author} {\bibfnamefont {K.}~\bibnamefont
  {Dasgupta}}, \ and\ \bibinfo {author} {\bibfnamefont {Z.}~\bibnamefont
  {Wang}},\ }\href {\doibase 10.1103/PhysRevD.102.063514} {\bibfield  {journal}
  {\bibinfo  {journal} {Phys. Rev. D}\ }\textbf {\bibinfo {volume} {102}},\
  \bibinfo {pages} {063514} (\bibinfo {year} {2020})},\ \Eprint
  {http://arxiv.org/abs/2007.01203} {arXiv:2007.01203 [hep-th]} \BibitemShut
  {NoStop}%
\bibitem [{\citenamefont {Bartolo}\ \emph {et~al.}(2019)\citenamefont
  {Bartolo}, \citenamefont {Bertacca}, \citenamefont {Matarrese}, \citenamefont
  {Peloso}, \citenamefont {Ricciardone}, \citenamefont {Riotto},\ and\
  \citenamefont {Tasinato}}]{Bartolo:2019oiq}%
  \BibitemOpen
  \bibfield  {author} {\bibinfo {author} {\bibfnamefont {N.}~\bibnamefont
  {Bartolo}}, \bibinfo {author} {\bibfnamefont {D.}~\bibnamefont {Bertacca}},
  \bibinfo {author} {\bibfnamefont {S.}~\bibnamefont {Matarrese}}, \bibinfo
  {author} {\bibfnamefont {M.}~\bibnamefont {Peloso}}, \bibinfo {author}
  {\bibfnamefont {A.}~\bibnamefont {Ricciardone}}, \bibinfo {author}
  {\bibfnamefont {A.}~\bibnamefont {Riotto}}, \ and\ \bibinfo {author}
  {\bibfnamefont {G.}~\bibnamefont {Tasinato}},\ }\href {\doibase
  10.1103/PhysRevD.100.121501} {\bibfield  {journal} {\bibinfo  {journal}
  {Phys. Rev. D}\ }\textbf {\bibinfo {volume} {100}},\ \bibinfo {pages}
  {121501} (\bibinfo {year} {2019})},\ \Eprint
  {http://arxiv.org/abs/1908.00527} {arXiv:1908.00527 [astro-ph.CO]}
  \BibitemShut {NoStop}%
\bibitem [{\citenamefont {Bartolo}\ \emph {et~al.}(2020)\citenamefont
  {Bartolo}, \citenamefont {Bertacca}, \citenamefont {De~Luca}, \citenamefont
  {Franciolini}, \citenamefont {Matarrese}, \citenamefont {Peloso},
  \citenamefont {Ricciardone}, \citenamefont {Riotto},\ and\ \citenamefont
  {Tasinato}}]{Bartolo:2019zvb}%
  \BibitemOpen
  \bibfield  {author} {\bibinfo {author} {\bibfnamefont {N.}~\bibnamefont
  {Bartolo}}, \bibinfo {author} {\bibfnamefont {D.}~\bibnamefont {Bertacca}},
  \bibinfo {author} {\bibfnamefont {V.}~\bibnamefont {De~Luca}}, \bibinfo
  {author} {\bibfnamefont {G.}~\bibnamefont {Franciolini}}, \bibinfo {author}
  {\bibfnamefont {S.}~\bibnamefont {Matarrese}}, \bibinfo {author}
  {\bibfnamefont {M.}~\bibnamefont {Peloso}}, \bibinfo {author} {\bibfnamefont
  {A.}~\bibnamefont {Ricciardone}}, \bibinfo {author} {\bibfnamefont
  {A.}~\bibnamefont {Riotto}}, \ and\ \bibinfo {author} {\bibfnamefont
  {G.}~\bibnamefont {Tasinato}},\ }\href {\doibase
  10.1088/1475-7516/2020/02/028} {\bibfield  {journal} {\bibinfo  {journal}
  {JCAP}\ }\textbf {\bibinfo {volume} {02}},\ \bibinfo {pages} {028} (\bibinfo
  {year} {2020})},\ \Eprint {http://arxiv.org/abs/1909.12619} {arXiv:1909.12619
  [astro-ph.CO]} \BibitemShut {NoStop}%
\bibitem [{\citenamefont {Bartolo}\ \emph {et~al.}(2022)\citenamefont {Bartolo}
  \emph {et~al.}}]{LISACosmologyWorkingGroup:2022kbp}%
  \BibitemOpen
  \bibfield  {author} {\bibinfo {author} {\bibfnamefont {N.}~\bibnamefont
  {Bartolo}} \emph {et~al.} (\bibinfo {collaboration} {LISA Cosmology Working
  Group}),\ }\href {\doibase 10.1088/1475-7516/2022/11/009} {\bibfield
  {journal} {\bibinfo  {journal} {JCAP}\ }\textbf {\bibinfo {volume} {11}},\
  \bibinfo {pages} {009} (\bibinfo {year} {2022})},\ \Eprint
  {http://arxiv.org/abs/2201.08782} {arXiv:2201.08782 [astro-ph.CO]}
  \BibitemShut {NoStop}%
\bibitem [{\citenamefont {Galloni}\ \emph {et~al.}(2022)\citenamefont
  {Galloni}, \citenamefont {Bartolo}, \citenamefont {Matarrese}, \citenamefont
  {Migliaccio}, \citenamefont {Ricciardone},\ and\ \citenamefont
  {Vittorio}}]{Galloni:2022rgg}%
  \BibitemOpen
  \bibfield  {author} {\bibinfo {author} {\bibfnamefont {G.}~\bibnamefont
  {Galloni}}, \bibinfo {author} {\bibfnamefont {N.}~\bibnamefont {Bartolo}},
  \bibinfo {author} {\bibfnamefont {S.}~\bibnamefont {Matarrese}}, \bibinfo
  {author} {\bibfnamefont {M.}~\bibnamefont {Migliaccio}}, \bibinfo {author}
  {\bibfnamefont {A.}~\bibnamefont {Ricciardone}}, \ and\ \bibinfo {author}
  {\bibfnamefont {N.}~\bibnamefont {Vittorio}},\ }\href {\doibase
  10.1088/1475-7516/2022/09/046} {\bibfield  {journal} {\bibinfo  {journal}
  {JCAP}\ }\textbf {\bibinfo {volume} {09}},\ \bibinfo {pages} {046} (\bibinfo
  {year} {2022})},\ \Eprint {http://arxiv.org/abs/2202.12858} {arXiv:2202.12858
  [astro-ph.CO]} \BibitemShut {NoStop}%
\bibitem [{\citenamefont {Schulze}\ \emph {et~al.}(2023)\citenamefont
  {Schulze}, \citenamefont {Valbusa~Dall'Armi}, \citenamefont {Lesgourgues},
  \citenamefont {Ricciardone}, \citenamefont {Bartolo}, \citenamefont
  {Bertacca}, \citenamefont {Fidler},\ and\ \citenamefont
  {Matarrese}}]{Schulze:2023ich}%
  \BibitemOpen
  \bibfield  {author} {\bibinfo {author} {\bibfnamefont {F.}~\bibnamefont
  {Schulze}}, \bibinfo {author} {\bibfnamefont {L.}~\bibnamefont
  {Valbusa~Dall'Armi}}, \bibinfo {author} {\bibfnamefont {J.}~\bibnamefont
  {Lesgourgues}}, \bibinfo {author} {\bibfnamefont {A.}~\bibnamefont
  {Ricciardone}}, \bibinfo {author} {\bibfnamefont {N.}~\bibnamefont
  {Bartolo}}, \bibinfo {author} {\bibfnamefont {D.}~\bibnamefont {Bertacca}},
  \bibinfo {author} {\bibfnamefont {C.}~\bibnamefont {Fidler}}, \ and\ \bibinfo
  {author} {\bibfnamefont {S.}~\bibnamefont {Matarrese}},\ }\href@noop {} {\
  (\bibinfo {year} {2023})},\ \Eprint {http://arxiv.org/abs/2305.01602}
  {arXiv:2305.01602 [gr-qc]} \BibitemShut {NoStop}%
\bibitem [{\citenamefont {Liu}(2023)}]{Liu:2023hte}%
  \BibitemOpen
  \bibfield  {author} {\bibinfo {author} {\bibfnamefont {J.}~\bibnamefont
  {Liu}},\ }\href@noop {} {\  (\bibinfo {year} {2023})},\ \Eprint
  {http://arxiv.org/abs/2305.15100} {arXiv:2305.15100 [astro-ph.CO]}
  \BibitemShut {NoStop}%
\bibitem [{\citenamefont {King}\ \emph {et~al.}(2023)\citenamefont {King},
  \citenamefont {Marfatia},\ and\ \citenamefont {Rahat}}]{King:2023cgv}%
  \BibitemOpen
  \bibfield  {author} {\bibinfo {author} {\bibfnamefont {S.~F.}\ \bibnamefont
  {King}}, \bibinfo {author} {\bibfnamefont {D.}~\bibnamefont {Marfatia}}, \
  and\ \bibinfo {author} {\bibfnamefont {M.~H.}\ \bibnamefont {Rahat}},\
  }\href@noop {} {\  (\bibinfo {year} {2023})},\ \Eprint
  {http://arxiv.org/abs/2306.05389} {arXiv:2306.05389 [hep-ph]} \BibitemShut
  {NoStop}%
\bibitem [{\citenamefont {Zu}\ \emph {et~al.}(2023)\citenamefont {Zu},
  \citenamefont {Zhang}, \citenamefont {Li}, \citenamefont {Gu}, \citenamefont
  {Tsai},\ and\ \citenamefont {Fan}}]{Zu:2023olm}%
  \BibitemOpen
  \bibfield  {author} {\bibinfo {author} {\bibfnamefont {L.}~\bibnamefont
  {Zu}}, \bibinfo {author} {\bibfnamefont {C.}~\bibnamefont {Zhang}}, \bibinfo
  {author} {\bibfnamefont {Y.-Y.}\ \bibnamefont {Li}}, \bibinfo {author}
  {\bibfnamefont {Y.-C.}\ \bibnamefont {Gu}}, \bibinfo {author} {\bibfnamefont
  {Y.-L.~S.}\ \bibnamefont {Tsai}}, \ and\ \bibinfo {author} {\bibfnamefont
  {Y.-Z.}\ \bibnamefont {Fan}},\ }\href@noop {} {\  (\bibinfo {year} {2023})},\
  \Eprint {http://arxiv.org/abs/2306.16769} {arXiv:2306.16769 [astro-ph.HE]}
  \BibitemShut {NoStop}%
\bibitem [{\citenamefont {Han}\ \emph {et~al.}(2023)\citenamefont {Han},
  \citenamefont {Xie}, \citenamefont {Yang},\ and\ \citenamefont
  {Zhang}}]{Han:2023olf}%
  \BibitemOpen
  \bibfield  {author} {\bibinfo {author} {\bibfnamefont {C.}~\bibnamefont
  {Han}}, \bibinfo {author} {\bibfnamefont {K.-P.}\ \bibnamefont {Xie}},
  \bibinfo {author} {\bibfnamefont {J.~M.}\ \bibnamefont {Yang}}, \ and\
  \bibinfo {author} {\bibfnamefont {M.}~\bibnamefont {Zhang}},\ }\href@noop {}
  {\  (\bibinfo {year} {2023})},\ \Eprint {http://arxiv.org/abs/2306.16966}
  {arXiv:2306.16966 [hep-ph]} \BibitemShut {NoStop}%
\bibitem [{\citenamefont {Lambiase}\ \emph {et~al.}(2023)\citenamefont
  {Lambiase}, \citenamefont {Mastrototaro},\ and\ \citenamefont
  {Visinelli}}]{Lambiase:2023pxd}%
  \BibitemOpen
  \bibfield  {author} {\bibinfo {author} {\bibfnamefont {G.}~\bibnamefont
  {Lambiase}}, \bibinfo {author} {\bibfnamefont {L.}~\bibnamefont
  {Mastrototaro}}, \ and\ \bibinfo {author} {\bibfnamefont {L.}~\bibnamefont
  {Visinelli}},\ }\href@noop {} {\  (\bibinfo {year} {2023})},\ \Eprint
  {http://arxiv.org/abs/2306.16977} {arXiv:2306.16977 [astro-ph.HE]}
  \BibitemShut {NoStop}%
\bibitem [{\citenamefont {Ellis}\ \emph
  {et~al.}(2023{\natexlab{a}})\citenamefont {Ellis}, \citenamefont {Fairbairn},
  \citenamefont {H\"utsi}, \citenamefont {Raidal}, \citenamefont {Urrutia},
  \citenamefont {Vaskonen},\ and\ \citenamefont {Veerm\"ae}}]{Ellis:2023dgf}%
  \BibitemOpen
  \bibfield  {author} {\bibinfo {author} {\bibfnamefont {J.}~\bibnamefont
  {Ellis}}, \bibinfo {author} {\bibfnamefont {M.}~\bibnamefont {Fairbairn}},
  \bibinfo {author} {\bibfnamefont {G.}~\bibnamefont {H\"utsi}}, \bibinfo
  {author} {\bibfnamefont {J.}~\bibnamefont {Raidal}}, \bibinfo {author}
  {\bibfnamefont {J.}~\bibnamefont {Urrutia}}, \bibinfo {author} {\bibfnamefont
  {V.}~\bibnamefont {Vaskonen}}, \ and\ \bibinfo {author} {\bibfnamefont
  {H.}~\bibnamefont {Veerm\"ae}},\ }\href@noop {} {\  (\bibinfo {year}
  {2023}{\natexlab{a}})},\ \Eprint {http://arxiv.org/abs/2306.17021}
  {arXiv:2306.17021 [astro-ph.CO]} \BibitemShut {NoStop}%
\bibitem [{\citenamefont {Guo}\ \emph {et~al.}(2023)\citenamefont {Guo},
  \citenamefont {Khlopov}, \citenamefont {Liu}, \citenamefont {Wu},
  \citenamefont {Wu},\ and\ \citenamefont {Zhu}}]{Guo:2023hyp}%
  \BibitemOpen
  \bibfield  {author} {\bibinfo {author} {\bibfnamefont {S.-Y.}\ \bibnamefont
  {Guo}}, \bibinfo {author} {\bibfnamefont {M.}~\bibnamefont {Khlopov}},
  \bibinfo {author} {\bibfnamefont {X.}~\bibnamefont {Liu}}, \bibinfo {author}
  {\bibfnamefont {L.}~\bibnamefont {Wu}}, \bibinfo {author} {\bibfnamefont
  {Y.}~\bibnamefont {Wu}}, \ and\ \bibinfo {author} {\bibfnamefont
  {B.}~\bibnamefont {Zhu}},\ }\href@noop {} {\  (\bibinfo {year} {2023})},\
  \Eprint {http://arxiv.org/abs/2306.17022} {arXiv:2306.17022 [hep-ph]}
  \BibitemShut {NoStop}%
\bibitem [{\citenamefont {Megias}\ \emph {et~al.}(2023)\citenamefont {Megias},
  \citenamefont {Nardini},\ and\ \citenamefont {Quiros}}]{Megias:2023kiy}%
  \BibitemOpen
  \bibfield  {author} {\bibinfo {author} {\bibfnamefont {E.}~\bibnamefont
  {Megias}}, \bibinfo {author} {\bibfnamefont {G.}~\bibnamefont {Nardini}}, \
  and\ \bibinfo {author} {\bibfnamefont {M.}~\bibnamefont {Quiros}},\
  }\href@noop {} {\  (\bibinfo {year} {2023})},\ \Eprint
  {http://arxiv.org/abs/2306.17071} {arXiv:2306.17071 [hep-ph]} \BibitemShut
  {NoStop}%
\bibitem [{\citenamefont {Fujikura}\ \emph {et~al.}(2023)\citenamefont
  {Fujikura}, \citenamefont {Girmohanta}, \citenamefont {Nakai},\ and\
  \citenamefont {Suzuki}}]{Fujikura:2023lkn}%
  \BibitemOpen
  \bibfield  {author} {\bibinfo {author} {\bibfnamefont {K.}~\bibnamefont
  {Fujikura}}, \bibinfo {author} {\bibfnamefont {S.}~\bibnamefont
  {Girmohanta}}, \bibinfo {author} {\bibfnamefont {Y.}~\bibnamefont {Nakai}}, \
  and\ \bibinfo {author} {\bibfnamefont {M.}~\bibnamefont {Suzuki}},\
  }\href@noop {} {\  (\bibinfo {year} {2023})},\ \Eprint
  {http://arxiv.org/abs/2306.17086} {arXiv:2306.17086 [hep-ph]} \BibitemShut
  {NoStop}%
\bibitem [{\citenamefont {Yang}\ \emph
  {et~al.}(2023{\natexlab{a}})\citenamefont {Yang}, \citenamefont {Xie},\ and\
  \citenamefont {Huang}}]{Yang:2023aak}%
  \BibitemOpen
  \bibfield  {author} {\bibinfo {author} {\bibfnamefont {J.}~\bibnamefont
  {Yang}}, \bibinfo {author} {\bibfnamefont {N.}~\bibnamefont {Xie}}, \ and\
  \bibinfo {author} {\bibfnamefont {F.~P.}\ \bibnamefont {Huang}},\ }\href@noop
  {} {\  (\bibinfo {year} {2023}{\natexlab{a}})},\ \Eprint
  {http://arxiv.org/abs/2306.17113} {arXiv:2306.17113 [hep-ph]} \BibitemShut
  {NoStop}%
\bibitem [{\citenamefont {Li}\ \emph {et~al.}(2023)\citenamefont {Li},
  \citenamefont {Zhang}, \citenamefont {Wang}, \citenamefont {Cui},
  \citenamefont {Tsai}, \citenamefont {Yuan},\ and\ \citenamefont
  {Fan}}]{Li:2023yaj}%
  \BibitemOpen
  \bibfield  {author} {\bibinfo {author} {\bibfnamefont {Y.}~\bibnamefont
  {Li}}, \bibinfo {author} {\bibfnamefont {C.}~\bibnamefont {Zhang}}, \bibinfo
  {author} {\bibfnamefont {Z.}~\bibnamefont {Wang}}, \bibinfo {author}
  {\bibfnamefont {M.}~\bibnamefont {Cui}}, \bibinfo {author} {\bibfnamefont
  {Y.-L.~S.}\ \bibnamefont {Tsai}}, \bibinfo {author} {\bibfnamefont
  {Q.}~\bibnamefont {Yuan}}, \ and\ \bibinfo {author} {\bibfnamefont {Y.-Z.}\
  \bibnamefont {Fan}},\ }\href@noop {} {\  (\bibinfo {year} {2023})},\ \Eprint
  {http://arxiv.org/abs/2306.17124} {arXiv:2306.17124 [astro-ph.HE]}
  \BibitemShut {NoStop}%
\bibitem [{\citenamefont {Deng}\ \emph {et~al.}(2023)\citenamefont {Deng},
  \citenamefont {B\'ecsy}, \citenamefont {Siemens}, \citenamefont {Cornish},\
  and\ \citenamefont {Madison}}]{Deng:2023btv}%
  \BibitemOpen
  \bibfield  {author} {\bibinfo {author} {\bibfnamefont {H.}~\bibnamefont
  {Deng}}, \bibinfo {author} {\bibfnamefont {B.}~\bibnamefont {B\'ecsy}},
  \bibinfo {author} {\bibfnamefont {X.}~\bibnamefont {Siemens}}, \bibinfo
  {author} {\bibfnamefont {N.~J.}\ \bibnamefont {Cornish}}, \ and\ \bibinfo
  {author} {\bibfnamefont {D.~R.}\ \bibnamefont {Madison}},\ }\href@noop {} {\
  (\bibinfo {year} {2023})},\ \Eprint {http://arxiv.org/abs/2306.17130}
  {arXiv:2306.17130 [gr-qc]} \BibitemShut {NoStop}%
\bibitem [{\citenamefont {Franciolini}\ \emph
  {et~al.}(2023{\natexlab{a}})\citenamefont {Franciolini}, \citenamefont
  {Racco},\ and\ \citenamefont {Rompineve}}]{Franciolini:2023wjm}%
  \BibitemOpen
  \bibfield  {author} {\bibinfo {author} {\bibfnamefont {G.}~\bibnamefont
  {Franciolini}}, \bibinfo {author} {\bibfnamefont {D.}~\bibnamefont {Racco}},
  \ and\ \bibinfo {author} {\bibfnamefont {F.}~\bibnamefont {Rompineve}},\
  }\href@noop {} {\  (\bibinfo {year} {2023}{\natexlab{a}})},\ \Eprint
  {http://arxiv.org/abs/2306.17136} {arXiv:2306.17136 [astro-ph.CO]}
  \BibitemShut {NoStop}%
\bibitem [{\citenamefont {Shen}\ \emph {et~al.}(2023)\citenamefont {Shen},
  \citenamefont {Yuan}, \citenamefont {Wang},\ and\ \citenamefont
  {Wang}}]{Shen:2023pan}%
  \BibitemOpen
  \bibfield  {author} {\bibinfo {author} {\bibfnamefont {Z.-Q.}\ \bibnamefont
  {Shen}}, \bibinfo {author} {\bibfnamefont {G.-W.}\ \bibnamefont {Yuan}},
  \bibinfo {author} {\bibfnamefont {Y.-Y.}\ \bibnamefont {Wang}}, \ and\
  \bibinfo {author} {\bibfnamefont {Y.-Z.}\ \bibnamefont {Wang}},\ }\href@noop
  {} {\  (\bibinfo {year} {2023})},\ \Eprint {http://arxiv.org/abs/2306.17143}
  {arXiv:2306.17143 [astro-ph.HE]} \BibitemShut {NoStop}%
\bibitem [{\citenamefont {Kitajima}\ \emph {et~al.}(2023)\citenamefont
  {Kitajima}, \citenamefont {Lee}, \citenamefont {Murai}, \citenamefont
  {Takahashi},\ and\ \citenamefont {Yin}}]{Kitajima:2023cek}%
  \BibitemOpen
  \bibfield  {author} {\bibinfo {author} {\bibfnamefont {N.}~\bibnamefont
  {Kitajima}}, \bibinfo {author} {\bibfnamefont {J.}~\bibnamefont {Lee}},
  \bibinfo {author} {\bibfnamefont {K.}~\bibnamefont {Murai}}, \bibinfo
  {author} {\bibfnamefont {F.}~\bibnamefont {Takahashi}}, \ and\ \bibinfo
  {author} {\bibfnamefont {W.}~\bibnamefont {Yin}},\ }\href@noop {} {\
  (\bibinfo {year} {2023})},\ \Eprint {http://arxiv.org/abs/2306.17146}
  {arXiv:2306.17146 [hep-ph]} \BibitemShut {NoStop}%
\bibitem [{\citenamefont {Ellis}\ \emph
  {et~al.}(2023{\natexlab{b}})\citenamefont {Ellis}, \citenamefont {Lewicki},
  \citenamefont {Lin},\ and\ \citenamefont {Vaskonen}}]{Ellis:2023tsl}%
  \BibitemOpen
  \bibfield  {author} {\bibinfo {author} {\bibfnamefont {J.}~\bibnamefont
  {Ellis}}, \bibinfo {author} {\bibfnamefont {M.}~\bibnamefont {Lewicki}},
  \bibinfo {author} {\bibfnamefont {C.}~\bibnamefont {Lin}}, \ and\ \bibinfo
  {author} {\bibfnamefont {V.}~\bibnamefont {Vaskonen}},\ }\href@noop {} {\
  (\bibinfo {year} {2023}{\natexlab{b}})},\ \Eprint
  {http://arxiv.org/abs/2306.17147} {arXiv:2306.17147 [astro-ph.CO]}
  \BibitemShut {NoStop}%
\bibitem [{\citenamefont {Franciolini}\ \emph
  {et~al.}(2023{\natexlab{b}})\citenamefont {Franciolini}, \citenamefont
  {Iovino}, \citenamefont {Vaskonen},\ and\ \citenamefont
  {Veermae}}]{Franciolini:2023pbf}%
  \BibitemOpen
  \bibfield  {author} {\bibinfo {author} {\bibfnamefont {G.}~\bibnamefont
  {Franciolini}}, \bibinfo {author} {\bibfnamefont {A.}~\bibnamefont {Iovino},
  \bibfnamefont {Junior.}}, \bibinfo {author} {\bibfnamefont {V.}~\bibnamefont
  {Vaskonen}}, \ and\ \bibinfo {author} {\bibfnamefont {H.}~\bibnamefont
  {Veermae}},\ }\href@noop {} {\  (\bibinfo {year} {2023}{\natexlab{b}})},\
  \Eprint {http://arxiv.org/abs/2306.17149} {arXiv:2306.17149 [astro-ph.CO]}
  \BibitemShut {NoStop}%
\bibitem [{\citenamefont {Wang}\ \emph
  {et~al.}(2023{\natexlab{a}})\citenamefont {Wang}, \citenamefont {Lei},
  \citenamefont {Jiao}, \citenamefont {Feng},\ and\ \citenamefont
  {Fan}}]{Wang:2023len}%
  \BibitemOpen
  \bibfield  {author} {\bibinfo {author} {\bibfnamefont {Z.}~\bibnamefont
  {Wang}}, \bibinfo {author} {\bibfnamefont {L.}~\bibnamefont {Lei}}, \bibinfo
  {author} {\bibfnamefont {H.}~\bibnamefont {Jiao}}, \bibinfo {author}
  {\bibfnamefont {L.}~\bibnamefont {Feng}}, \ and\ \bibinfo {author}
  {\bibfnamefont {Y.-Z.}\ \bibnamefont {Fan}},\ }\href@noop {} {\  (\bibinfo
  {year} {2023}{\natexlab{a}})},\ \Eprint {http://arxiv.org/abs/2306.17150}
  {arXiv:2306.17150 [astro-ph.HE]} \BibitemShut {NoStop}%
\bibitem [{\citenamefont {Ghoshal}\ and\ \citenamefont
  {Strumia}(2023)}]{Ghoshal:2023fhh}%
  \BibitemOpen
  \bibfield  {author} {\bibinfo {author} {\bibfnamefont {A.}~\bibnamefont
  {Ghoshal}}\ and\ \bibinfo {author} {\bibfnamefont {A.}~\bibnamefont
  {Strumia}},\ }\href@noop {} {\  (\bibinfo {year} {2023})},\ \Eprint
  {http://arxiv.org/abs/2306.17158} {arXiv:2306.17158 [astro-ph.CO]}
  \BibitemShut {NoStop}%
\bibitem [{\citenamefont {Bai}\ \emph {et~al.}(2023)\citenamefont {Bai},
  \citenamefont {Chen},\ and\ \citenamefont {Korwar}}]{Bai:2023cqj}%
  \BibitemOpen
  \bibfield  {author} {\bibinfo {author} {\bibfnamefont {Y.}~\bibnamefont
  {Bai}}, \bibinfo {author} {\bibfnamefont {T.-K.}\ \bibnamefont {Chen}}, \
  and\ \bibinfo {author} {\bibfnamefont {M.}~\bibnamefont {Korwar}},\
  }\href@noop {} {\  (\bibinfo {year} {2023})},\ \Eprint
  {http://arxiv.org/abs/2306.17160} {arXiv:2306.17160 [hep-ph]} \BibitemShut
  {NoStop}%
\bibitem [{\citenamefont {Addazi}\ \emph {et~al.}(2023)\citenamefont {Addazi},
  \citenamefont {Cai}, \citenamefont {Marciano},\ and\ \citenamefont
  {Visinelli}}]{Addazi:2023jvg}%
  \BibitemOpen
  \bibfield  {author} {\bibinfo {author} {\bibfnamefont {A.}~\bibnamefont
  {Addazi}}, \bibinfo {author} {\bibfnamefont {Y.-F.}\ \bibnamefont {Cai}},
  \bibinfo {author} {\bibfnamefont {A.}~\bibnamefont {Marciano}}, \ and\
  \bibinfo {author} {\bibfnamefont {L.}~\bibnamefont {Visinelli}},\ }\href@noop
  {} {\  (\bibinfo {year} {2023})},\ \Eprint {http://arxiv.org/abs/2306.17205}
  {arXiv:2306.17205 [astro-ph.CO]} \BibitemShut {NoStop}%
\bibitem [{\citenamefont {Athron}\ \emph {et~al.}(2023)\citenamefont {Athron},
  \citenamefont {Fowlie}, \citenamefont {Lu}, \citenamefont {Morris},
  \citenamefont {Wu}, \citenamefont {Wu},\ and\ \citenamefont
  {Xu}}]{Athron:2023mer}%
  \BibitemOpen
  \bibfield  {author} {\bibinfo {author} {\bibfnamefont {P.}~\bibnamefont
  {Athron}}, \bibinfo {author} {\bibfnamefont {A.}~\bibnamefont {Fowlie}},
  \bibinfo {author} {\bibfnamefont {C.-T.}\ \bibnamefont {Lu}}, \bibinfo
  {author} {\bibfnamefont {L.}~\bibnamefont {Morris}}, \bibinfo {author}
  {\bibfnamefont {L.}~\bibnamefont {Wu}}, \bibinfo {author} {\bibfnamefont
  {Y.}~\bibnamefont {Wu}}, \ and\ \bibinfo {author} {\bibfnamefont
  {Z.}~\bibnamefont {Xu}},\ }\href@noop {} {\  (\bibinfo {year} {2023})},\
  \Eprint {http://arxiv.org/abs/2306.17239} {arXiv:2306.17239 [hep-ph]}
  \BibitemShut {NoStop}%
\bibitem [{\citenamefont {Oikonomou}(2023{\natexlab{c}})}]{Oikonomou:2023qfz}%
  \BibitemOpen
  \bibfield  {author} {\bibinfo {author} {\bibfnamefont {V.~K.}\ \bibnamefont
  {Oikonomou}},\ }\href@noop {} {\  (\bibinfo {year} {2023}{\natexlab{c}})},\
  \Eprint {http://arxiv.org/abs/2306.17351} {arXiv:2306.17351 [astro-ph.CO]}
  \BibitemShut {NoStop}%
\bibitem [{\citenamefont {Kitajima}\ and\ \citenamefont
  {Nakayama}(2023)}]{Kitajima:2023vre}%
  \BibitemOpen
  \bibfield  {author} {\bibinfo {author} {\bibfnamefont {N.}~\bibnamefont
  {Kitajima}}\ and\ \bibinfo {author} {\bibfnamefont {K.}~\bibnamefont
  {Nakayama}},\ }\href@noop {} {\  (\bibinfo {year} {2023})},\ \Eprint
  {http://arxiv.org/abs/2306.17390} {arXiv:2306.17390 [hep-ph]} \BibitemShut
  {NoStop}%
\bibitem [{\citenamefont {Huang}\ \emph {et~al.}(2023)\citenamefont {Huang},
  \citenamefont {Cai}, \citenamefont {Jiang}, \citenamefont {Zhang},\ and\
  \citenamefont {Piao}}]{Huang:2023chx}%
  \BibitemOpen
  \bibfield  {author} {\bibinfo {author} {\bibfnamefont {H.-L.}\ \bibnamefont
  {Huang}}, \bibinfo {author} {\bibfnamefont {Y.}~\bibnamefont {Cai}}, \bibinfo
  {author} {\bibfnamefont {J.-Q.}\ \bibnamefont {Jiang}}, \bibinfo {author}
  {\bibfnamefont {J.}~\bibnamefont {Zhang}}, \ and\ \bibinfo {author}
  {\bibfnamefont {Y.-S.}\ \bibnamefont {Piao}},\ }\href@noop {} {\  (\bibinfo
  {year} {2023})},\ \Eprint {http://arxiv.org/abs/2306.17577} {arXiv:2306.17577
  [gr-qc]} \BibitemShut {NoStop}%
\bibitem [{\citenamefont {Eichhorn}\ \emph {et~al.}(2023)\citenamefont
  {Eichhorn}, \citenamefont {Lino~dos Santos},\ and\ \citenamefont
  {Miqueleto}}]{Eichhorn:2023gat}%
  \BibitemOpen
  \bibfield  {author} {\bibinfo {author} {\bibfnamefont {A.}~\bibnamefont
  {Eichhorn}}, \bibinfo {author} {\bibfnamefont {R.~R.}\ \bibnamefont {Lino~dos
  Santos}}, \ and\ \bibinfo {author} {\bibfnamefont {J.~a.~L.}\ \bibnamefont
  {Miqueleto}},\ }\href@noop {} {\  (\bibinfo {year} {2023})},\ \Eprint
  {http://arxiv.org/abs/2306.17718} {arXiv:2306.17718 [gr-qc]} \BibitemShut
  {NoStop}%
\bibitem [{\citenamefont {Buchmueller}\ \emph {et~al.}(2023)\citenamefont
  {Buchmueller}, \citenamefont {Ellis},\ and\ \citenamefont
  {Schneider}}]{Buchmueller:2023nll}%
  \BibitemOpen
  \bibfield  {author} {\bibinfo {author} {\bibfnamefont {O.}~\bibnamefont
  {Buchmueller}}, \bibinfo {author} {\bibfnamefont {J.}~\bibnamefont {Ellis}},
  \ and\ \bibinfo {author} {\bibfnamefont {U.}~\bibnamefont {Schneider}},\
  }\href@noop {} {\  (\bibinfo {year} {2023})},\ \Eprint
  {http://arxiv.org/abs/2306.17726} {arXiv:2306.17726 [astro-ph.CO]}
  \BibitemShut {NoStop}%
\bibitem [{\citenamefont {Lazarides}\ \emph {et~al.}(2023)\citenamefont
  {Lazarides}, \citenamefont {Maji},\ and\ \citenamefont
  {Shafi}}]{Lazarides:2023ksx}%
  \BibitemOpen
  \bibfield  {author} {\bibinfo {author} {\bibfnamefont {G.}~\bibnamefont
  {Lazarides}}, \bibinfo {author} {\bibfnamefont {R.}~\bibnamefont {Maji}}, \
  and\ \bibinfo {author} {\bibfnamefont {Q.}~\bibnamefont {Shafi}},\
  }\href@noop {} {\  (\bibinfo {year} {2023})},\ \Eprint
  {http://arxiv.org/abs/2306.17788} {arXiv:2306.17788 [hep-ph]} \BibitemShut
  {NoStop}%
\bibitem [{\citenamefont {Broadhurst}\ \emph {et~al.}(2023)\citenamefont
  {Broadhurst}, \citenamefont {Chen}, \citenamefont {Liu},\ and\ \citenamefont
  {Zheng}}]{Broadhurst:2023tus}%
  \BibitemOpen
  \bibfield  {author} {\bibinfo {author} {\bibfnamefont {T.}~\bibnamefont
  {Broadhurst}}, \bibinfo {author} {\bibfnamefont {C.}~\bibnamefont {Chen}},
  \bibinfo {author} {\bibfnamefont {T.}~\bibnamefont {Liu}}, \ and\ \bibinfo
  {author} {\bibfnamefont {K.-F.}\ \bibnamefont {Zheng}},\ }\href@noop {} {\
  (\bibinfo {year} {2023})},\ \Eprint {http://arxiv.org/abs/2306.17821}
  {arXiv:2306.17821 [astro-ph.HE]} \BibitemShut {NoStop}%
\bibitem [{\citenamefont {Cai}\ \emph {et~al.}(2023)\citenamefont {Cai},
  \citenamefont {He}, \citenamefont {Ma}, \citenamefont {Yan},\ and\
  \citenamefont {Yuan}}]{Cai:2023dls}%
  \BibitemOpen
  \bibfield  {author} {\bibinfo {author} {\bibfnamefont {Y.-F.}\ \bibnamefont
  {Cai}}, \bibinfo {author} {\bibfnamefont {X.-C.}\ \bibnamefont {He}},
  \bibinfo {author} {\bibfnamefont {X.}~\bibnamefont {Ma}}, \bibinfo {author}
  {\bibfnamefont {S.-F.}\ \bibnamefont {Yan}}, \ and\ \bibinfo {author}
  {\bibfnamefont {G.-W.}\ \bibnamefont {Yuan}},\ }\href@noop {} {\  (\bibinfo
  {year} {2023})},\ \Eprint {http://arxiv.org/abs/2306.17822} {arXiv:2306.17822
  [gr-qc]} \BibitemShut {NoStop}%
\bibitem [{\citenamefont {Yang}\ \emph
  {et~al.}(2023{\natexlab{b}})\citenamefont {Yang}, \citenamefont {Ma},
  \citenamefont {Jiang},\ and\ \citenamefont {Huang}}]{Yang:2023qlf}%
  \BibitemOpen
  \bibfield  {author} {\bibinfo {author} {\bibfnamefont {A.}~\bibnamefont
  {Yang}}, \bibinfo {author} {\bibfnamefont {J.}~\bibnamefont {Ma}}, \bibinfo
  {author} {\bibfnamefont {S.}~\bibnamefont {Jiang}}, \ and\ \bibinfo {author}
  {\bibfnamefont {F.~P.}\ \bibnamefont {Huang}},\ }\href@noop {} {\  (\bibinfo
  {year} {2023}{\natexlab{b}})},\ \Eprint {http://arxiv.org/abs/2306.17827}
  {arXiv:2306.17827 [hep-ph]} \BibitemShut {NoStop}%
\bibitem [{\citenamefont {Blasi}\ \emph
  {et~al.}(2023{\natexlab{b}})\citenamefont {Blasi}, \citenamefont {Mariotti},
  \citenamefont {Rase},\ and\ \citenamefont {Sevrin}}]{Blasi:2023sej}%
  \BibitemOpen
  \bibfield  {author} {\bibinfo {author} {\bibfnamefont {S.}~\bibnamefont
  {Blasi}}, \bibinfo {author} {\bibfnamefont {A.}~\bibnamefont {Mariotti}},
  \bibinfo {author} {\bibfnamefont {A.}~\bibnamefont {Rase}}, \ and\ \bibinfo
  {author} {\bibfnamefont {A.}~\bibnamefont {Sevrin}},\ }\href@noop {} {\
  (\bibinfo {year} {2023}{\natexlab{b}})},\ \Eprint
  {http://arxiv.org/abs/2306.17830} {arXiv:2306.17830 [hep-ph]} \BibitemShut
  {NoStop}%
\bibitem [{\citenamefont {Inomata}\ \emph {et~al.}(2023)\citenamefont
  {Inomata}, \citenamefont {Kohri},\ and\ \citenamefont
  {Terada}}]{Inomata:2023zup}%
  \BibitemOpen
  \bibfield  {author} {\bibinfo {author} {\bibfnamefont {K.}~\bibnamefont
  {Inomata}}, \bibinfo {author} {\bibfnamefont {K.}~\bibnamefont {Kohri}}, \
  and\ \bibinfo {author} {\bibfnamefont {T.}~\bibnamefont {Terada}},\
  }\href@noop {} {\  (\bibinfo {year} {2023})},\ \Eprint
  {http://arxiv.org/abs/2306.17834} {arXiv:2306.17834 [astro-ph.CO]}
  \BibitemShut {NoStop}%
\bibitem [{\citenamefont {Depta}\ \emph {et~al.}(2023)\citenamefont {Depta},
  \citenamefont {Schmidt-Hoberg},\ and\ \citenamefont
  {Tasillo}}]{Depta:2023qst}%
  \BibitemOpen
  \bibfield  {author} {\bibinfo {author} {\bibfnamefont {P.~F.}\ \bibnamefont
  {Depta}}, \bibinfo {author} {\bibfnamefont {K.}~\bibnamefont
  {Schmidt-Hoberg}}, \ and\ \bibinfo {author} {\bibfnamefont {C.}~\bibnamefont
  {Tasillo}},\ }\href@noop {} {\  (\bibinfo {year} {2023})},\ \Eprint
  {http://arxiv.org/abs/2306.17836} {arXiv:2306.17836 [astro-ph.CO]}
  \BibitemShut {NoStop}%
\bibitem [{\citenamefont {Gouttenoire}\ and\ \citenamefont
  {Vitagliano}(2023)}]{Gouttenoire:2023ftk}%
  \BibitemOpen
  \bibfield  {author} {\bibinfo {author} {\bibfnamefont {Y.}~\bibnamefont
  {Gouttenoire}}\ and\ \bibinfo {author} {\bibfnamefont {E.}~\bibnamefont
  {Vitagliano}},\ }\href@noop {} {\  (\bibinfo {year} {2023})},\ \Eprint
  {http://arxiv.org/abs/2306.17841} {arXiv:2306.17841 [gr-qc]} \BibitemShut
  {NoStop}%
\bibitem [{\citenamefont {Borah}\ \emph
  {et~al.}(2023{\natexlab{b}})\citenamefont {Borah}, \citenamefont
  {Jyoti~Das},\ and\ \citenamefont {Samanta}}]{Borah:2023sbc}%
  \BibitemOpen
  \bibfield  {author} {\bibinfo {author} {\bibfnamefont {D.}~\bibnamefont
  {Borah}}, \bibinfo {author} {\bibfnamefont {S.}~\bibnamefont {Jyoti~Das}}, \
  and\ \bibinfo {author} {\bibfnamefont {R.}~\bibnamefont {Samanta}},\
  }\href@noop {} {\  (\bibinfo {year} {2023}{\natexlab{b}})},\ \Eprint
  {http://arxiv.org/abs/2307.00537} {arXiv:2307.00537 [hep-ph]} \BibitemShut
  {NoStop}%
\bibitem [{\citenamefont {Wang}\ \emph
  {et~al.}(2023{\natexlab{b}})\citenamefont {Wang}, \citenamefont {Zhao},
  \citenamefont {Li},\ and\ \citenamefont {Zhu}}]{Wang:2023ost}%
  \BibitemOpen
  \bibfield  {author} {\bibinfo {author} {\bibfnamefont {S.}~\bibnamefont
  {Wang}}, \bibinfo {author} {\bibfnamefont {Z.-C.}\ \bibnamefont {Zhao}},
  \bibinfo {author} {\bibfnamefont {J.-P.}\ \bibnamefont {Li}}, \ and\ \bibinfo
  {author} {\bibfnamefont {Q.-H.}\ \bibnamefont {Zhu}},\ }\href@noop {} {\
  (\bibinfo {year} {2023}{\natexlab{b}})},\ \Eprint
  {http://arxiv.org/abs/2307.00572} {arXiv:2307.00572 [astro-ph.CO]}
  \BibitemShut {NoStop}%
\bibitem [{\citenamefont {Murai}\ and\ \citenamefont
  {Yin}(2023)}]{Murai:2023gkv}%
  \BibitemOpen
  \bibfield  {author} {\bibinfo {author} {\bibfnamefont {K.}~\bibnamefont
  {Murai}}\ and\ \bibinfo {author} {\bibfnamefont {W.}~\bibnamefont {Yin}},\
  }\href@noop {} {\  (\bibinfo {year} {2023})},\ \Eprint
  {http://arxiv.org/abs/2307.00628} {arXiv:2307.00628 [hep-ph]} \BibitemShut
  {NoStop}%
\bibitem [{\citenamefont {Datta}(2023)}]{Datta:2023vbs}%
  \BibitemOpen
  \bibfield  {author} {\bibinfo {author} {\bibfnamefont {S.}~\bibnamefont
  {Datta}},\ }\href@noop {} {\  (\bibinfo {year} {2023})},\ \Eprint
  {http://arxiv.org/abs/2307.00646} {arXiv:2307.00646 [hep-ph]} \BibitemShut
  {NoStop}%
\bibitem [{\citenamefont {Barman}\ \emph {et~al.}(2023)\citenamefont {Barman},
  \citenamefont {Borah}, \citenamefont {Jyoti~Das},\ and\ \citenamefont
  {Saha}}]{Barman:2023fad}%
  \BibitemOpen
  \bibfield  {author} {\bibinfo {author} {\bibfnamefont {B.}~\bibnamefont
  {Barman}}, \bibinfo {author} {\bibfnamefont {D.}~\bibnamefont {Borah}},
  \bibinfo {author} {\bibfnamefont {S.}~\bibnamefont {Jyoti~Das}}, \ and\
  \bibinfo {author} {\bibfnamefont {I.}~\bibnamefont {Saha}},\ }\href@noop {}
  {\  (\bibinfo {year} {2023})},\ \Eprint {http://arxiv.org/abs/2307.00656}
  {arXiv:2307.00656 [hep-ph]} \BibitemShut {NoStop}%
\bibitem [{\citenamefont {Bi}\ \emph {et~al.}(2023)\citenamefont {Bi},
  \citenamefont {Wu}, \citenamefont {Chen},\ and\ \citenamefont
  {Huang}}]{Bi:2023tib}%
  \BibitemOpen
  \bibfield  {author} {\bibinfo {author} {\bibfnamefont {Y.-C.}\ \bibnamefont
  {Bi}}, \bibinfo {author} {\bibfnamefont {Y.-M.}\ \bibnamefont {Wu}}, \bibinfo
  {author} {\bibfnamefont {Z.-C.}\ \bibnamefont {Chen}}, \ and\ \bibinfo
  {author} {\bibfnamefont {Q.-G.}\ \bibnamefont {Huang}},\ }\href@noop {} {\
  (\bibinfo {year} {2023})},\ \Eprint {http://arxiv.org/abs/2307.00722}
  {arXiv:2307.00722 [astro-ph.CO]} \BibitemShut {NoStop}%
\bibitem [{\citenamefont {Lu}\ and\ \citenamefont {Chiang}(2023)}]{Lu:2023mcz}%
  \BibitemOpen
  \bibfield  {author} {\bibinfo {author} {\bibfnamefont {B.-Q.}\ \bibnamefont
  {Lu}}\ and\ \bibinfo {author} {\bibfnamefont {C.-W.}\ \bibnamefont
  {Chiang}},\ }\href@noop {} {\  (\bibinfo {year} {2023})},\ \Eprint
  {http://arxiv.org/abs/2307.00746} {arXiv:2307.00746 [hep-ph]} \BibitemShut
  {NoStop}%
\bibitem [{\citenamefont {Xiao}\ \emph {et~al.}(2023)\citenamefont {Xiao},
  \citenamefont {Yang},\ and\ \citenamefont {Zhang}}]{Xiao:2023dbb}%
  \BibitemOpen
  \bibfield  {author} {\bibinfo {author} {\bibfnamefont {Y.}~\bibnamefont
  {Xiao}}, \bibinfo {author} {\bibfnamefont {J.~M.}\ \bibnamefont {Yang}}, \
  and\ \bibinfo {author} {\bibfnamefont {Y.}~\bibnamefont {Zhang}},\
  }\href@noop {} {\  (\bibinfo {year} {2023})},\ \Eprint
  {http://arxiv.org/abs/2307.01072} {arXiv:2307.01072 [hep-ph]} \BibitemShut
  {NoStop}%
\bibitem [{\citenamefont {Li}\ and\ \citenamefont {Xie}(2023)}]{Li:2023bxy}%
  \BibitemOpen
  \bibfield  {author} {\bibinfo {author} {\bibfnamefont {S.-P.}\ \bibnamefont
  {Li}}\ and\ \bibinfo {author} {\bibfnamefont {K.-P.}\ \bibnamefont {Xie}},\
  }\href@noop {} {\  (\bibinfo {year} {2023})},\ \Eprint
  {http://arxiv.org/abs/2307.01086} {arXiv:2307.01086 [hep-ph]} \BibitemShut
  {NoStop}%
\bibitem [{\citenamefont {Zhang}\ \emph {et~al.}(2023)\citenamefont {Zhang},
  \citenamefont {Dai}, \citenamefont {Gao}, \citenamefont {Gong}, \citenamefont
  {Jiang},\ and\ \citenamefont {Lu}}]{Zhang:2023lzt}%
  \BibitemOpen
  \bibfield  {author} {\bibinfo {author} {\bibfnamefont {C.}~\bibnamefont
  {Zhang}}, \bibinfo {author} {\bibfnamefont {N.}~\bibnamefont {Dai}}, \bibinfo
  {author} {\bibfnamefont {Q.}~\bibnamefont {Gao}}, \bibinfo {author}
  {\bibfnamefont {Y.}~\bibnamefont {Gong}}, \bibinfo {author} {\bibfnamefont
  {T.}~\bibnamefont {Jiang}}, \ and\ \bibinfo {author} {\bibfnamefont
  {X.}~\bibnamefont {Lu}},\ }\href@noop {} {\  (\bibinfo {year} {2023})},\
  \Eprint {http://arxiv.org/abs/2307.01093} {arXiv:2307.01093 [gr-qc]}
  \BibitemShut {NoStop}%
\bibitem [{\citenamefont {Anchordoqui}\ \emph {et~al.}(2023)\citenamefont
  {Anchordoqui}, \citenamefont {Antoniadis},\ and\ \citenamefont
  {Lust}}]{Anchordoqui:2023tln}%
  \BibitemOpen
  \bibfield  {author} {\bibinfo {author} {\bibfnamefont {L.~A.}\ \bibnamefont
  {Anchordoqui}}, \bibinfo {author} {\bibfnamefont {I.}~\bibnamefont
  {Antoniadis}}, \ and\ \bibinfo {author} {\bibfnamefont {D.}~\bibnamefont
  {Lust}},\ }\href@noop {} {\  (\bibinfo {year} {2023})},\ \Eprint
  {http://arxiv.org/abs/2307.01100} {arXiv:2307.01100 [hep-ph]} \BibitemShut
  {NoStop}%
\bibitem [{\citenamefont {Liu}\ \emph {et~al.}(2023)\citenamefont {Liu},
  \citenamefont {Chen},\ and\ \citenamefont {Huang}}]{Liu:2023ymk}%
  \BibitemOpen
  \bibfield  {author} {\bibinfo {author} {\bibfnamefont {L.}~\bibnamefont
  {Liu}}, \bibinfo {author} {\bibfnamefont {Z.-C.}\ \bibnamefont {Chen}}, \
  and\ \bibinfo {author} {\bibfnamefont {Q.-G.}\ \bibnamefont {Huang}},\
  }\href@noop {} {\  (\bibinfo {year} {2023})},\ \Eprint
  {http://arxiv.org/abs/2307.01102} {arXiv:2307.01102 [astro-ph.CO]}
  \BibitemShut {NoStop}%
\bibitem [{\citenamefont {Konoplya}\ and\ \citenamefont
  {Zhidenko}(2023)}]{Konoplya:2023fmh}%
  \BibitemOpen
  \bibfield  {author} {\bibinfo {author} {\bibfnamefont {R.~A.}\ \bibnamefont
  {Konoplya}}\ and\ \bibinfo {author} {\bibfnamefont {A.}~\bibnamefont
  {Zhidenko}},\ }\href@noop {} {\  (\bibinfo {year} {2023})},\ \Eprint
  {http://arxiv.org/abs/2307.01110} {arXiv:2307.01110 [gr-qc]} \BibitemShut
  {NoStop}%
\bibitem [{\citenamefont {Chowdhury}\ \emph {et~al.}(2023)\citenamefont
  {Chowdhury}, \citenamefont {Tasinato},\ and\ \citenamefont
  {Zavala}}]{Chowdhury:2023opo}%
  \BibitemOpen
  \bibfield  {author} {\bibinfo {author} {\bibfnamefont {D.}~\bibnamefont
  {Chowdhury}}, \bibinfo {author} {\bibfnamefont {G.}~\bibnamefont {Tasinato}},
  \ and\ \bibinfo {author} {\bibfnamefont {I.}~\bibnamefont {Zavala}},\
  }\href@noop {} {\  (\bibinfo {year} {2023})},\ \Eprint
  {http://arxiv.org/abs/2307.01188} {arXiv:2307.01188 [hep-th]} \BibitemShut
  {NoStop}%
\bibitem [{\citenamefont {Niu}\ and\ \citenamefont
  {Rahat}(2023)}]{Niu:2023bsr}%
  \BibitemOpen
  \bibfield  {author} {\bibinfo {author} {\bibfnamefont {X.}~\bibnamefont
  {Niu}}\ and\ \bibinfo {author} {\bibfnamefont {M.~H.}\ \bibnamefont
  {Rahat}},\ }\href@noop {} {\  (\bibinfo {year} {2023})},\ \Eprint
  {http://arxiv.org/abs/2307.01192} {arXiv:2307.01192 [hep-ph]} \BibitemShut
  {NoStop}%
\end{thebibliography}%

\end{document}